\documentclass[twocolumn,trackchanges]{aastex6}
\usepackage{amsmath}
\usepackage{gensymb}
\usepackage{paralist}
\usepackage{hyperref}
\usepackage[all]{hypcap}

\begin{document}
\title{Smooth H\,{\scriptsize i} low column density outskirts in nearby galaxies}
\author{R. Ianjamasimanana\altaffilmark{1}, Fabian 
Walter\altaffilmark{1}, W.J.G. de  Blok\altaffilmark{2,3,4}, George
  H. Heald\altaffilmark{5,4}, Elias Brinks\altaffilmark{6}} 
  \altaffiltext{1}{Max-Planck Institut f$\rm{\ddot{u}}$r Astronomie, 
  K\"onigstuhl 17, 69117, Heidelberg, Germany}
    \altaffiltext{2}{Netherlands Institute for
  Radio Astronomy (ASTRON), Postbus 2, 7990 AA Dwingeloo, The  Netherlands}
    \altaffiltext{3}{Astrophysics, Cosmology and Gravity Centre, 
  Department of Astronomy, University of Cape
  Town, Private Bag X3, Rondebosch 7701, South Africa}  
    \altaffiltext{4}{Kapteyn Astronomical Institute, 
    University of Groningen, PO Box 800, 9700 AV, Groningen, The Netherlands}
    \altaffiltext{5}{CSIRO Astronomy and Space Science, 26 Dick Perry Avenue, Kensington WA 6151, Australia}
    \altaffiltext{6}{Centre for Astrophysics Research, University of Hertfordshire, College Lane, Hatfield, AL10 9AB, UK}
   \email{ianjamasimanana26@gmail.com} 
   \email{walter@mpia.de}
   \email{blok@astron.nl}
   \email{George.Heald@csiro.au}
   \email{e.brinks@herts.ac.uk}
\accepted{for Publication in the Astronomical Journal}
\begin{abstract}
    The low column density gas at the outskirts of galaxies as
    traced by the 21 cm hydrogen line emission (H\,{\sc i}) 
    represents the interface between galaxies and the intergalactic medium, i.e., 
    where galaxies are believed to get their supply of gas to
    fuel future episodes of star formation. Photoionization models predict a 
    break in the radial profiles of H\,{\sc i} at a column density of $\rm{\sim~5~\times~10^{19}~cm^{-2}}$ 
    due to the lack of self-shielding against extragalactic ionizing photons. To investigate the 
    prevalence of such breaks in galactic disks and to characterize what
    determines the potential \textit{edge} of the H\,{\sc i} disks, we study the 
    azimuthally-averaged H\,{\sc i} column density profiles of 17 nearby galaxies from The H\,{\sc i}
    Nearby Galaxy Survey (THINGS) and supplemented in two cases with published Hydrogen Accretion in
    LOcal GAlaxieS (HALOGAS) data. To detect potential faint H\,{\sc i} emission that would
    otherwise be undetected using conventional moment map analysis, we line up individual profiles to the same 
    reference velocity and average them azimuthally to derive stacked radial profiles. 
    To do so, we use model velocity fields created from a simple 
    extrapolation of the rotation curves to align the profiles in velocity at radii beyond 
    the extent probed with the sensitivity of traditional 
    integrated H\,{\sc i} maps. With this method, 
    we improve our sensitivity to outer-disk H\,{\sc i} emission by up to an order of magnitude. 
    Except for a few disturbed galaxies, none show evidence for a sudden change in the slope of the H\,{\sc i} radial
    profiles, the alleged signature of ionization by the extragalactic background.

\end{abstract}


\section{Introduction} \label{sec:intro}
The 21 cm atomic hydrogen (H\,{\sc i}) line can trace the interstellar medium (ISM) out 
to large radii in galaxies \citep[e.g.,][]{bosma81}, 
typically well beyond the optical disk. 
Detailed studies of the shapes of the radial H\,{\sc i}  
distribution are important to understand the physical conditions of the 
outer disk of galaxies, e.g., constrain models of the 
H\,{\sc i} to H\,{\sc ii} transition, to trace the influence of 
environment on H\,{\sc i}, and to use the H\,{\sc i} gas as a tracer of the (dark) 
matter distribution.
In general, the radial H\,{\sc i} distribution 
tends to be approximately flat out to about the edge of the 
optical disk (with a central depression for many spiral galaxies), 
followed by a gradual decline 
\citep[e.g., early work by][]{sancisi83, broeilsetal94} at a column density of 
about a few times  $10^{20}$ cm$^{-2}$. A second break in the radial H\,{\sc i} 
profiles has been reported using high sensitivity observations 
of M33 \citep{corbellietal89} and NGC 3198 \citep{vangorkom93}. 
In these studies, a break was reported 
at a column density of a few times $10^{19}$ cm$^{-2}$. 
Such a behavior has been interpreted in the context of the 
existence of a critical column density below which H\,{\sc i} is ionized by the 
extragalactic radiation field \citep{corbellisalpeter93, maloney93, doveshull94}. 
However, as mentioned by \citet{irwin95}, 
the degree of steepness of the radial H\,{\sc i} fall-off predicted by 
ionization models depends on many factors such 
as the vertical structure of the H\,{\sc i} disk, the H\,{\sc i} 
mass distribution, and more importantly the flux of the 
ionizing photons, which are not well 
constrained by observations \citep[see also][]{doveshull94, fumagallietal2017}.  

Although there have been many discussions in the literature 
explaining the edge of the H\,{\sc i} disk from theoretical 
viewpoints \citep{sunyaev69, bergerongunn77, 
bosma81, corbellisalpeter93, maloney93, doveshull94}, 
detailed observations of the radial extent of the outer H\,{\sc i} disk in 
a sample of galaxies are lacking. 

Using single-dish observations with the Arecibo telescope, 
\citet{corbellietal89} examined the outer radial H\,{\sc i} 
distribution of the northern part of the major axis of M33 
down to a (3$\sigma$ detection limit) column 
density of $\sim$ 2 $\times$ 10$^{18}$ cm$^{-2}$. The column density is 
$\sim$ 2 $\times$ 10$^{21}$ cm$^{-2}$ in the central disk and the value drops 
below $\sim$ 2 $\times$ 10$^{20}$ cm$^{-2}$ just outside the optical disk, which then 
gradually declined to $\sim$ 3 $\times$ 10$^{19}$ cm$^{-2}$. 
A second break in the radial 
profile was observed at $\sim$ 3 $\times$ 10$^{19}$ cm$^{-2}$, below which the value 
decreased to $\sim$ 2 $\times$ 10$^{18}$ cm$^{-2}$ within $\sim$ 1 kpc. 
The NRAO\footnote{The National Radio Astronomy 
Observatory is a facility of the National Science 
Foundation operated under cooperative agreement by Associated Universities, Inc.} 
Very Large Array (VLA) observation of NGC 3198 by \citet{vangorkom93}, with a 
3$\sigma$ column density limit of 4 $\times$ 10$^{18}$ cm$^{-2}$, 
also revealed the existence of 
a radial break of the H\,{\sc i} disk below $\sim$ 2 $\times$ 10$^{19}$ cm$^{-2}$. 
The VLA data showed that the H\,{\sc i} surface density of NGC 3198 decreased 
by an order of magnitude from $\sim$ 30 to $\sim$ 33 kpc. 
The abrupt decline in NGC 3198 was, however, not resolved within 
the synthesized beam of the observation ($\sim$ 2.7 kpc).
Recently \citet{healdetal16} studied the H\,{\sc i} edge in M83 using interferometric 
observations with the Karoo Array Telescope \citep[KAT-7, a 7 
dish radio telescope in South Africa, precursor to the MeerKAT telescope;][]{carignanetal13}. 
These observations reached 
a 3$\sigma$ column density sensitivity 
of $5.6\times10^{18}~\rm{cm}^{-2}$. Based on the 
analysis of the H\,{\sc i} column density map 
of M83, they concluded that this galaxy also 
has an H\,{\sc i} 
edge similar to what was found for 
NGC 3198 and M33. \citet{healdetal16} considered 
ram pressure or ionization by the intergalactic 
medium as a possible cause of the H\,{\sc i} edge. 
There are also other observations targeting the H\,{\sc i} outskirts  
of nearby galaxies \citep[e.g.,][]{Irwinetal09} but 
we restrict ourselves to those that specifically address 
the edge of the H\,{\sc i} disk.                          

In this paper, we study the shapes of the radial H\,{\sc i} 
column density profiles for a large 
sample of nearby galaxies and investigate whether the presence of 
an edge is a common feature of the H\,{\sc i} disk. 
We do this by employing a new method to extract faint H\,{\sc i} emission at 
large galactocentric radii from H\,{\sc i} data cubes. 
Our paper is organized as follows. 
In Section \ref{sec:data}, we present our data and 
sample galaxies. In Section \ref{sec:methodology}, 
we describe our new stacking approach to derive the 
radial H\,{\sc i} column density profiles. 
In Section \ref{sec:results}, we discuss our results. 
In Section \ref{sec:summary}, we present our summary and conclusions. 
\section{Data and sample}\label{sec:data}
Analyzing the distribution of the H\,{\sc i} emission at 
large radii requires high sensitivity observations. 
With the limitation of current state-of-the-art telescopes, 
only observations of nearby galaxies 
provide enough spatial resolution and sensitivity to resolve the H\,{\sc i} edge 
and investigate the H\,{\sc i} distribution at large radii. 
Based on the model by \citet{maloney93} and the previous study of NGC 3198 by \citet{vangorkom93}, 
a column density sensitivity limit of $\rm{\sim10^{19}~cm^{-2}}$ and a spatial resolution better than 2.7 kpc 
are required to detect and resolve the H\,{\sc i} edge. 
Using a stacking method that will be described later, we satisfy this criteria 
using data from The H\,{\sc i} Nearby Galaxy Survey \citep[THINGS;][]{walteretal08}. 
THINGS provides VLA 21-cm line maps of 34 nearby 
spiral and dwarf galaxies at an average angular 
resolution of $\sim11\arcsec$ (linear resolution is given in Table~\ref{tab:sample}). 
\twocolumngrid
{\renewcommand{\arraystretch}{1}\begin{deluxetable}{l c c c c c}
\centering
\tablewidth{0pt}
\tablecaption{The Sample galaxies \label{tab:sample}}
\tablehead{
\multicolumn{1}{c}{Galaxy} & incl & Dist. & 
\multicolumn{1}{c}{$\rm{B_{min}}$}&$\rm{B_{maj}}$ & Lin. res.\\ 
& $[\degree]$&[Mpc]& [\arcsec]&[\arcsec]&[pc]\\
\multicolumn{1}{c}{1}&\multicolumn{1}{c}{2}&\multicolumn{1}{c}{3}
& 4 & 5 & \multicolumn{1}{c}{6}
}
\startdata
IC 2574 & 53 & 4.0 & 11.9 & 12.8 & 240 \\
NGC 3621 & 65 & 6.6 & 10.2 & 15.9 & 419 \\
NGC 2366 & 64 & 3.4 & 11.8 & 13.1 & 206 \\
DDO 154 & 66 & 4.3 & 12.6 & 14.1 & 278 \\
NGC 6946 & 33 & 5.9 & 5.6 & 6.0 & 167 \\
NGC 4736 & 41 & 4.7 & 9.1 & 10.2 & 220 \\
NGC 3521 & 73 & 10.7 & 11.2 & 14.1 & 656 \\
NGC 4826 & 65 & 7.5 & 9.3 & 12.2 & 391 \\
NGC 3627 & 62 & 9.3 & 8.8 & 10.6 & 438 \\
NGC 925 & 66 & 9.2 & 5.7 & 5.9 & 260 \\
NGC 3198 & 72 & 9.4 & 11.6 & 13.0 & 560 \\
NGC 7331 & 76 & 14.7 & 5.6 & 6.1 & 418 \\
NGC 2841 & 74 & 14.1 & 9.4 & 11.1 & 698 \\
NGC 2403 & 63 & 3.2 & 7.7 & 8.8 & 127 \\
NGC 2903 & 65 & 8.9 & 13.3 & 15.3 & 617 \\
NGC 7793 & 50 & 3.9 & 10.8 & 15.6 & 250 \\
NGC 2976 & 65 & 3.6 & 6.4 & 7.4 & 121 
\enddata
\tablecomments{Column 1: name of galaxy; 
Column 2: inclination; Column 3: distance; Column 4\&5: 
minor and major axes of synthesized beam in arcsec using a natural weighting; 
Column 6: average linear resolution.}
\end{deluxetable}}
In this work, we use the high resolution rotation curves derived by \citet{debloketal08} 
(see Section \ref{sec:methodology}). Thus we require that our galaxies are part of the 
\citet{debloketal08} sample. However, we exclude NGC 5055 as this galaxy is strongly warped. We also exclude 
M81 as its H\,{\sc i} emission is heavily disturbed, due to its interaction with M82 and NGC 3077. 
Thus, our sample is reduced to 17 spiral and dwarf galaxies. 
Their observational properties and names are 
shown for reference in Table \ref{tab:sample} \citep[see also][]{walteretal08}. 

In this paper, we use the THINGS natural-weighted data and do not 
apply any residual flux correction nor blanking so that 
we preserve the noise properties of the data.  
However, to remove the flux discrepancy due to 
the mismatch between the shapes of the dirty beam and the restoring clean beam 
\citep[for a full discussion, see][]{walteretal08}, 
we clean the data close to the noise level 
(1.5$\sigma$, as opposed to the 2.5$\sigma$ of the standard THINGS data release).
When cleaning to this depth, we find better agreement between 
the flux derived from the natural non-residual scaled cubes and the flux from the 
standard THINGS moment zero maps, which were derived 
from residual-scaled cubes \citep[see also][]{Ianjamasimananaetal17}. 
For completeness, we also present in this paper the 
radial H\,{\sc i} column density profiles 
derived from the standard THINGS moment zero maps.

A few of the THINGS galaxies have also been observed as 
part of the Hydrogen Accretion in LOcal GAlaxieS Survey 
\citep[HALOGAS;][]{healdetal11}. For two galaxies (NGC 3198 and NGC 2403) 
which overlap with the HALOGAS sample and 
fulfill our sample selection criterion, we compare our results from the THINGS data 
with those from the HALOGAS data. HALOGAS is a deep H\,{\sc i} survey of 
24 nearby galaxies, with 23 galaxies observed by the 
WSRT\footnote{Westerbork Synthesis Radio Telescope} 
\citep{oosterlooetal07, healdetal11} and one 
by the VLA \citep{fraternalietal02}. The WSRT data by \citet{healdetal11} have a 
typical ($5\sigma$ per 4.1 $\rm{km~s^{-1}}$ channel) 
column density sensitivity limit of $\sim$5 
$\times$ 10$^{18}$ cm$^{-2}$ at 30\arcsec~resolution using 
a Brigg's \citep{brigs} robustness parameter of 0 with a Gaussian taper. 
The VLA data (NGC 2403) by \citet{fraternalietal02} reached a 
($5\sigma$ per 5.2 $\rm{km~s^{-1}}$ channel) column density 
sensitivity of $\sim5\times$ 10$^{19}$ cm$^{-2}$ 
using a Robust (robust = 0.2) or Briggs weighting scheme 
at 15$\arcsec$ (230 pc) angular resolution.  
\section{Methodology}\label{sec:methodology}

We aim to derive H\,{\sc i} column density values as a function of 
radius, beyond the extent probed by conventional pixel-by-pixel moment map 
analysis. To do so, we divide each galaxy into concentric elliptical annuli, 
each of them with a width along the major axis 
similar to the size of the synthesized beam of the observations. 
For THINGS and the VLA data from \citet{fraternalietal02}, 
we adopt a common width of 15$\arcsec$; for the 
HALOGAS data of NGC 3198, we use a 30$\arcsec$ width. Using a stacking technique that 
will be described in the next section, we derive the 
average H\,{\sc i} column density within each annulus to derive the radial profiles. 
Since we aim to recover 
H\,{\sc i} flux out to large radii, correction for
sensitivity bias due to the response of the primary beam of the telescope is
important. We therefore apply a primary beam correction in the analysis that follows.

\subsection{Stacking of \rm{H\,{\sc i}} \textit{spectra}}

Detecting H\,{\sc i} emission in the extreme outskirts of galaxies is 
challenging due to the low signal-to-noise (S/N) of the 
individual profiles.   
We here aim to push the sensitivity limit of the available H\,{\sc i} data 
to much deeper levels. To do so, we stack and sum individual profiles within the 
elliptical annuli mentioned in the previous section to get high 
S/N azimuthally averaged H\,{\sc i} profiles. 
Individual profiles at different positions within a data cube peak at different 
velocities predominantly due to the rotation 
of the gas within the galaxy disk. Consequently, 
they cannot be simply summed or averaged but individual 
spectra need to be aligned to a common 
reference velocity prior to the averaging. 
Faint H\,{\sc i} 
emission that would otherwise be undetected 
(or ``blanked'') using conventional pixel-by-pixel moment 
map analysis can thus be recovered. 
We therefore shift individual profiles 
by the amount suggested by the galaxy's velocity 
field to attempt to align them to the same velocity 
\citep{ianjamasimananaetal12}. This technique is implemented in the GIPSY\footnote{The Groningen Image
Processing System, \url{https://www.astro.rug.nl/~gipsy/articles/gipsypaper.html}}
task SHUFFLE. The stacking is  
straightforward in the bright inner disk of galaxies, 
where the velocity fields are well determined 
\citep[for a full discussion of the derivation of the 
THINGS velocity fields, see][]{debloketal08}. However, in the faint outskirts of the 
disks, where the peak emission of the individual profiles falls below 3 
times the rms noise, direct 
measurements of the velocity field cannot be derived. 
At such large radii, we will use extrapolations 
of the rotation curves derived in the inner disk 
by \citet{debloketal08} to create model velocity fields. 
As most galaxies reach the flat part of the rotation 
curve at large radii, we here simply assume that the rotation curve will 
continue to stay flat beyond the last point where the 
rotation curves can be directly measured. 
We also assume that the position angle and inclination remain constant as well, i.e., 
there is no warp beyond the outermost radius directly detected in H\,{\sc i} emission. 
These assumptions are the key elements of our stacking procedure. 
As we will see below, 
stacked H\,{\sc i} signal is recovered beyond the measured 
rotation curves, which implies that the assumption of flat, 
extended rotation curves holds.

\subsection{Effects of noisy profiles} 
 
Since we are stacking profiles azimuthally, we are summing both low and high S/N profiles.
We thus assign a weight, $w_{i}$, to each individual 
profile or spectrum, $S_{i}$, before the stacking. 
We assume 
that the noise in the data cubes is uniform before the primary beam correction. 
This is a good assumption for our data cubes. 
After the primary beam correction, the noise is no longer uniform and we 
define our stacked profiles as
\begin{equation}
S=\dfrac{\sum\limits_{i=1}^{N} w_{i}S_{i}}{\sum\limits_{i=1}^{N} w_{i}},
\label{eq:spectra}
\end{equation}
 where $w_{i}=\dfrac{p_{i}^{2}}{\sigma_{i}^{2}}$ is the weight assigned 
 to spectrum $S_{i}$, with $p_{i}$ being the associated primary beam 
correction factor (we adopt a Gaussian shape, with $p_{i}$ = 1 at the center and 
0.5 at 15.4\arcmin~and 17.0\arcmin~from the pointing center of the VLA and WSRT, respectively) 
and $\sigma_{i}$ is the rms noise before the correction. 
Since $\sigma_{i}$ is assumed to be equal to $\sigma$ for all 
profiles in a data cube, Equation \ref{eq:spectra} becomes 
\begin{equation}
S=\dfrac{\sum\limits_{i=1}^{N}p_{i}^{2}S_{i}}{\sum\limits_{i=1}^{N} p_{i}^{2}}.
\end{equation} 
We use a simple propagation of 
errors to define the noise of the stacked profiles 
in each elliptical annulus. Some stacked profiles show 
negative bowls due to unsampled ($u,~v$) data (i.e., ``missing short spacings''). 
We correct for this by fitting a single or double Gaussian function with a 
polynomial background to the stacked profiles. 
We then convert the integrated flux of the fitted Gaussian 
components with respect to the fitted baseline to a column density. Finally, to correct for the orientation of the galaxies, 
we multiply the results by the cosine of the inclination of the galaxies to derive the actual surface density 
profiles. The adopted inclination values are listed in 
Table~\ref{tab:sample}. 

\begin{figure*}
    \begin{tabular}{c c}
    \includegraphics[scale=0.46]{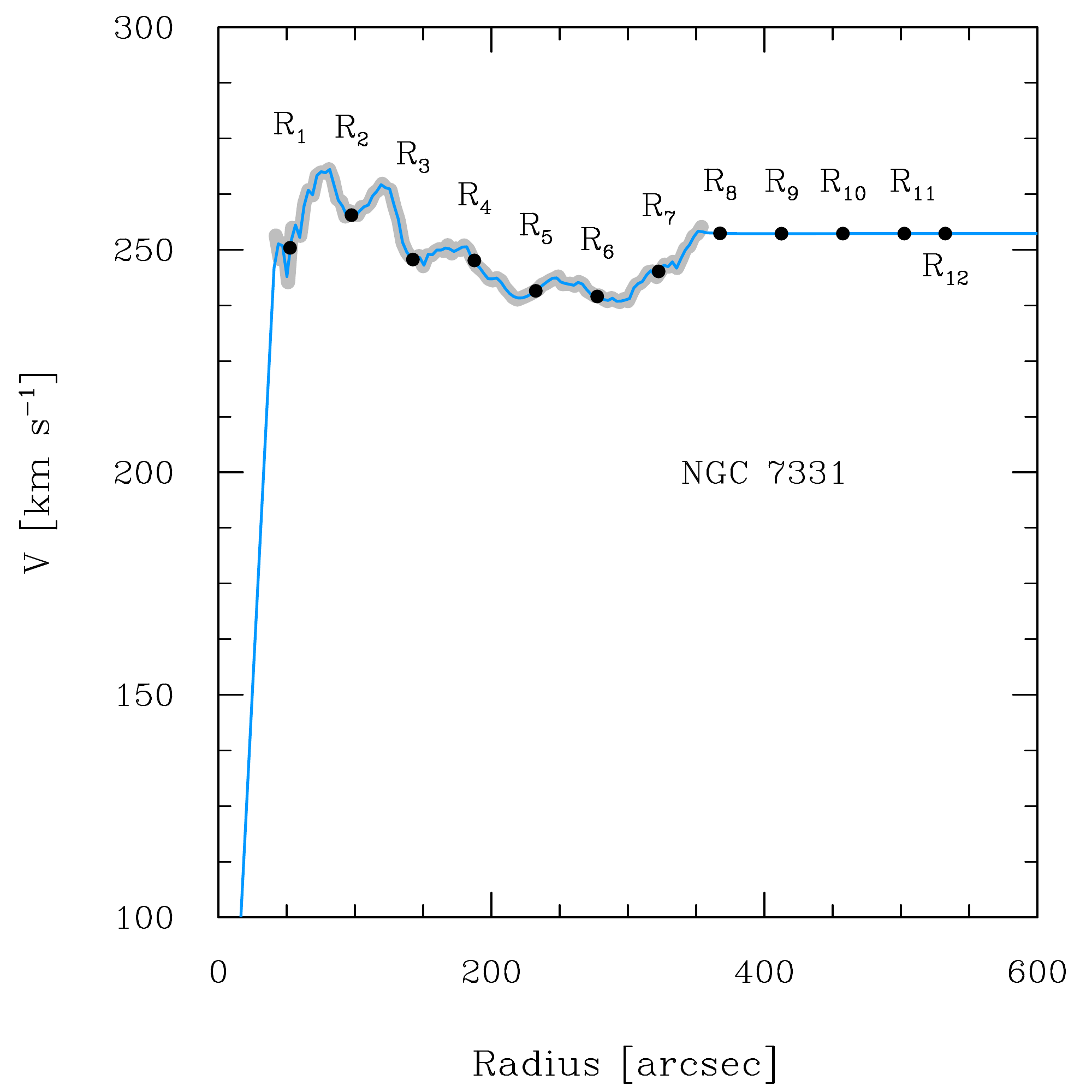} &
    \includegraphics[scale=0.46]{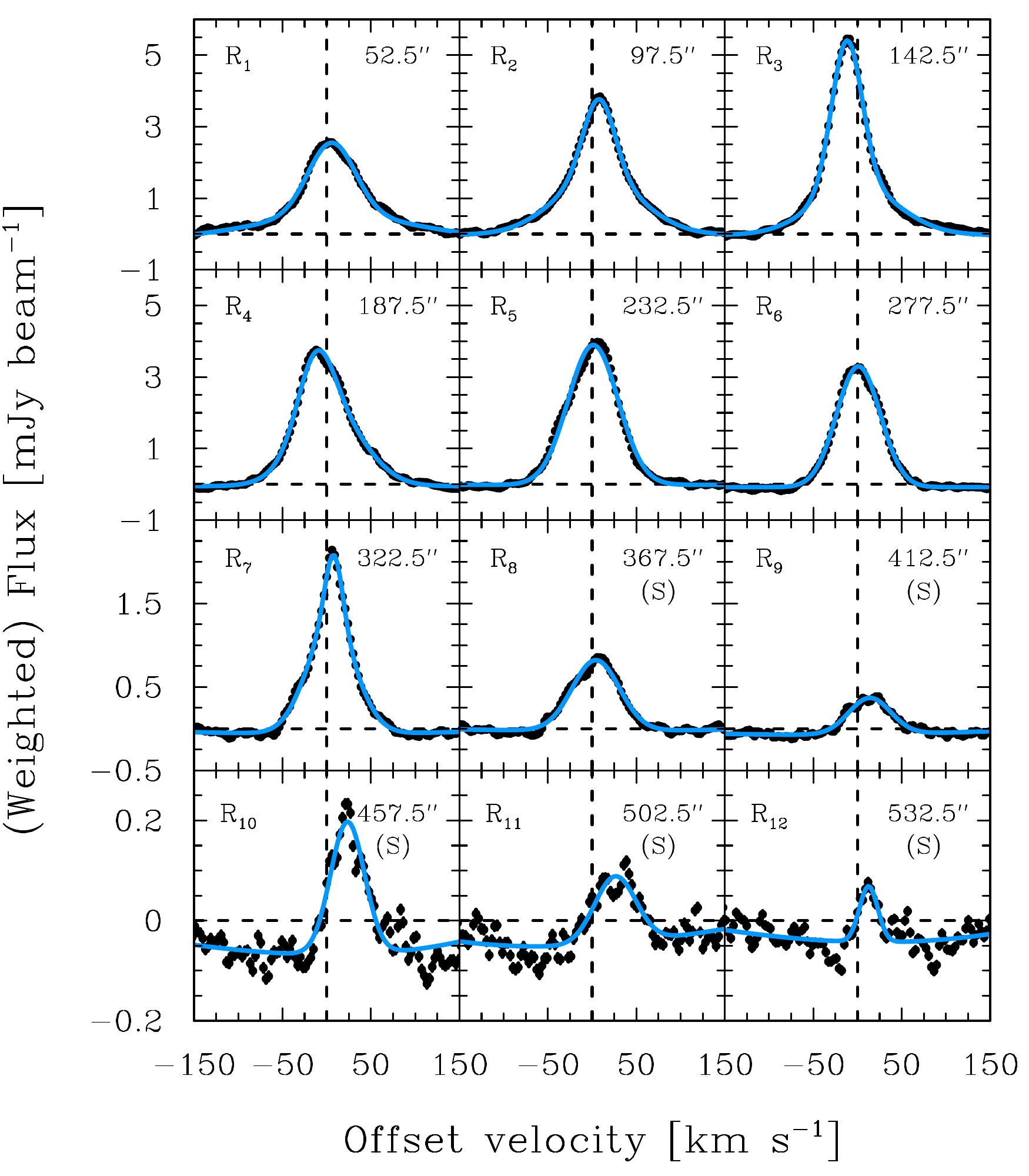}
    \end{tabular}
    \caption{Left: the 
    observed rotation curve of NGC 7331 \citep[thick gray line,][]{debloketal08} 
    and our interpolation and extrapolation of the rotation curve (blue solid line). 
    Right: Azimuthally averaged stacked profiles of NGC 7331 
    (black circle symbols). The solid blue lines are the 
    fit to the data. The vertical and horizontal 
    dashed lines indicate the flux and offset velocity 
    with respect to the model velocity field to guide the eyes. 
    The (S) letters in some panels represent the `stacked radii', i.e., the 
    spectra at a radius larger than 
    the extent traced directly by the THINGS moment zero map.}  
    \label{fig:shape} 
\end{figure*}   

\begin{figure*}
\centering
    \begin{tabular}{l l l }
      \includegraphics[scale= 0.31]{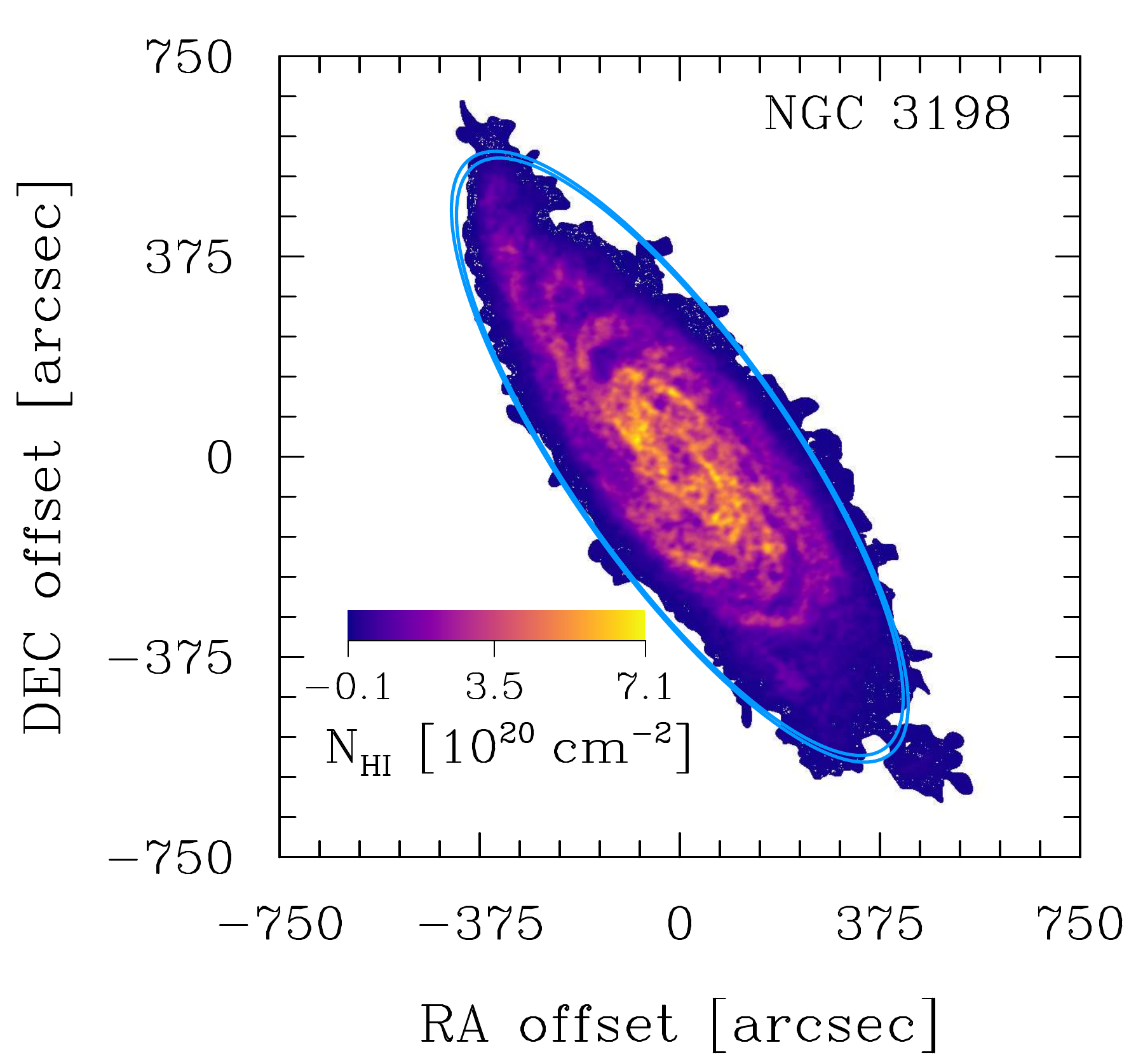} &
      \includegraphics[scale= 0.31]{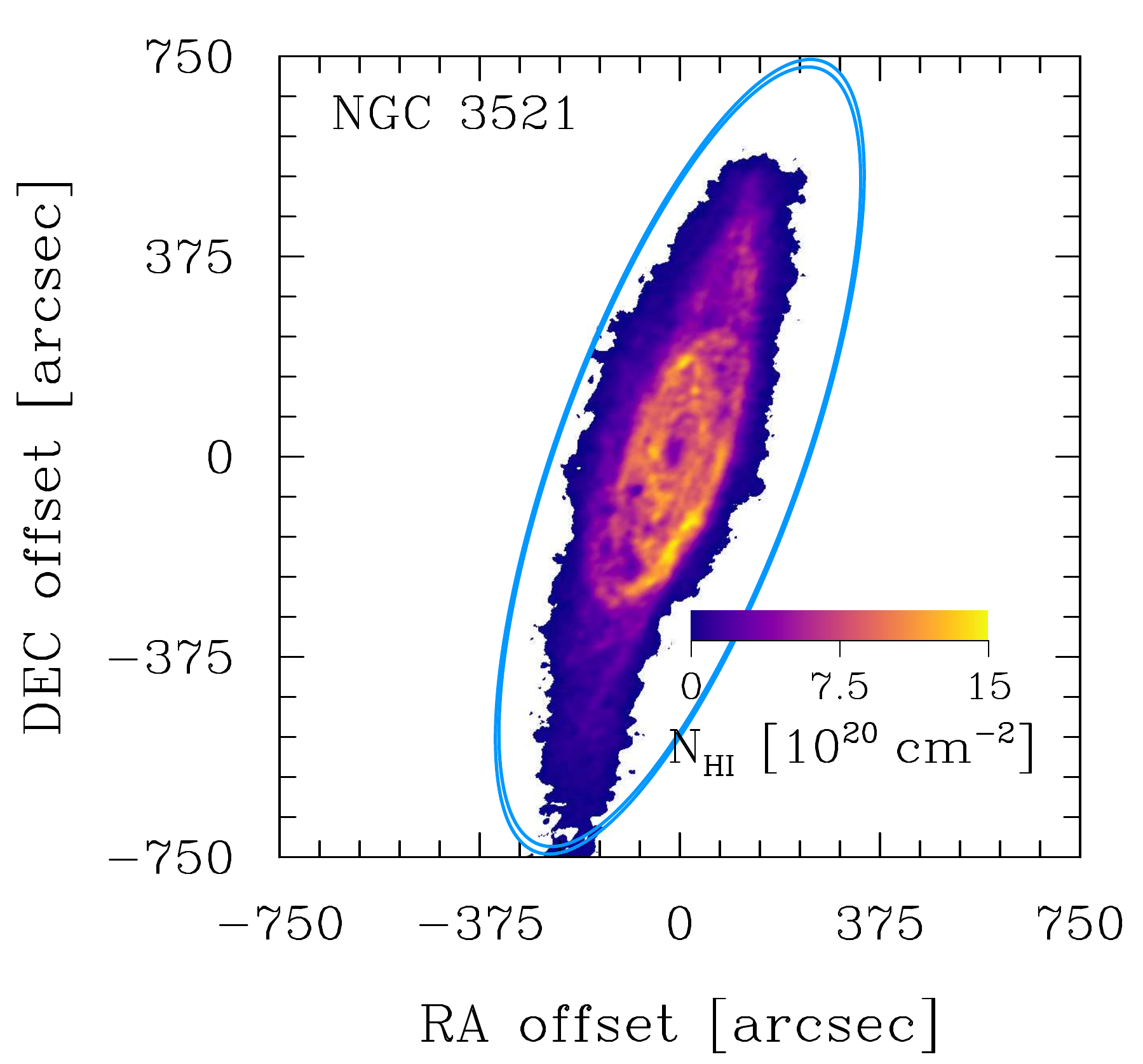} &
      \includegraphics[scale= 0.31]{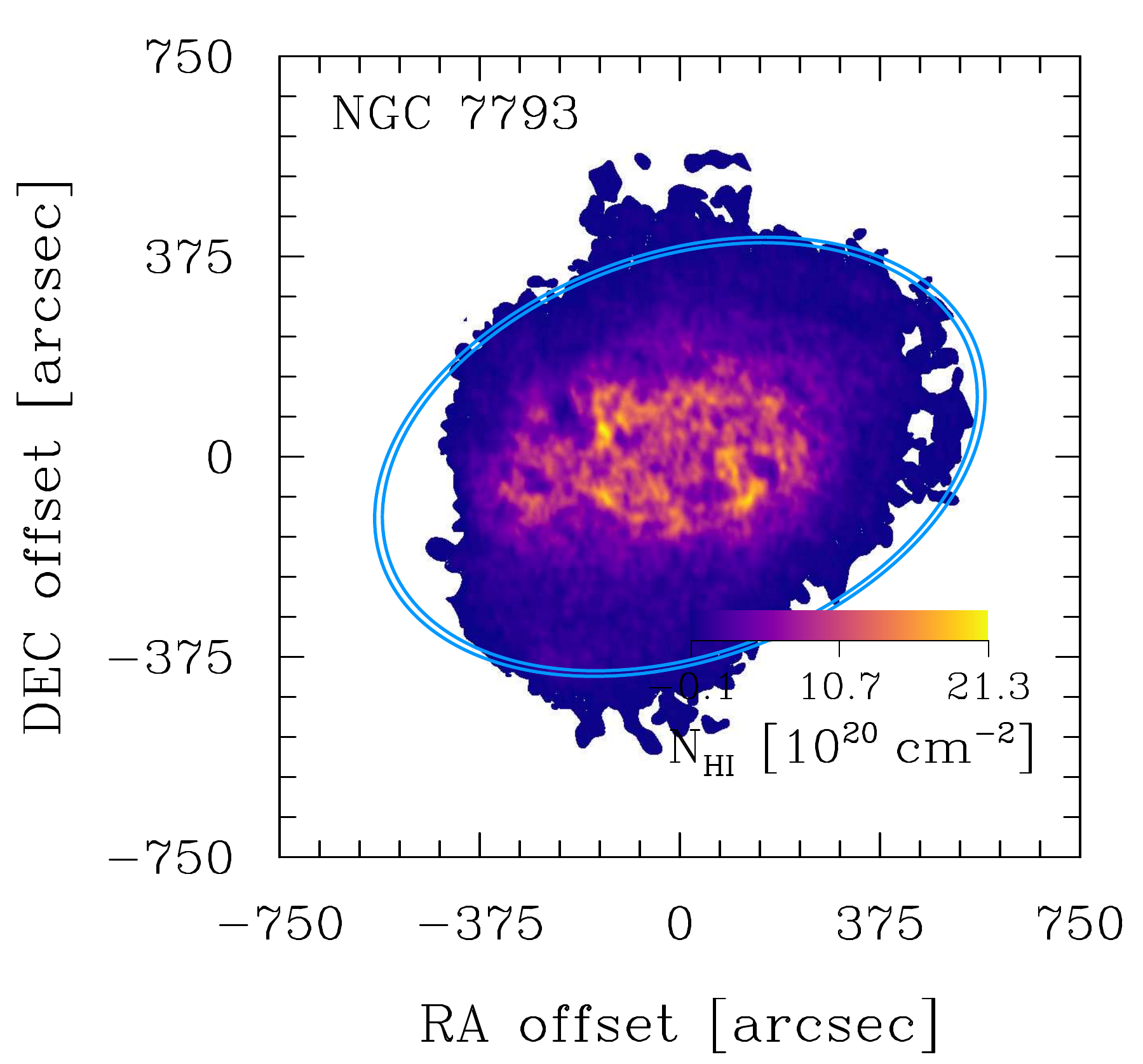} \\
      \includegraphics[scale= 0.3]{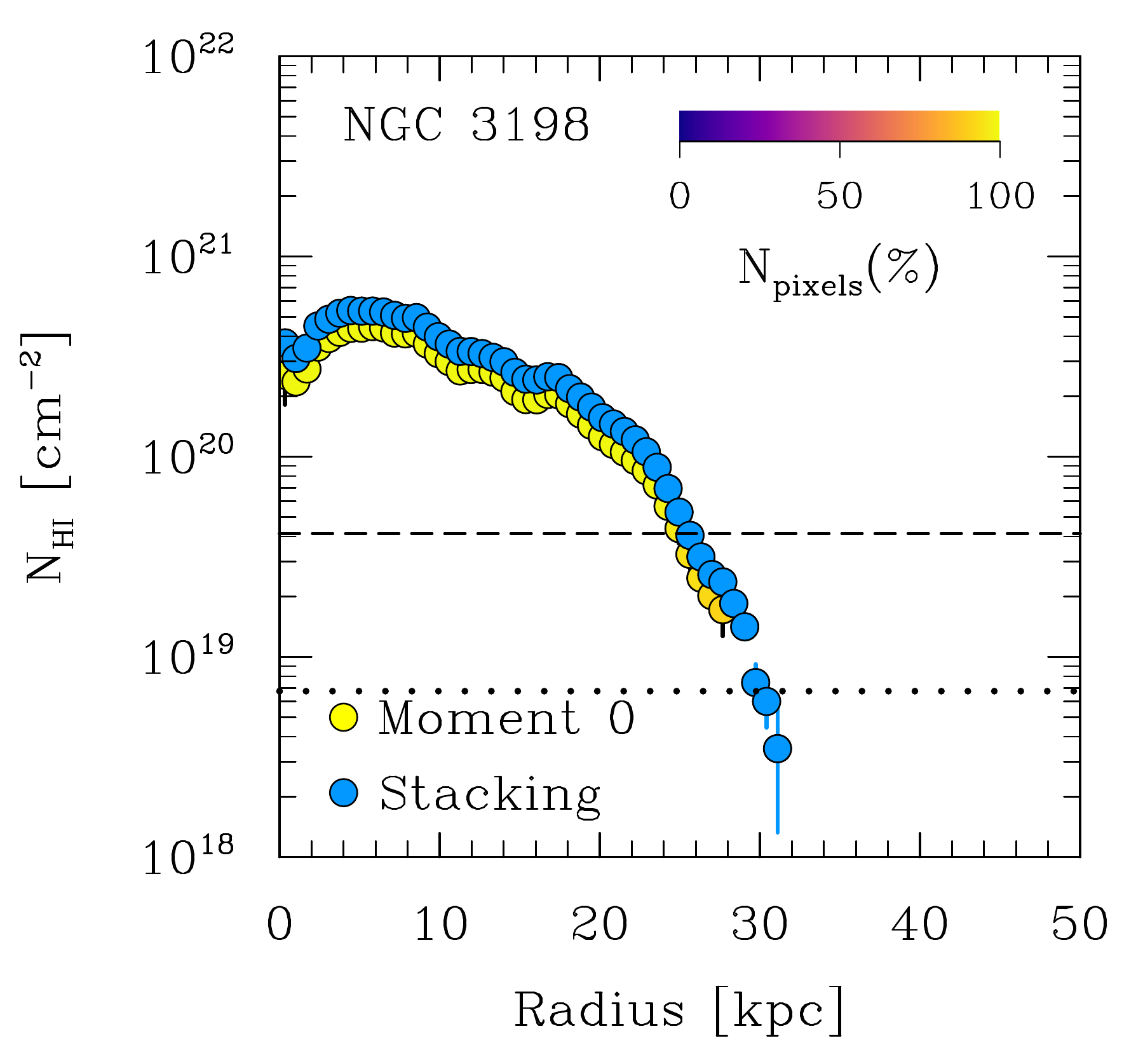} & 
      \includegraphics[scale= 0.3]{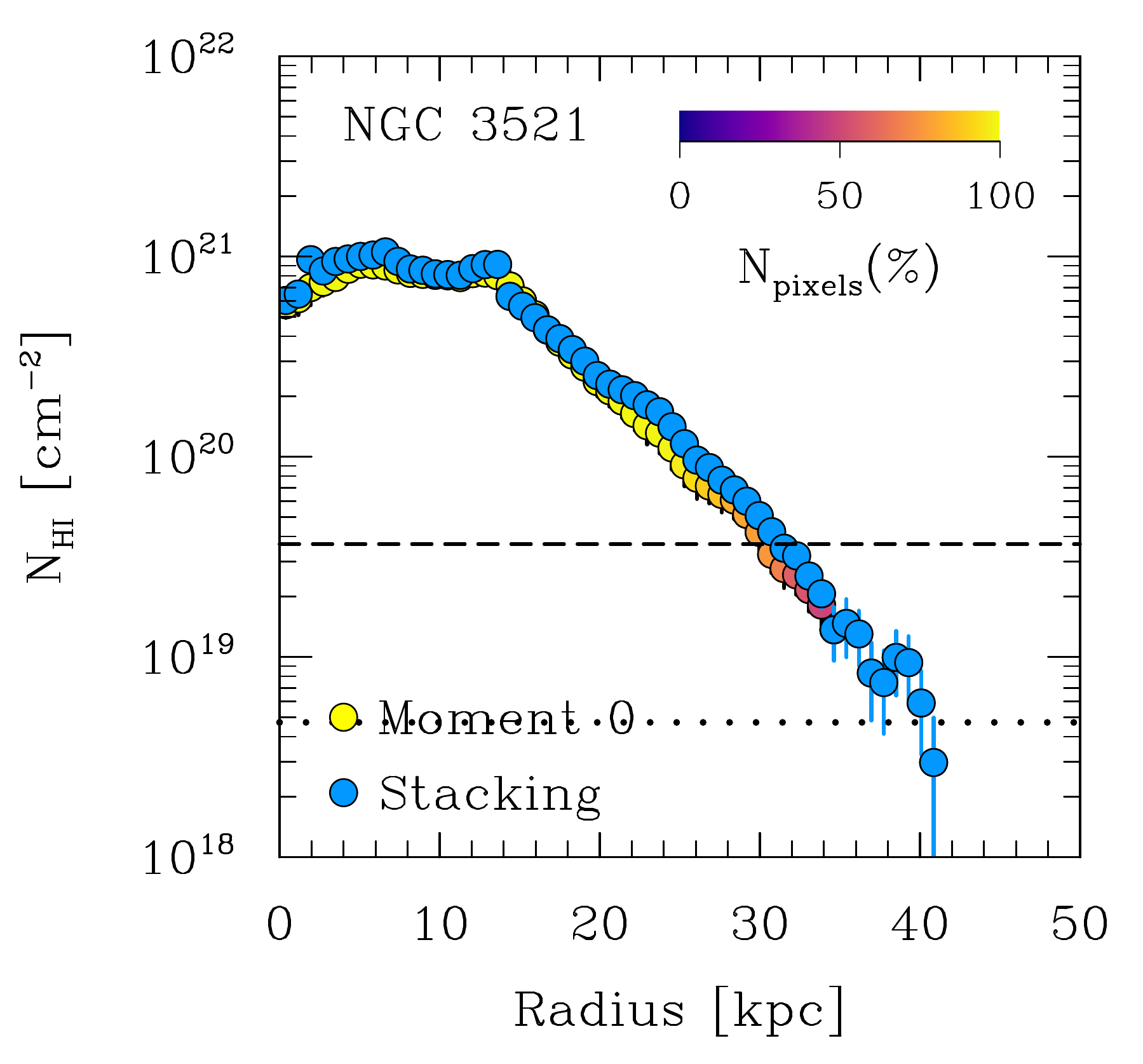}&
      \includegraphics[scale= 0.3]{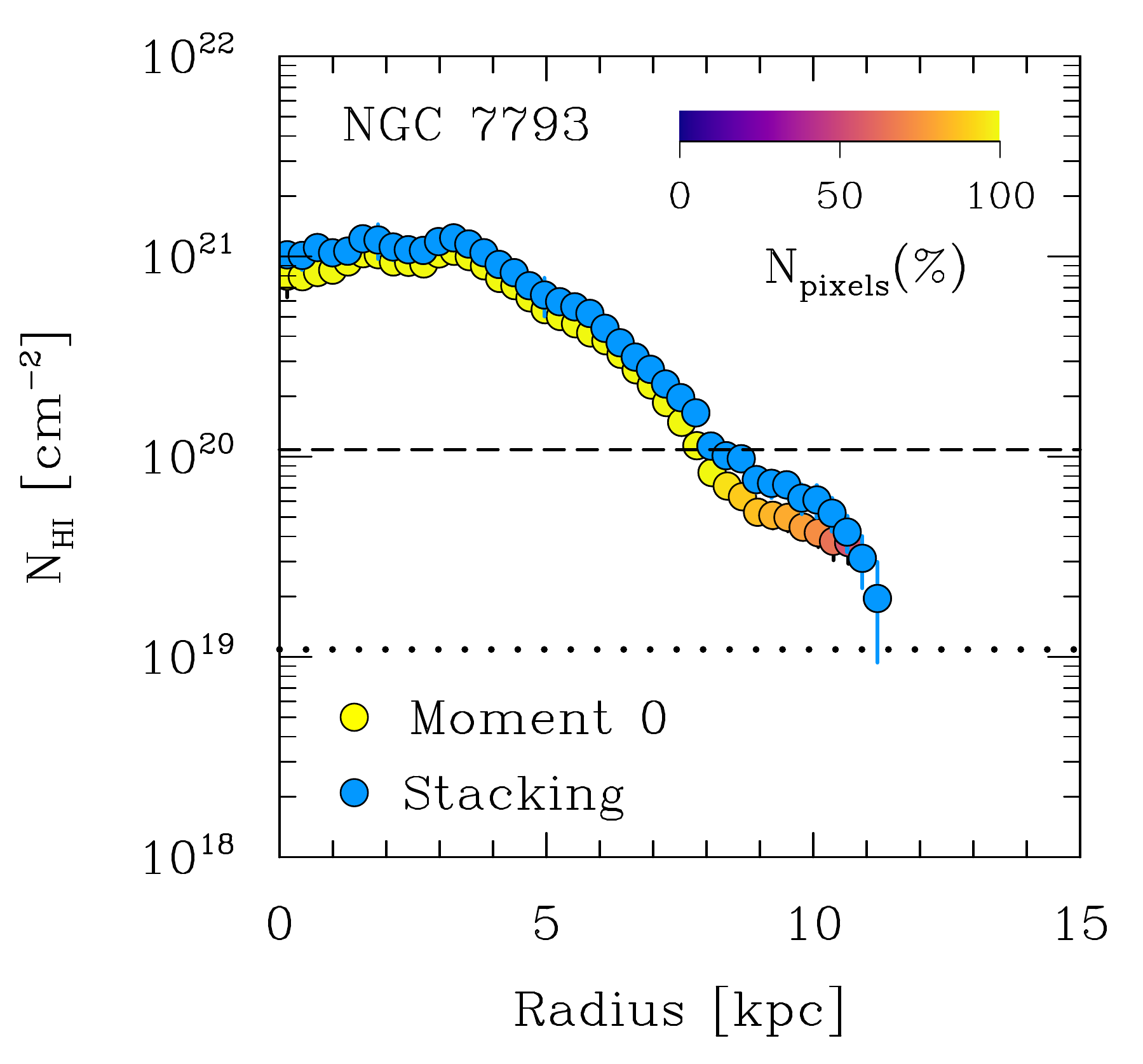}\\
      \includegraphics[scale= 0.31]{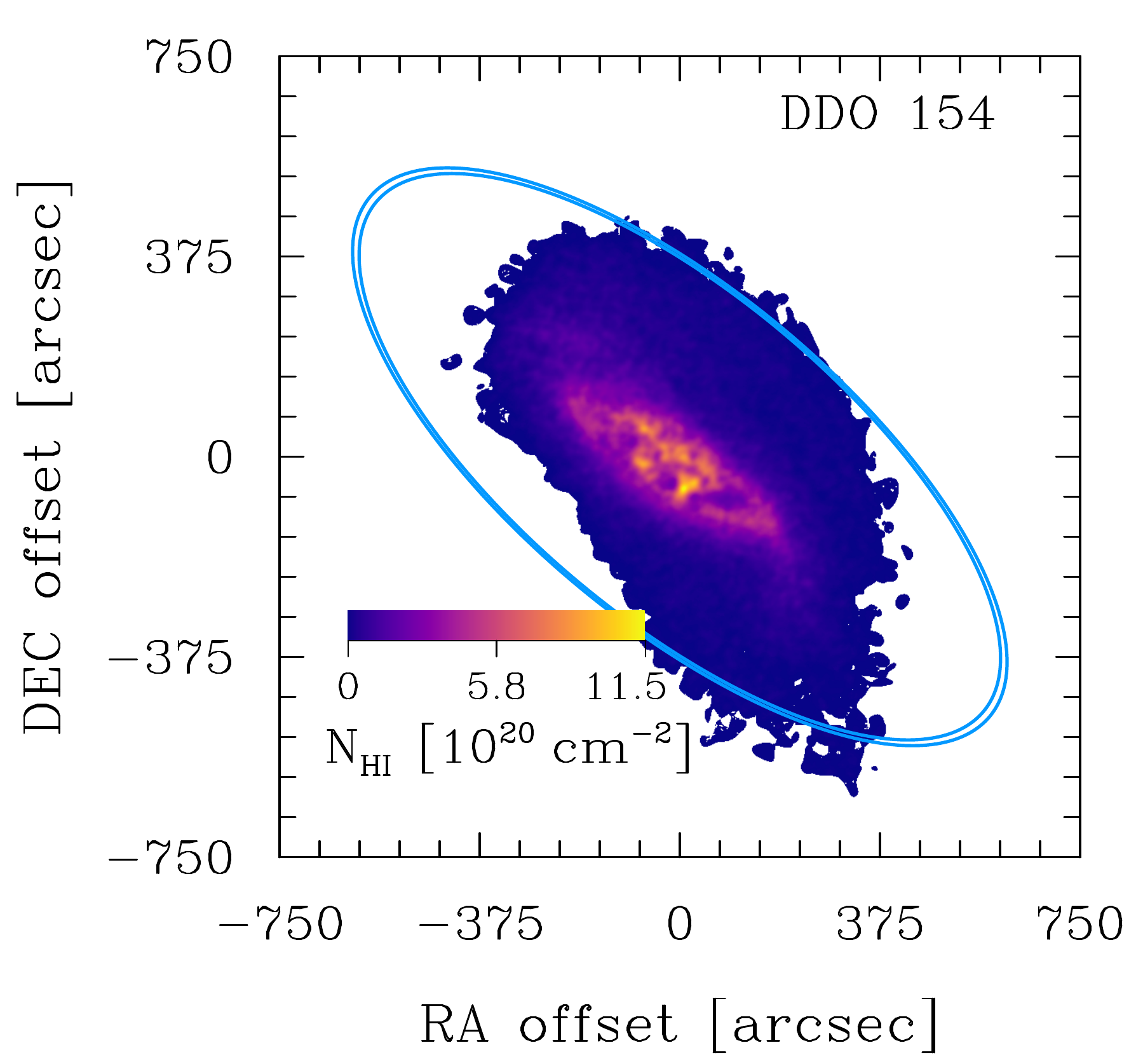} &
      \includegraphics[scale= 0.31]{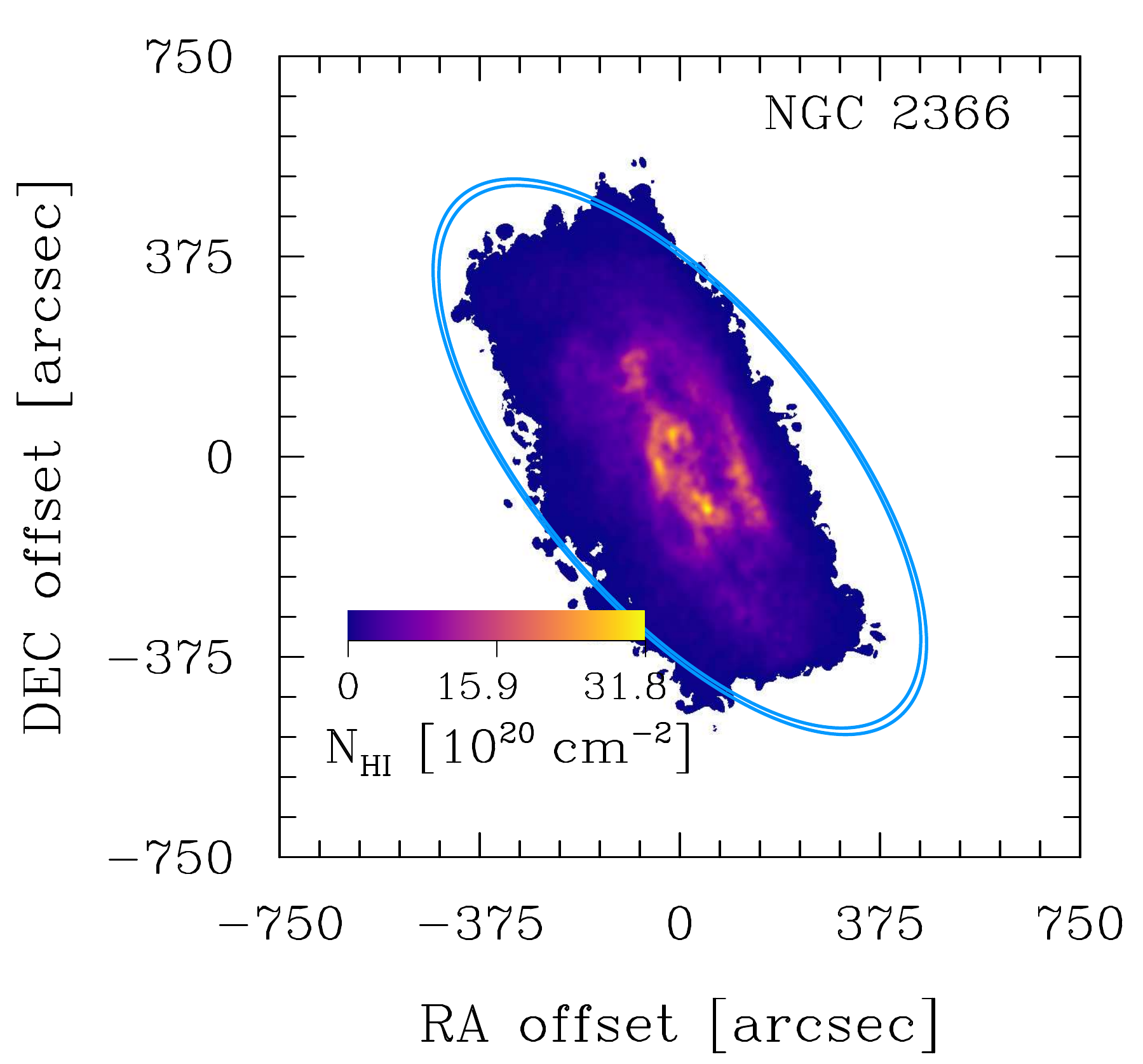} &
      \includegraphics[scale= 0.31]{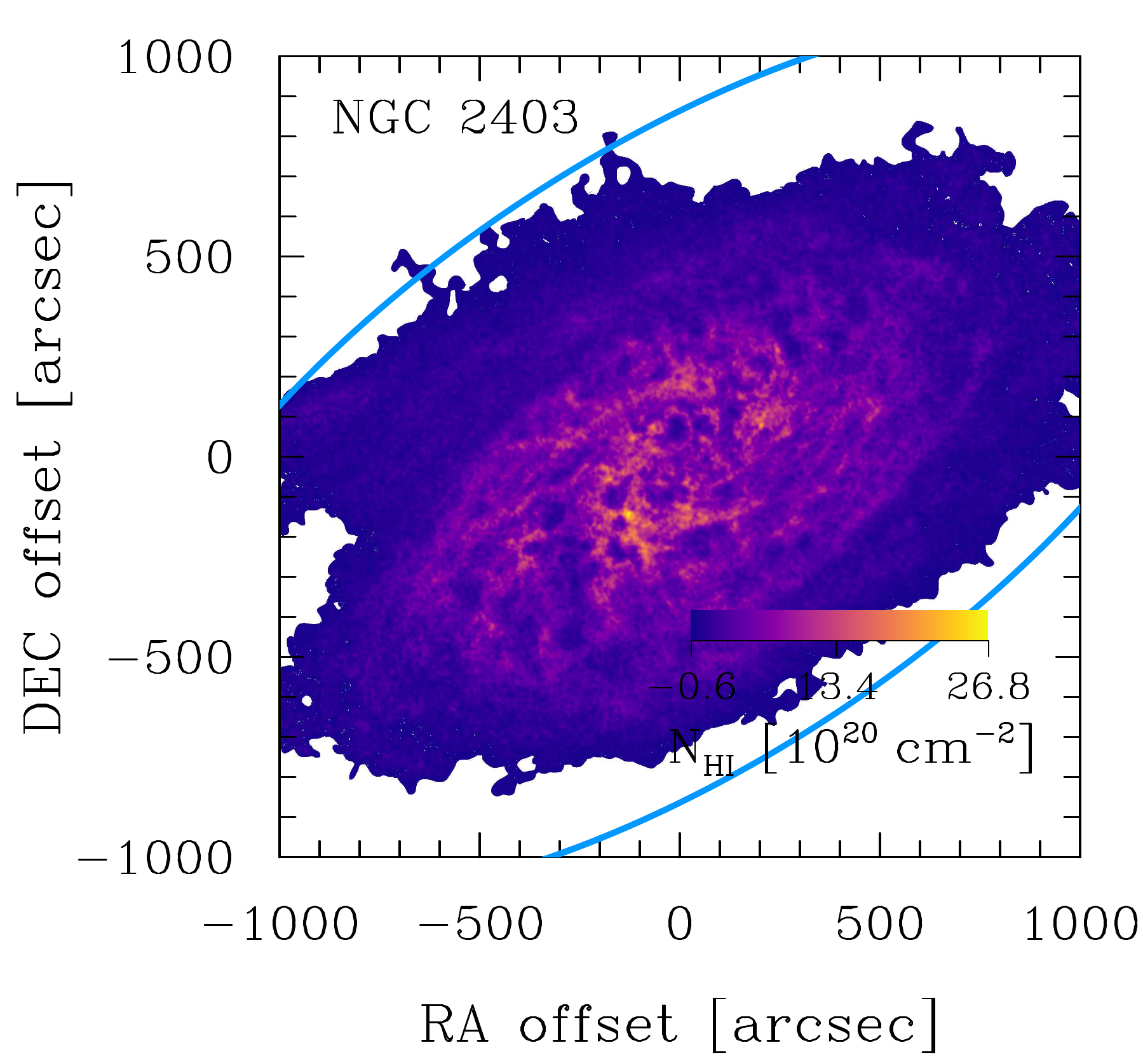} \\
      \includegraphics[scale= 0.31]{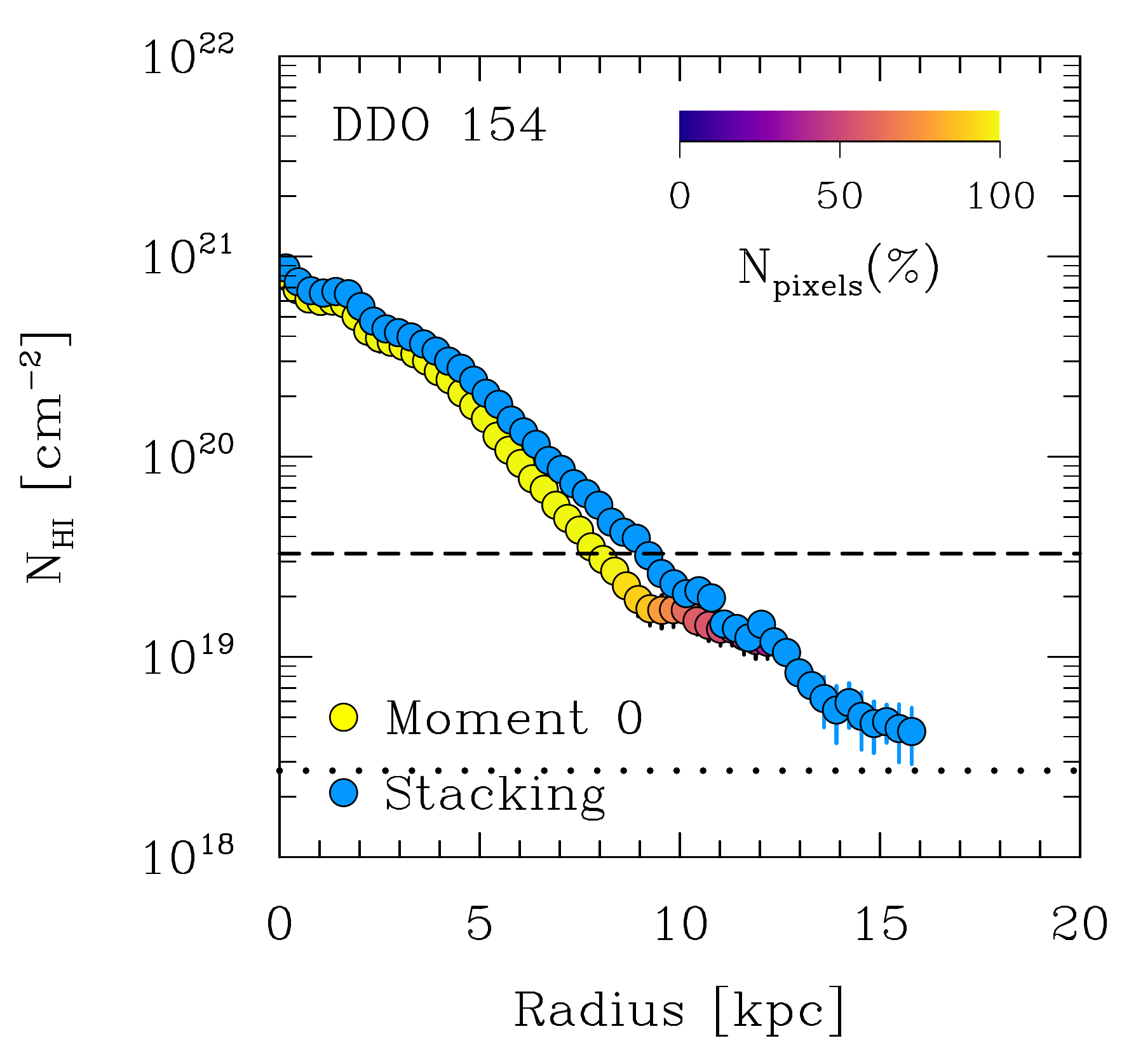}&
      \includegraphics[scale= 0.3]{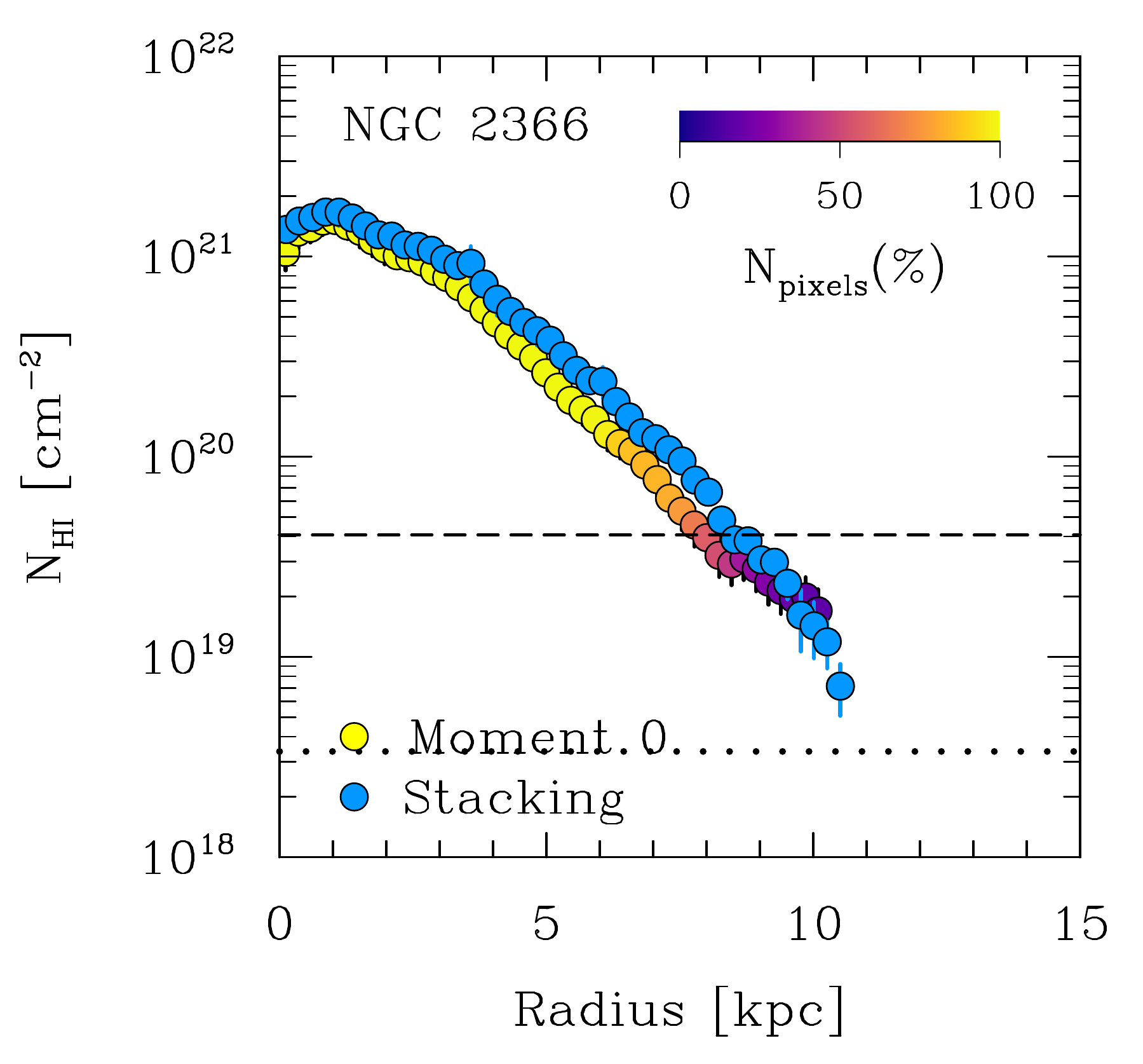}&
        \includegraphics[scale= 0.3]{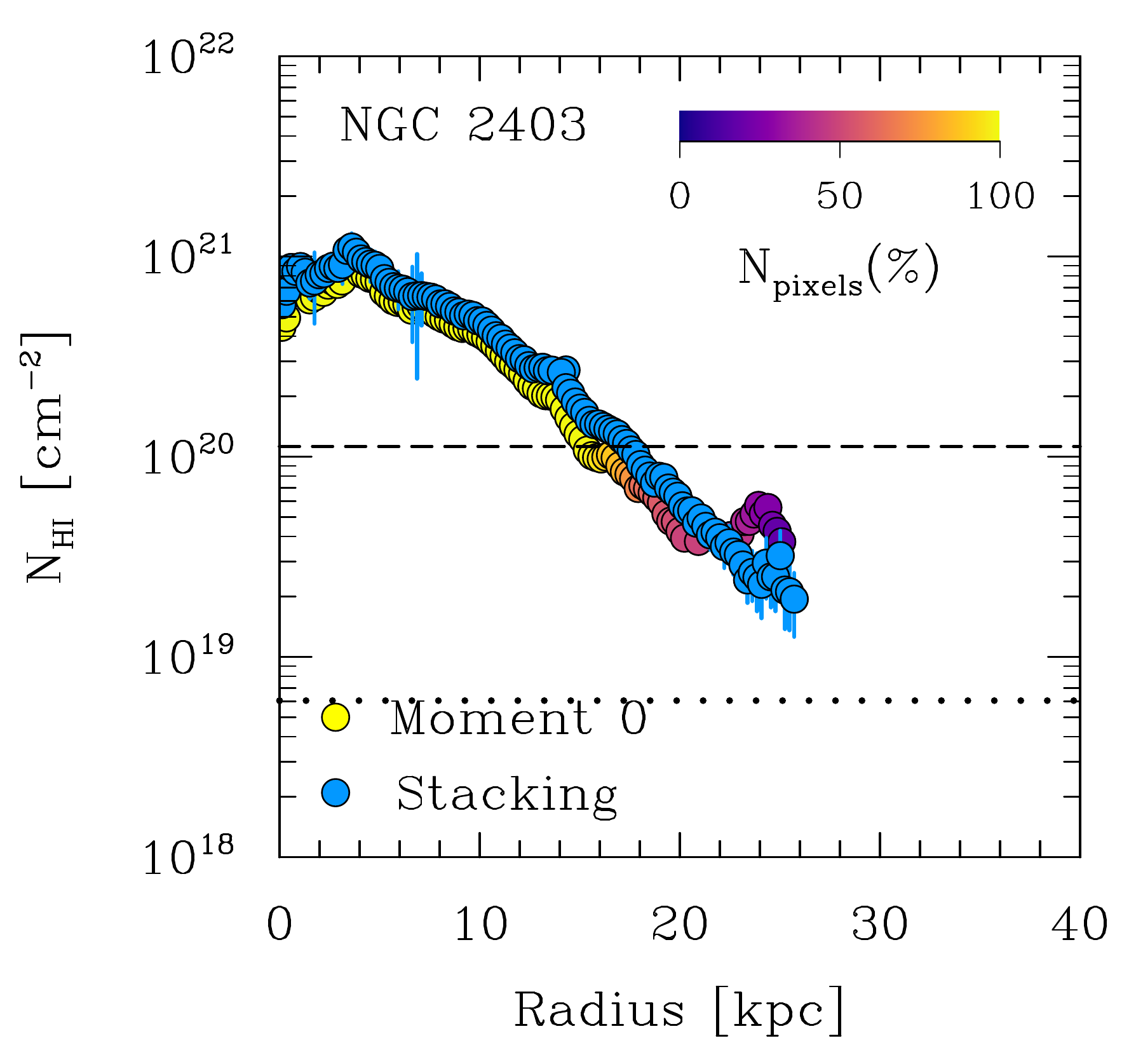}
    \end{tabular}
    \caption{Odd rows: H\,{\sc i} column density maps overlaid with blue ellipses 
    showing the annuli used to derive the radial profiles (only the outermost annulus is shown). 
    Even rows: the radial 
    H\,{\sc i} column density profiles of our sample galaxies. 
    The blue solid circle symbols represent the radial 
    profiles derived from our stacking method. 
    The color-coded (the color represents the number of pixels inside each annulus in \%) solid circle symbols 
    show the azimuthally averaged radial profiles derived from the THINGS moment zero maps. 
    The horizontal dashed lines correspond to the sensitivity of the 
    THINGS moment zero maps (3 times the rms noise). The dotted lines represent the 
    sensitivity of the azimuthally stacked profiles corresponding to 
    3 times the rms noise.} 
    \label{fig:radprof} 
\end{figure*} 
\begin{figure*}
\setcounter{figure}{1}
\centering
    \begin{tabular}{l l l}
      \includegraphics[scale= 0.31]{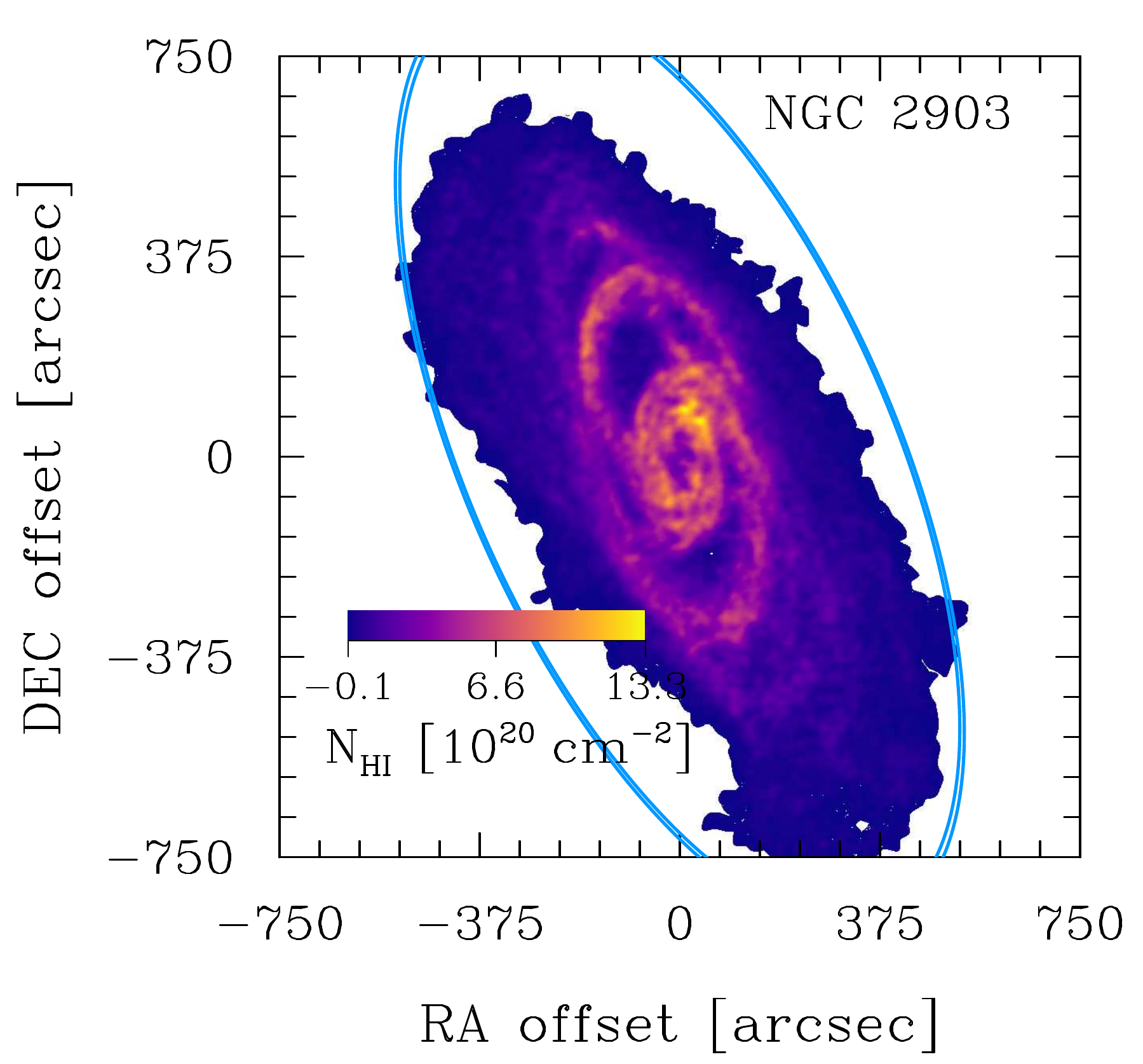}&
      \includegraphics[scale= 0.31]{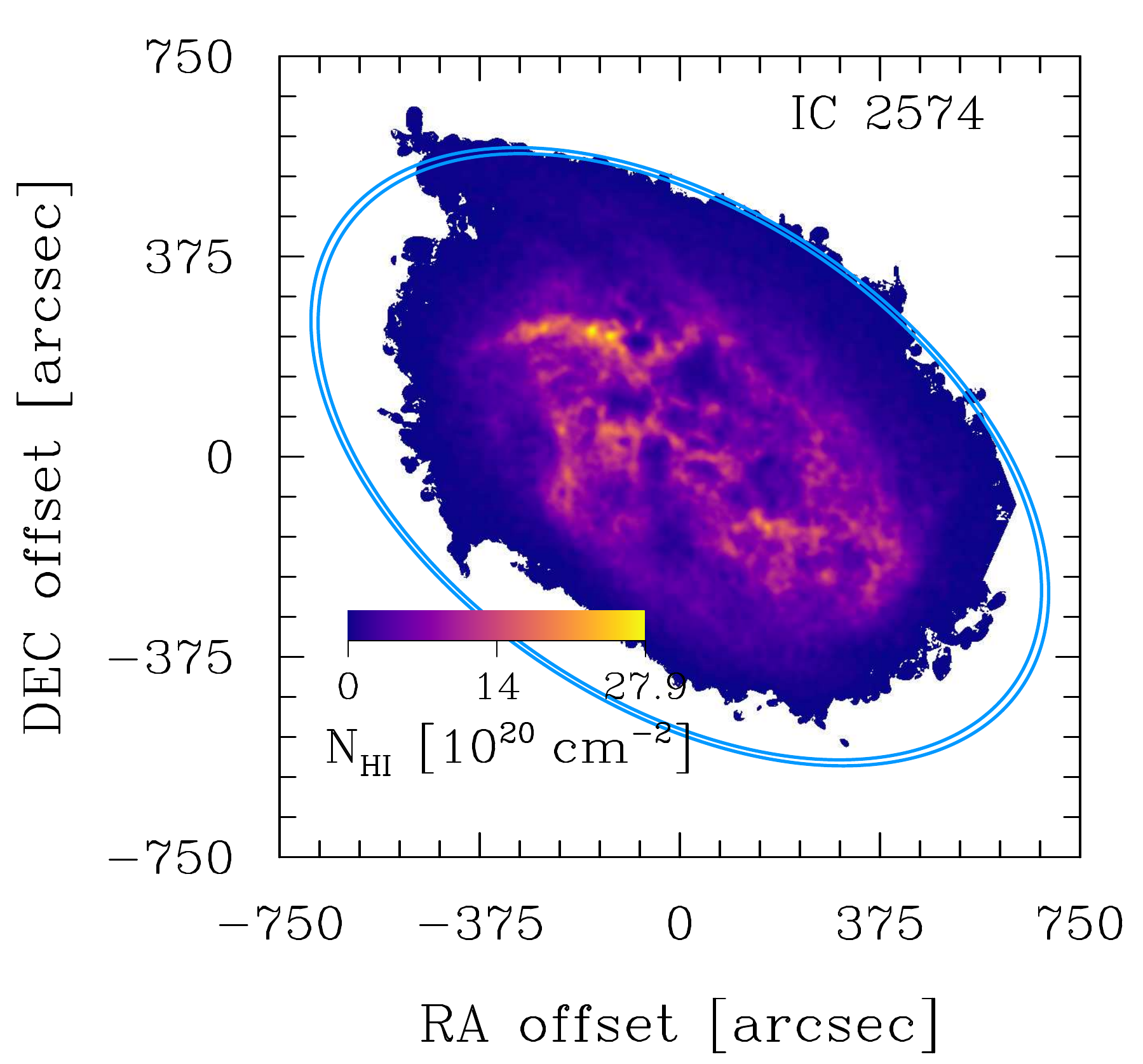}&
      \includegraphics[scale= 0.31]{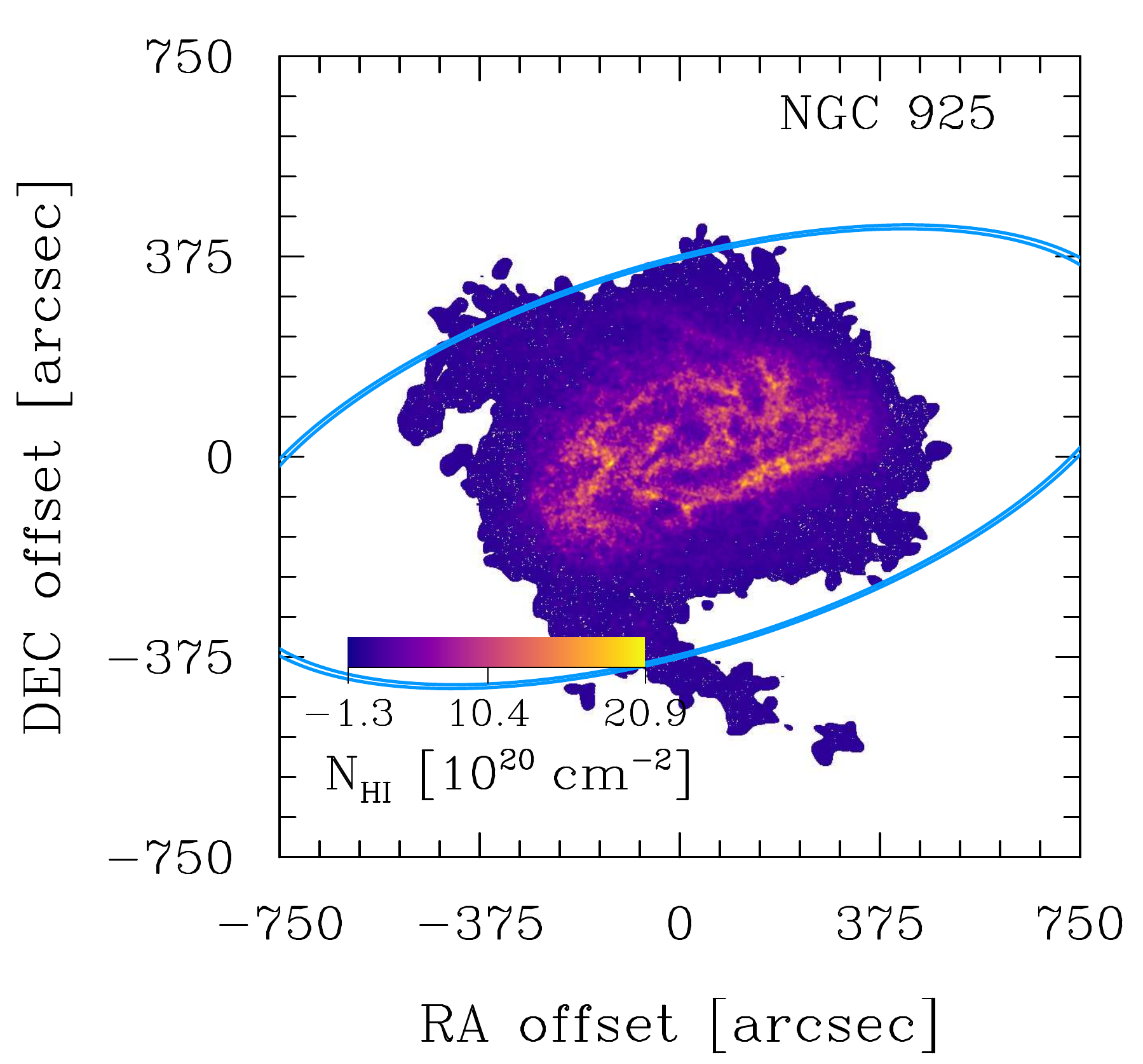}\\
      \includegraphics[scale= 0.3]{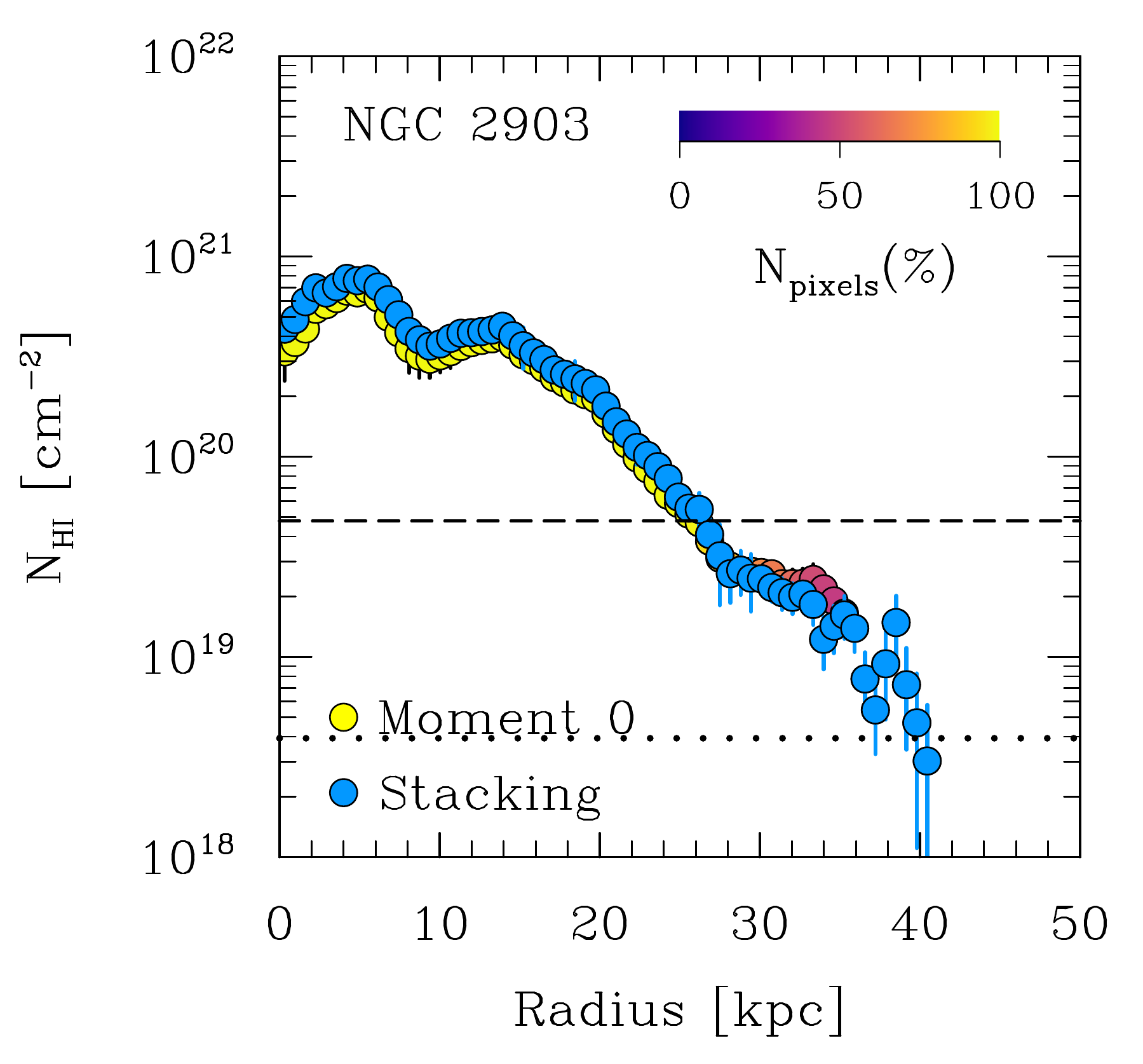}&
      \includegraphics[scale= 0.3]{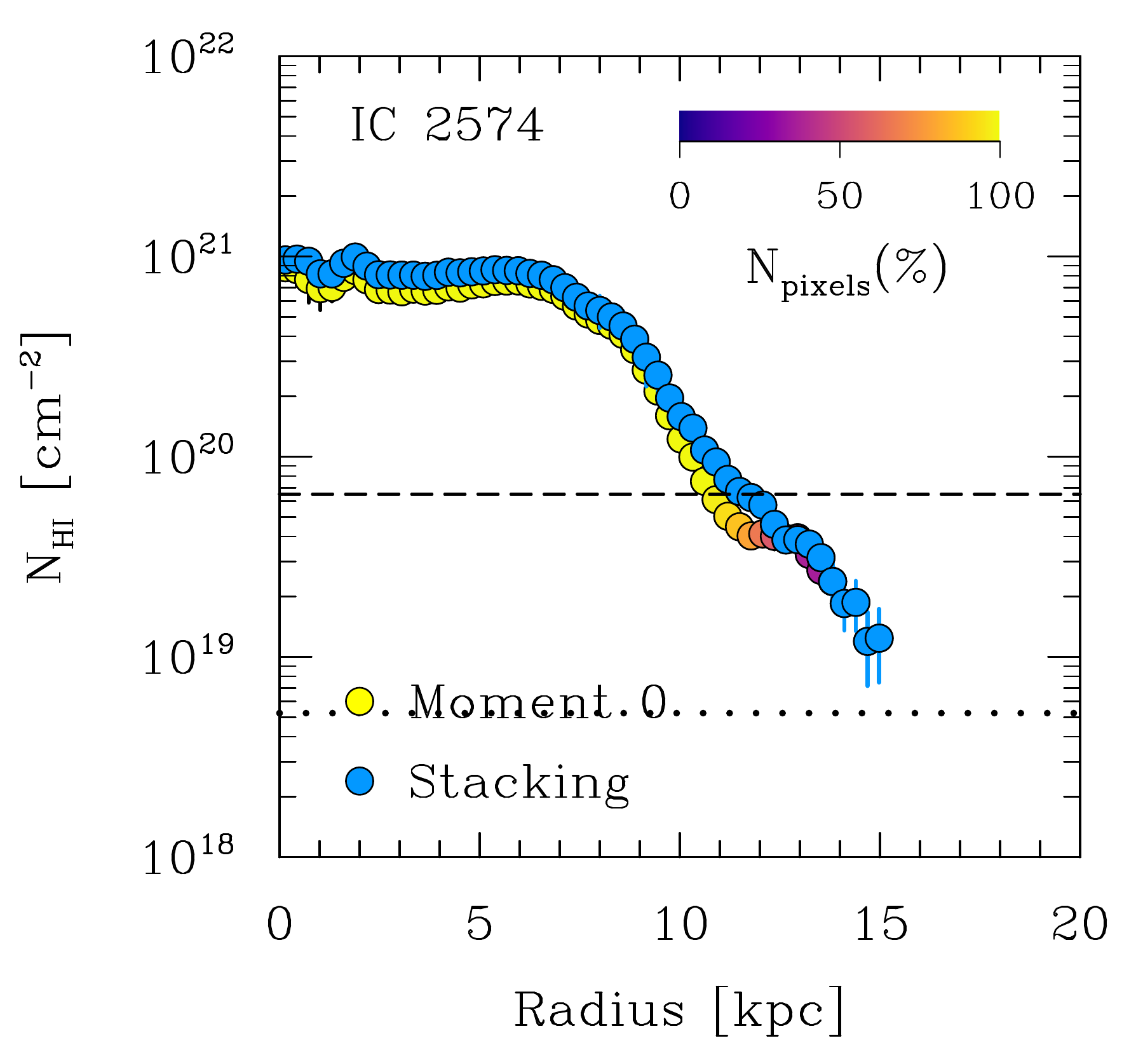}&
      \includegraphics[scale= 0.31]{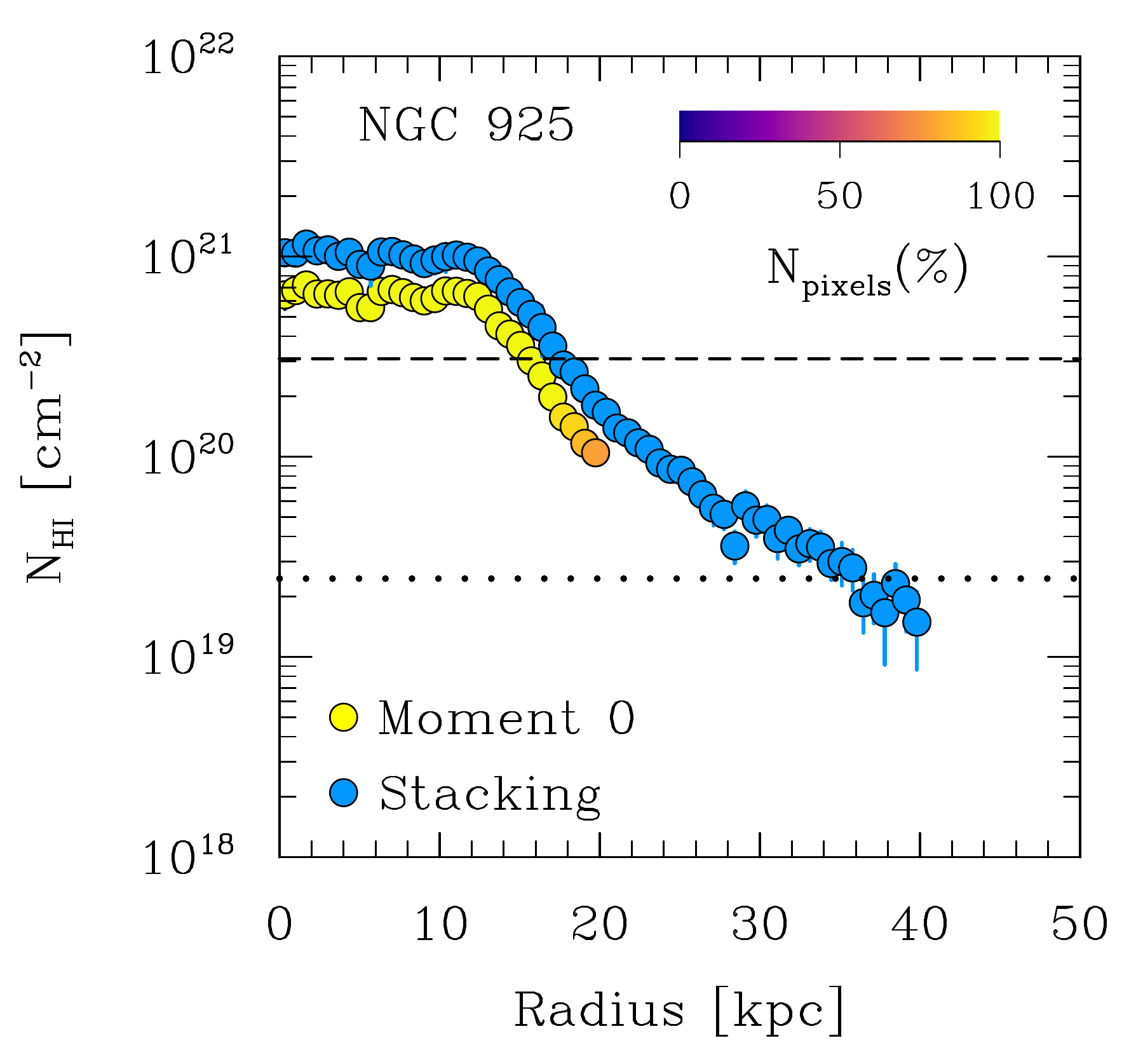}\\
      \includegraphics[scale= 0.31]{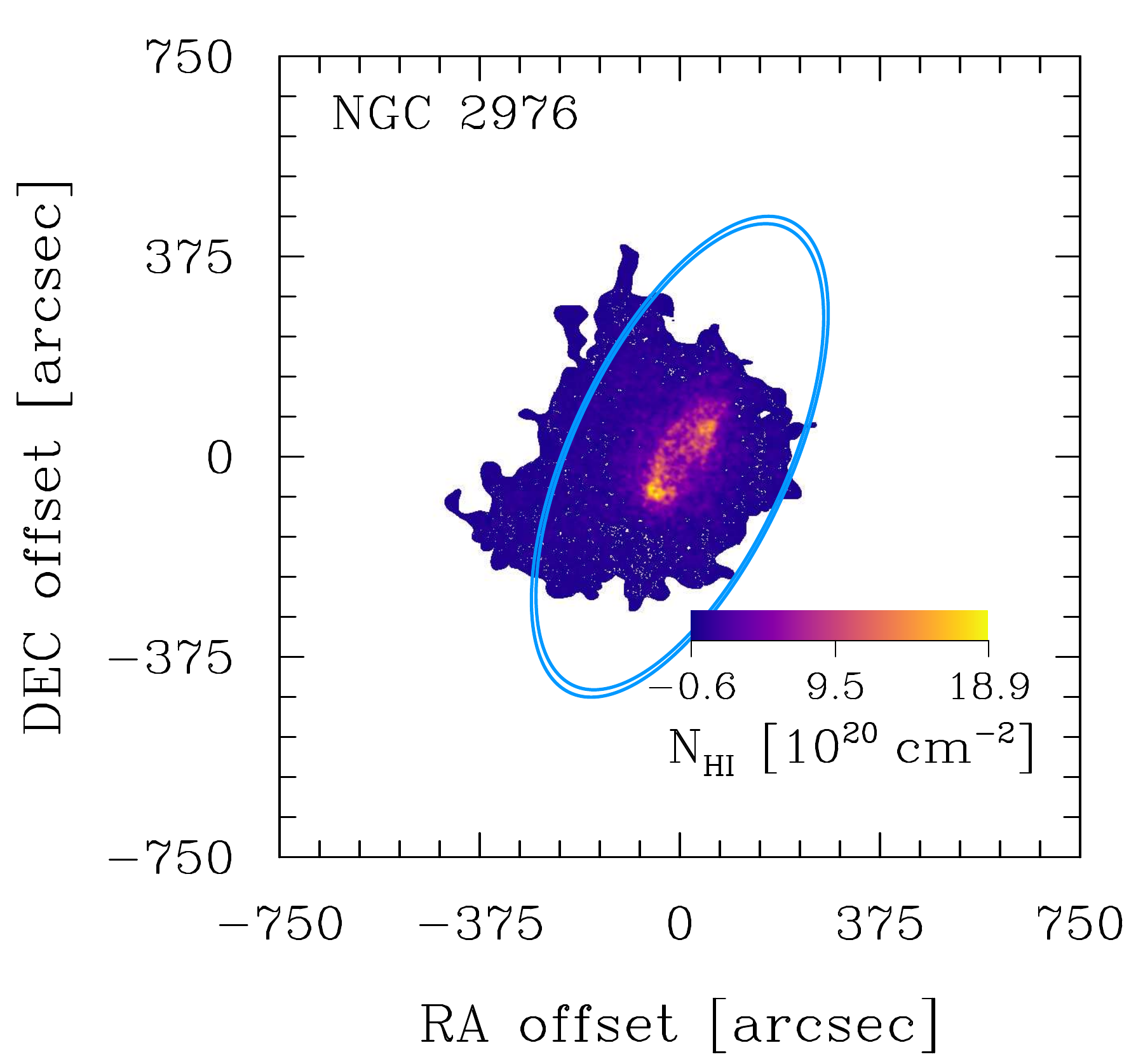} &
      \includegraphics[scale= 0.31]{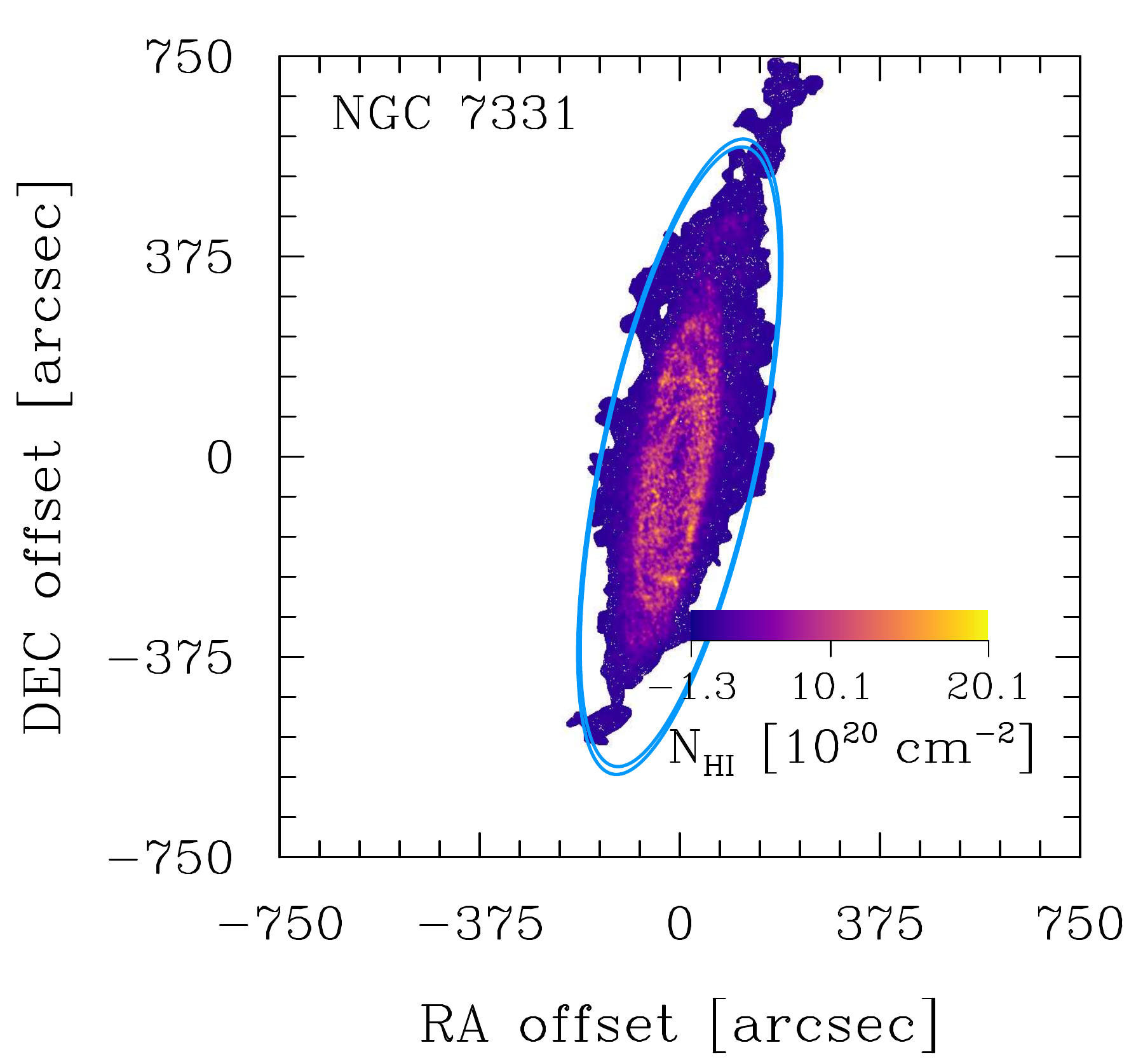} &
      \includegraphics[scale= 0.31]{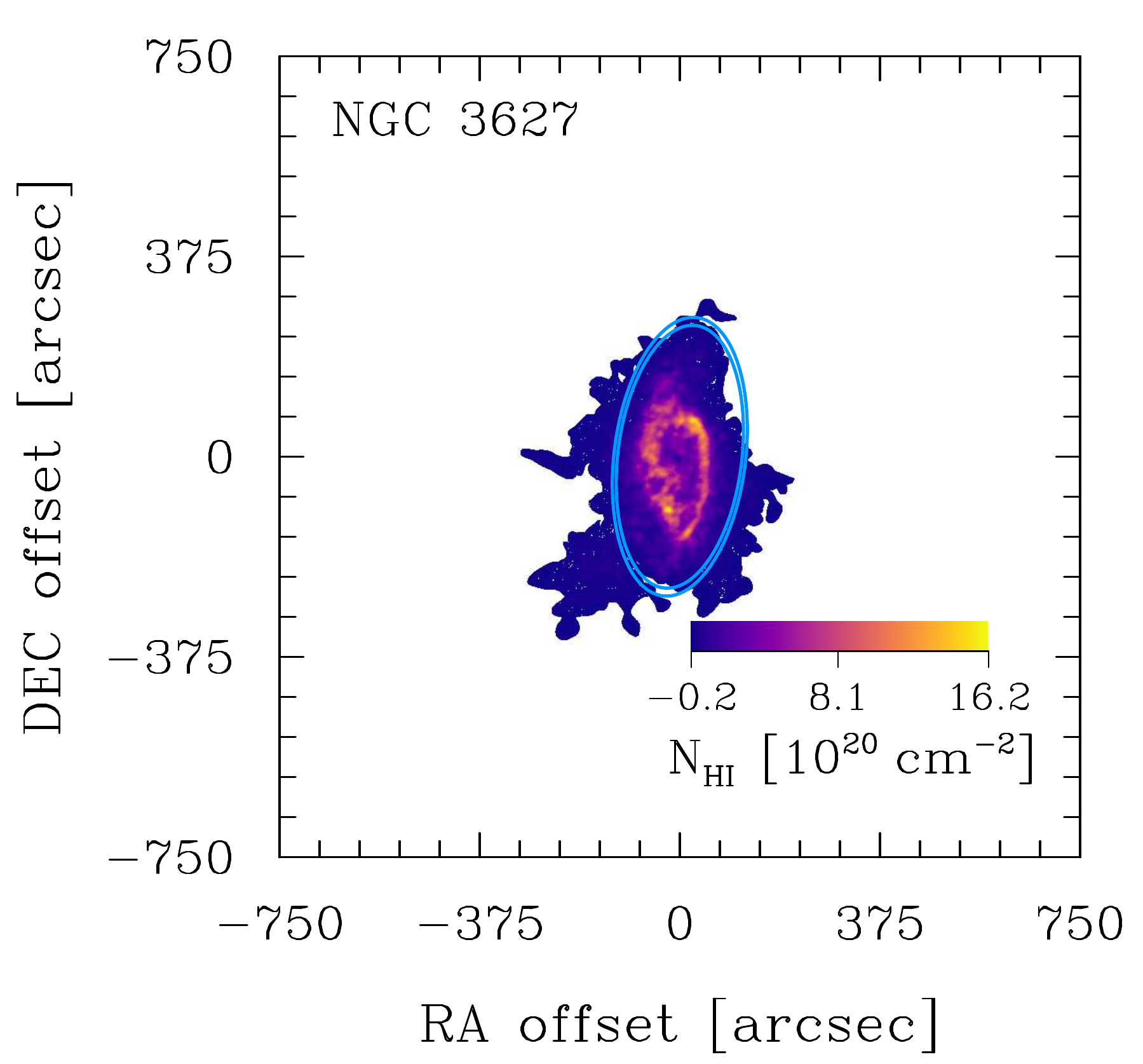} \\
      \includegraphics[scale= 0.3]{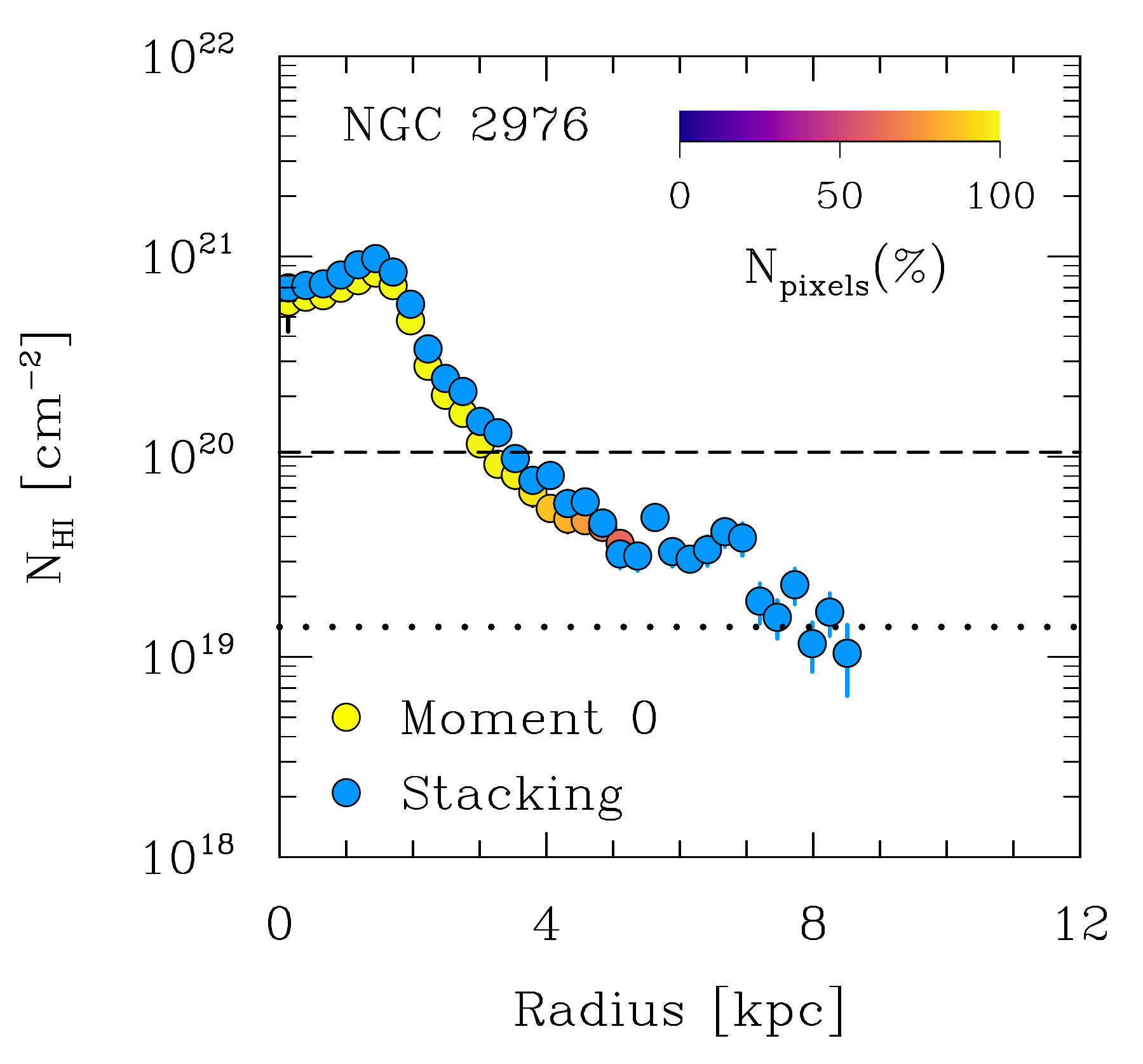}&
      \includegraphics[scale= 0.3]{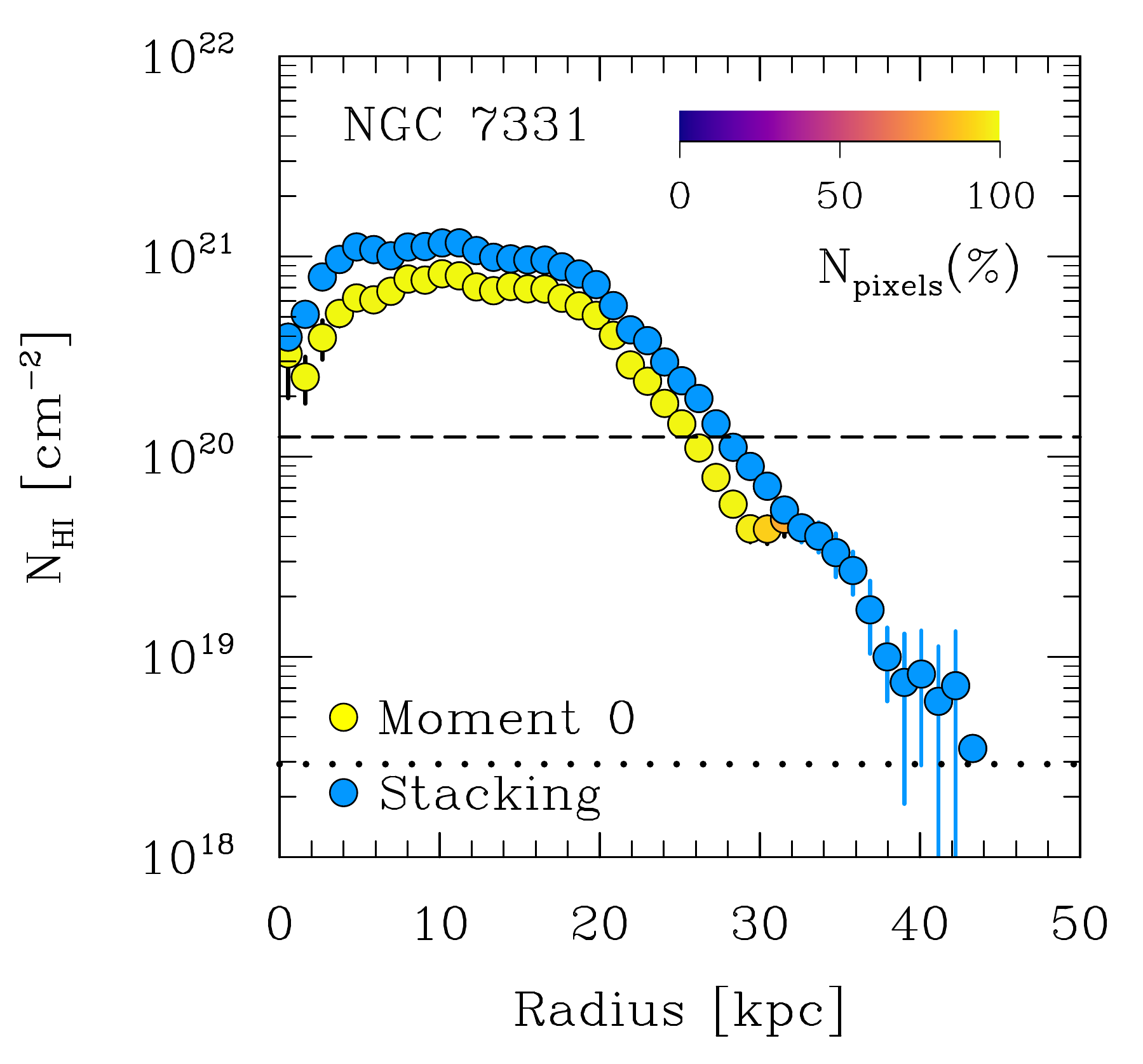}&
      \includegraphics[scale= 0.3]{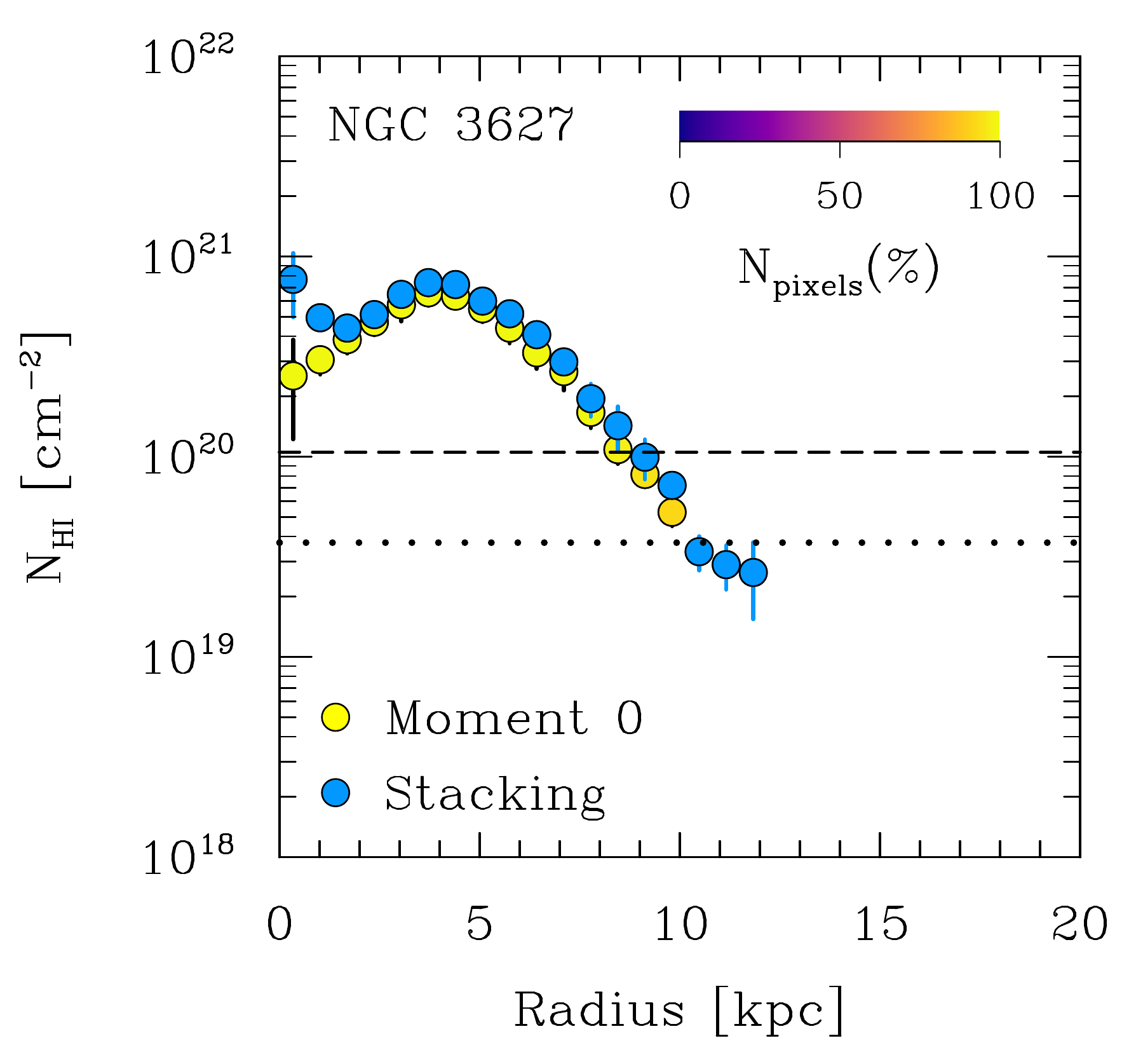}
    \end{tabular}
\caption{Continued. \label{fig:radprof}}
\end{figure*} 
\begin{figure*}
\setcounter{figure}{1}
\centering
    \begin{tabular}{l l l}
      \includegraphics[scale= 0.31]{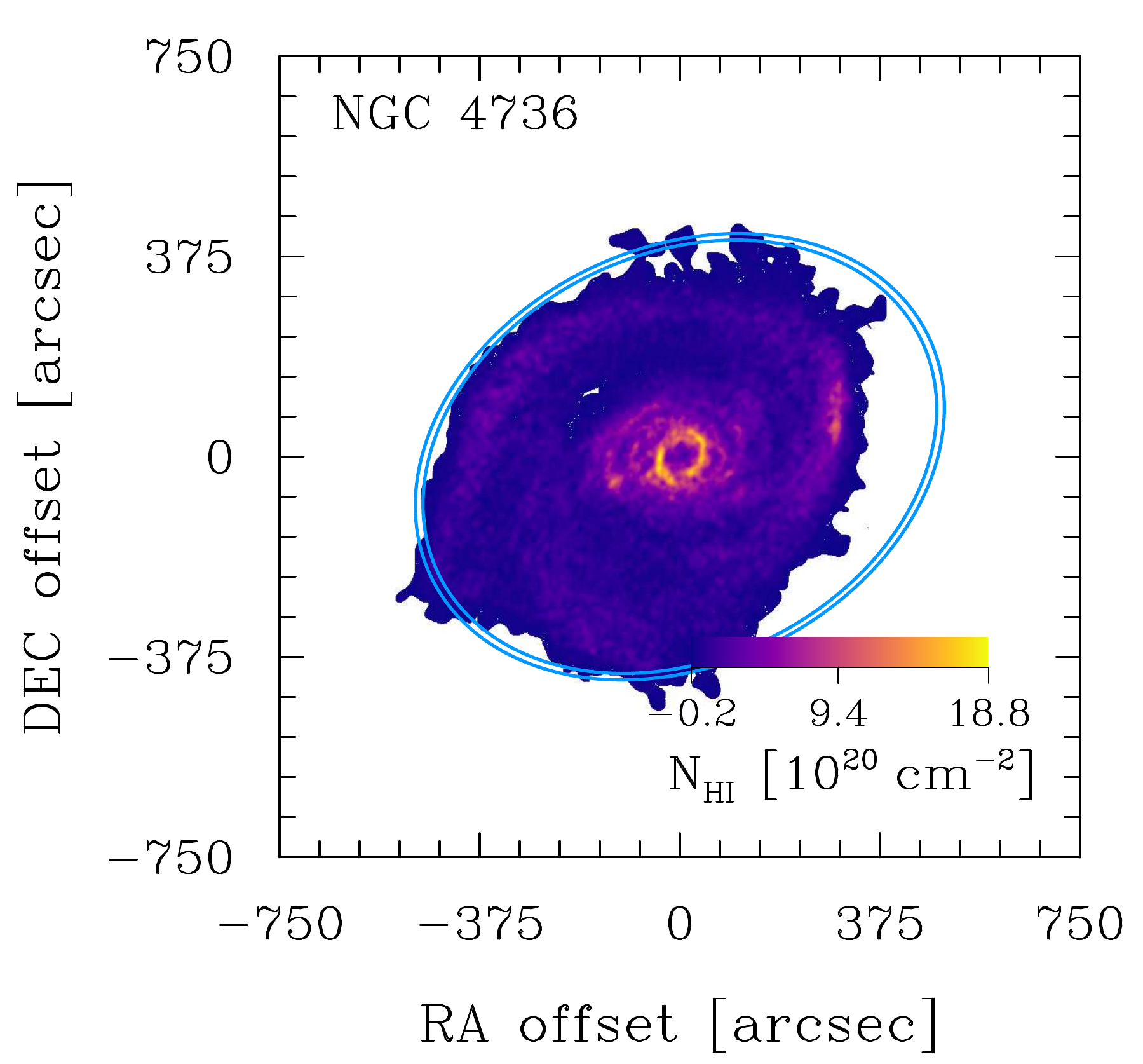} &
      \includegraphics[scale= 0.31]{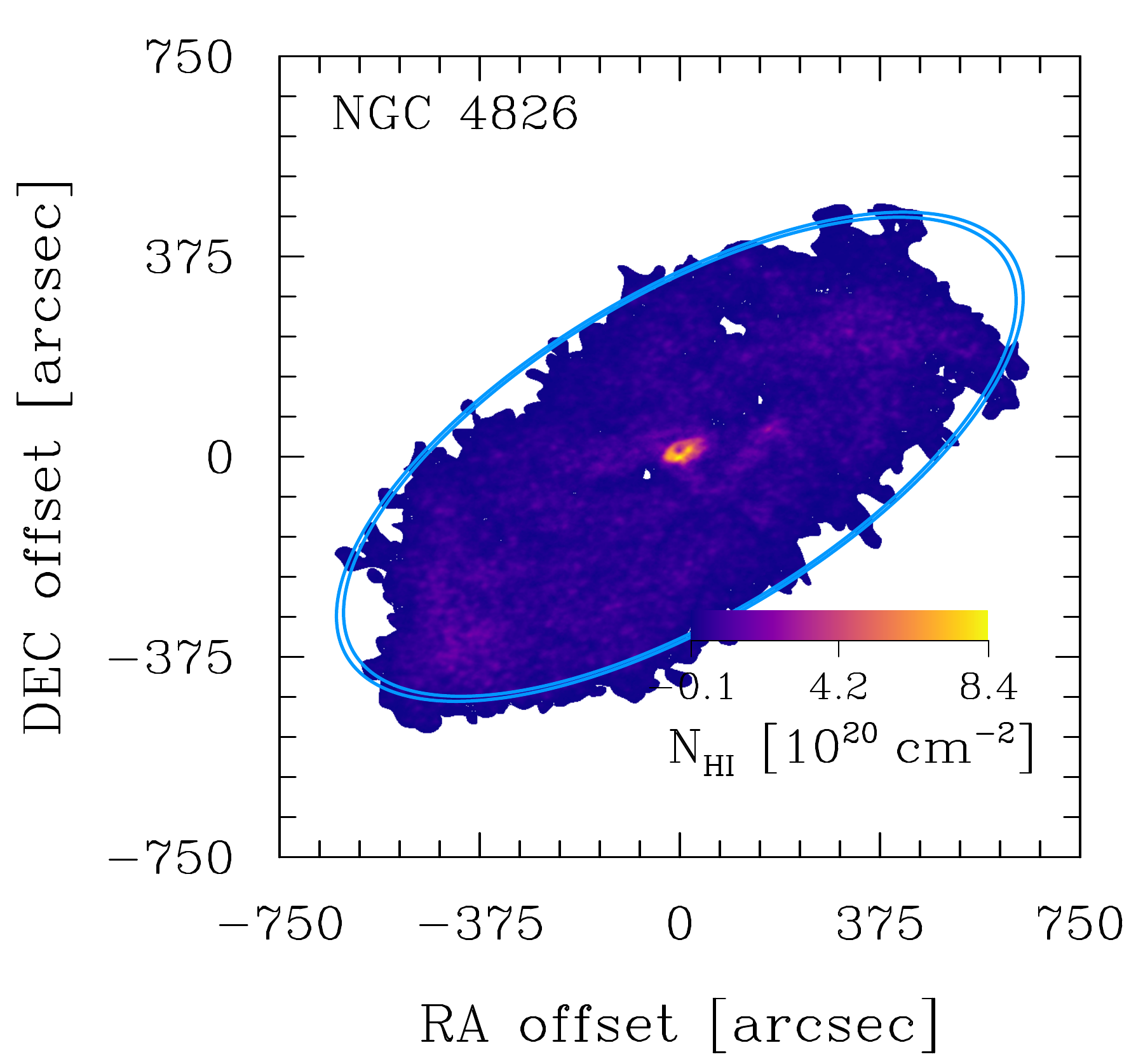}&
      \includegraphics[scale= 0.31]{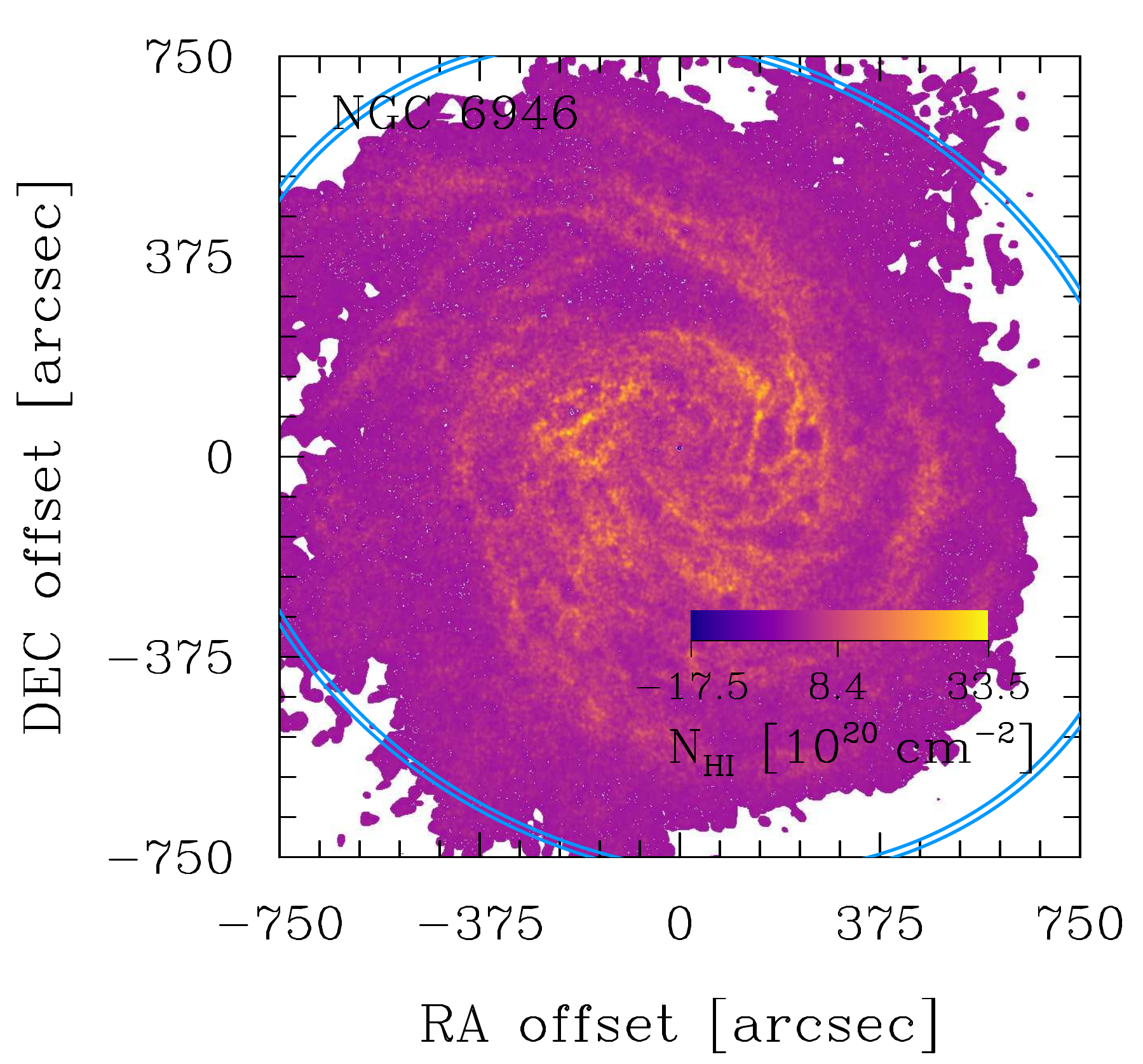}\\
      \includegraphics[scale= 0.3]{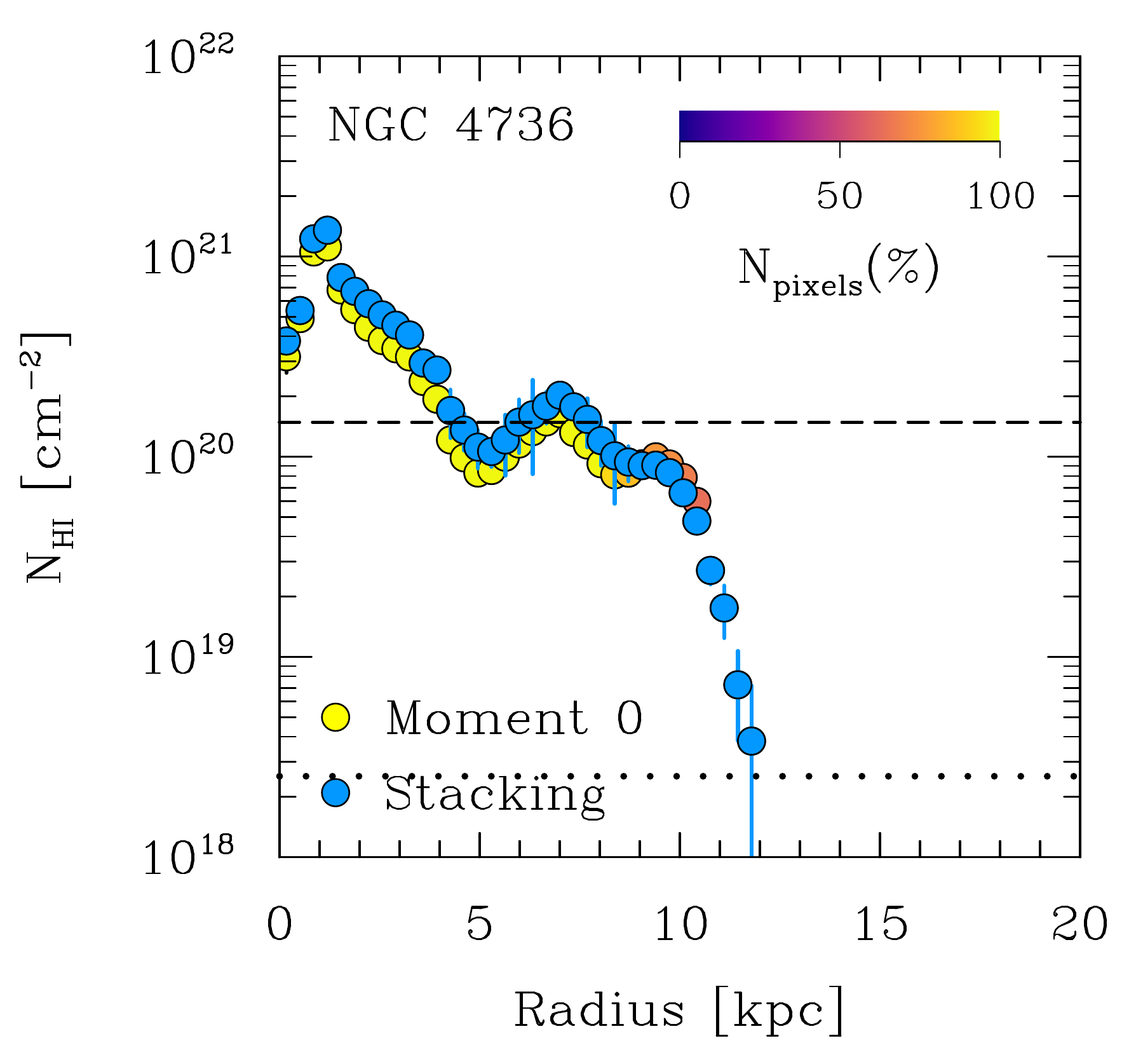}&
      \includegraphics[scale= 0.3]{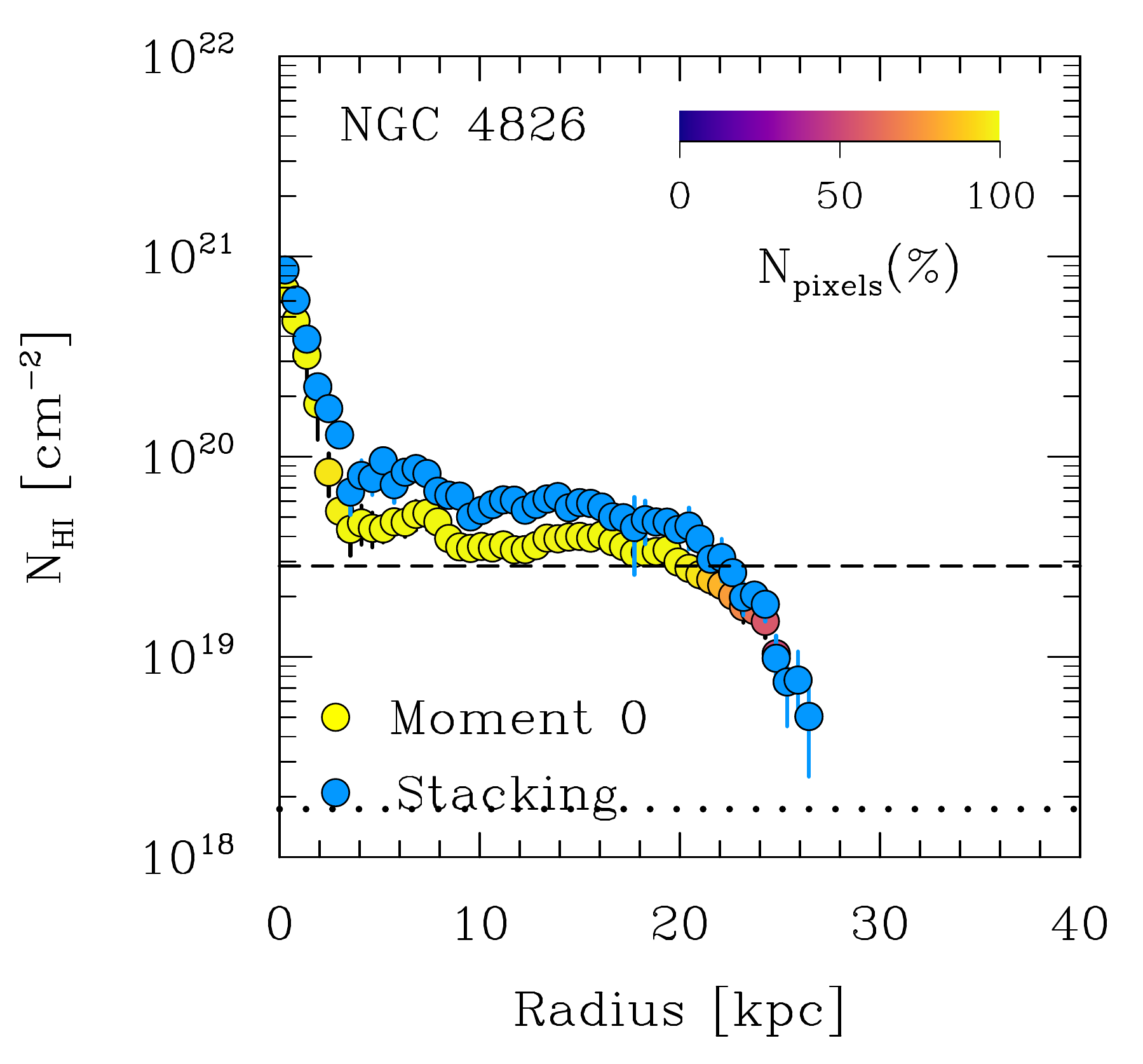}&
      \includegraphics[scale= 0.3]{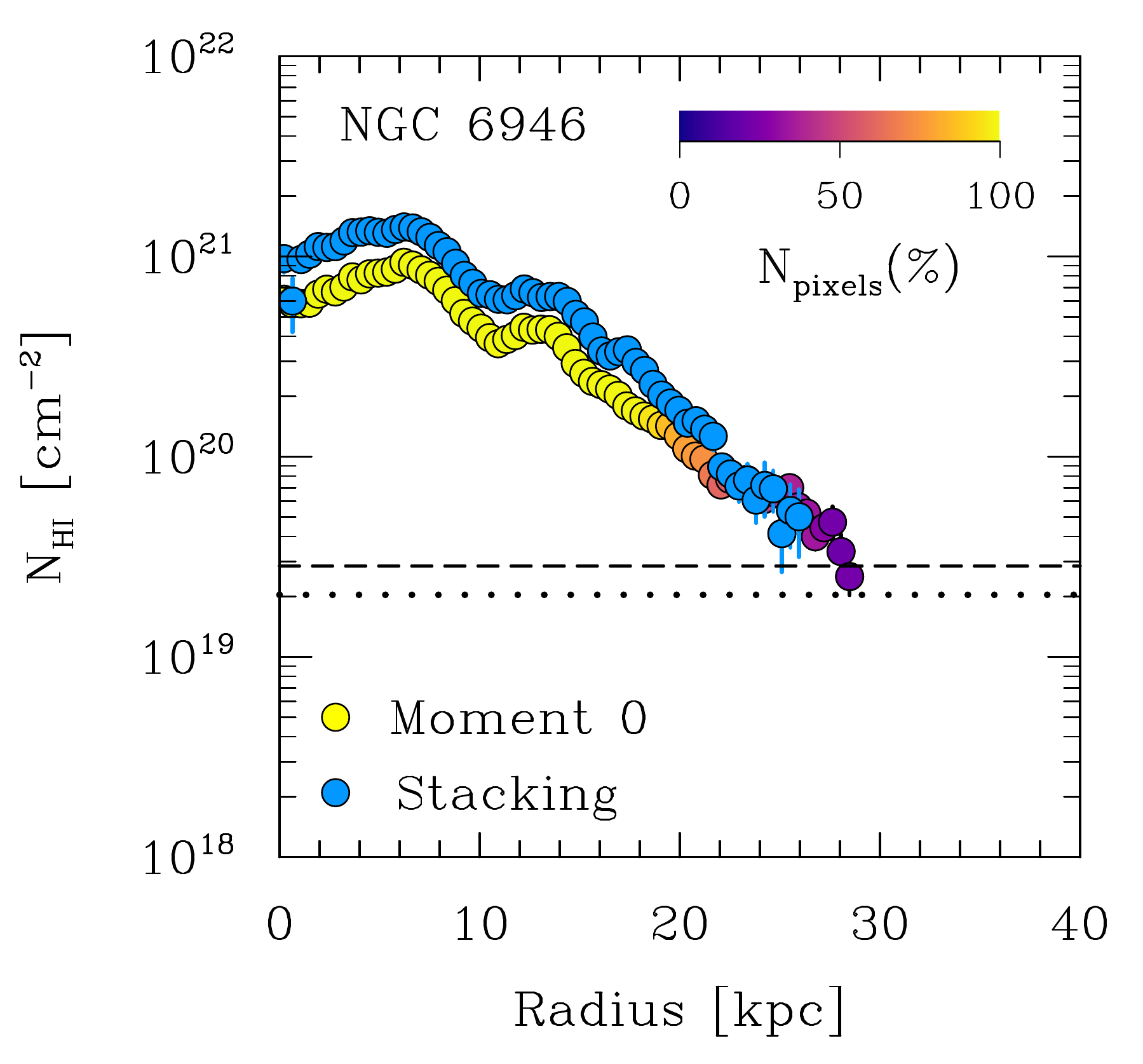}\\
      \includegraphics[scale= 0.31]{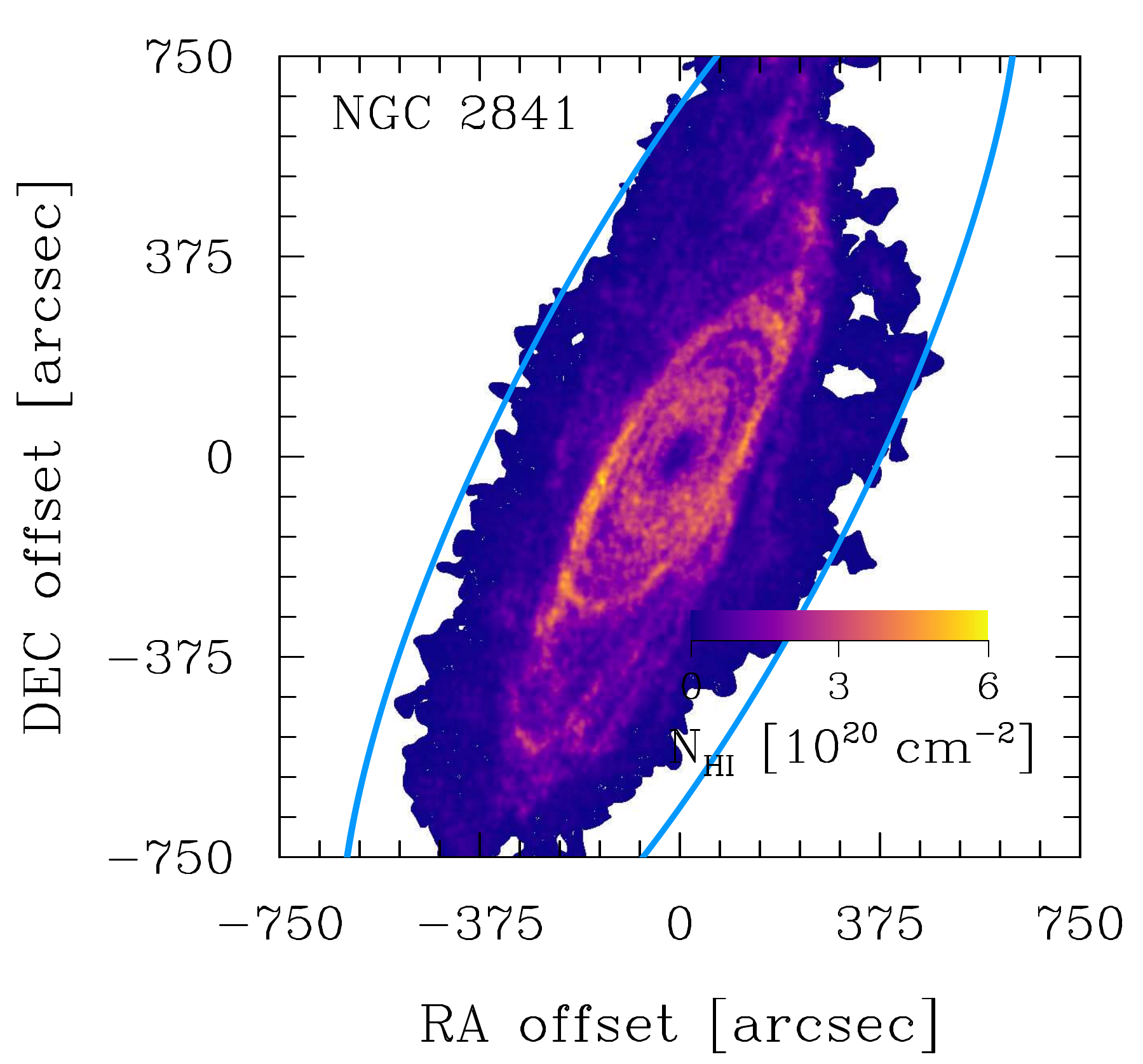} &
      \includegraphics[scale= 0.31]{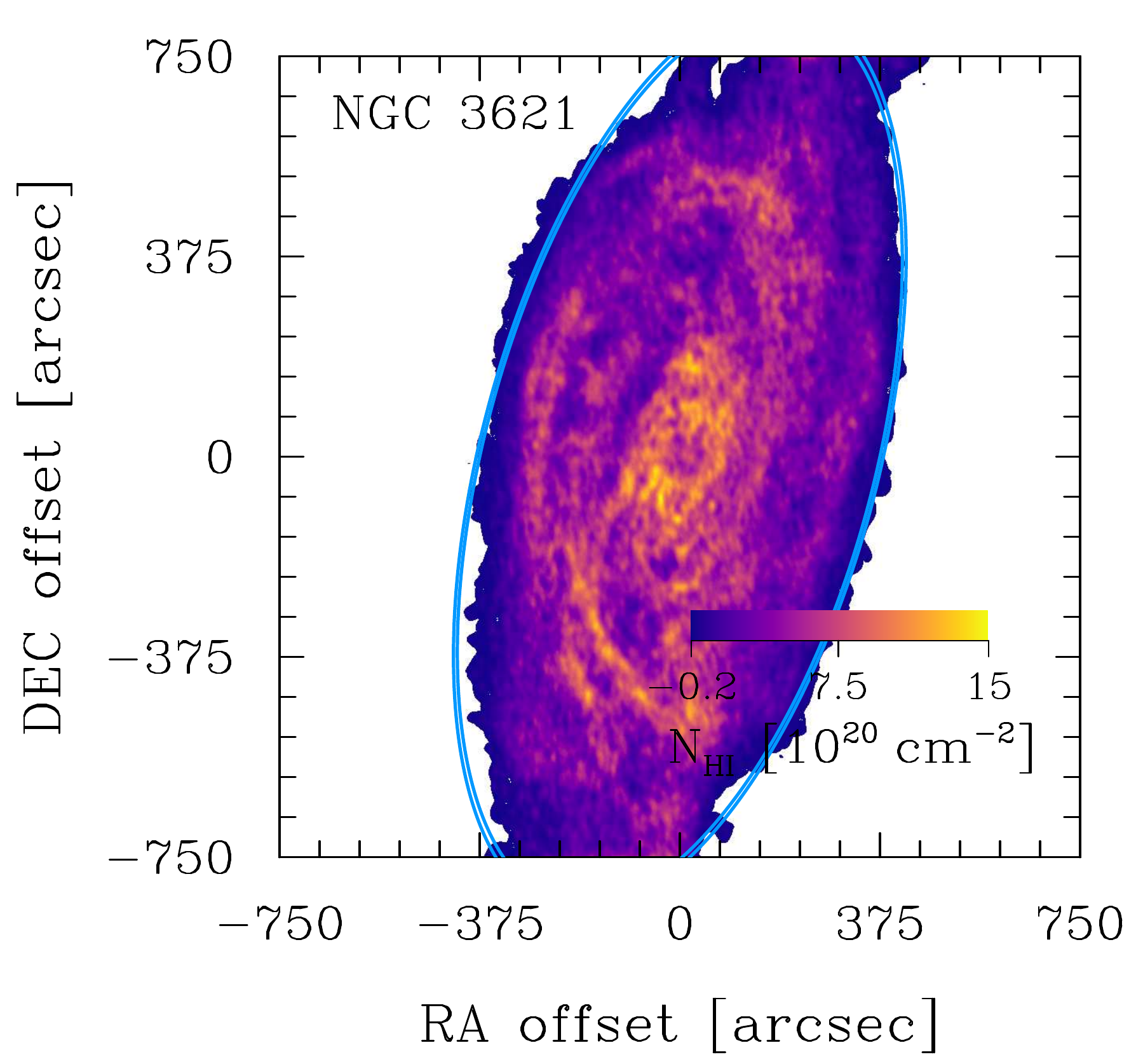}& 
      \\
     \includegraphics[scale= 0.3]{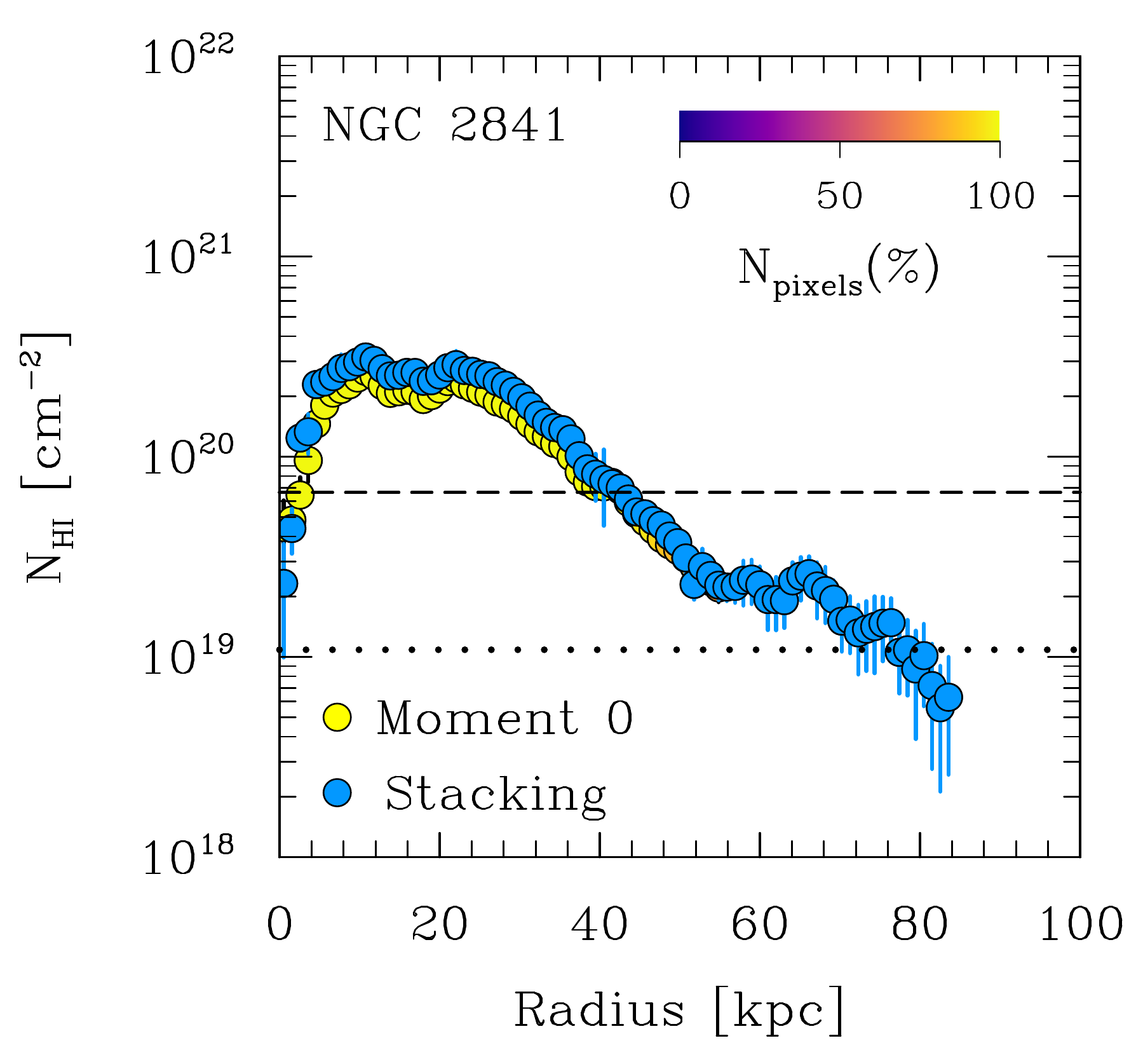}&
     \includegraphics[scale= 0.3]{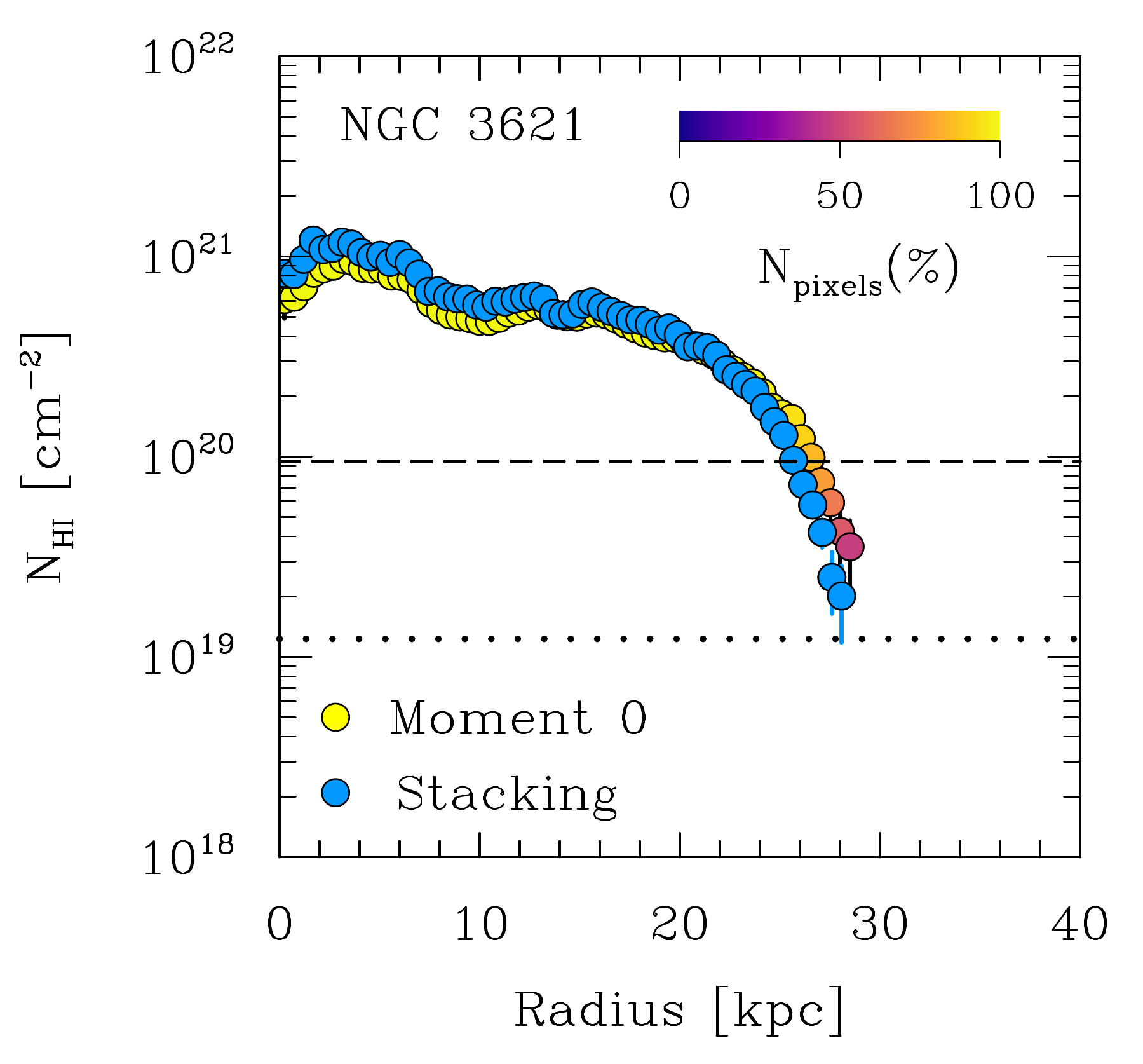}&
    \end{tabular}
\caption{Continued. \label{fig:radprof}}
\end{figure*} 
\begin{figure}
    \begin{tabular}{l}
      \includegraphics[scale= 0.32]{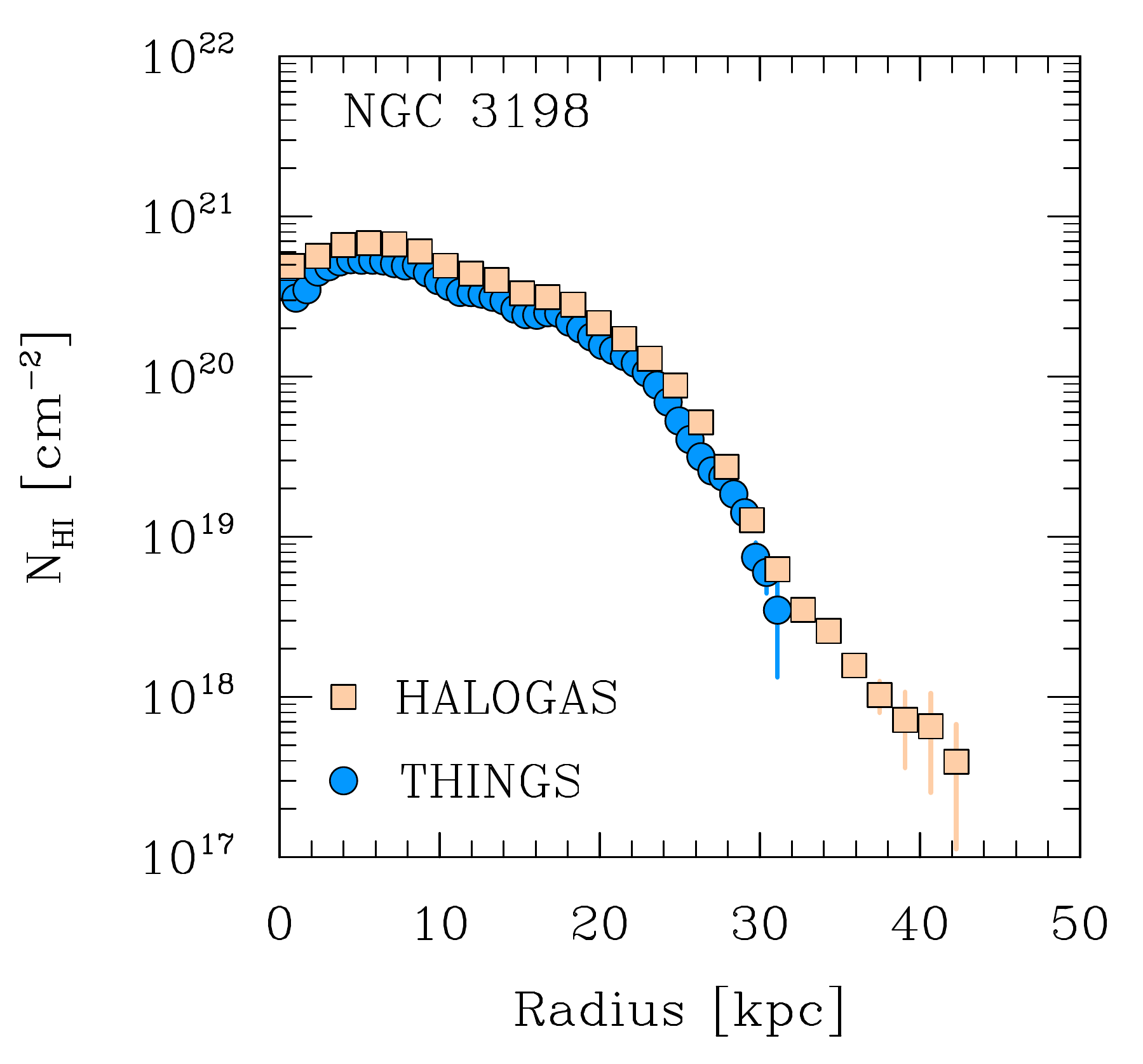}\\ 
      \includegraphics[scale= 0.32]{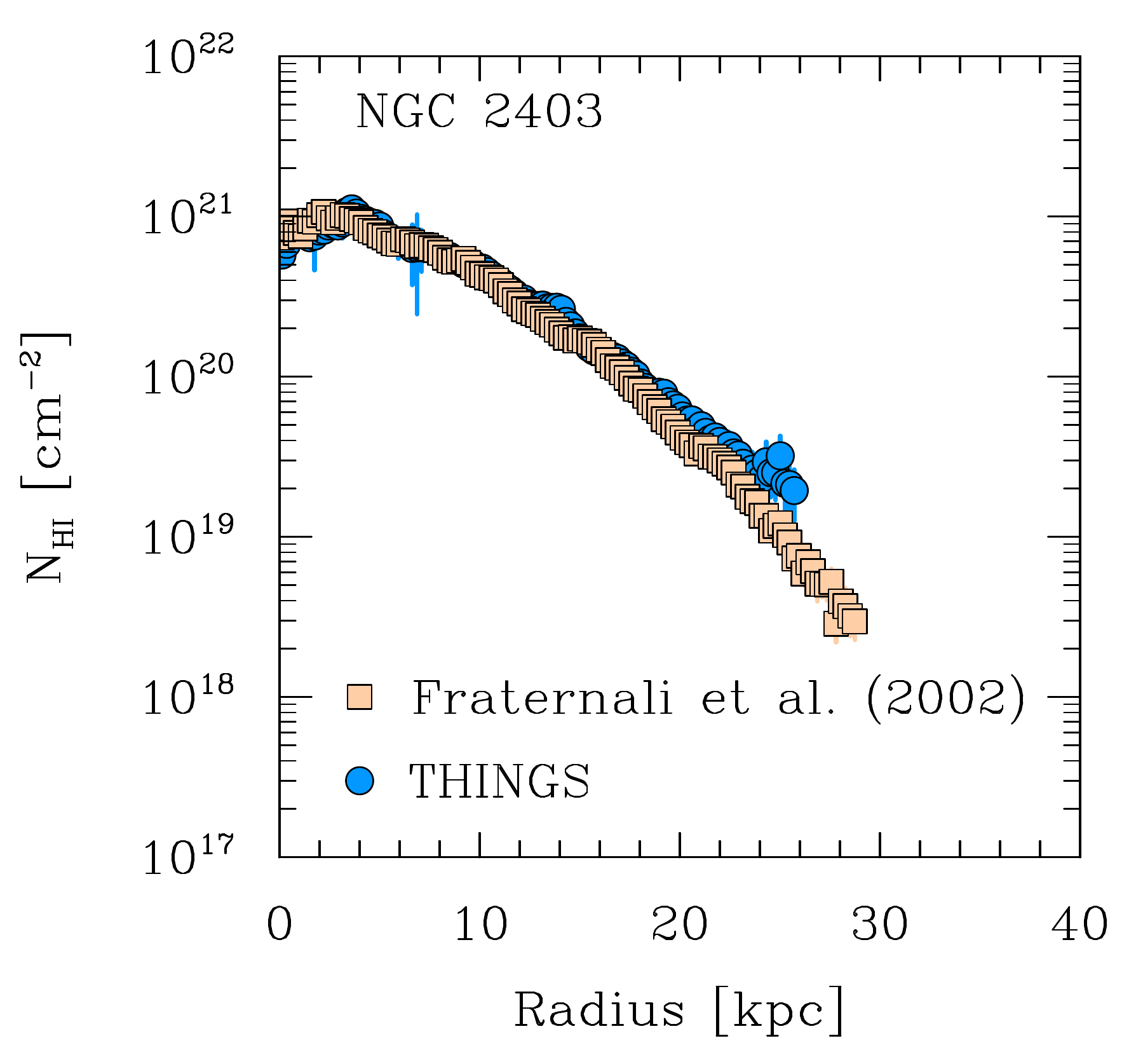} 
    \end{tabular}
    \caption{Radial H\,{\sc i} column density profiles of 
    NGC 3198 and NGC 2403, showing both the results based on the THINGS and HALOGAS data. 
    We find good agreement between the two independent methods to analyze the data.} 
    \label{fig:halo}
\end{figure}
\section{Results}\label{sec:results} 
\subsection{Examples of azimuthally stacked spectra}  
In Figure~\ref{fig:shape} we show examples of the 
azimuthally averaged stacked profiles of NGC 7331, derived using the model velocity field 
obtained from the model rotation curve shown as the blue solid line in the figure. 
We show the profiles for all sample galaxies in Figure~\ref{fig:stackedspectra} of the Appendix. 
In the region where the rotation curve is defined, we fit the rotation curve 
with a spline function. 
Beyond this, we assume a flat rotation curve. 
We derive extended model velocity fields from 
the model rotation curves and use them as inputs 
for the stacking method implemented in the 
GIPSY task SHUFFLE. 
Note that if our estimate of 
the rotation curves significantly deviates from the \textit{true} rotation curves, 
we would not be able to detect a signal.  
Figure~\ref{fig:shape} demonstrates that we are 
indeed able to detect signal beyond what is reached by 
conventional pixel-by-pixel analysis. The outermost point of the observed rotation curve 
measured by \citet{debloketal08} is at $\sim$350$\arcsec$ (using a 3 $\sigma$ cut-off) 
but we are able to still 
detect stacked H\,{\sc i} emission out to a radius of $\sim$530$\arcsec.$ 
The fairly constant width of the profiles indicates 
that our assumption of a flat rotation curve is reasonable. 
The profiles remain narrow and 
we do not see multiple peaked or broad profiles as would be expected from an incorrect 
shift of the individual velocity profiles. 
There is a shift away from `zero' velocities, though. This shift is the consequence of 
the northern side of NGC 7331 having slightly more H\,{\sc i} emission 
than the southern side. 
 
\subsection{Radial profiles for all galaxies}
In Figure~\ref{fig:radprof}, we show the radial column density profiles of the 
sample galaxies as derived using our stacking method. 
Out of the 17 galaxies studied, we reach a column density sensitivity limit of $\sim$ $10^{19}$~cm$^{-2}$ 
for 9 galaxies, and below $\sim$ $10^{19}$~cm$^{-2}$ for 8 galaxies. 
As a check, we also show the radial column density profiles 
derived from the standard THINGS 
moment-zero maps in Figure~\ref{fig:radprof}. There is still a difference between the column densities 
derived from the moment-zero maps and from the stacking method despite the deeper cleaning of our data cubes. 
This small difference is due to the stacking of residual uncleaned emission in the data cubes, leading to 
somewhat higher fluxes. This effect can be corrected for in the moment maps, but less easily so for stacked profiles. 
We refer to \citet{Ianjamasimananaetal17} for an extensive discussion. The small offset, 
however, does not affect the conclusions of this paper. 
The moment-zero values can only be reliably determined above 3 times the local 
rms noise. Note that the rms noise in a moment zero map 
is not the same as the rms noise in 
a data cube channel map since a variable number of channels, $N$, contribute 
to any given pixel in the moment map. 
We therefore calculate the rms noise in the moment maps as the rms noise in one channel 
map multiplied by the square root of the number of the contributing channels. 
This assumes channels are independent, which is a valid 
assumption as the THINGS data are Hanning-smoothed \citep[see][]{walteretal08}.
We indicate the (moment-zero) column density sensitivity corresponding to 3 times the 
rms noise in Figure~\ref{fig:radprof}. We also show in the same figure the 3 times 
rms noise column density sensitivity of the azimuthally stacked profiles. On average, 
the stacking method offers an order of magnitude 
sensitivity improvement compared to the traditional moment-zero analysis. 
Close inspection of Figure~\ref{fig:radprof} shows that the stacked 
profiles do not always reach the S/N = 3 limit. In these cases 
we are limited by the presence of low-level artefacts in the 
data (mostly due to deconvolution residuals) that prevent us 
from reaching lower limits. Nevertheless, the fact that with the 
current data set stacking already yields limits much lower than 
traditional moment-zero limits, clearly shows the promise of the method. 

The most surprising finding from Figure~\ref{fig:radprof} is that most galaxies show no evidence of a break. 
The profiles are mostly flat in the inner disk and then decline smoothly down to the sensitivity limit 
of our (stacked) data. 
Our assumption of a flat rotation curve 
 cannot be the reason for this absence of an H\,{\sc i} break. 
 Any incorrect estimates of the rotation curves would 
 result in stacked profiles that are systematically 
 lower in amplitude but broader in width and this 
 would not remove the signature of a break. 
 A major velocity offset between real and assumed velocities 
 would result in non-detections, which would mimic an 
 artificially strong break in the profiles. 
 
 A possible cause of the absence of a break is that the H\,{\sc i} 
 surface density distribution is azimuthally asymmetric. Our sample galaxies 
 exhibit this asymmetry. These include, among others, NGC 7331, NGC 3198, NGC 2903, NGC 3621. 
 We also note that the break in M33 observed by \citet{corbellietal89} is based on a single radial slice 
 along the major axis of the galaxy. In addition, the break in NGC 3198 observed by 
 \citet {vangorkom93} \citep[see also][]{healdetal16, kametal17} is seen in only one side of the galaxy.   
 Thus the stacking method may smooth out any potential break as we are averaging 
 over the full azimuthal angle. To explore this possibility, we also divide the galaxies into 
 30 degree wide ``pie-wedge-shaped'' sectors and do the stacking inside annular sectors instead of the 
 full azimuthal annuli. The results are shown in Figure~\ref{fig:sectors} of the Appendix. Although the 
 radial profiles are steeper in some sectors than in others, the profiles are smooth without a clear break. 
 We thus conclude that the absence of a break in the (full) azimuthally averaged profiles are not caused 
 by the presence of an asymmetric H\,{\sc i} distribution. 
\subsection{Exceptions}
 A close inspection of the individual galaxies reveals that there are a 
 few exceptions to the conclusion that radial breaks 
 are not present in this sample. These are  discussed in the following.
 The radial profiles of NGC 4736 and NGC 3621 decline steeply 
 around $10^{20}$~cm$^{-2}$ and for NGC 4826 at $\sim5\times10^{19}$~cm$^{-2}$. 
 These three galaxies, however, have some structural peculiarities. 
 NGC 4736 has a ring-like H\,{\sc i} morphology which manifests 
 itself as bumps in the H\,{\sc i} column density 
 profiles shown in Figure~\ref{fig:radprof}. Earlier 
 lower resolution H\,{\sc i} observations 
 by \citet{bosmaetal77} also revealed this pattern. The authors 
 classified the galaxies as 
 consisting of 5 different zones in which a faint outer 
 ring surrounds the main body of the galaxy. 
 The H\,{\sc i} edge around 7 kpc corresponds to the transition from the main disk of the 
 galaxy and the faint outer rings identified by \citet{bosmaetal77}. 
 The origin of the ring 
 is not clear but could be related to a past accretion event. 
 Due to the high column-density 
 at which it occurs ($\sim10^{20}$~cm$^{-2}$), 
 it is unlikely that the H\,{\sc i} break is related to the breaks driven by ionization. 
 NGC 3621, on the other hand, has an extended H\,{\sc i} distribution. 
 The integrated H\,{\sc i} intensity map shows a tail or stream of 
 gas extending along the major axis, 
 which could be the result of past interactions or signatures of gas accretion. 
 The break around $10^{20}$~cm$^{-2}$ for this galaxy therefore corresponds to the
 transition between the main disk of NGC 3621 and the stream or tail (See also Figure~\ref{fig:sectors}). 
 The profile of NGC 4826 declines steeply around 5$\times10^{19}$~cm$^{-2}$.
  However, it is unlikely that we are tracing effects due to ionization 
 by extragalactic photons. Kinematical studies 
 by \citet{braunetal94} and by \citet{rixetal95} 
 revealed that the inner disk is counter-rotating with respect to the outer disk. 
 As reported by \citet{rixetal95}, this counter-rotation 
 may indicate that the outer disk of NGC 4826 has been built from the merging of 
 gas rich dwarf galaxies or from the infall of H\,{\sc i} gas with a retrograde orbit. 
 Regardless of the exact cause, it is conceivable that we are tracing the 
 relics of a merging event or episodic gas 
 infall rather than the effects of photoionization. In the 
 discussion that follows we do not 
 include the three galaxies discussed in this section.          
\subsection{Comparison with HALOGAS}
For two galaxies we have both THINGS and HALOGAS data available. 
This provides us with the opportunity to check for possible systematics. 
In Figure~\ref{fig:halo}, we overplot the azimuthally stacked THINGS 
and HALOGAS H\,{\sc i} 
column density profiles for NGC 2403 and NGC 3198. 
In both cases, we find excellent agreement in the measurements 
(where they overlap). Also, neither galaxies column density 
show any indication for a break in H\,{\sc i}, but feature rather smooth profiles.
\subsection{Identification of any possible breaks and comparison with models}
\begin{figure*}
    \begin{tabular}{l l}
      \includegraphics[scale= 0.48]{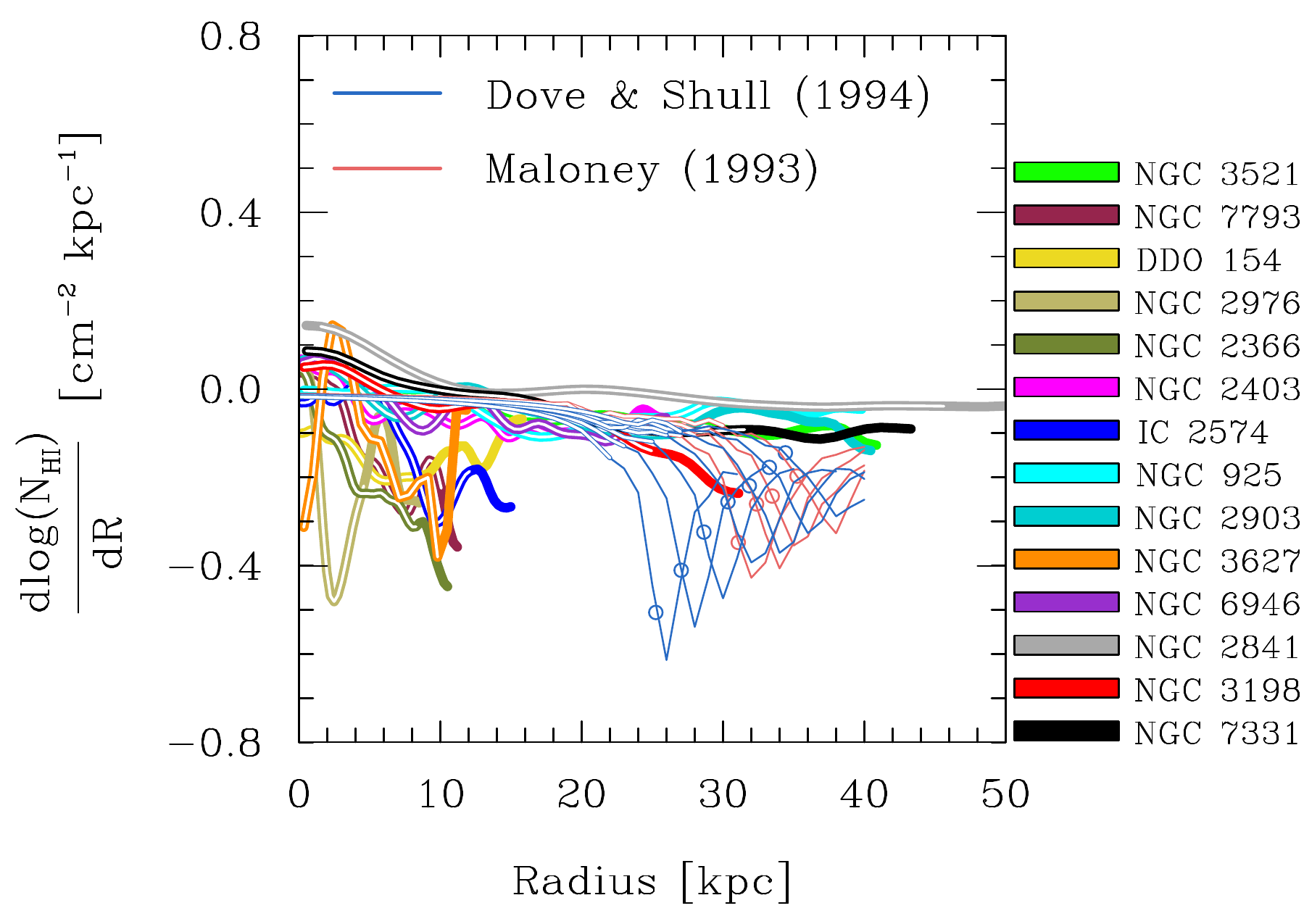} &
      \includegraphics[scale=0.5]{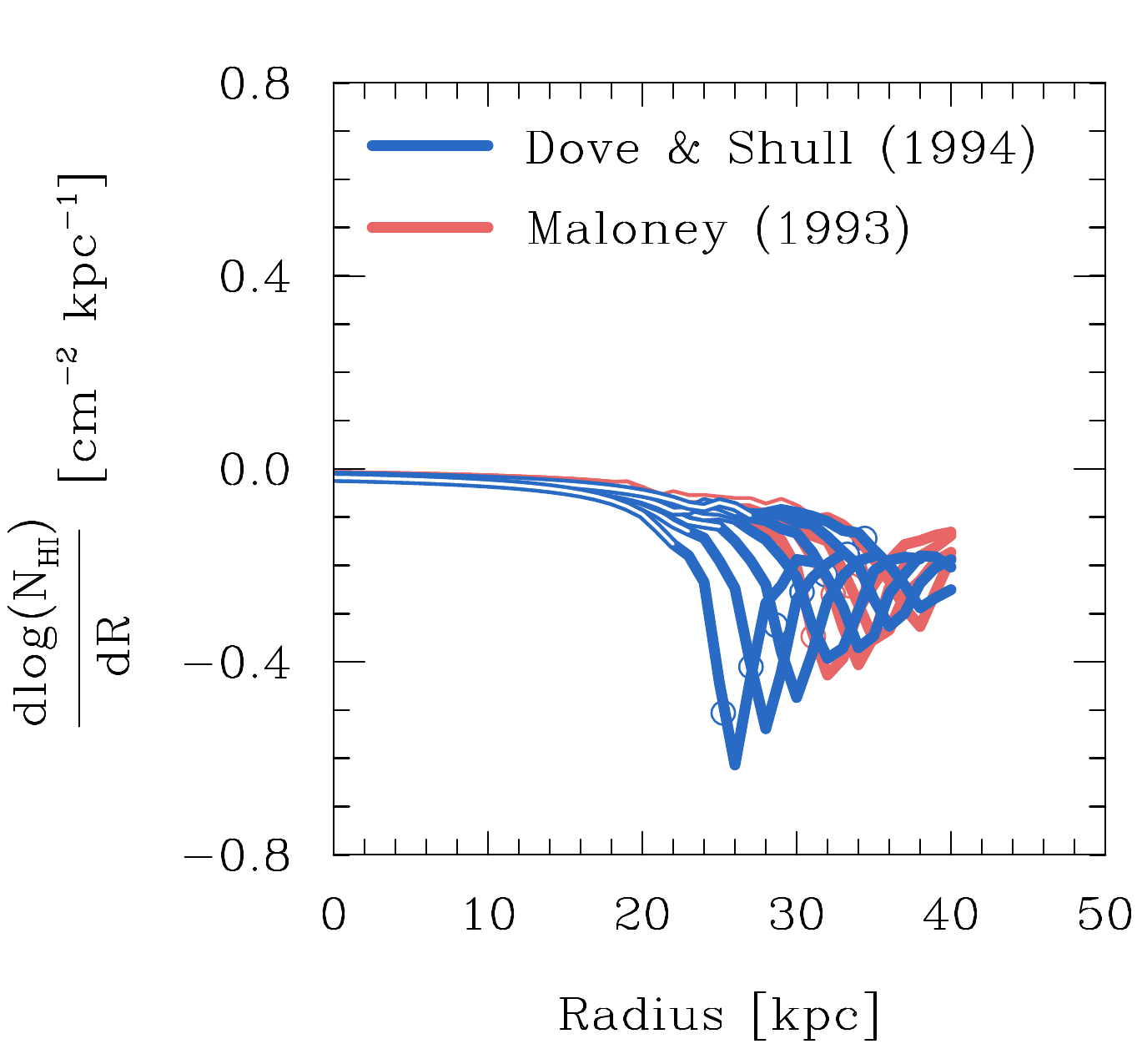}
    \end{tabular}
    \caption{Left: first logarithmic derivatives of the radial column density profiles of 
    our sample galaxies (thick solid lines) overplotted on the 
    photoionization models (thin solid lines) 
    by \citet{maloney93} and by \citet{doveshull94}. 
    Right: The photoionization models shown as thin lines in the left panel are 
    re-plotted for better visualization. The open lines indicate 
    column densities above 5$\times10^{19}$~cm$^{-2}$. 
    The open circle symbols indicate at which radius the 
    column densities drop below $10^{19}$~cm$^{-2}$.} 
    \label{fig:allder1} 
\end{figure*}

To better identify possible breaks in the azimuthally stacked profiles, we measure 
the change in slope of the profiles by calculating their first logarithmic derivatives. 
We also use the first logarithmic derivatives to gauge whether any break seen 
in the data resembles that of the photoionization models of \citet{maloney93} 
and \citet{doveshull94}. The derivative will be zero for a constant profile, 
negative if it is decreasing and positive if the 
profile is increasing. 
As illustrated in Figure \ref{fig:allder1}, 
the change in slope is close 
to zero for many galaxies and does not resemble that of the photoionization models. 
Note that the models 
by \citet{maloney93} and by \citet{doveshull94} only show 
$\rm{N_{H\,{\scriptscriptstyle I}}(R)}$ for 
R $\geq$ 20 kpc. Purely for illustrative purposes, we thus extrapolated 
the models to R = 0 kpc. 
The clear difference between the observed and model profiles shows that 
with our present data, we do not find convincing evidence that would suggest the 
presence of a break caused by photoionization 
in our column density profiles.        

\begin{figure*}
    \begin{tabular}{l l}
      \includegraphics[scale= 0.376]{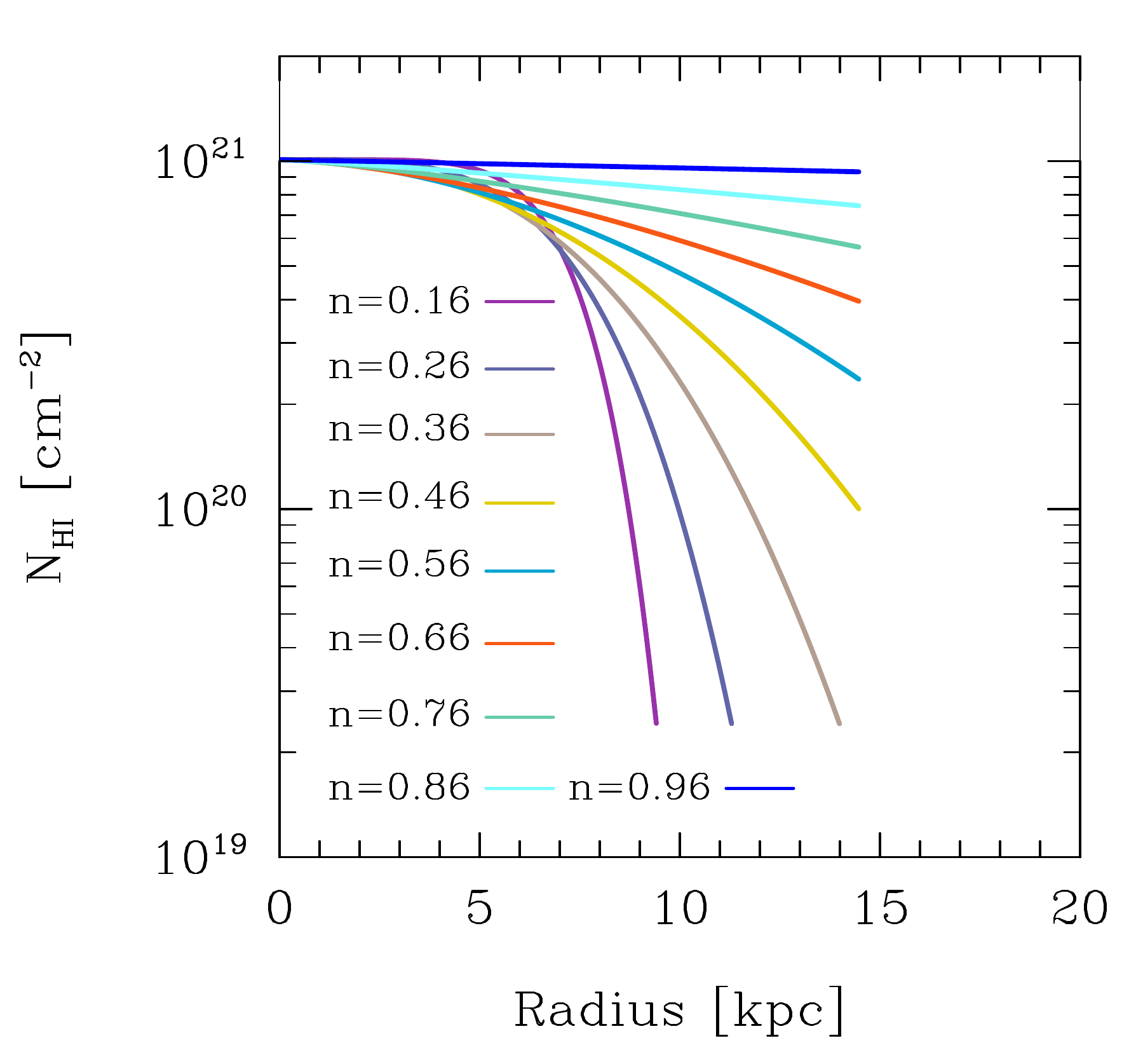} &
      \includegraphics[scale= 0.376]{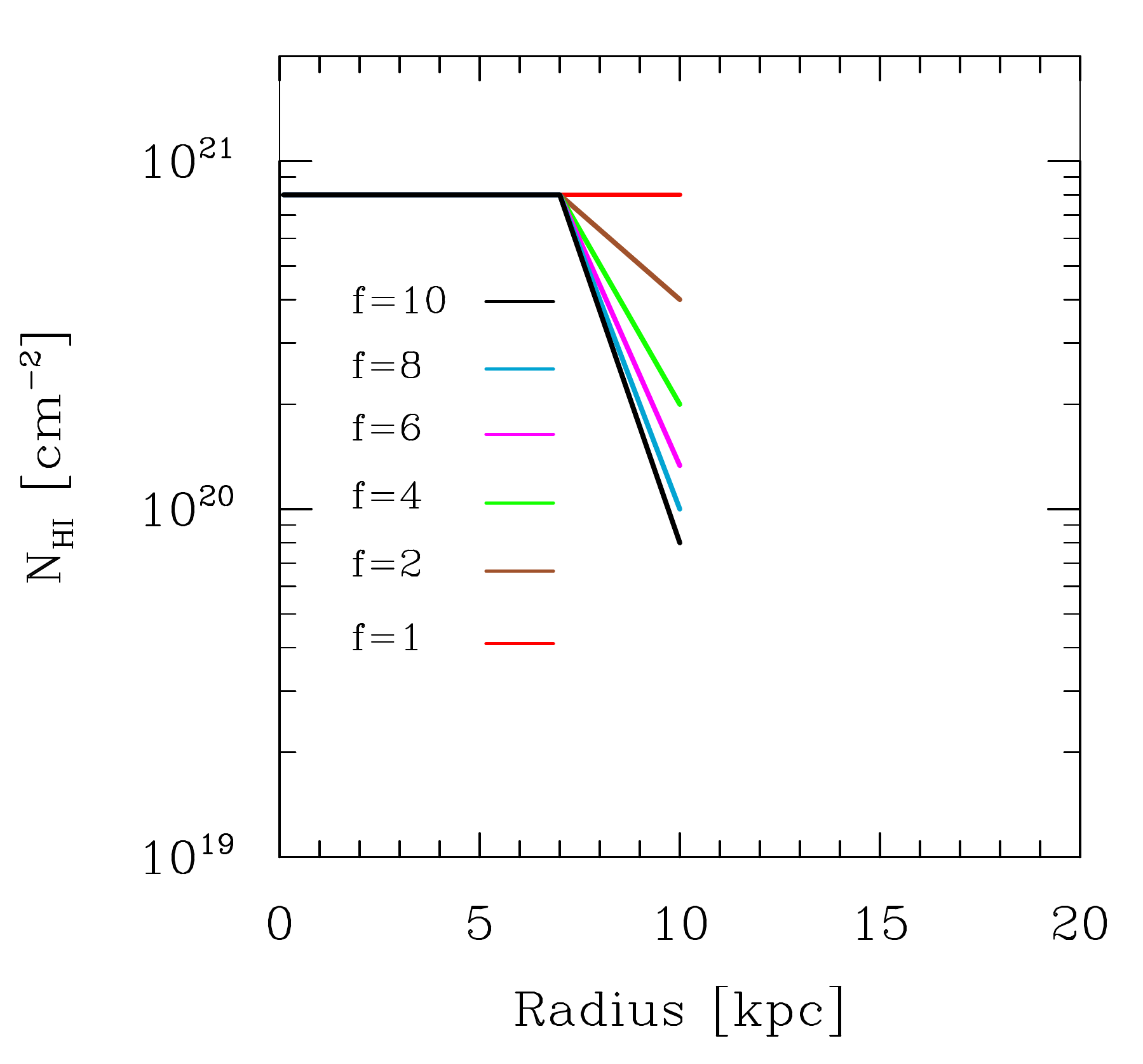} \\
      \includegraphics[scale= 0.376]{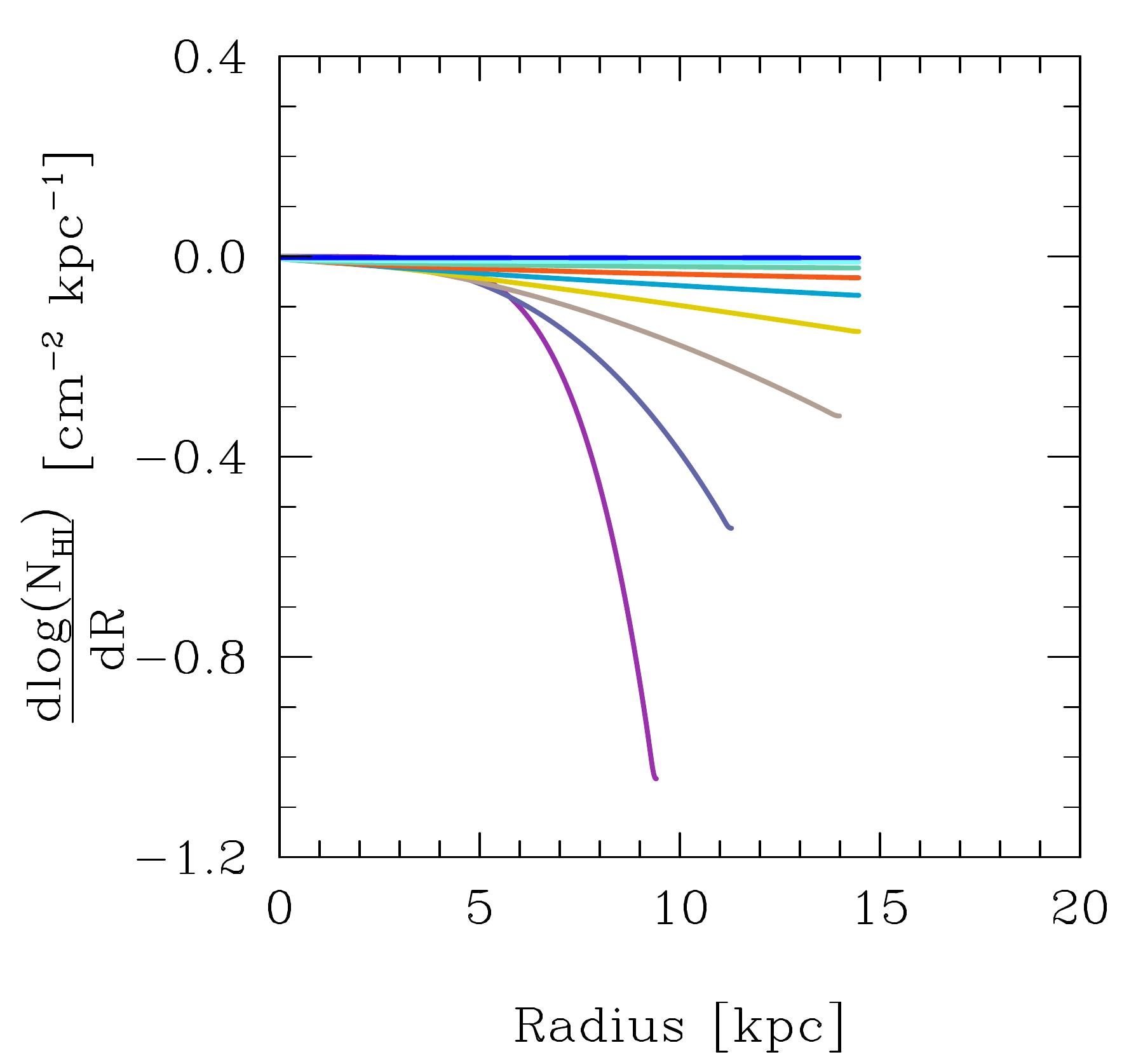} &
      \includegraphics[scale= 0.376]{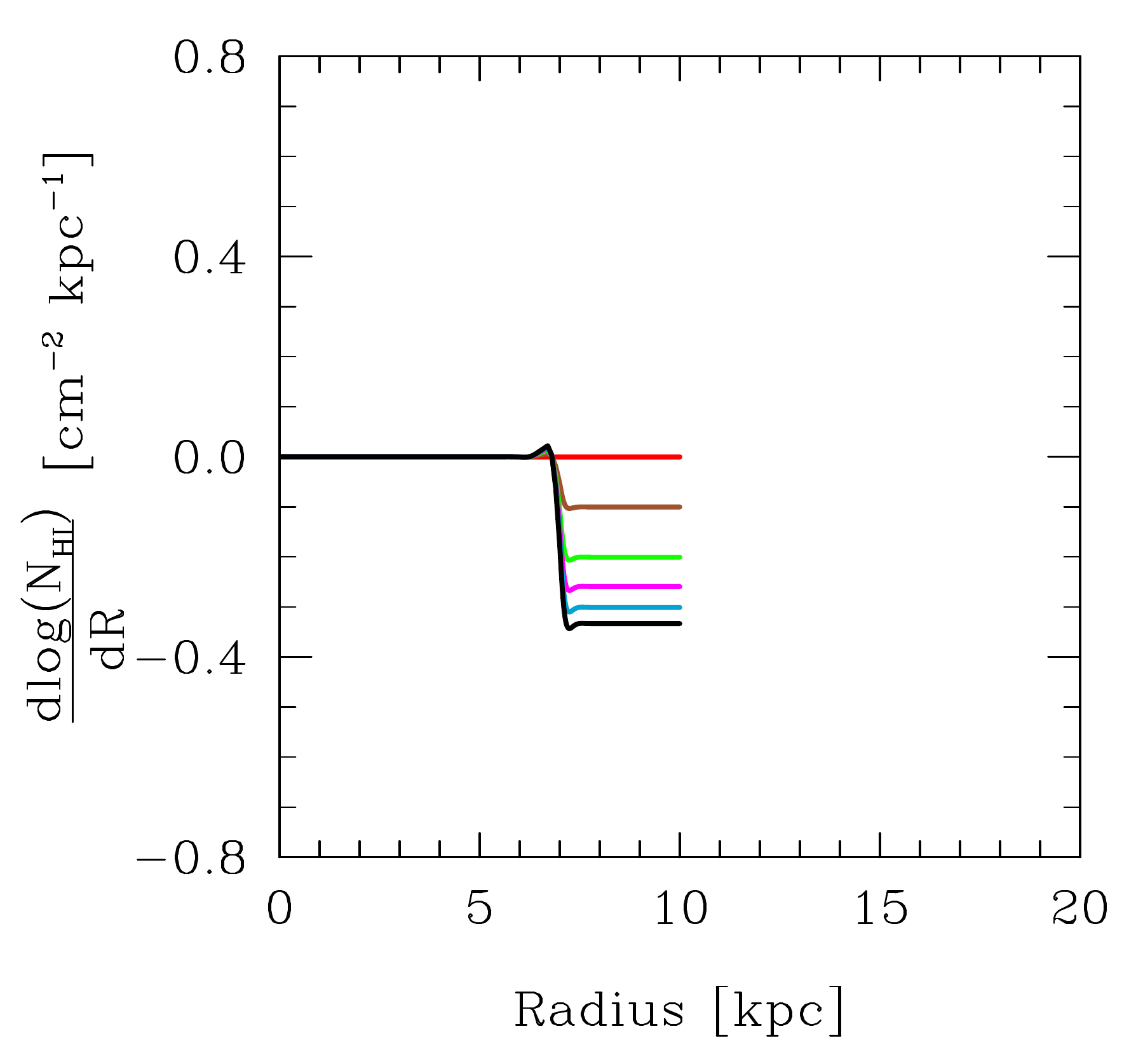} 
    \end{tabular}
    \caption{Solid lines: synthetic column density profiles (top panels) 
    and their respective first logarithmic derivatives (bottom panels).} 
    \label{fig:derivative}
\end{figure*}

To quantify the amount of change in slope, 
we fit the profiles using S\'{e}rsic profiles \citep{sersic} 
with different slope indexes, $n$. 
We here use the formalism introduced by \citet{Portas} who, for $n < 1 $, 
defines the S\'{e}rsic profile in terms of a break radius, $R_{b}$, instead of the 
half-light radius that is mostly used in the literature. 
Thus, following \citet{Portas}, we define the S\'{e}rsic profile as: \begin{equation} 
I(R) = I(0)\,\exp\left\{-(1-n)\left(\dfrac{R}{R_{b}}\right)^{1/n}\right\}, \end{equation} 
where I(0) is the central intensity. We vary the slope 
indexes, $n$, but fix $I(0)$ and $R_{b}$ and 
derive the first logarithmic derivatives of the resulting profiles. 
As seen in Figure \ref{fig:derivative}, a Sersic index of 0.46 roughly corresponds to a 
derivative value of $-0.2$ dex cm$^{-2}$ kpc$^{-1}$. 
We create a second set of synthetic profiles consisting of a 
straight line without a break and straight lines that break at a fixed radius, $R_{b}$, 
with varying slopes. 
We illustrate those results in Figure \ref{fig:derivative}. 
Here an order of magnitude decrease within 3 kpc 
(corresponding to f=10 in the Figure) corresponds to a 
derivative value change of $-0.3$ dex cm$^{-2}$ kpc$^{-1}$.
Note that the values of the derivatives of the real 
data are typically $\gtrsim$ $-0.2$ dex cm$^{-2}$ kpc$^{-1}$. 
The photoionization 
models by \citet{maloney93} and \citet{doveshull94}, on the other hand, have derivative 
values $\lesssim$ $-0.3$ dex cm$^{-2}$ kpc$^{-1}$. 
We conclude that, down to our sensitivity limit, 
a radial column density break is not a common feature of disk galaxies. 
More high sensitivity H\,{\sc i} observations are needed to investigate 
the H\,{\sc i} distribution at even lower column density values.  
\subsection{NGC 3198: Comparison with \citet{vangorkom93}}
\begin{figure*}
    \begin{tabular}{l}
      \includegraphics[scale= 0.5]{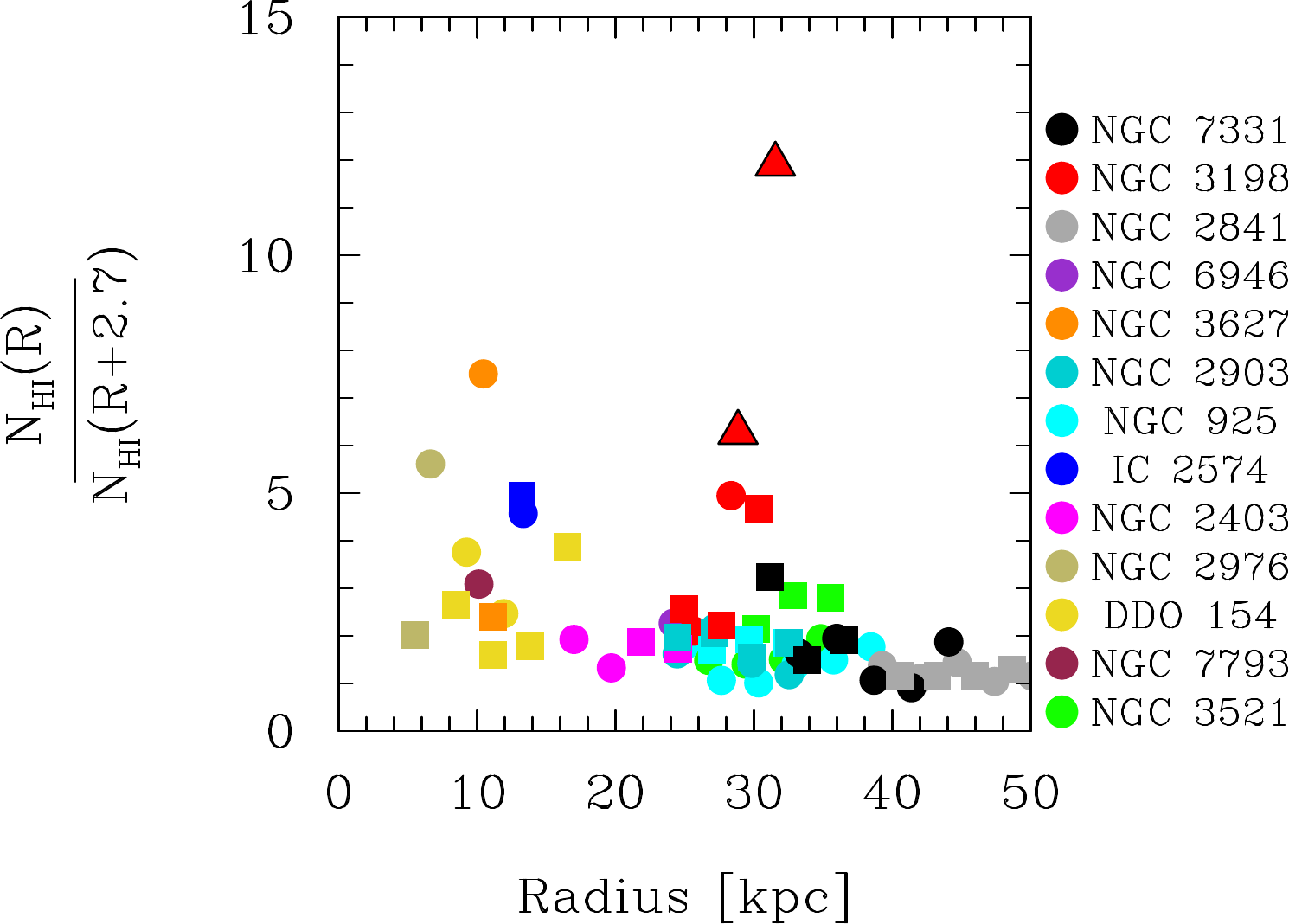} 
    \end{tabular}
    \caption{Slopes of the approaching and receding radial column density profiles 
    within 2.7 kpc (see text for motivation). Only the outer-radii 
    corresponding to $\rm{N_{H\,{\scriptscriptstyle I}}(R)} 
    \leq 10^{20}~cm^{-2}$ are shown. Circle symbols: approaching 
    sides; square symbols: receding sides. Triangle symbols: slopes derived from the 
    \citet{vangorkom93} data (approaching side).} 
    \label{fig:slopego}
\end{figure*}

NGC 3198 is a galaxy for which we do not find a global 
break in neither the THINGS nor the HALOGAS data. However, \citet{vangorkom93} reported a 
break toward the approaching 
side of this galaxy. She showed that the 
column density of NGC 3198 decreased by an order of 
magnitude within a synthesized beam of 2.7 kpc. 
When we redo the analysis for the approaching 
side of NGC 3198, we did not find a noticeable break as reported by 
\citet{vangorkom93}. This prompted us to redo the analysis 
for the approaching and receding sides of all galaxies. 
To allow a 
better comparison of our data with that of \citet{vangorkom93}, 
we divide the radial profiles of the approaching and receding sides of 
the galaxies in 2.7 kpc bins, thus using the same definition for the slope as 
\citet{vangorkom93}. We then calculate the slopes of 
the outer $\rm{N_{H\,{\scriptscriptstyle I}}(R)} \leq 10^{20}~cm^{-2}$ as 
$\rm{N_{H\,{\scriptscriptstyle I}}(R)}/\rm{N_{H\,{\scriptscriptstyle I}}(R+2.7)}$ (see Figure~\ref{fig:slopego}). 
From this analysis, we do not find evidence that an H\,{\sc i} break associated with ionization is a common feature in galaxies. 
\section{Summary and conclusion}\label{sec:summary}    
We studied the radial H\,{\sc i} column density profiles of 17 nearby galaxies. 
We used data from the THINGS and, for two galaxies, the HALOGAS survey.
To push the sensitivity limit of our data to deeper levels and trace the H\,{\sc i} 
distribution to previously unexplored radii, we stacked individual profiles azimuthally 
using a stacking technique. 
Model velocity fields derived from a simple (flat) 
extrapolation of the rotation curves were used 
to stack individual profiles at large radii. In general the radial column density profiles 
are flat in the inner disk and then smoothly decline down 
to the sensitivity limit of the data. 
Thus, our present 
data do not confirm the prediction of photoionization models that an H\,{\sc i} 
break at a column density of $\sim 5\times 10^{19}$~cm$^{-2}$ is a 
common feature of H\,{\sc i} disks. The absence of an H\,{\sc i} 
break may indicate that ionization by extragalactic photons is not the limiting 
factor of the extent of the H\,{\sc i} disk. Thus, the outskirts of the H\,{\sc i} 
disk may instead correspond to the transition to a low column density gas accreted from 
the cosmic web at later evolutionary stages of disk formation. 
However, in order to map the morphology and kinematics of 
the low column density outskirts, 
we require the next generation of radio telescopes \citep[e.g., MeerKAT,][]{meerblok}. 
Until then we can only start to trace the average properties at the outermost radii, 
as done in this paper.
\acknowledgments
The authors thank the anonymous referee for excellent comments that helped improve the
presentation of this paper. 
R.I. acknowledges funding from the Alexander von Humboldt foundation 
through the Georg Forster Research Fellowship 
programme. 
EB acknowledges support from the UK Science and 
Technology Facilities Council [grant number ST/M001008/1].  
We thank the HALOGAS team for providing us with the NGC 3198 and NGC 2403 data.  
\appendix 
\section{azimuthally stacked spectra}
In this Section, we show the model rotation curves used to derive the radial surface density profiles of our sample galaxies
(see Figure~\ref{fig:stackedspectra}). 
We also show examples of the azimuthally averaged stacked spectra at different radius (see Section~\ref{sec:methodology} for details). 
\begin{figure*}
    \begin{tabular}{l l}
      \includegraphics[scale= 0.376]{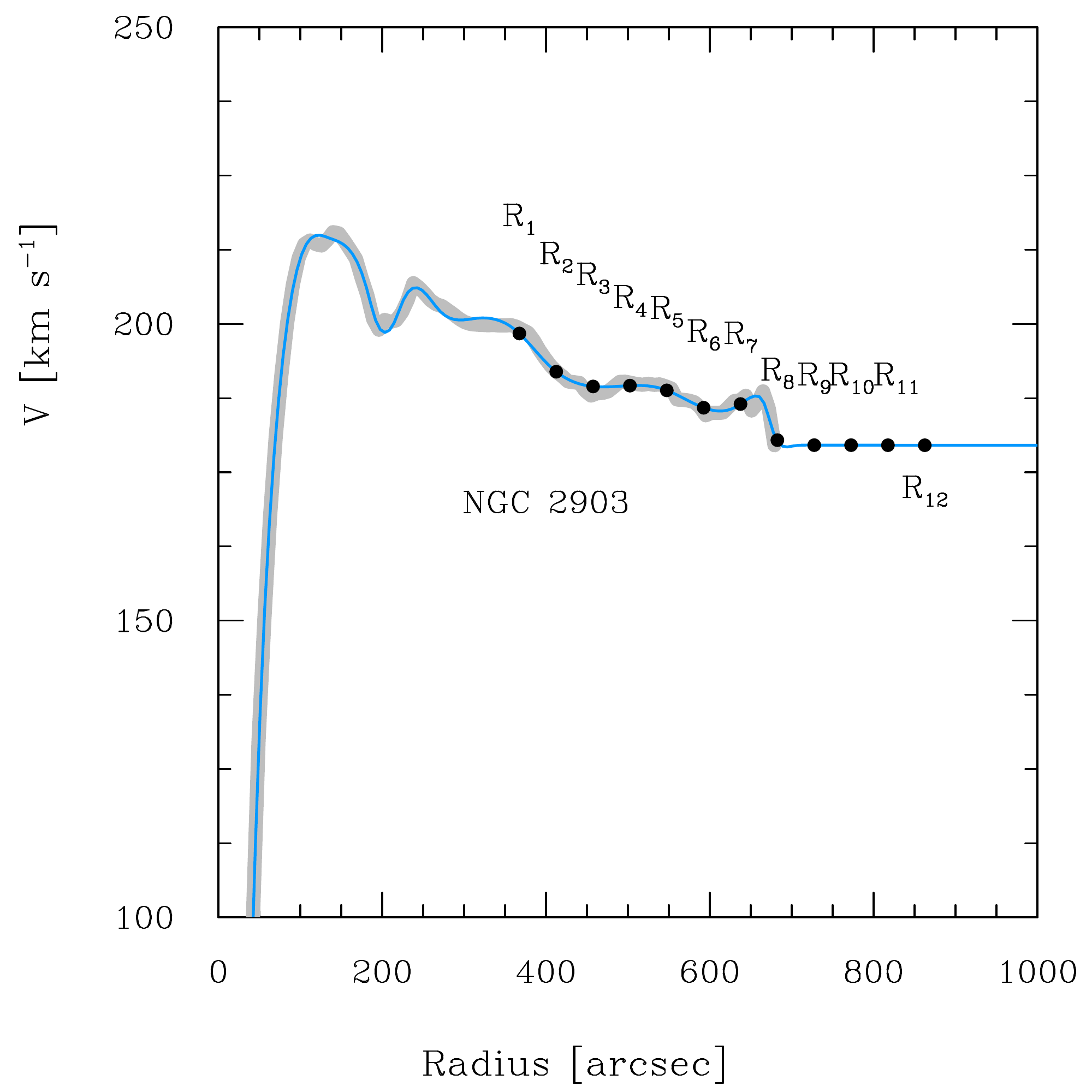} &
      \vspace{0.5cm}
      \includegraphics[scale= 0.376]{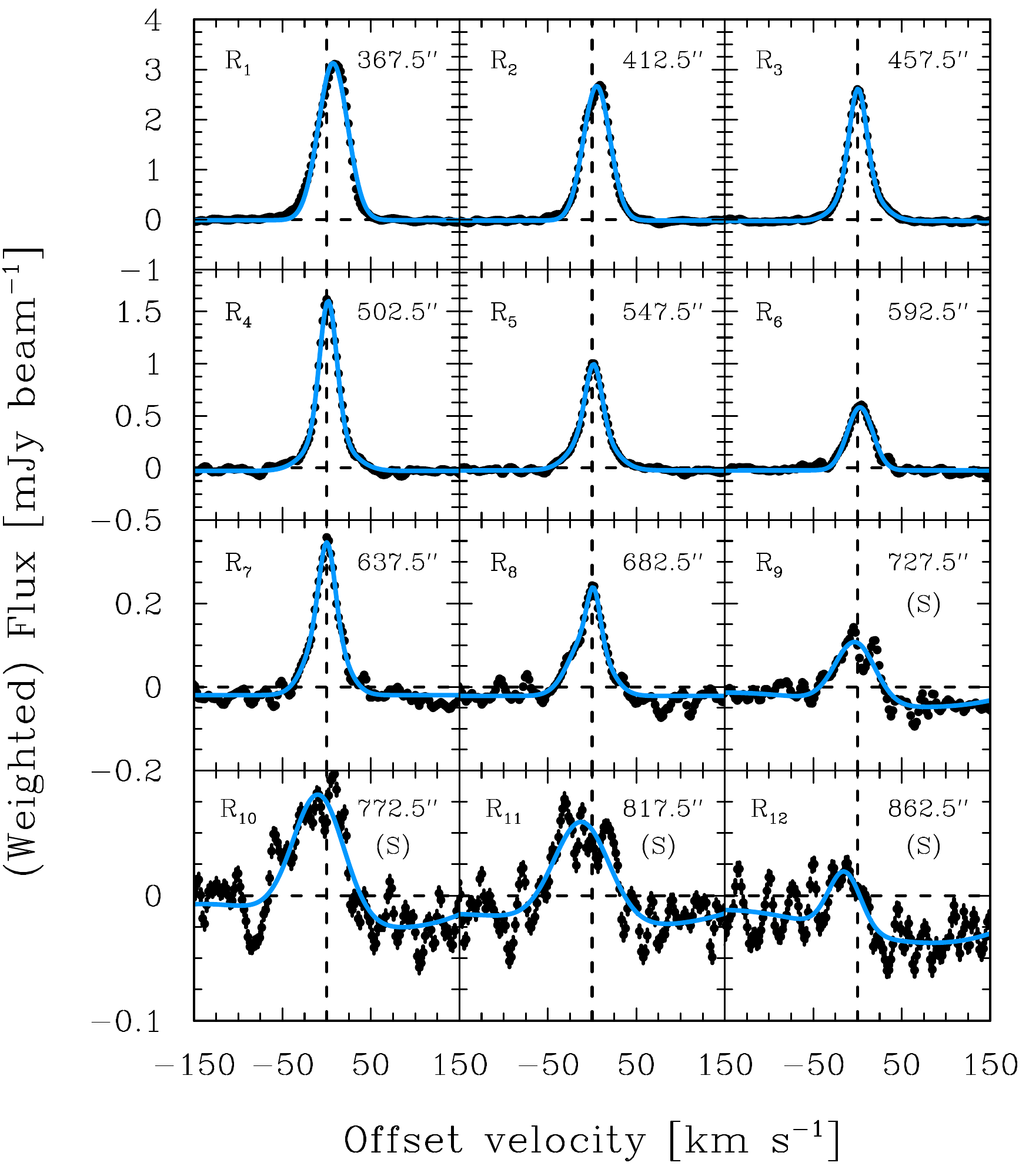} \\
      \includegraphics[scale= 0.376]{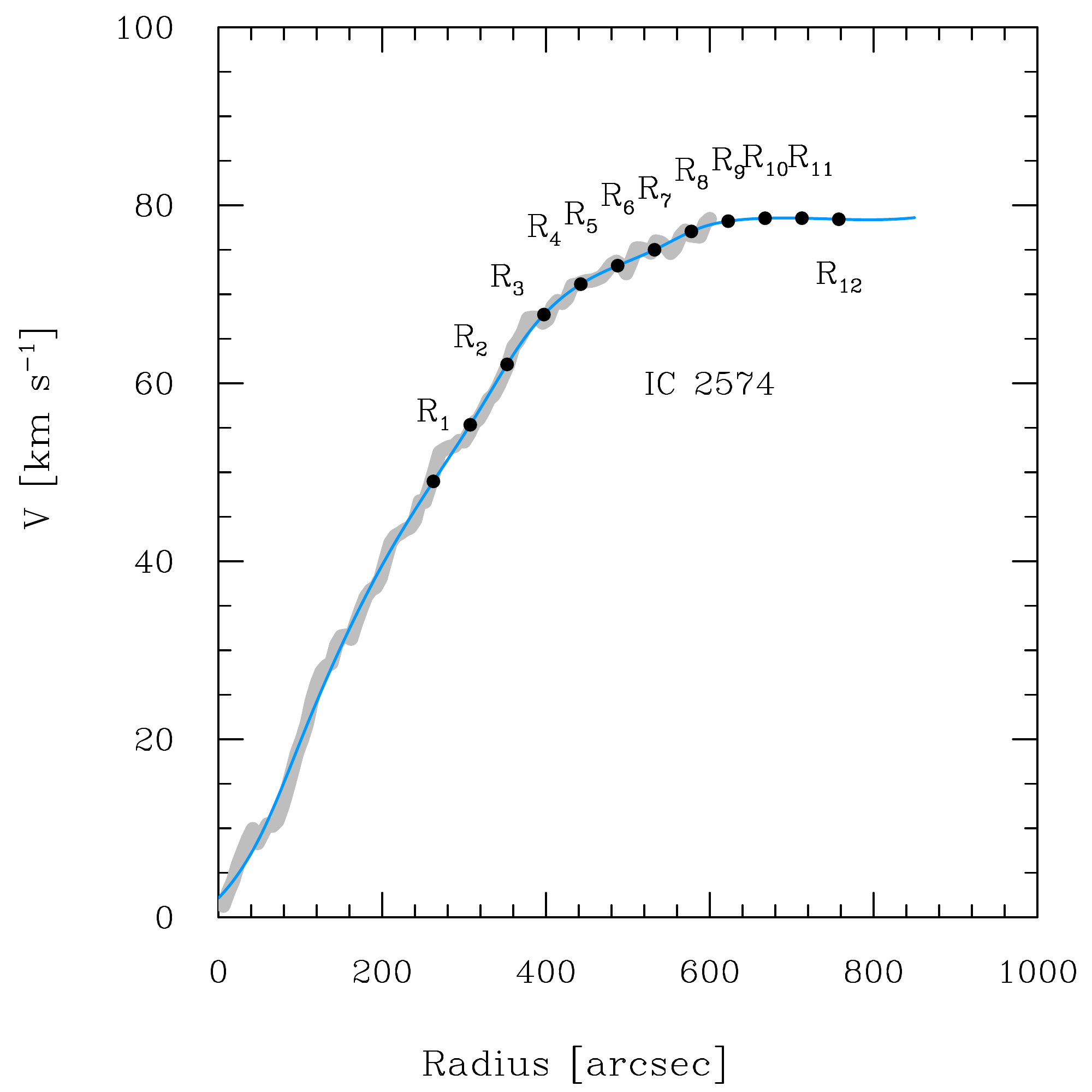} &
      \vspace{0.5cm}
      \includegraphics[scale= 0.376]{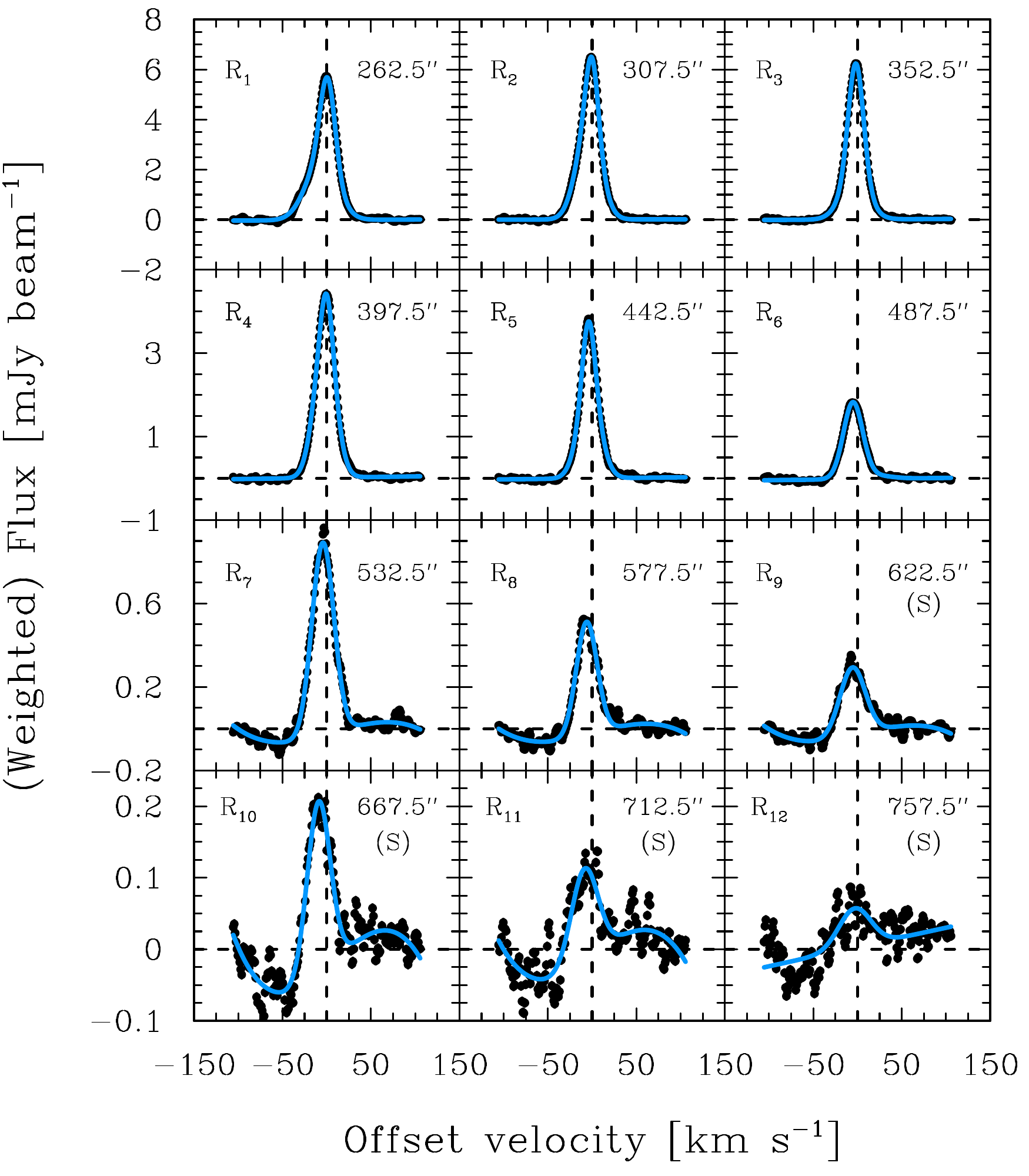} 
    \end{tabular}
    \caption{Left: the observed rotation curve by \citet{debloketal08} (thick gray line) 
    and our interpolation and extrapolation of the rotation curve (blue solid line). 
    Right: Azimuthally averaged stacked profiles (black circle symbols). The solid blue lines are the fit to the data. 
    The vertical and horizontal dashed lines indicate the flux and offset velocity 
    with respect to the model velocity field to guide the eyes. 
    The (S) letters in some panels represent the `stacked radii', i.e., the 
    spectra at a radius larger than the extent traced directly with the THINGS moment zero map.} 
    \label{fig:stackedspectra}
\end{figure*}
\setcounter{figure}{0}
\begin{figure*}
    \begin{tabular}{l l}
      \includegraphics[scale= 0.376]{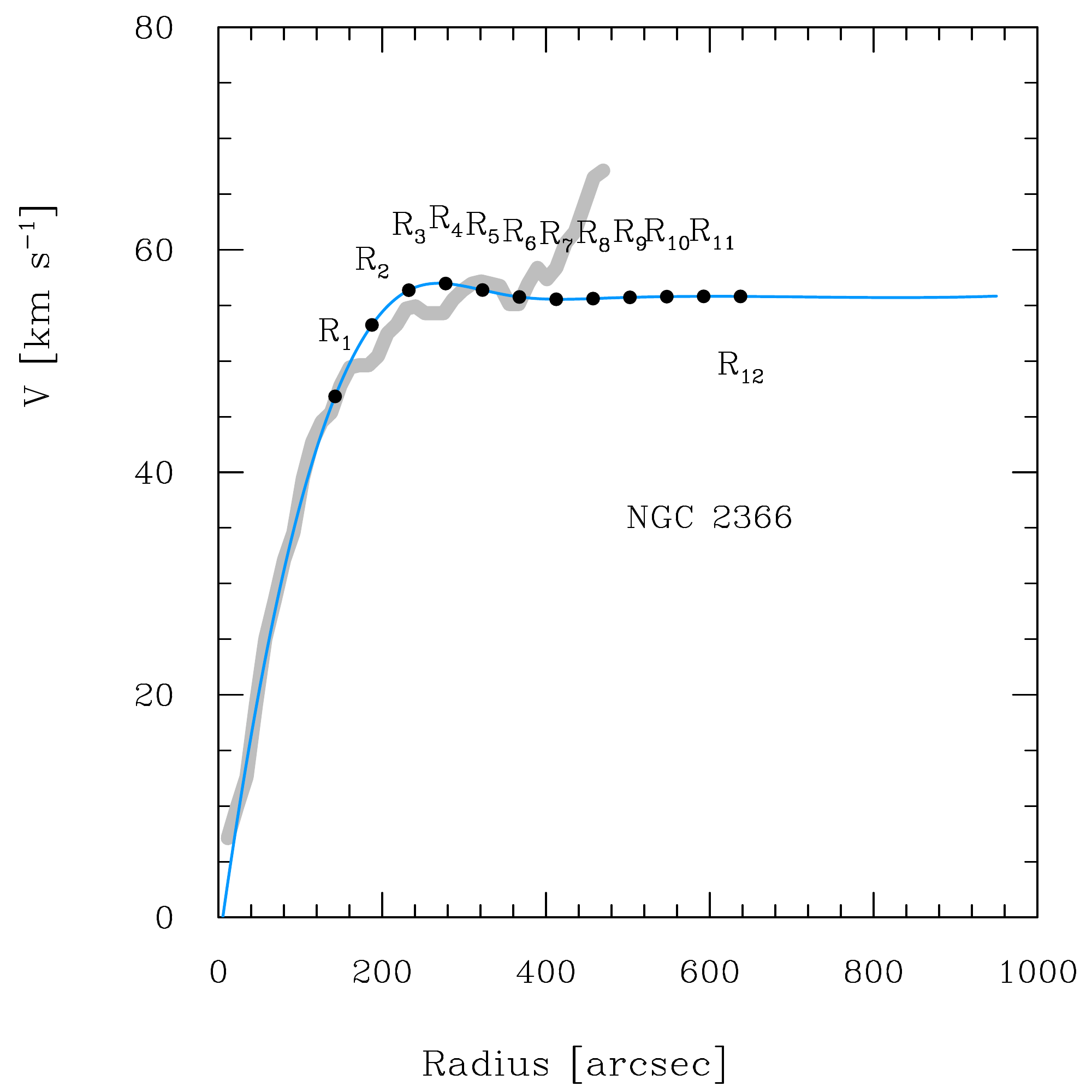} &
      \includegraphics[scale= 0.376]{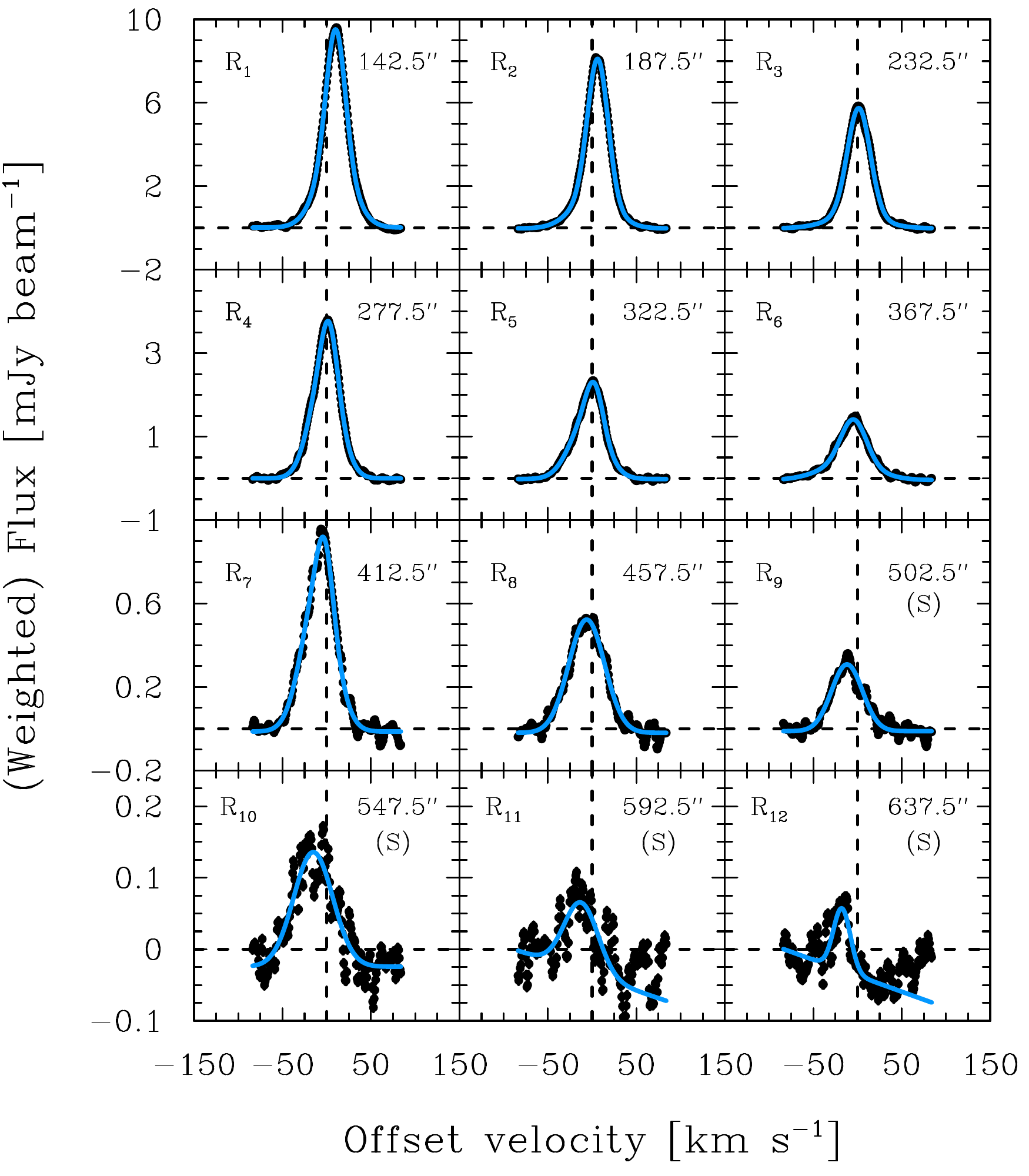} \\
      \includegraphics[scale= 0.376]{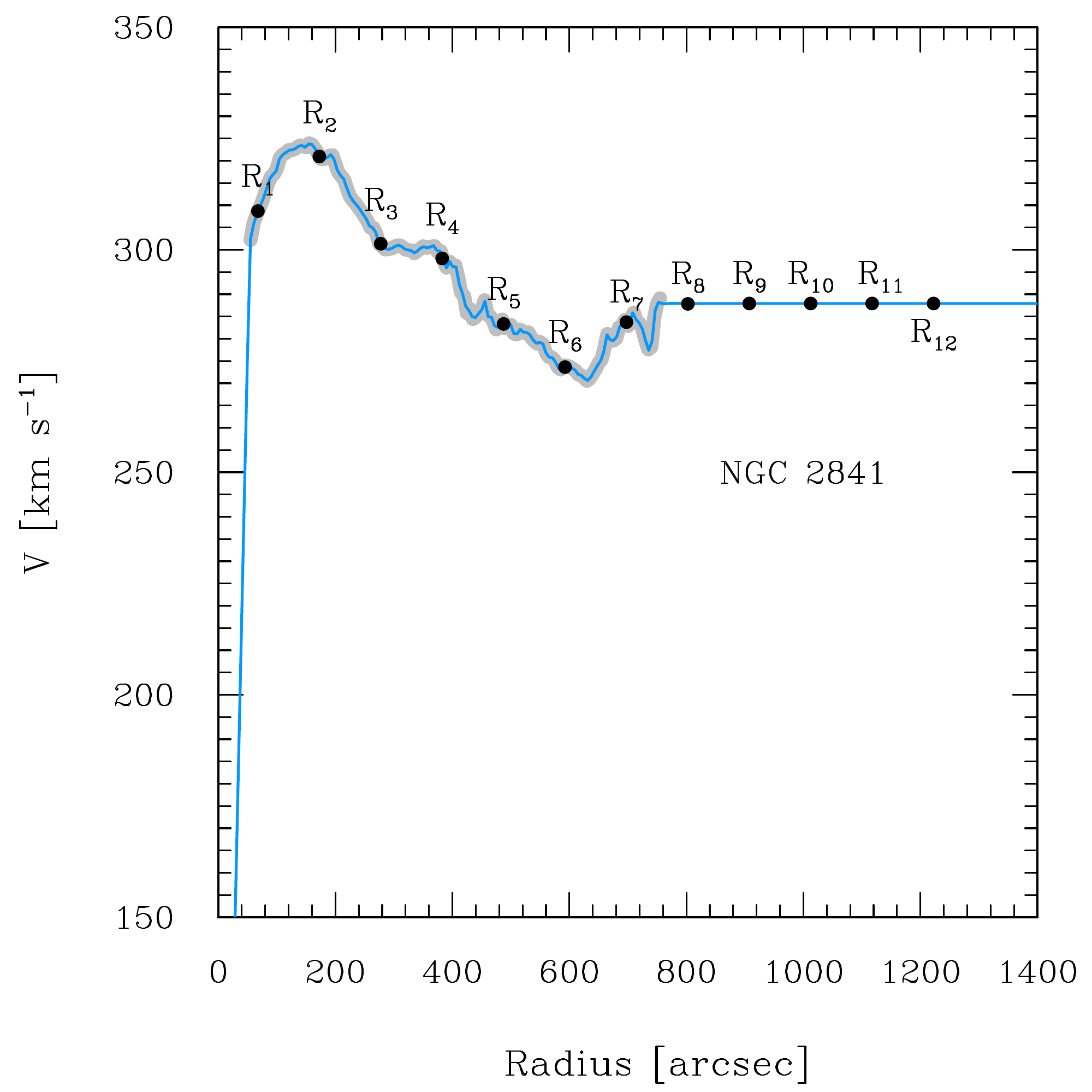} &
      \includegraphics[scale= 0.376]{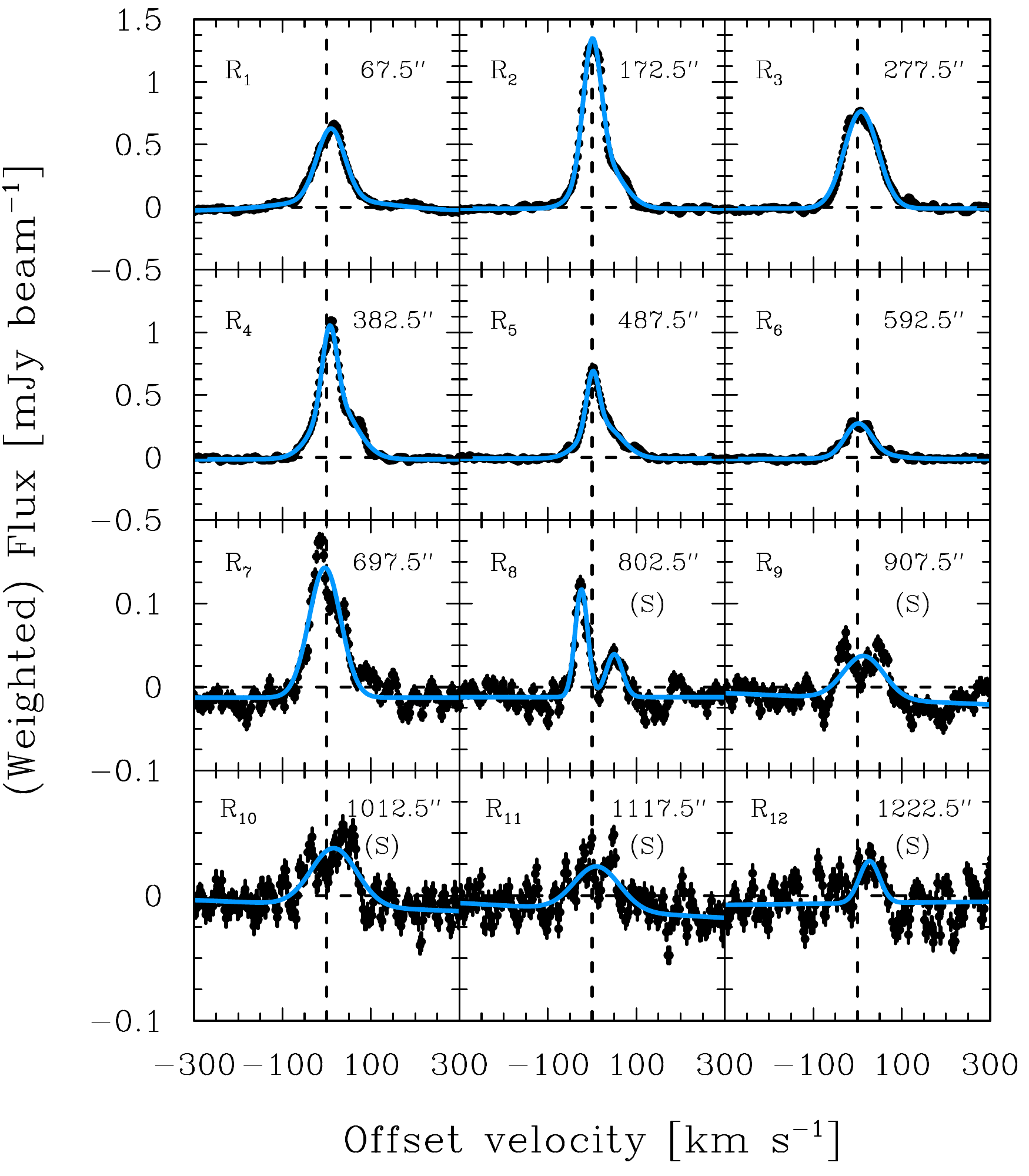}\\ 
      \includegraphics[scale= 0.376]{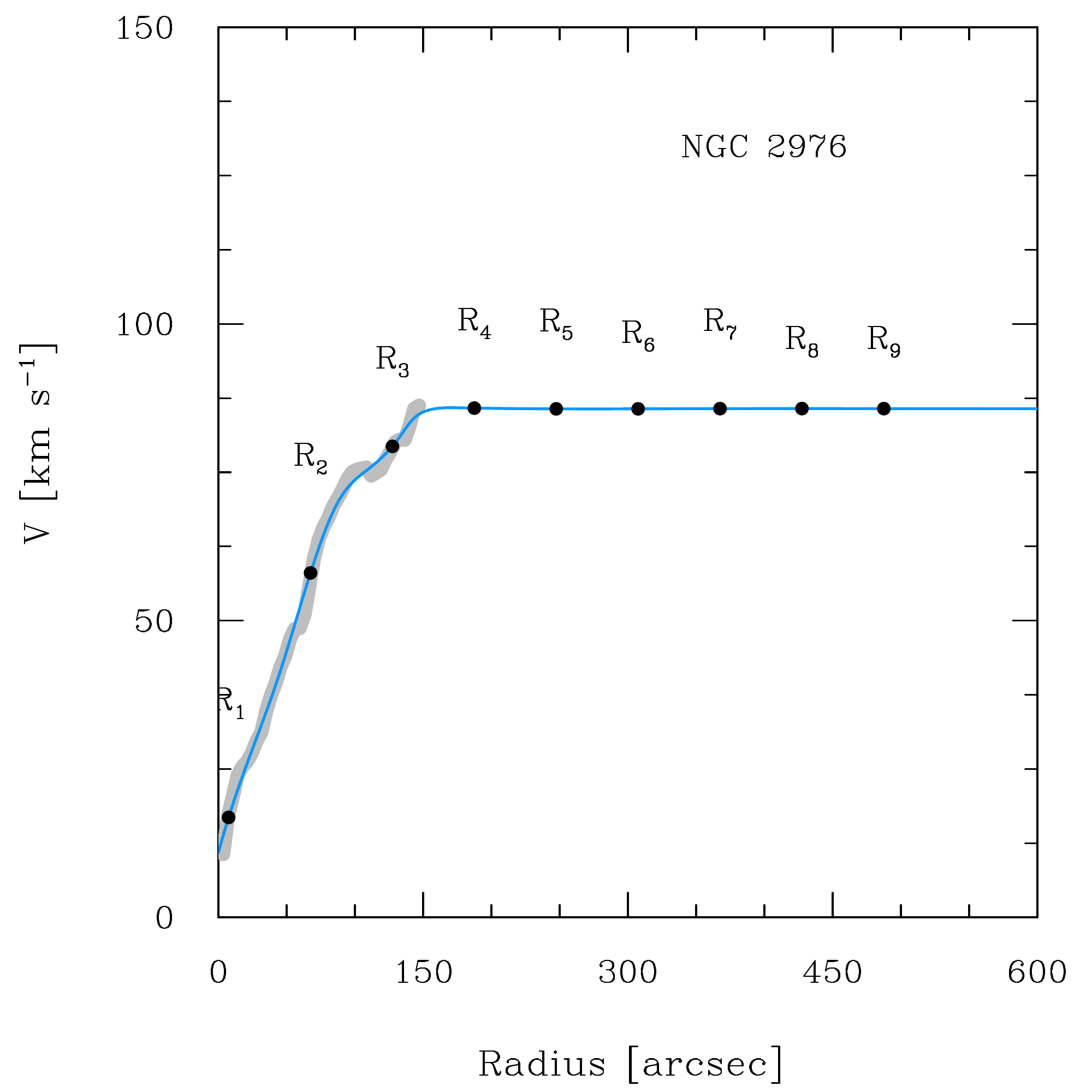} &
      \includegraphics[scale= 0.376]{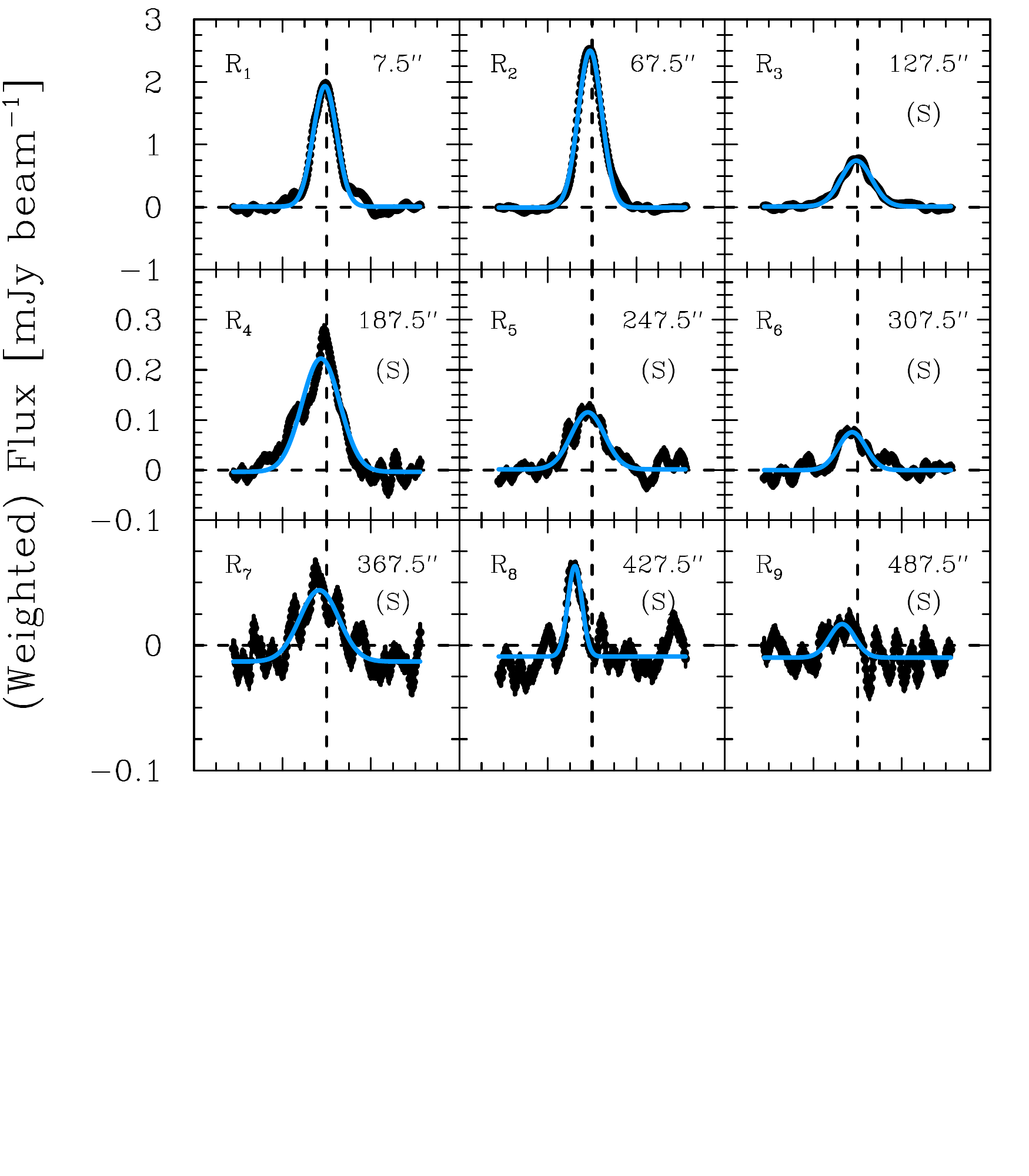}

    \end{tabular}
    \caption{ Continued. \label{stackedspectra}} 
\end{figure*}
\setcounter{figure}{0}
\begin{figure*}
    \begin{tabular}{l l}
      \includegraphics[scale= 0.376]{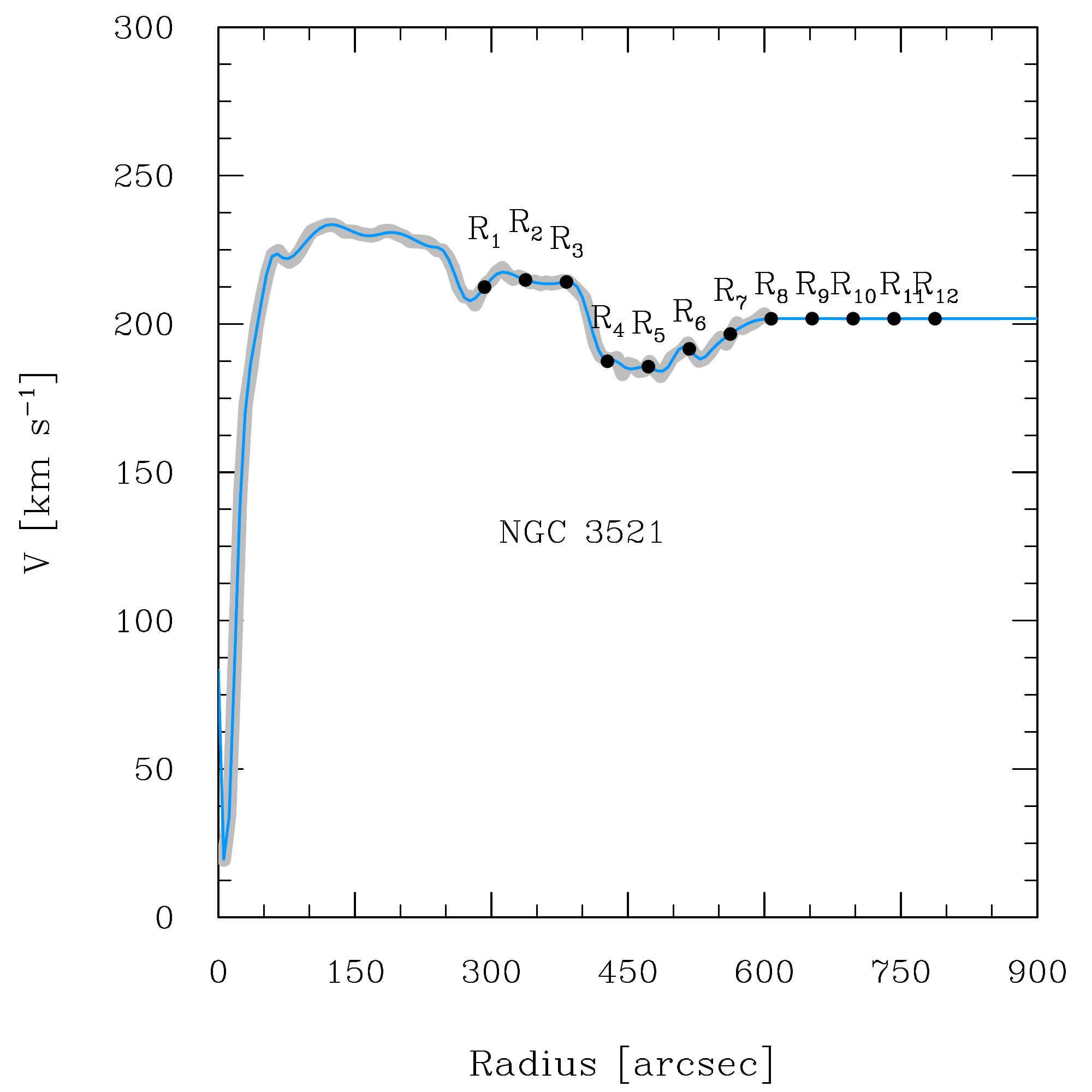} &
      \includegraphics[scale= 0.376]{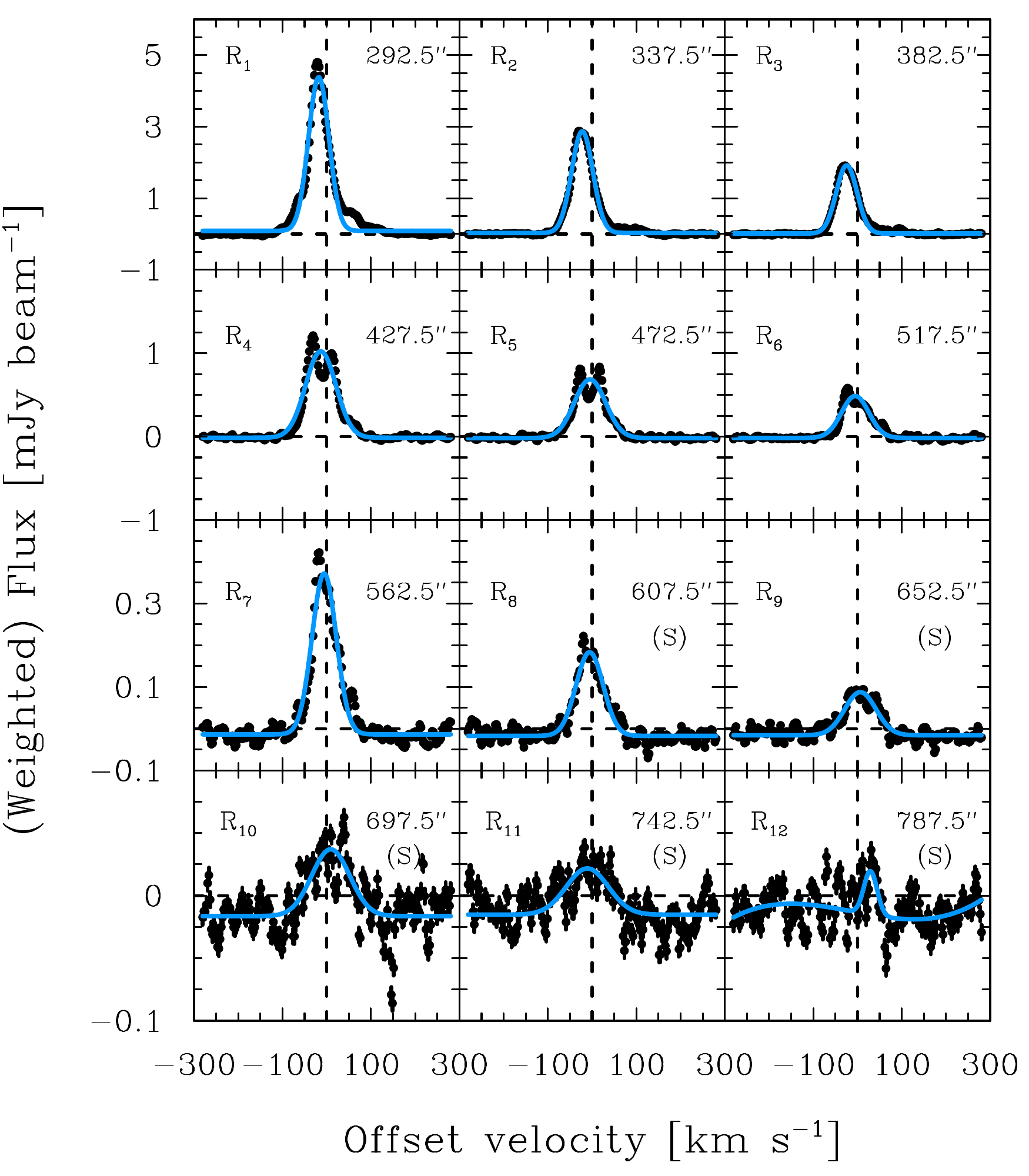}\\ 
      \includegraphics[scale= 0.376]{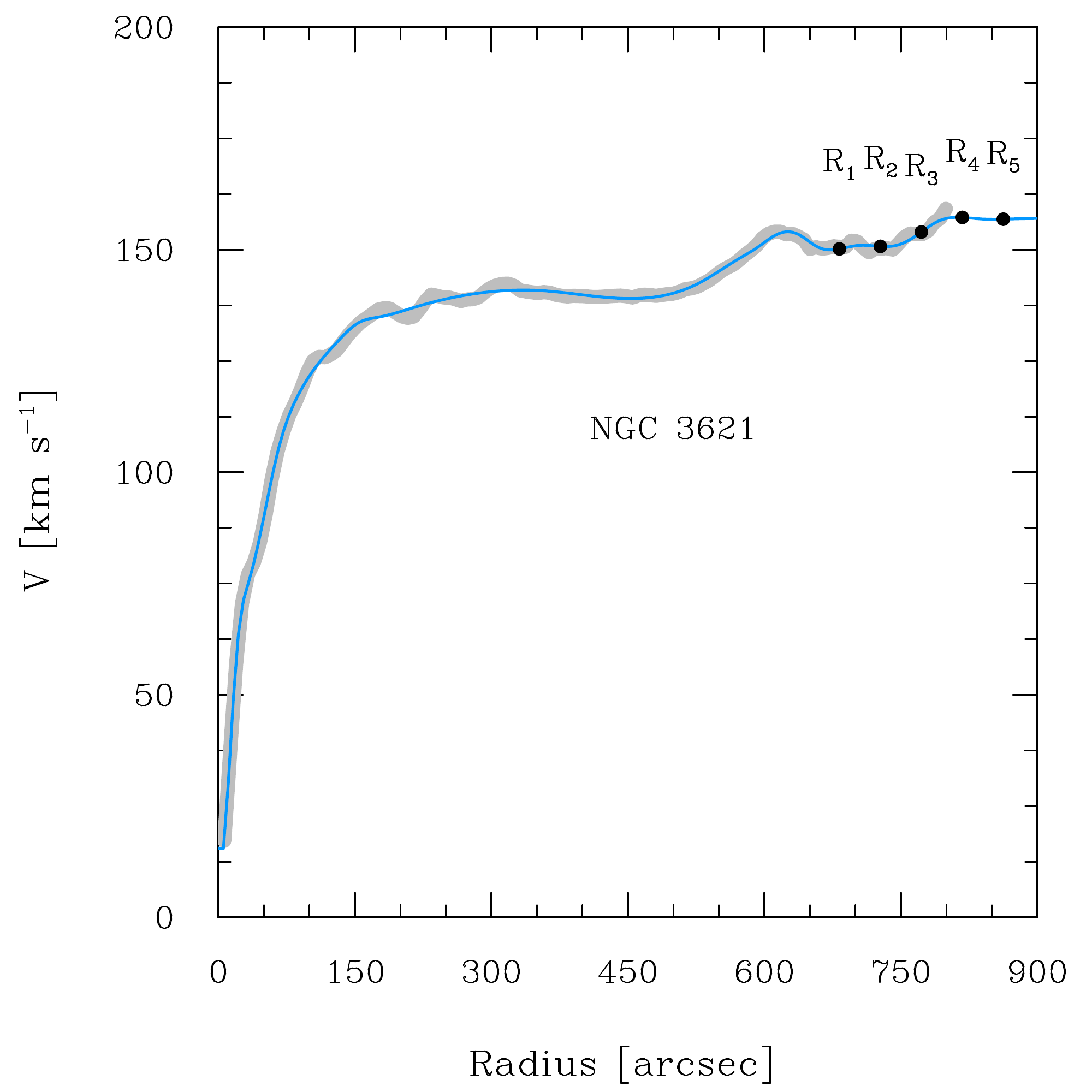} &
      \includegraphics[scale= 0.376]{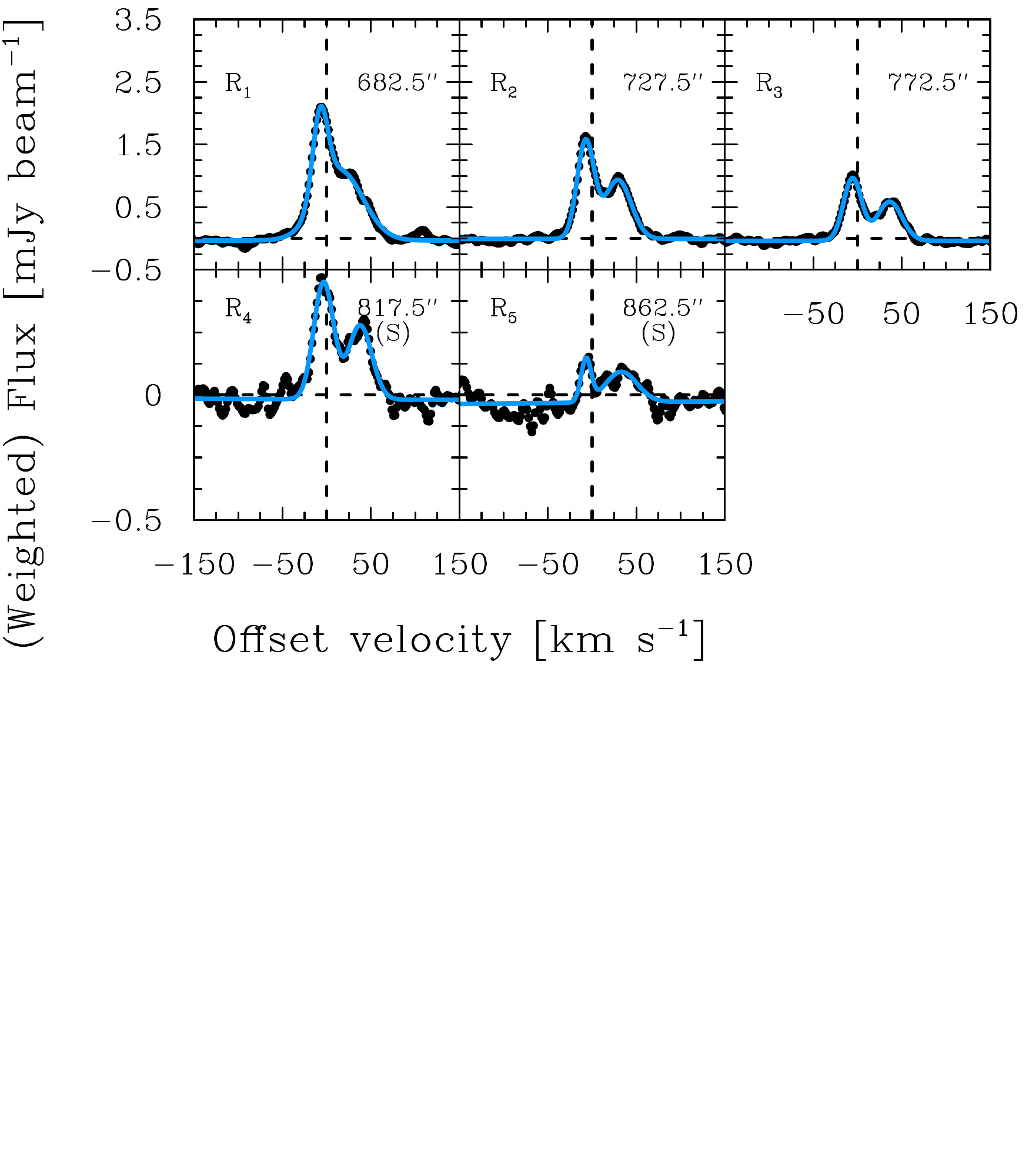} \\
      \includegraphics[scale= 0.376]{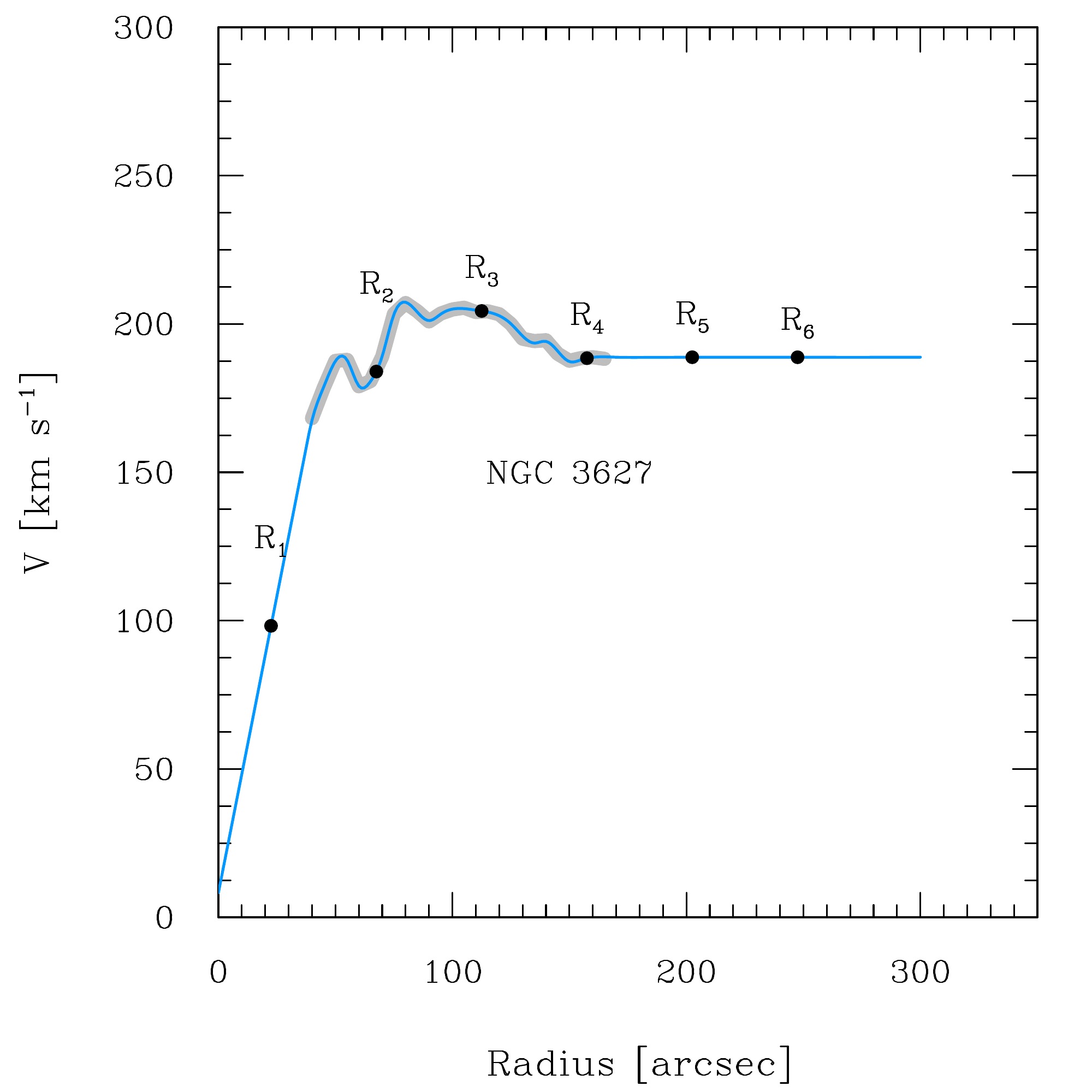} &
      \includegraphics[scale= 0.376]{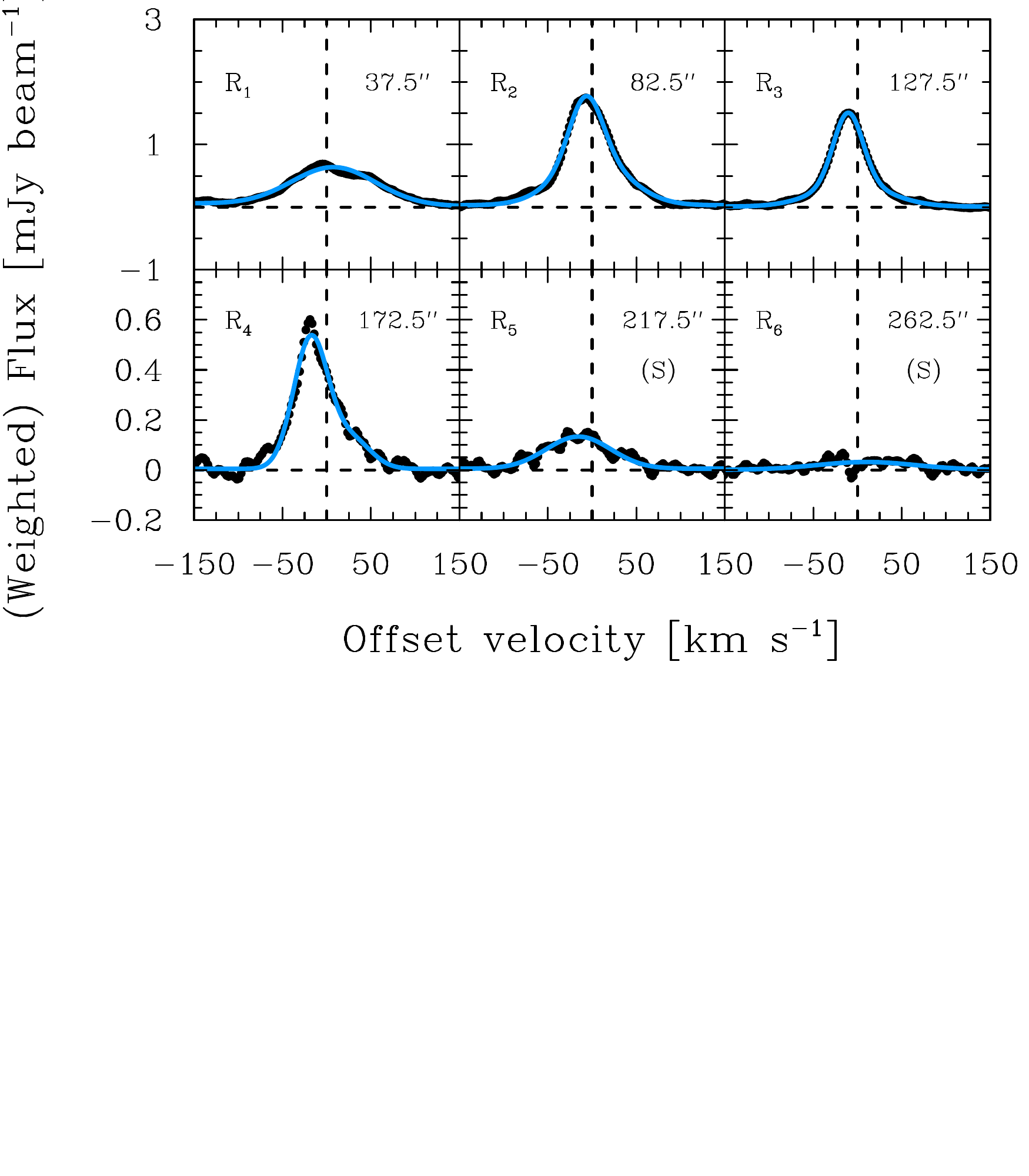} 
    \end{tabular}
    \caption{Continued. \label{stackedspectra}} 
\end{figure*}
\setcounter{figure}{0}
\begin{figure*}
    \begin{tabular}{l l}
      \includegraphics[scale= 0.376]{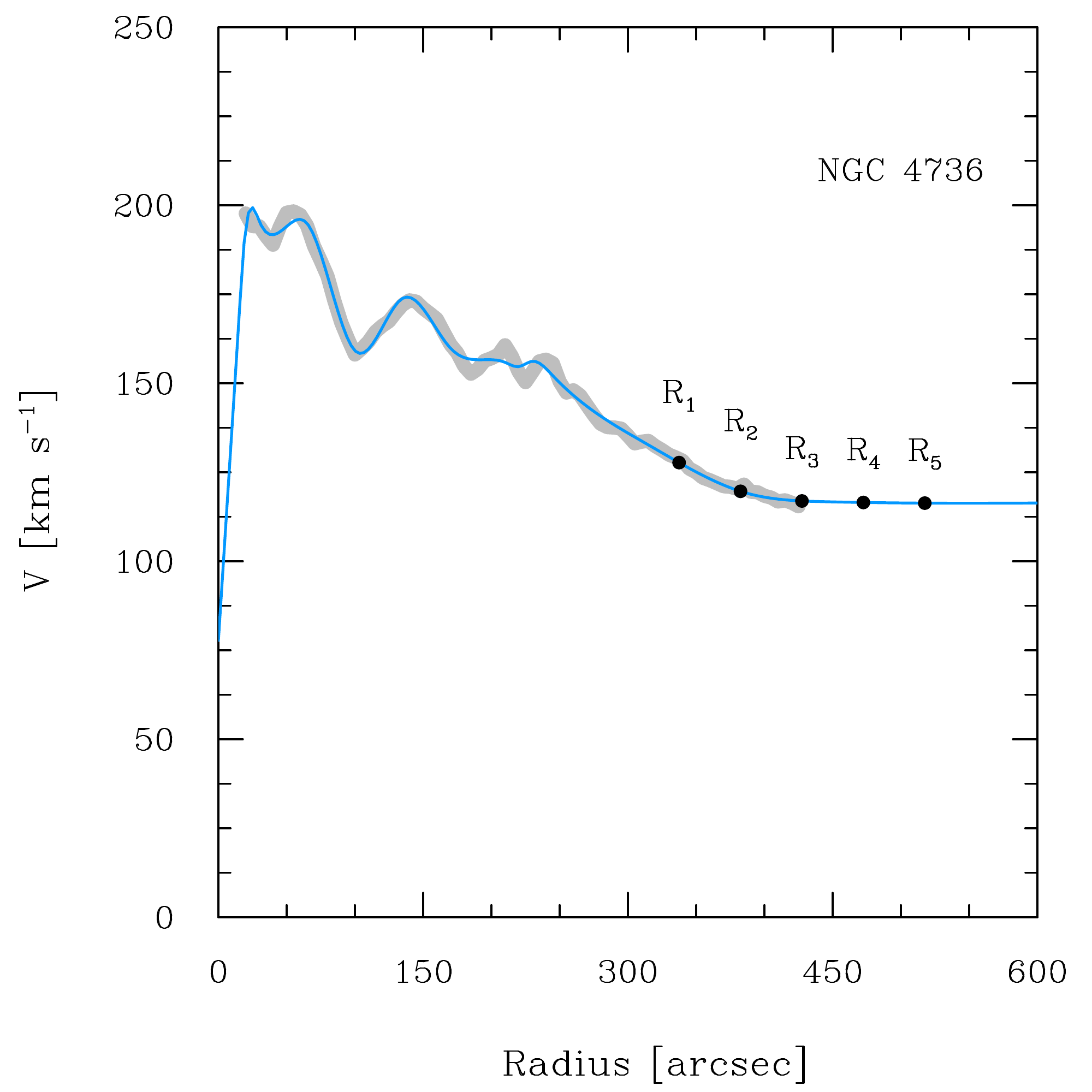} &
      \includegraphics[scale= 0.376]{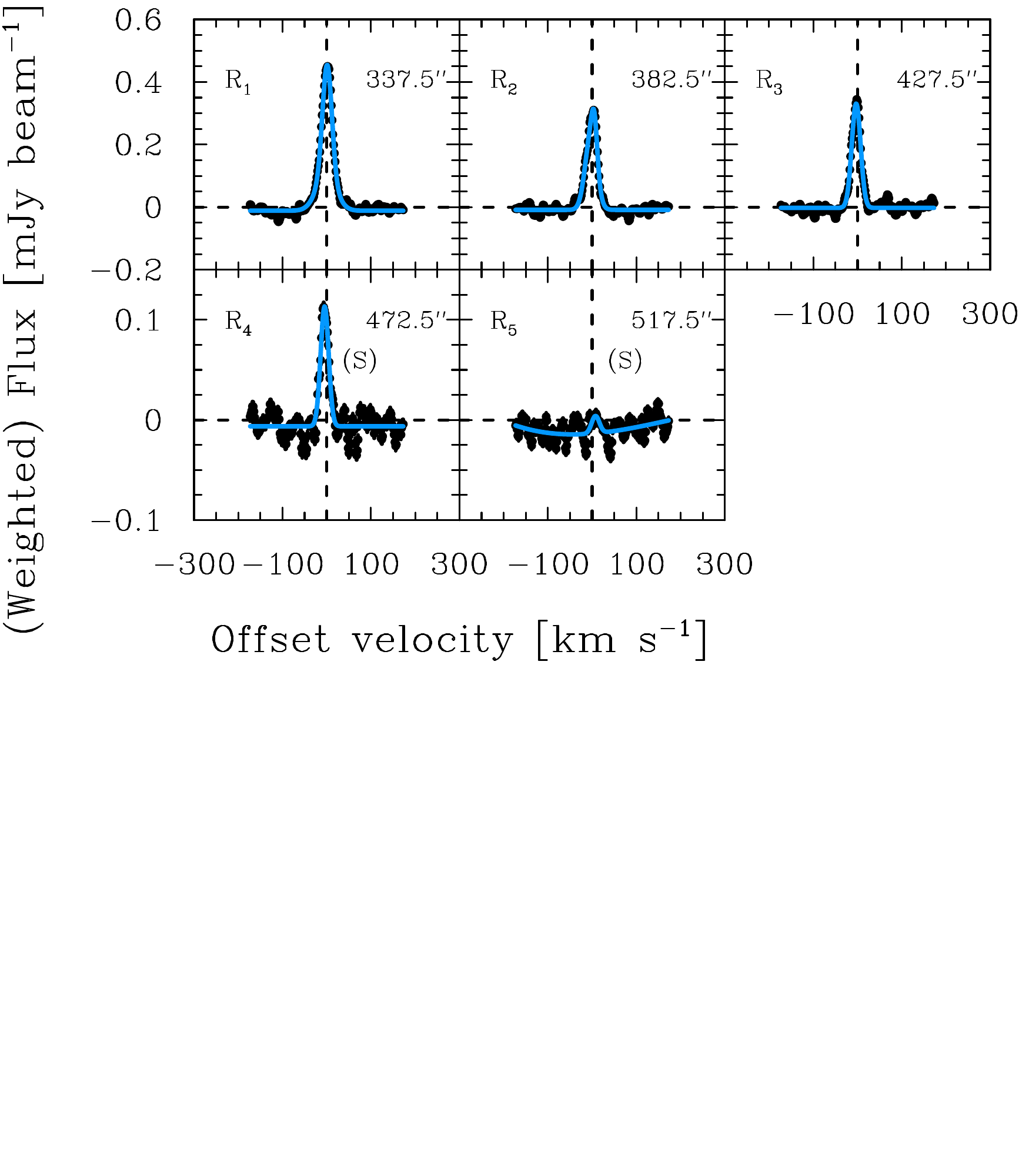} \\
      \includegraphics[scale= 0.376]{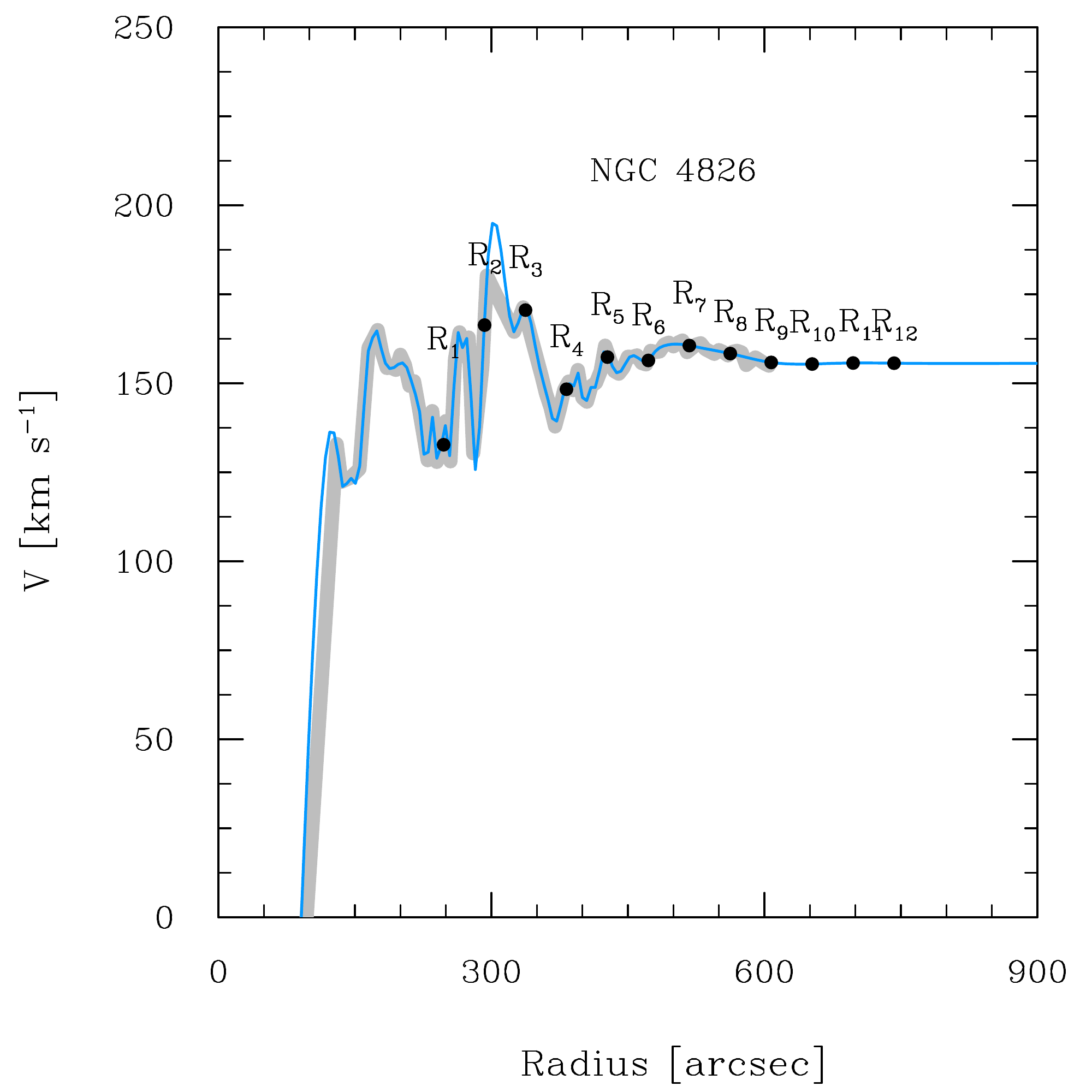} &
      \includegraphics[scale= 0.376]{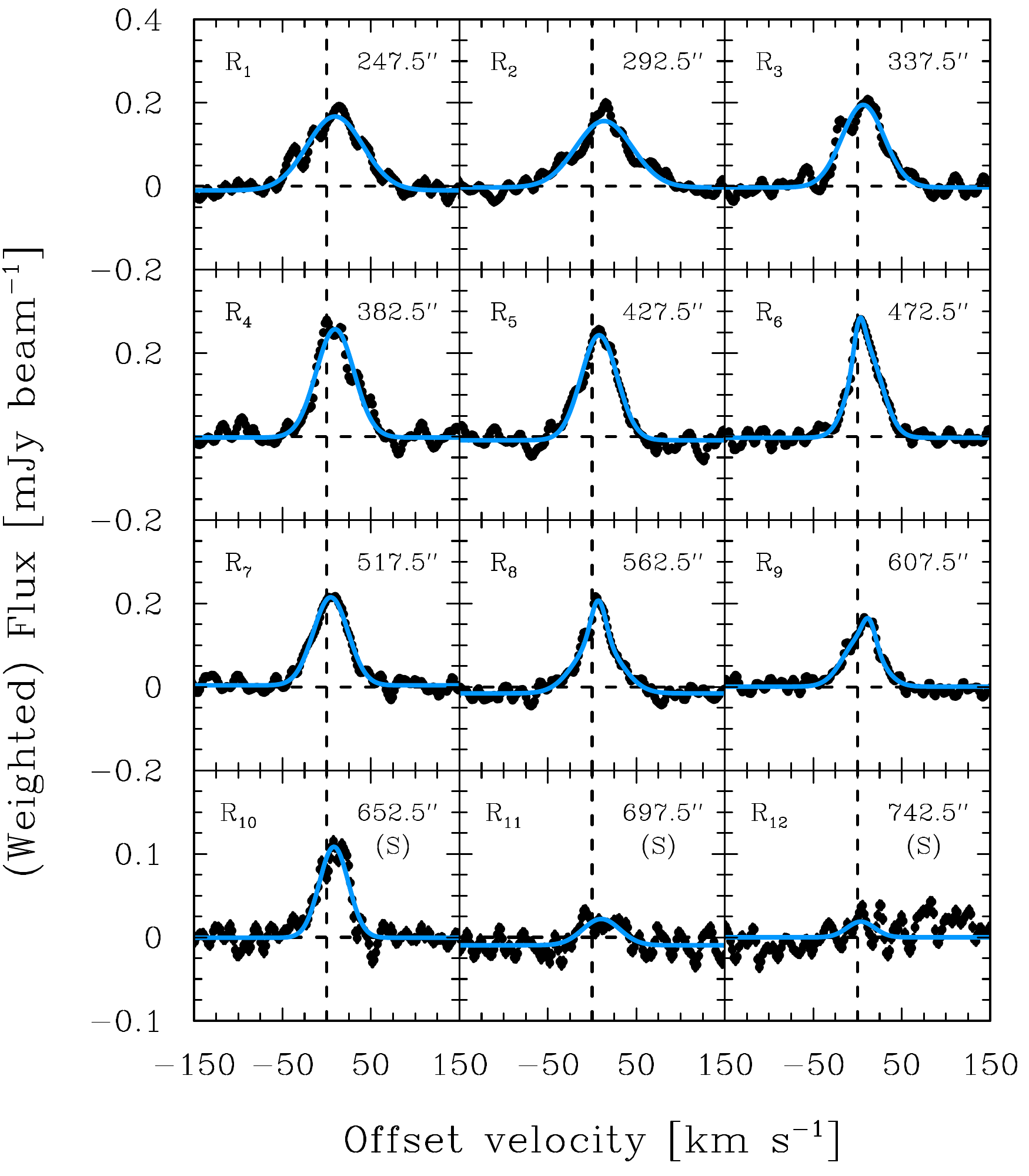}\\ 
      \includegraphics[scale= 0.376]{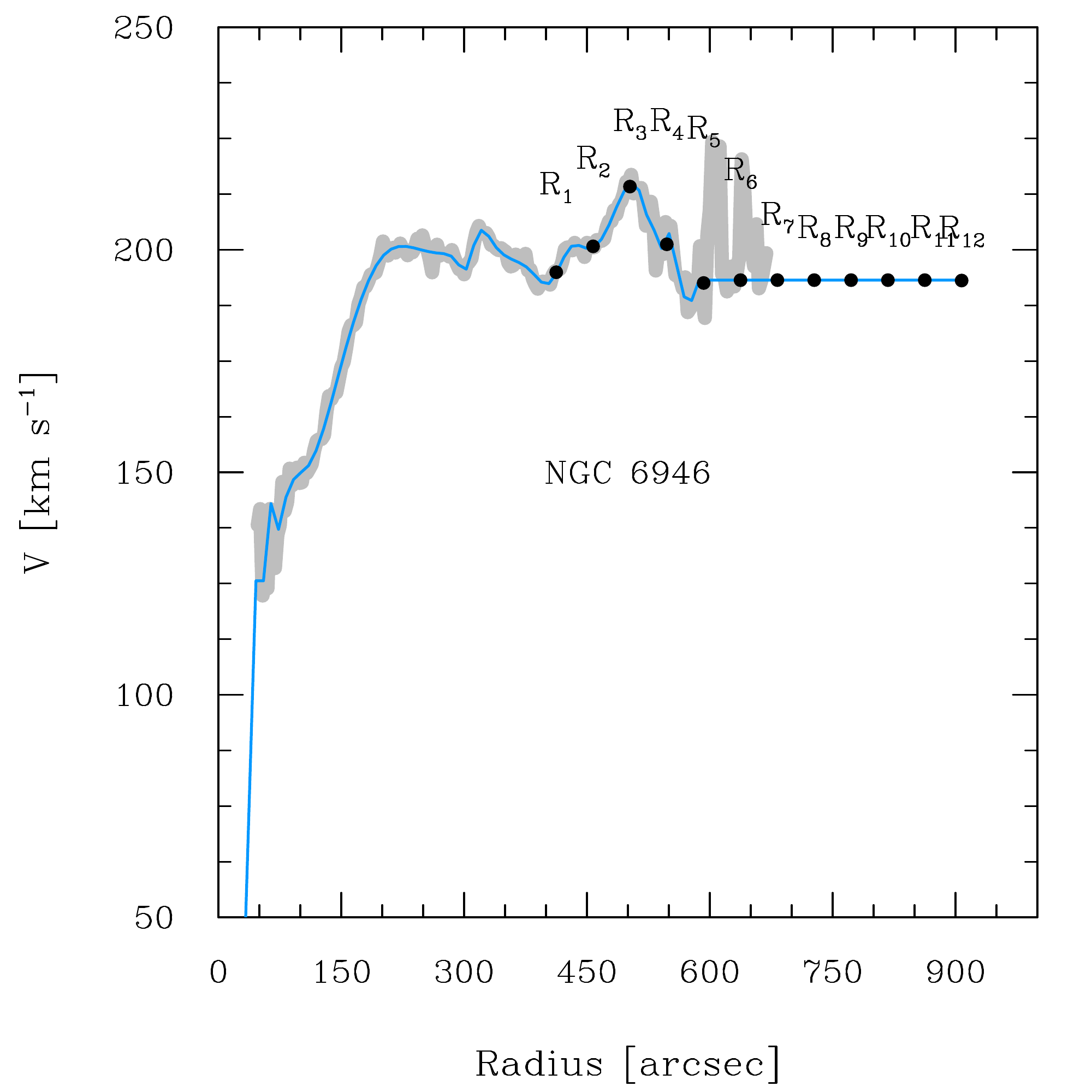} &
      \includegraphics[scale= 0.376]{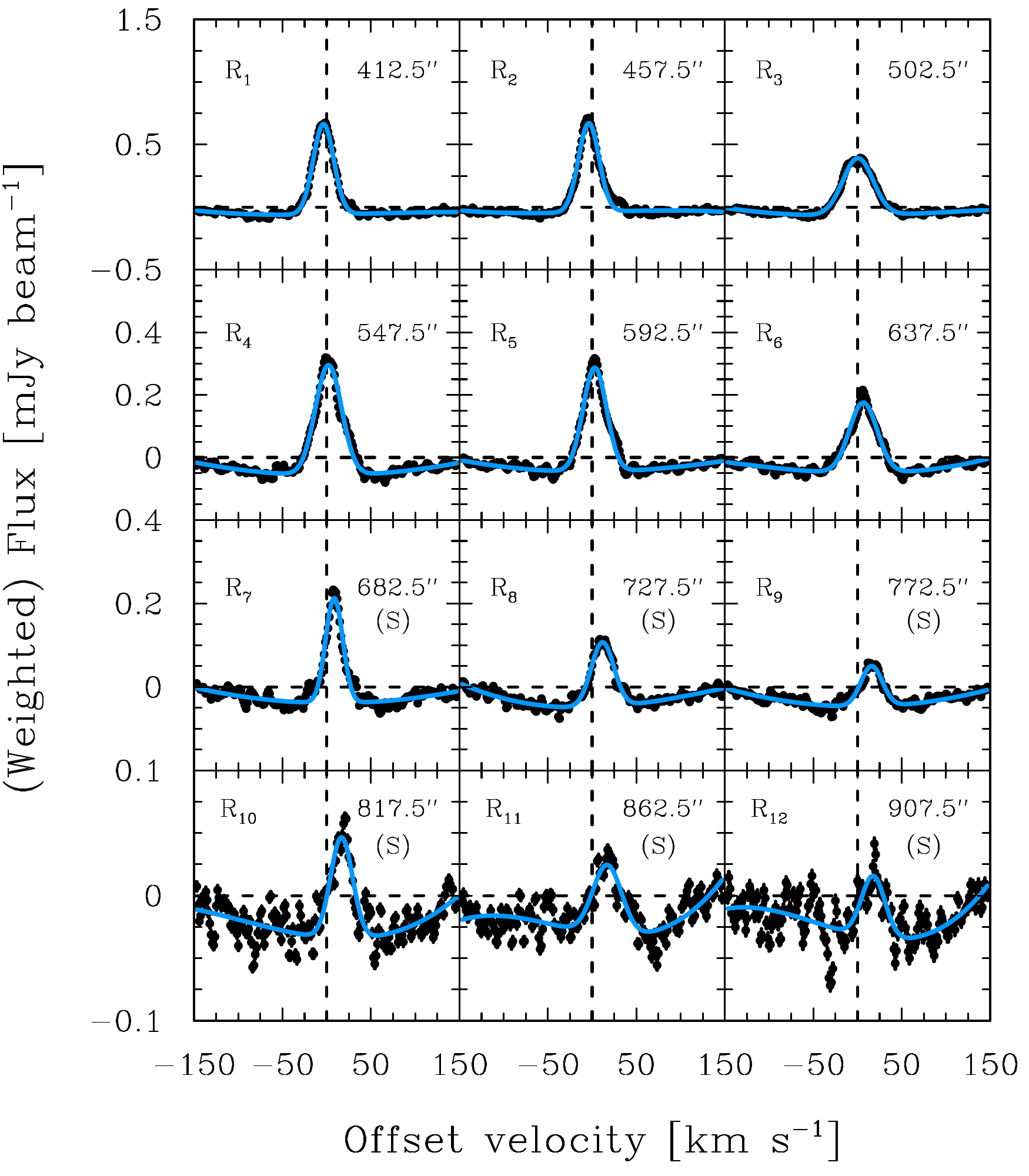} 
    \end{tabular}
    \caption{Continued. \label{stackedspectra}} 
\end{figure*}
\setcounter{figure}{0}
\begin{figure*}
    \begin{tabular}{l l}
      \includegraphics[scale= 0.376]{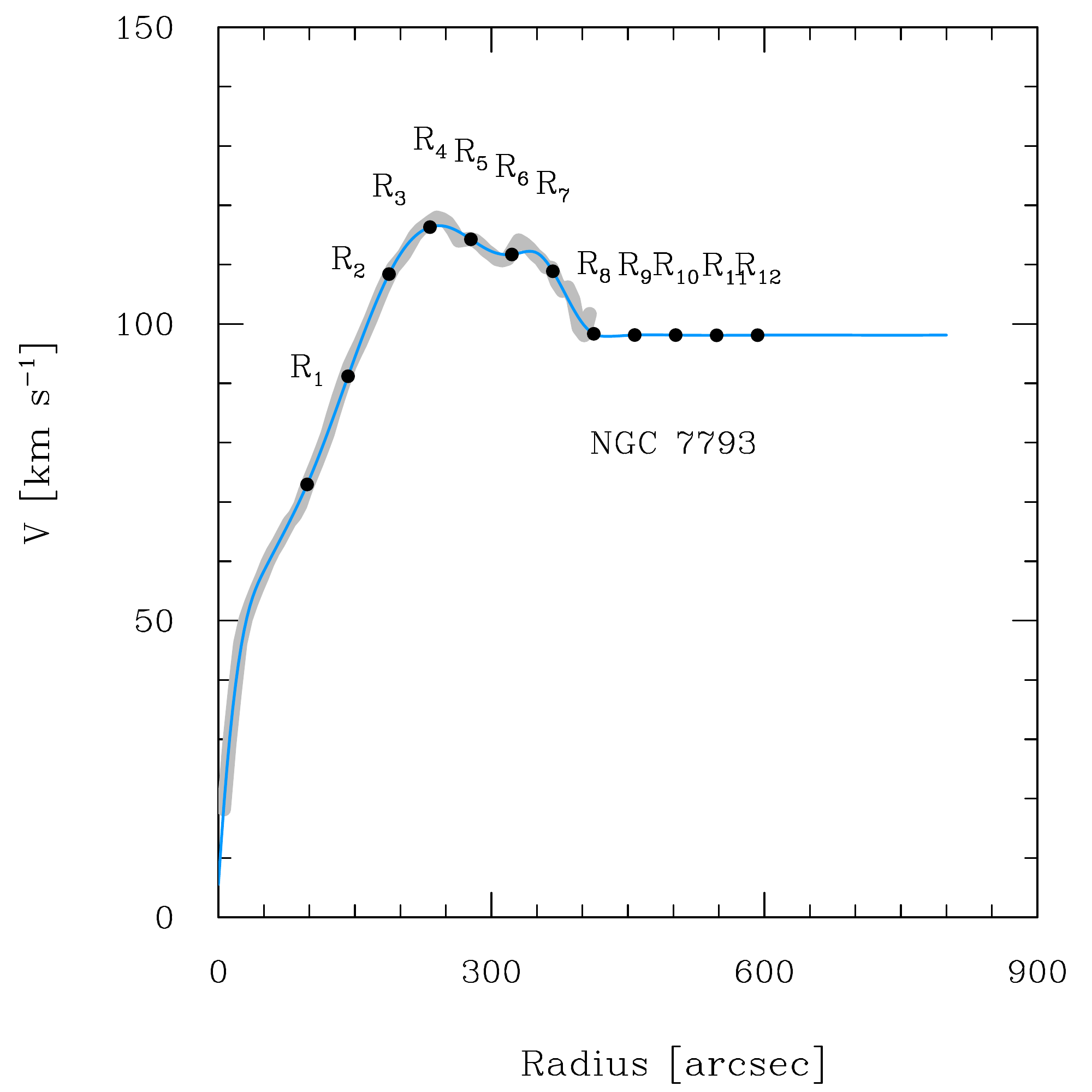} &
      \includegraphics[scale= 0.376]{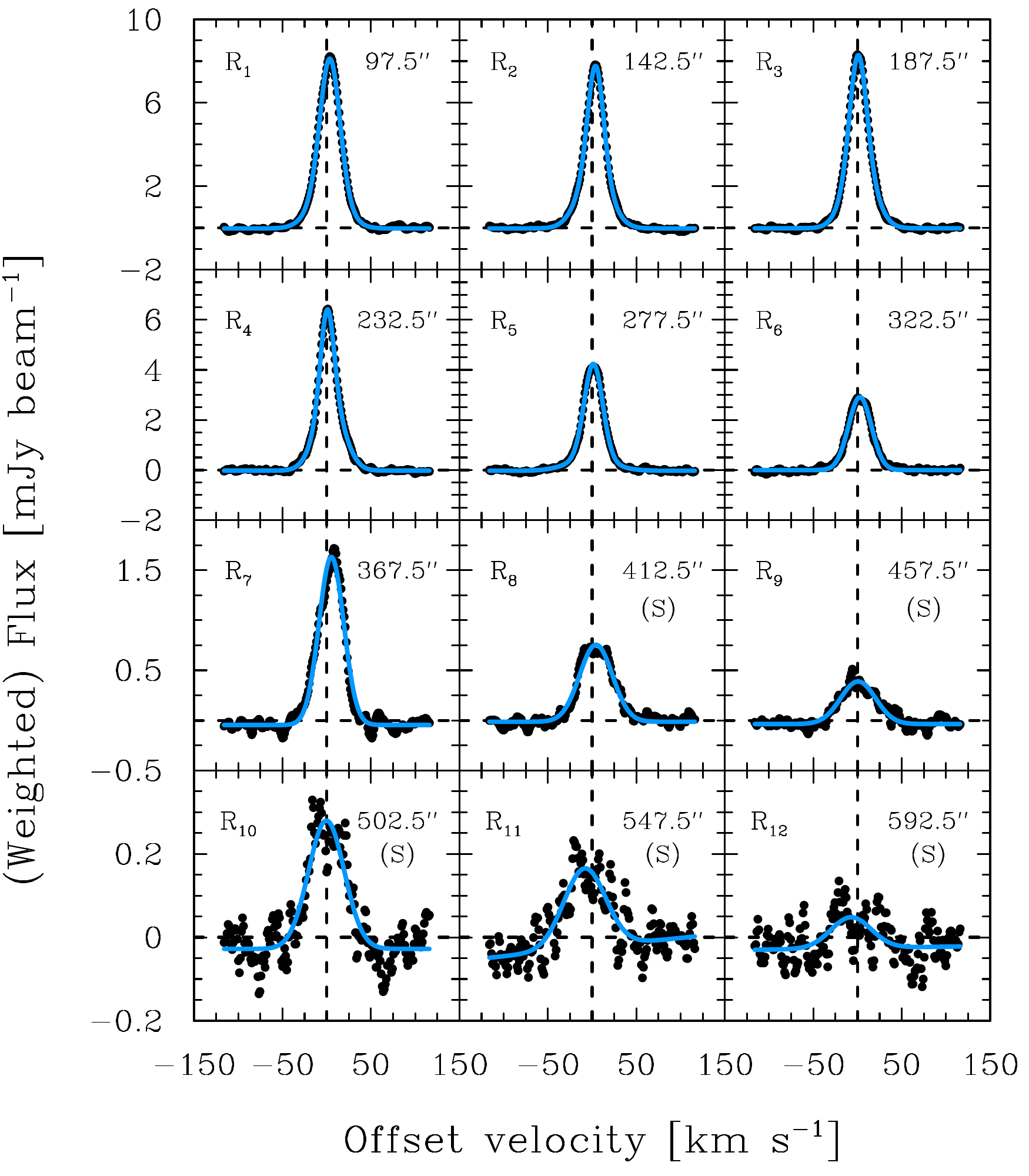} \\
      \includegraphics[scale= 0.376]{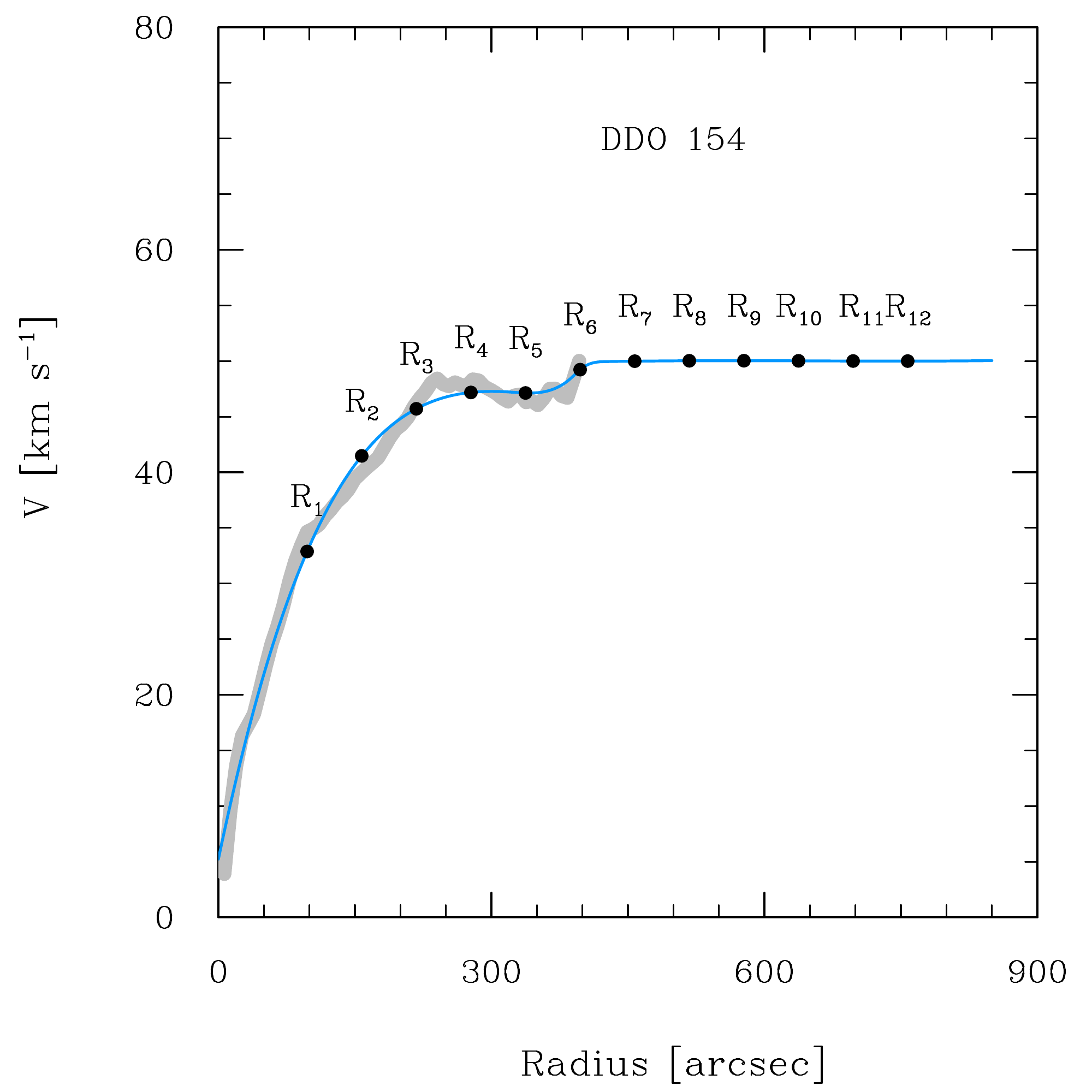} &
      \includegraphics[scale= 0.376]{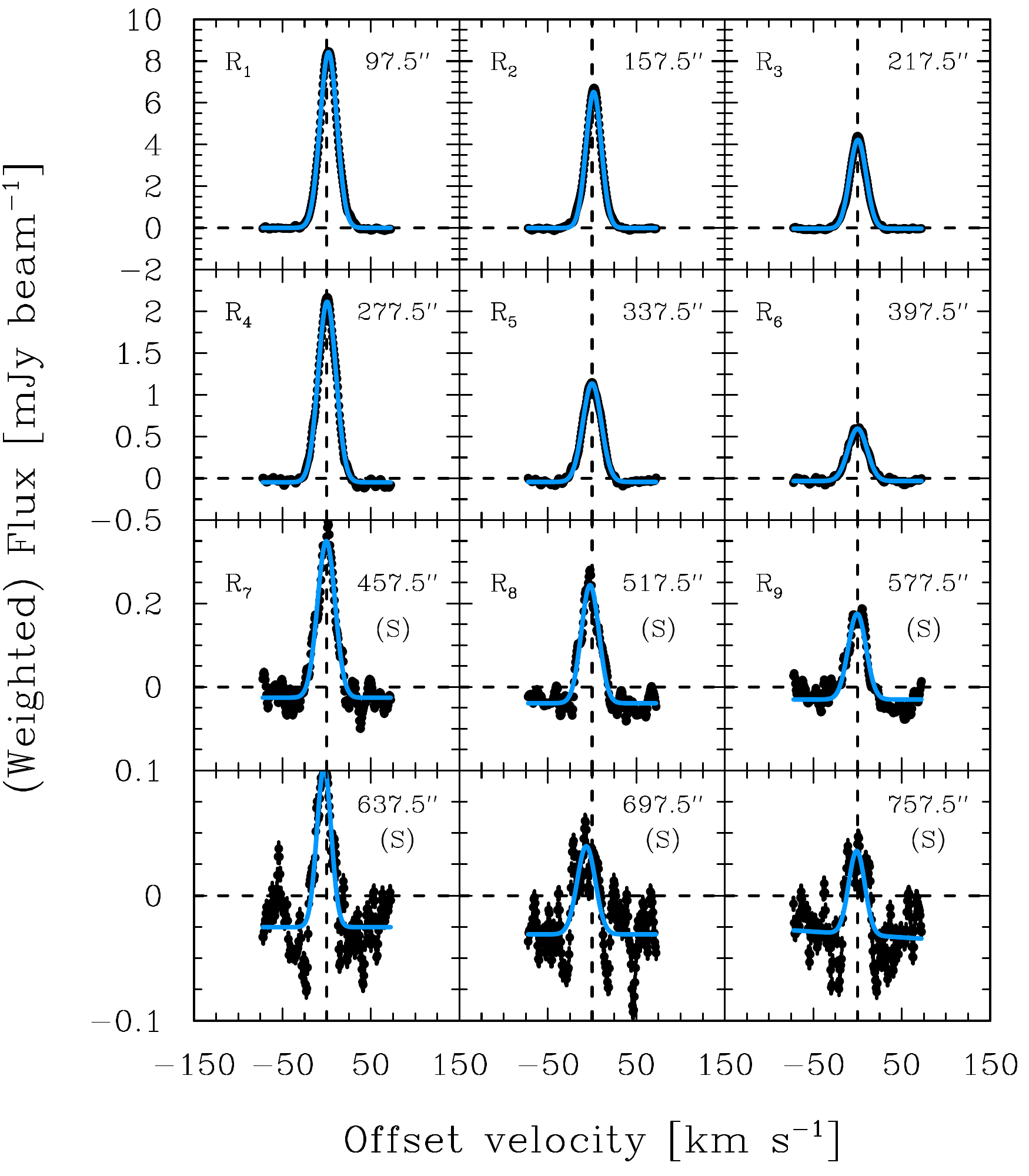} \\
      \includegraphics[scale= 0.376]{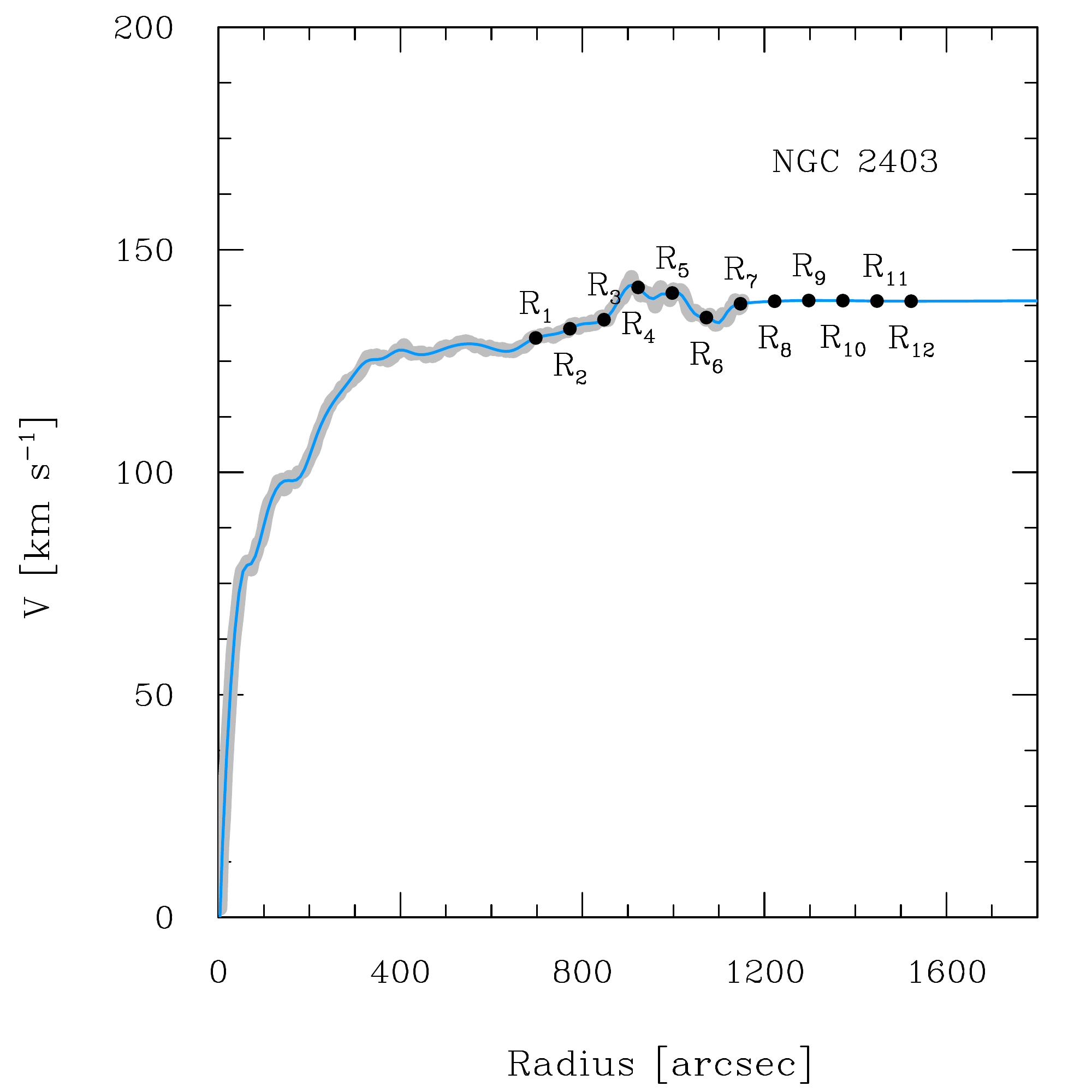} &
      \includegraphics[scale= 0.376]{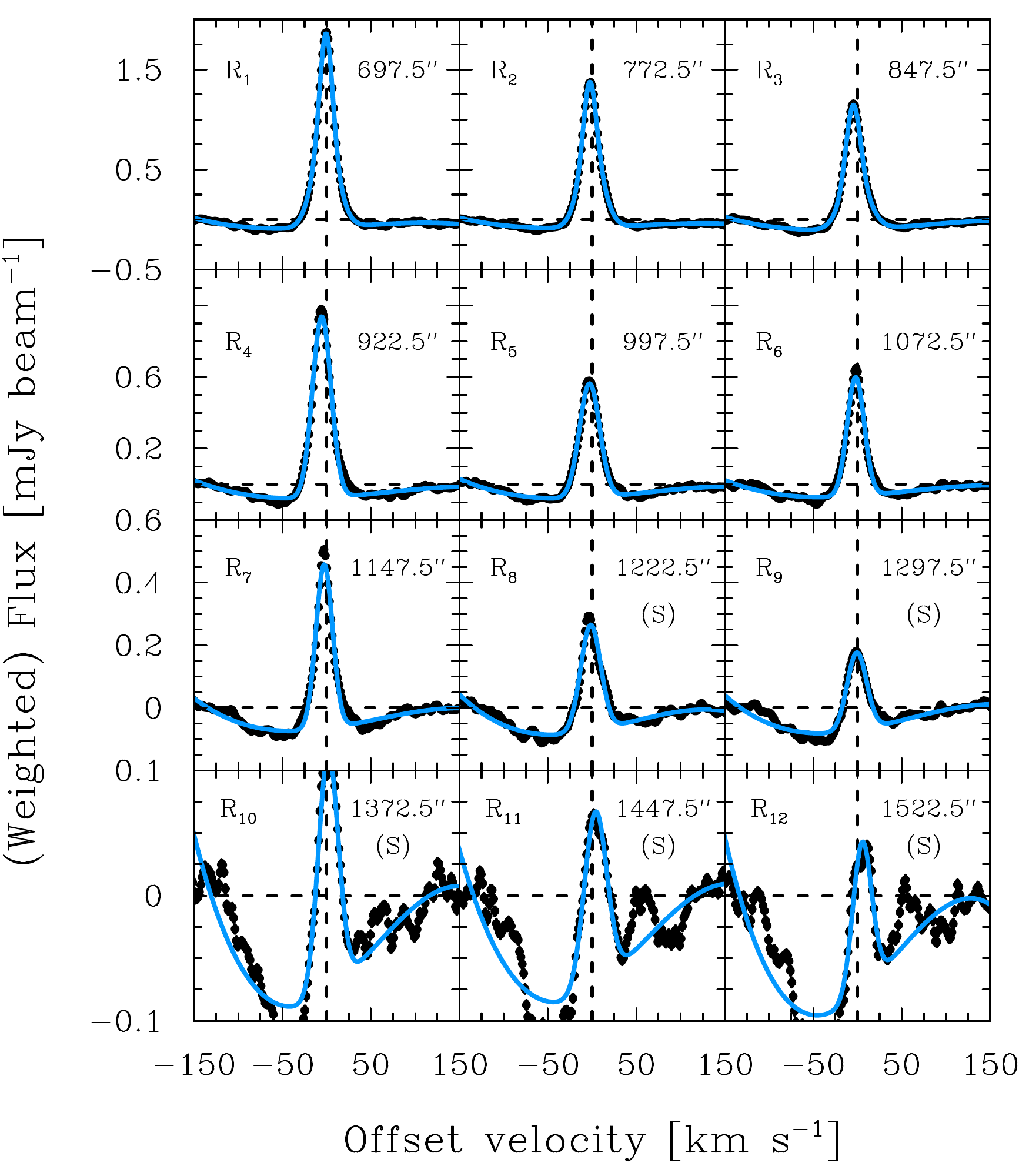} 
    \end{tabular}
    \caption{Continued. \label{stackedspectra}} 
\end{figure*}
\setcounter{figure}{0}
\begin{figure*}
     \begin{tabular}{l l}
      \includegraphics[scale= 0.376]{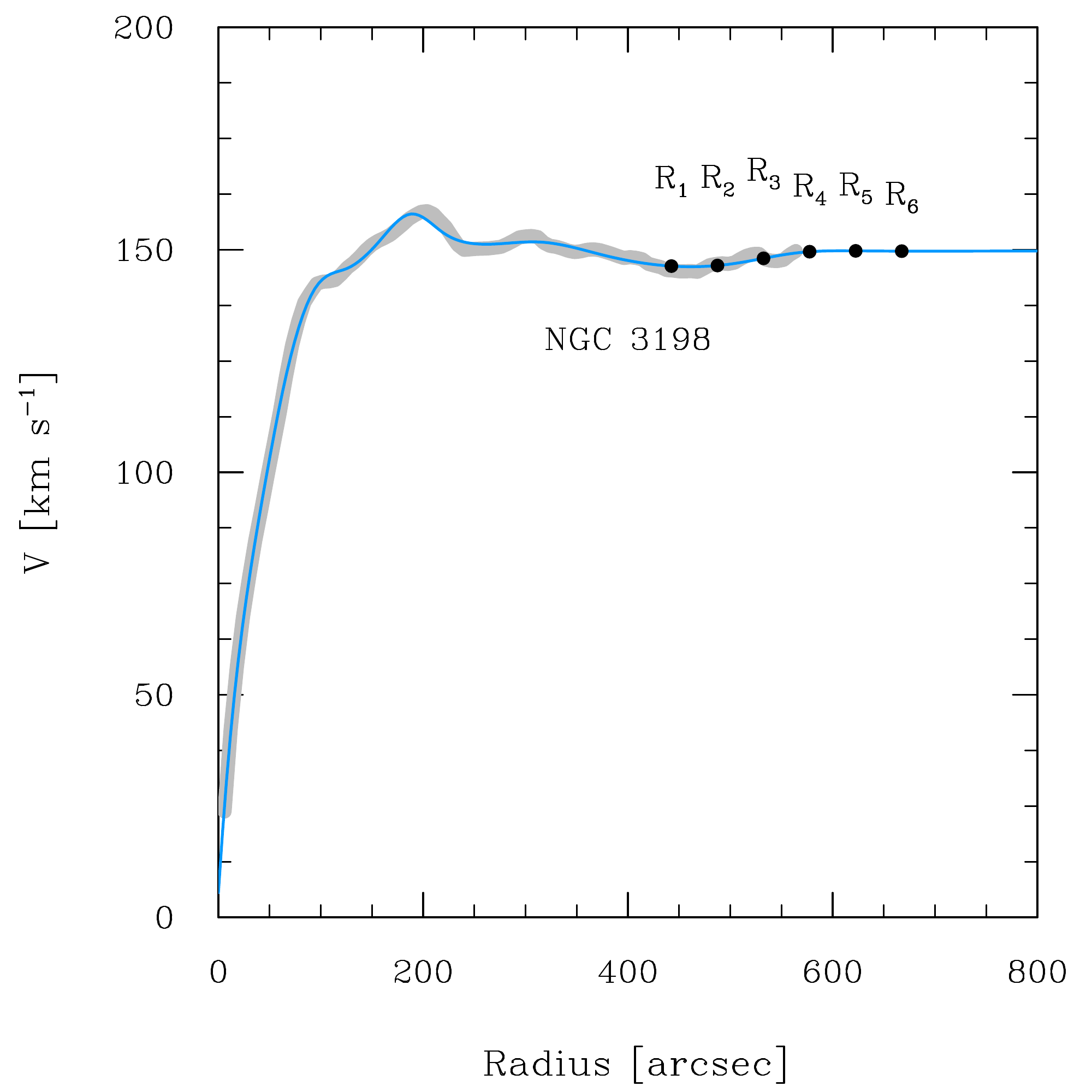} &
      \includegraphics[scale= 0.376]{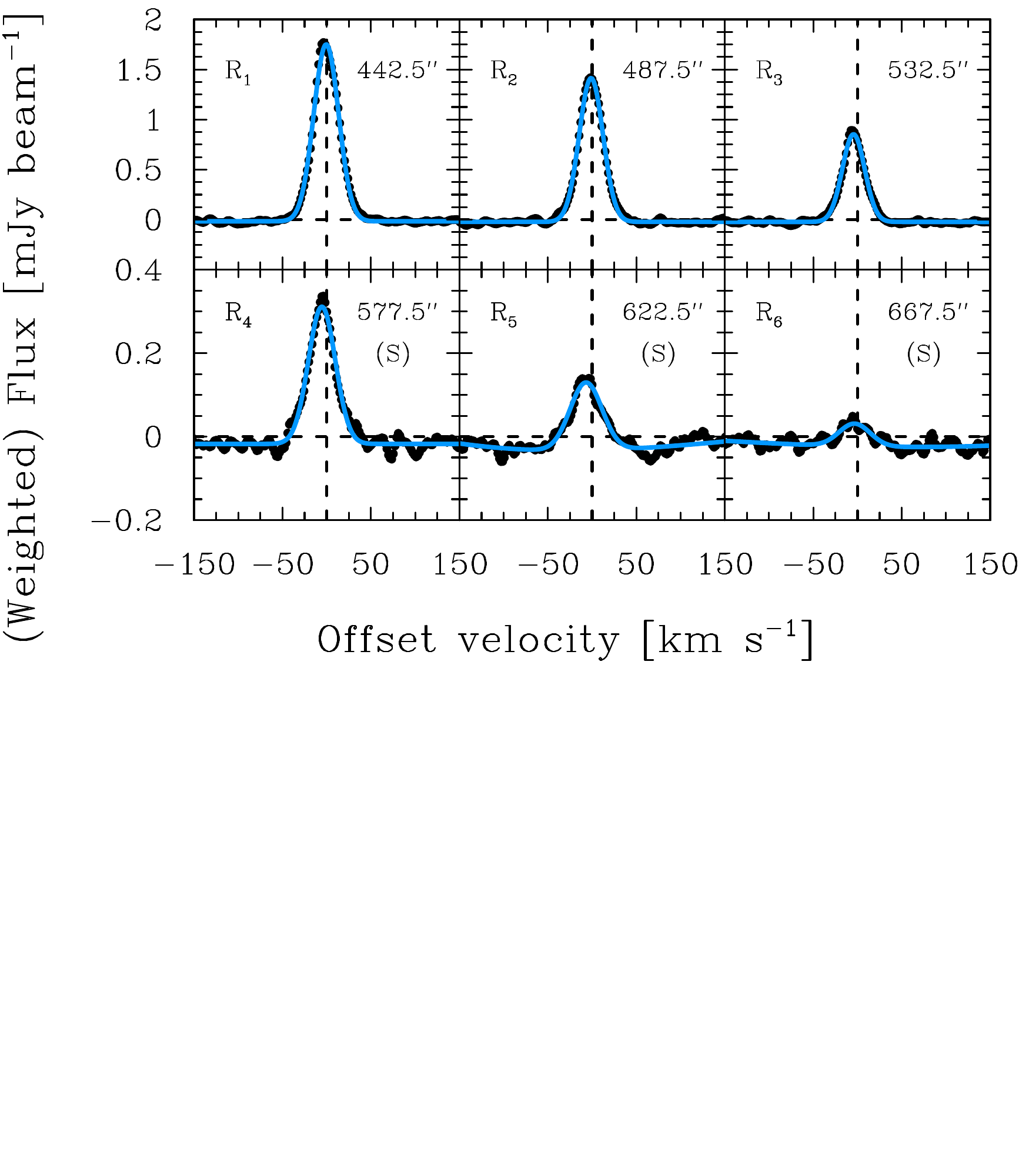} \\
      \includegraphics[scale= 0.376]{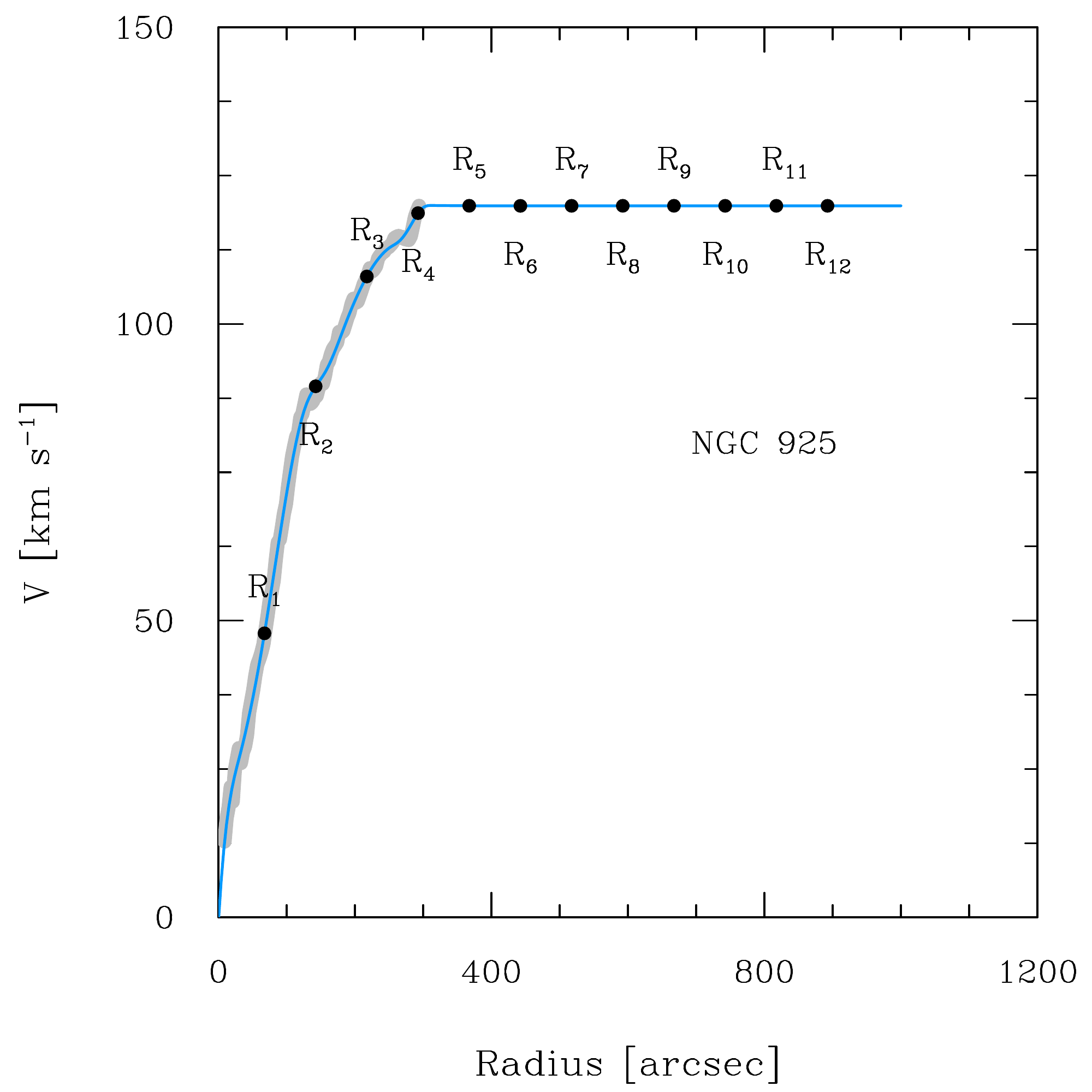} &
      \includegraphics[scale= 0.376]{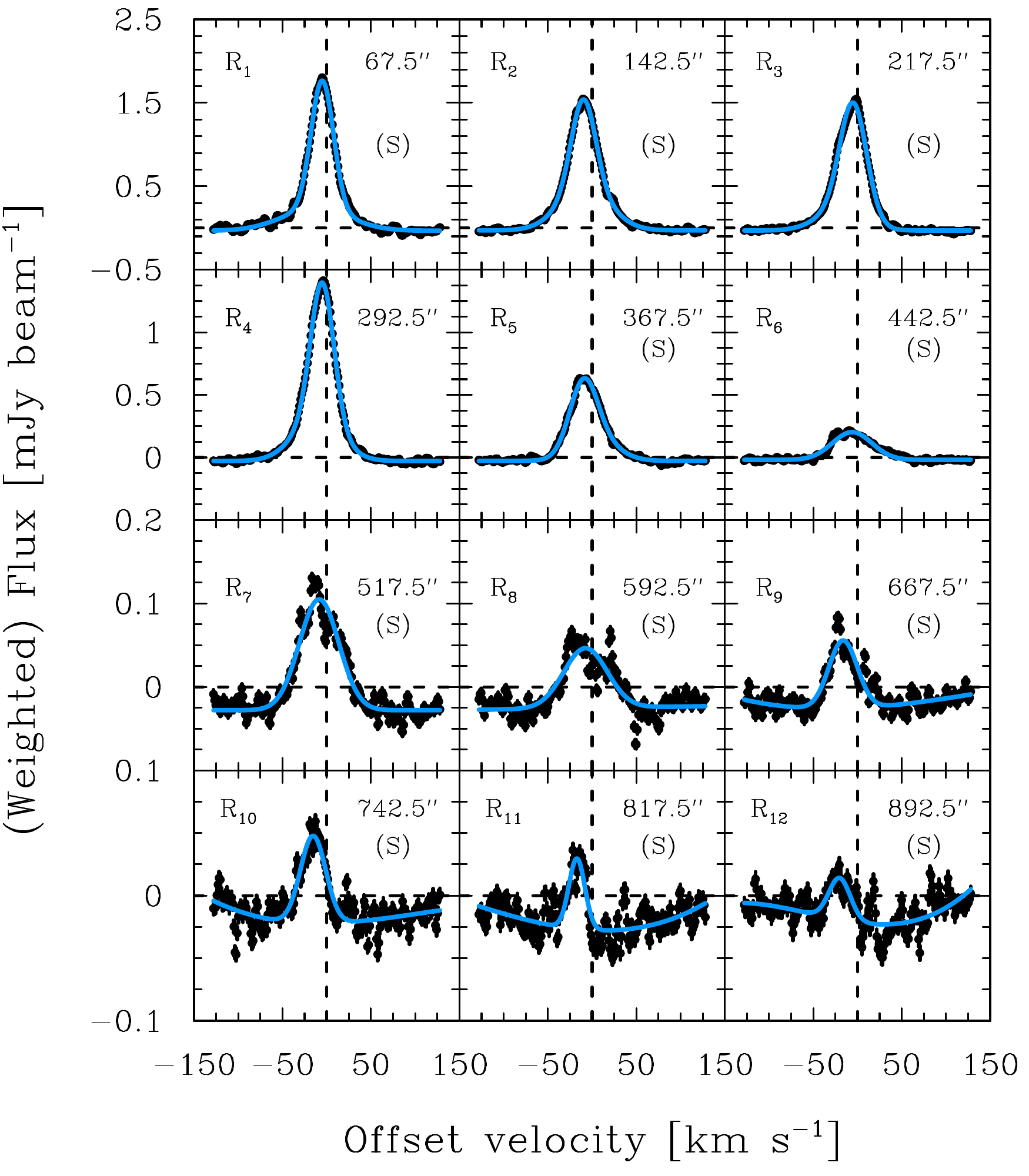}\\
      \includegraphics[scale= 0.376]{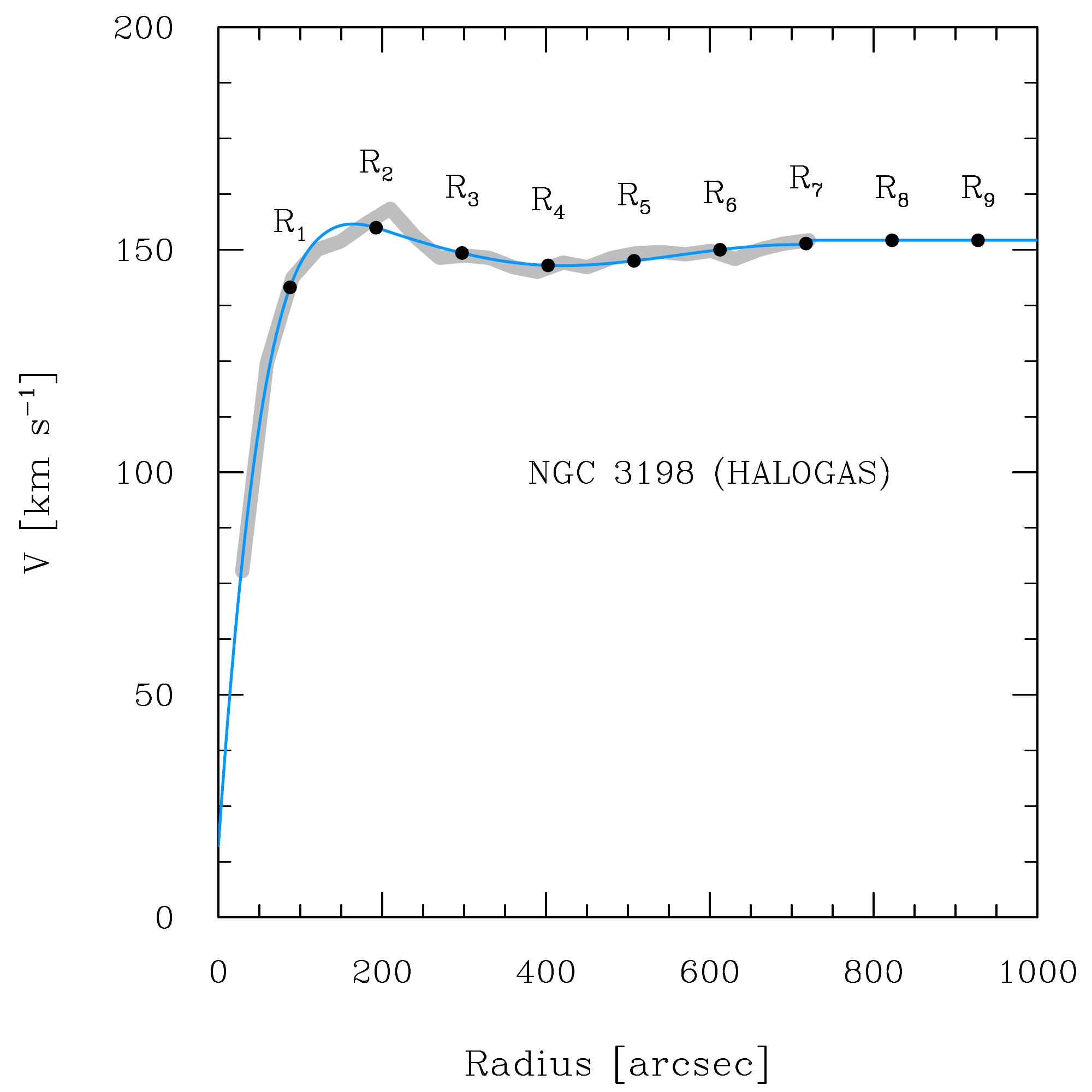} &
      \includegraphics[scale= 0.376]{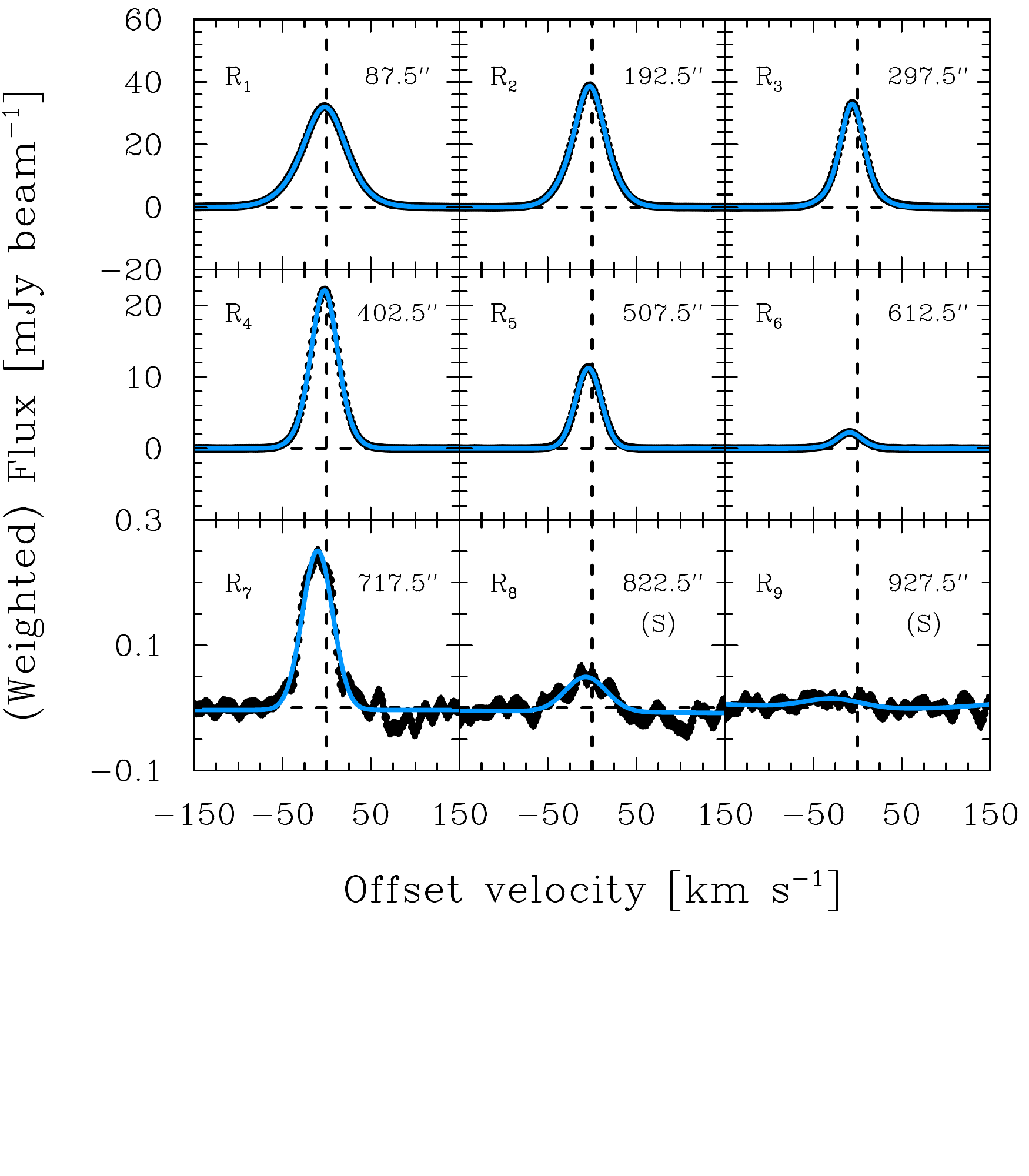} 
    \end{tabular}
    \caption{Continued. \label{stackedspectra}} 
\end{figure*}
\setcounter{figure}{0}
\begin{figure*}
    \begin{tabular}{l l}
      \includegraphics[scale= 0.376]{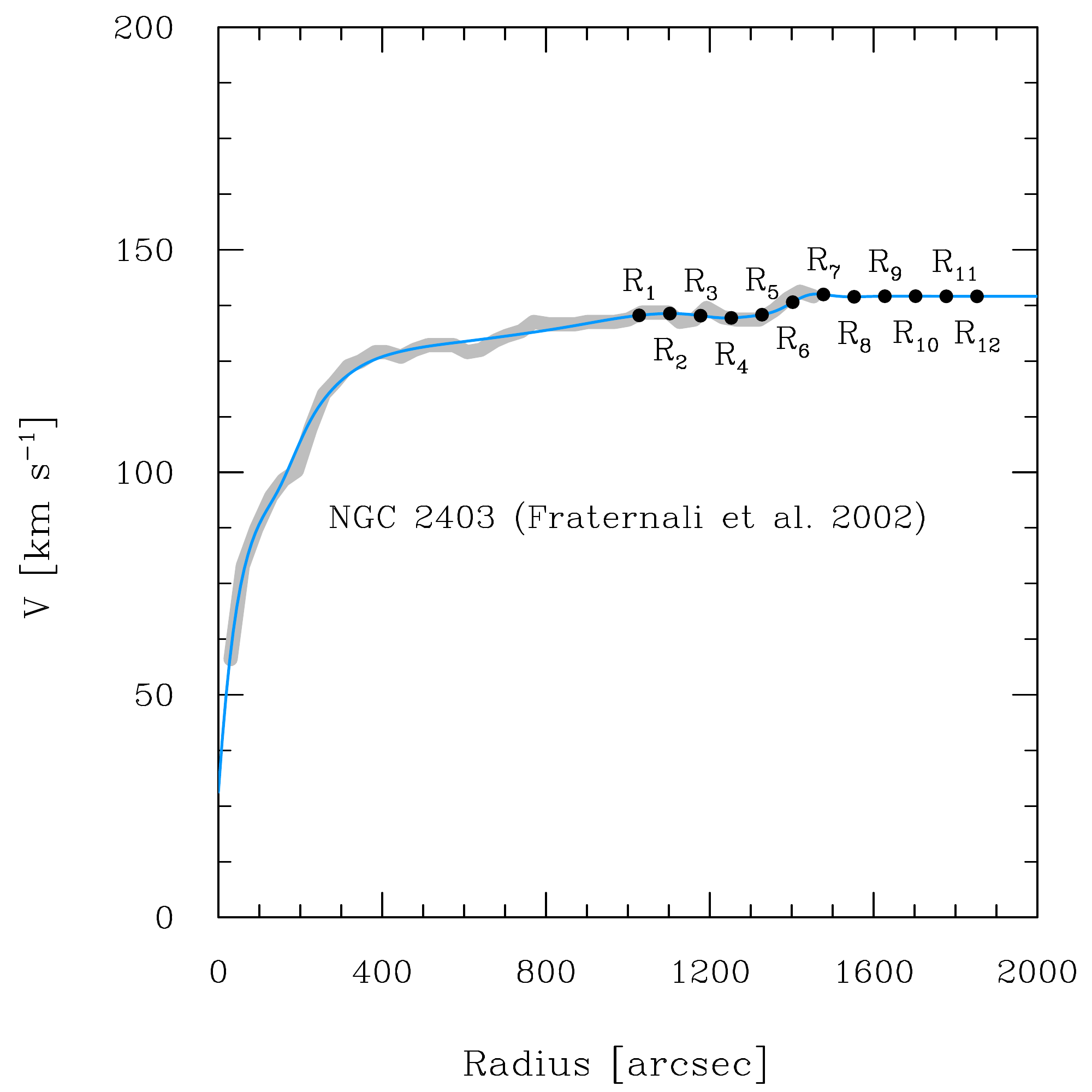} &
      \includegraphics[scale= 0.376]{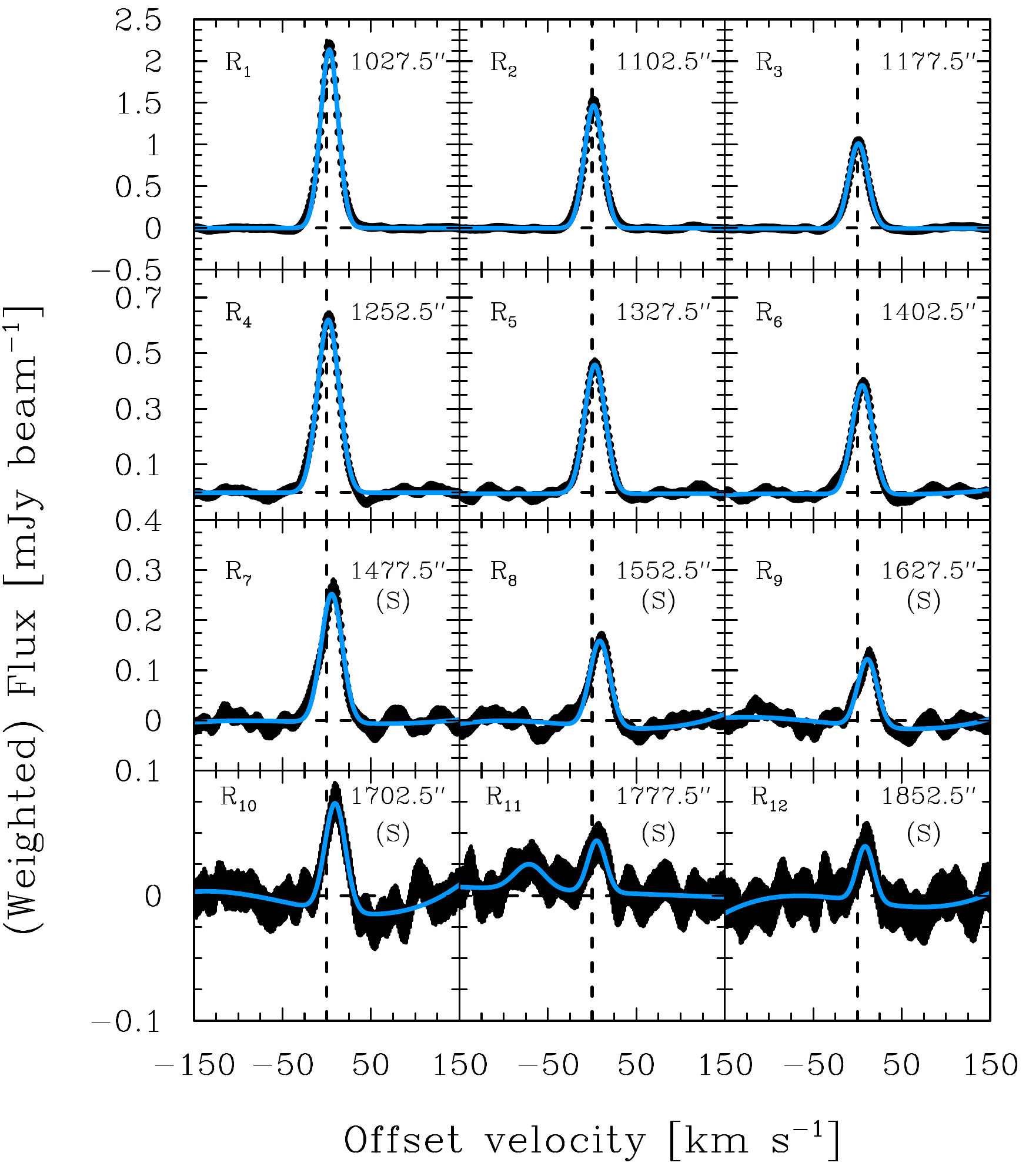} 
    \end{tabular}
    \caption{Continued. \label{stackedspectra}} 
\end{figure*}
\section{Radial column density profiles as a function of azimuthal positions}
Here we present the radial profiles derived inside the annular sectors described in Section~\ref{sec:results}. 
In summary, the galaxies are divided into concentric rings with widths close to the synthesized beam of the observations. 
All rings are then split into 30 degree wide sectors, resulting in annular sectors as shown in Figure~\ref{fig:sectors} 
(only the outer-most annular sectors are shown). The stacking of individual profiles at a given radius are 
done inside the annular sectors to obtain the radial column density profiles shown in 
Figure~\ref{fig:sectors}.
\setcounter{figure}{0}
\begin{figure*}
    \begin{tabular}{l l l}
     \includegraphics[scale= 0.31]{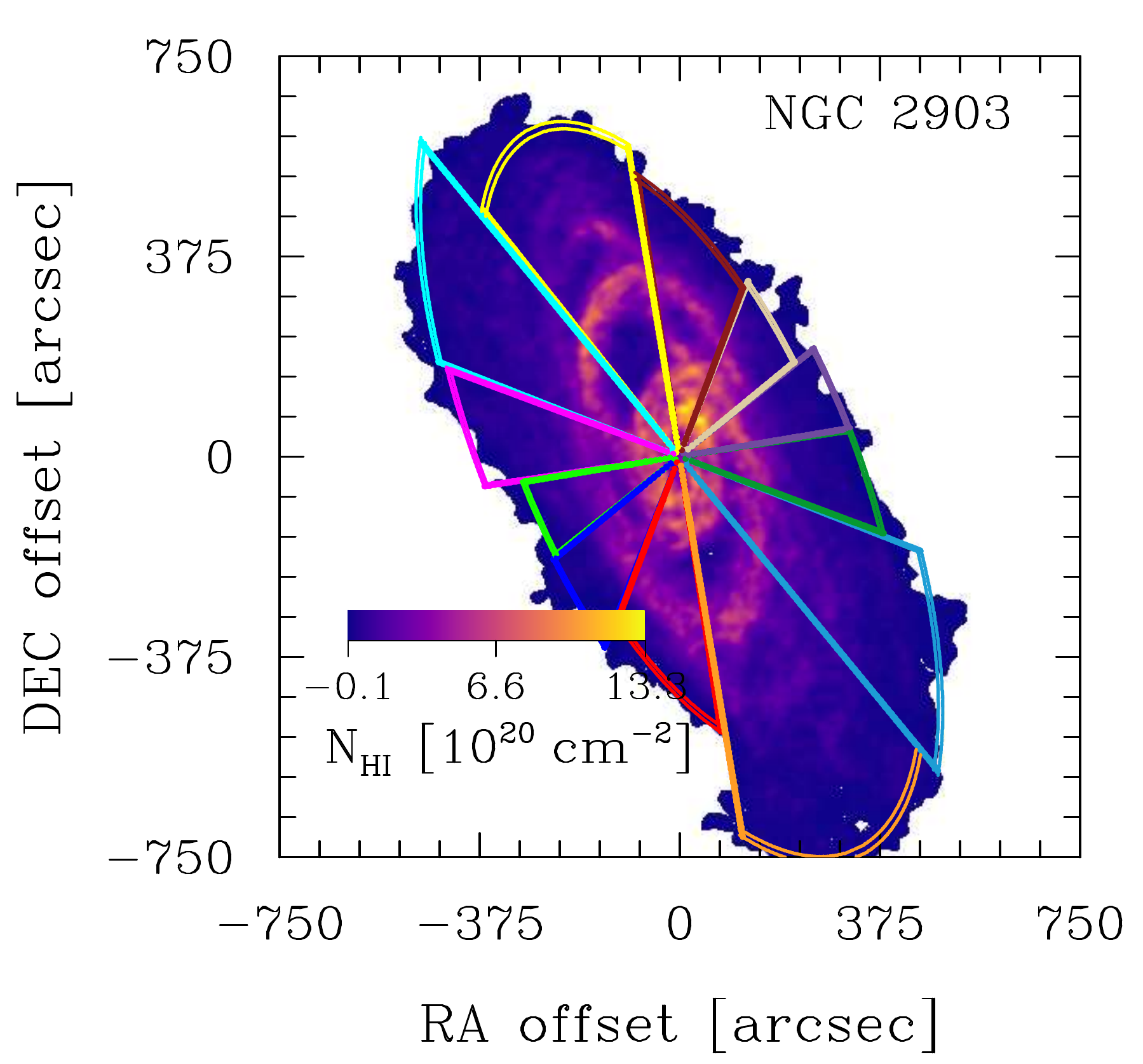} &
     \includegraphics[scale= 0.31]{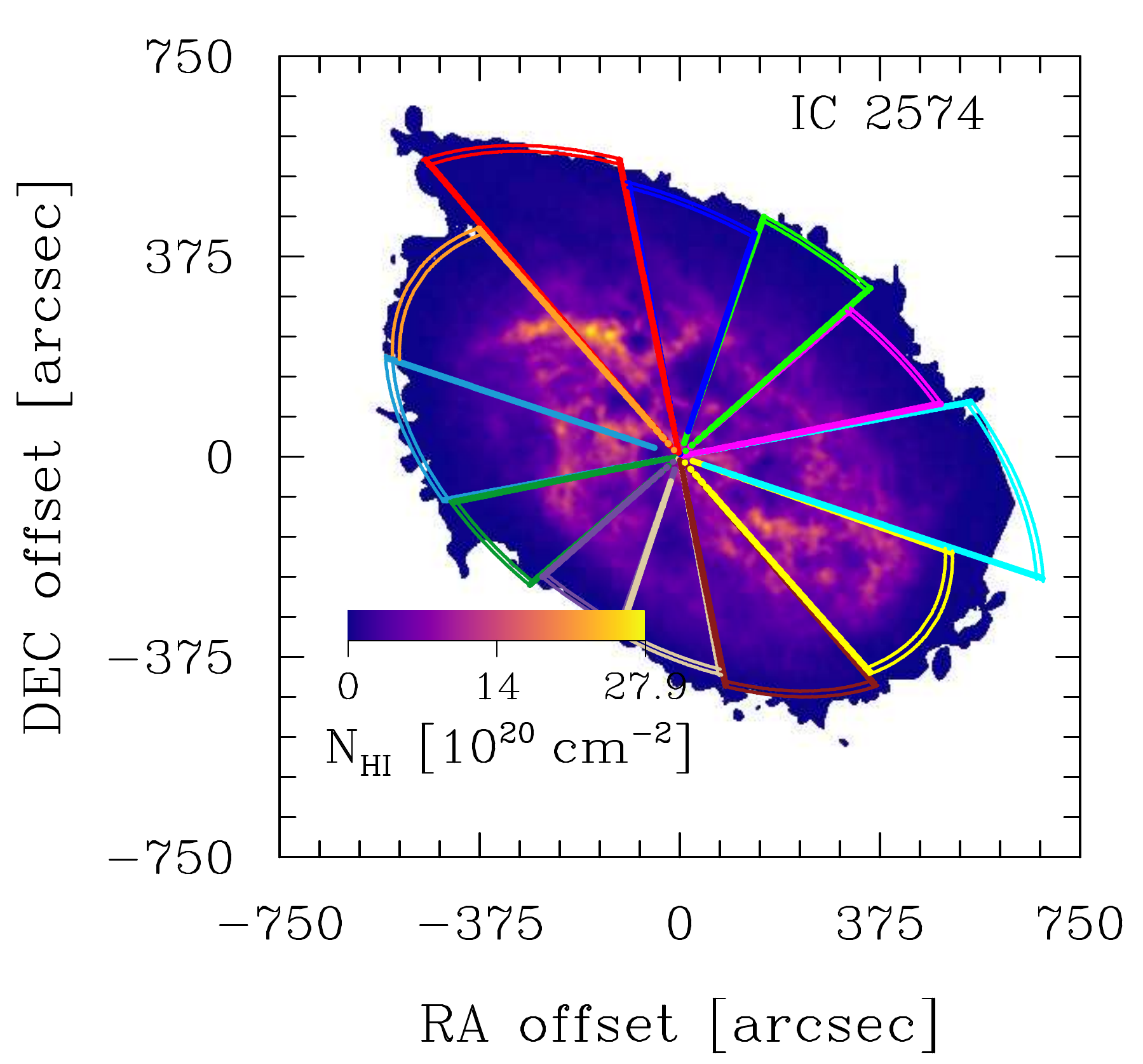} &
     \includegraphics[scale= 0.31]{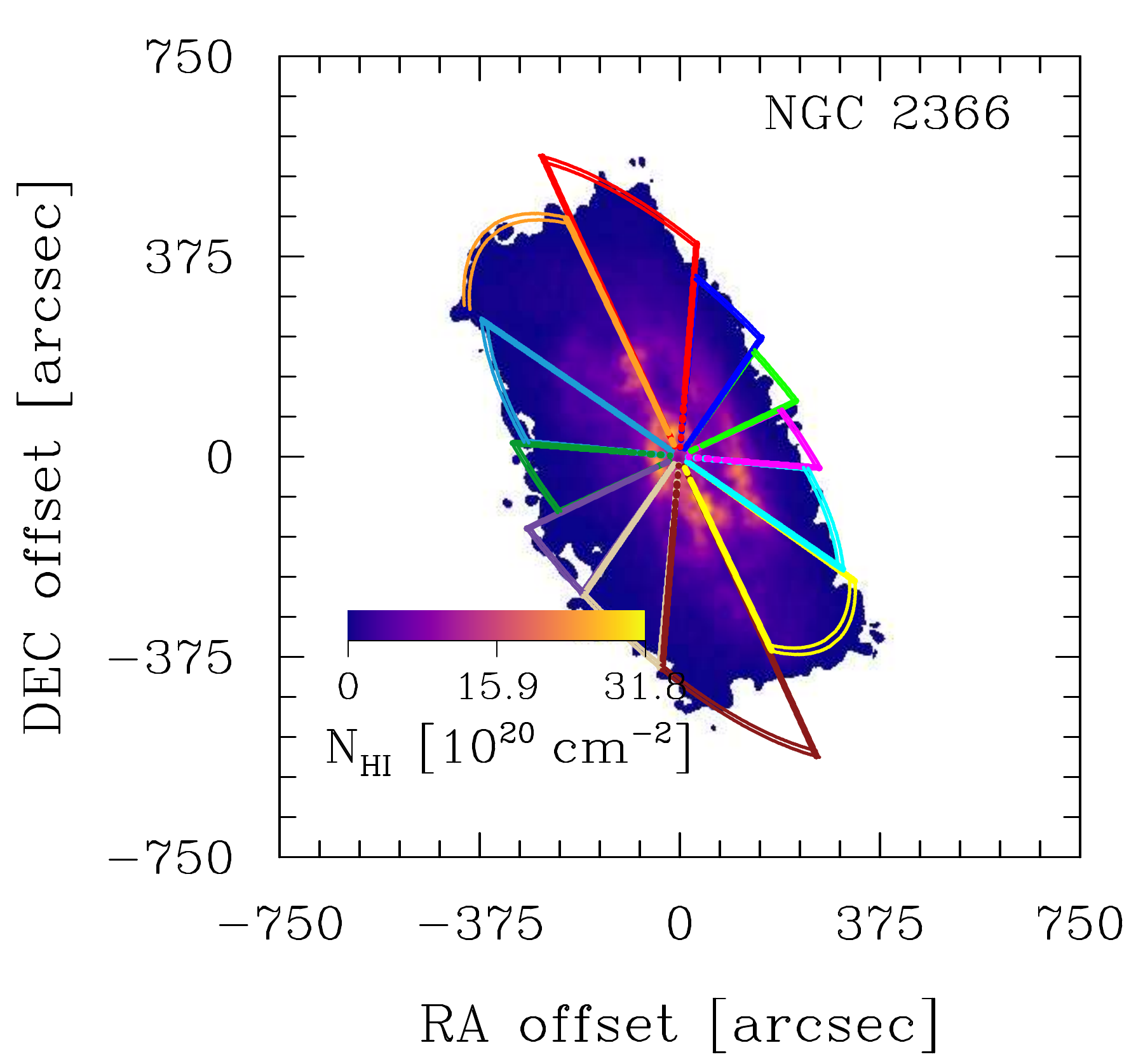} \\
     \includegraphics[scale= 0.31]{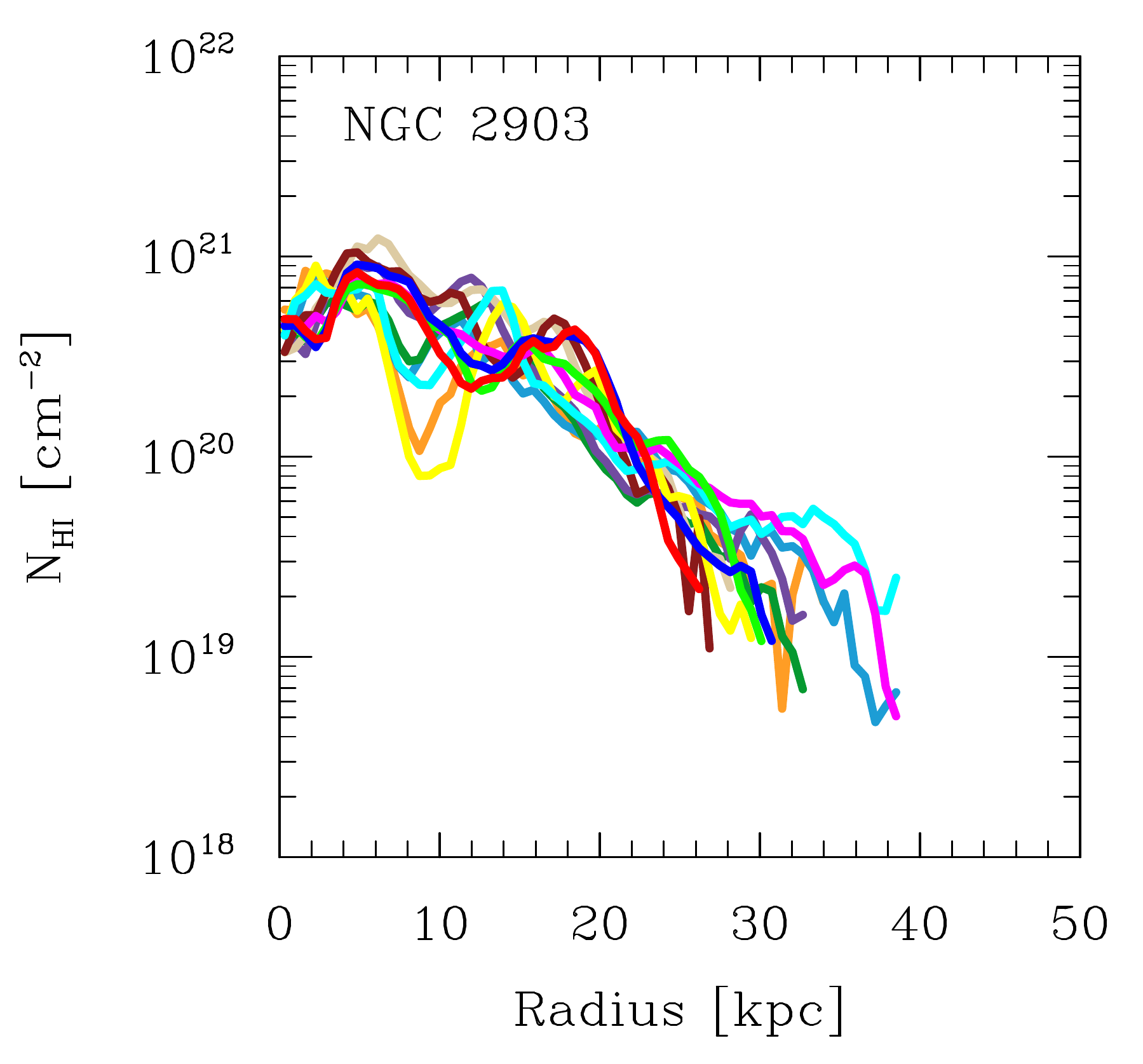} & 
     \includegraphics[scale= 0.31]{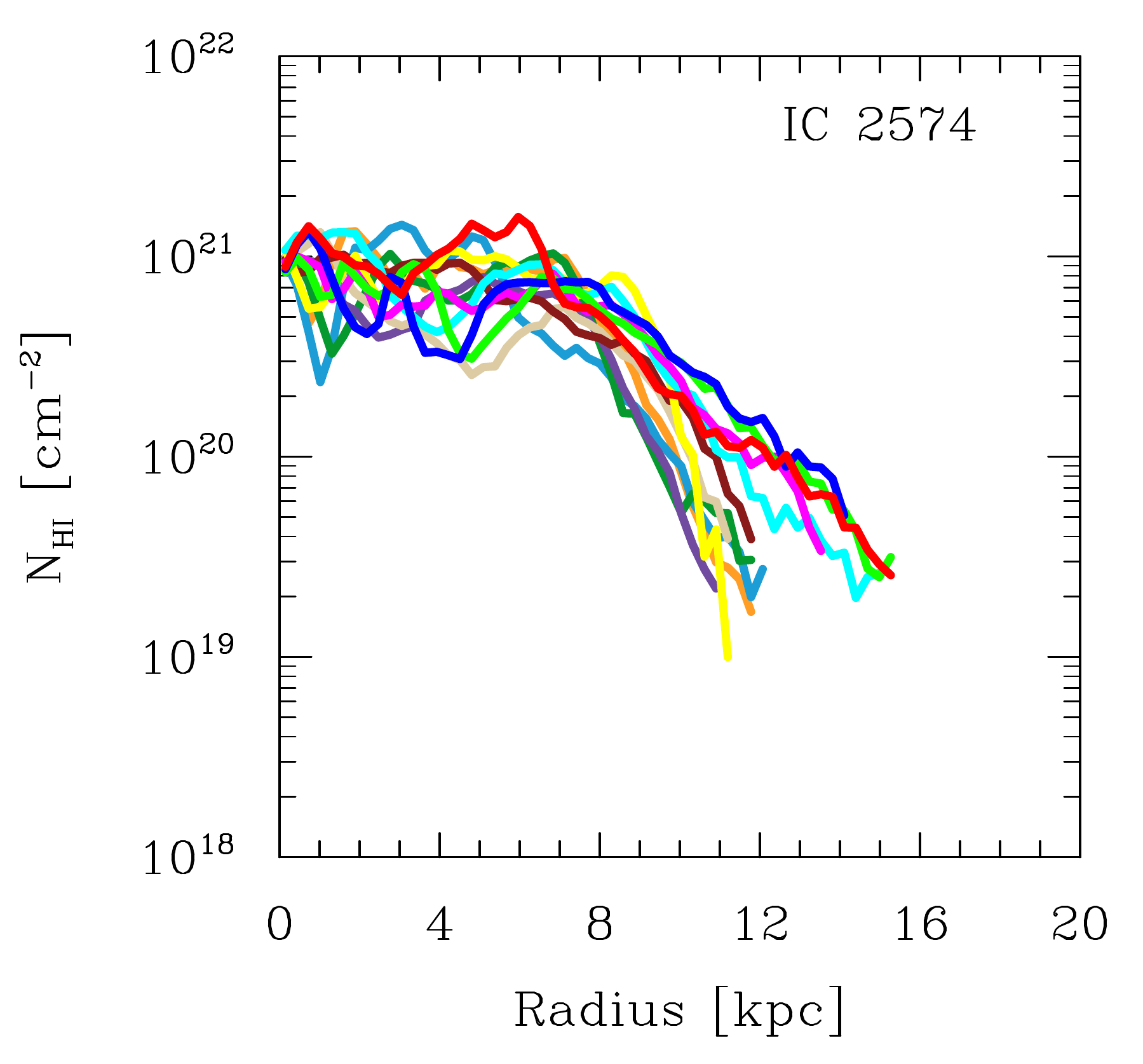} &
      \includegraphics[scale= 0.31]{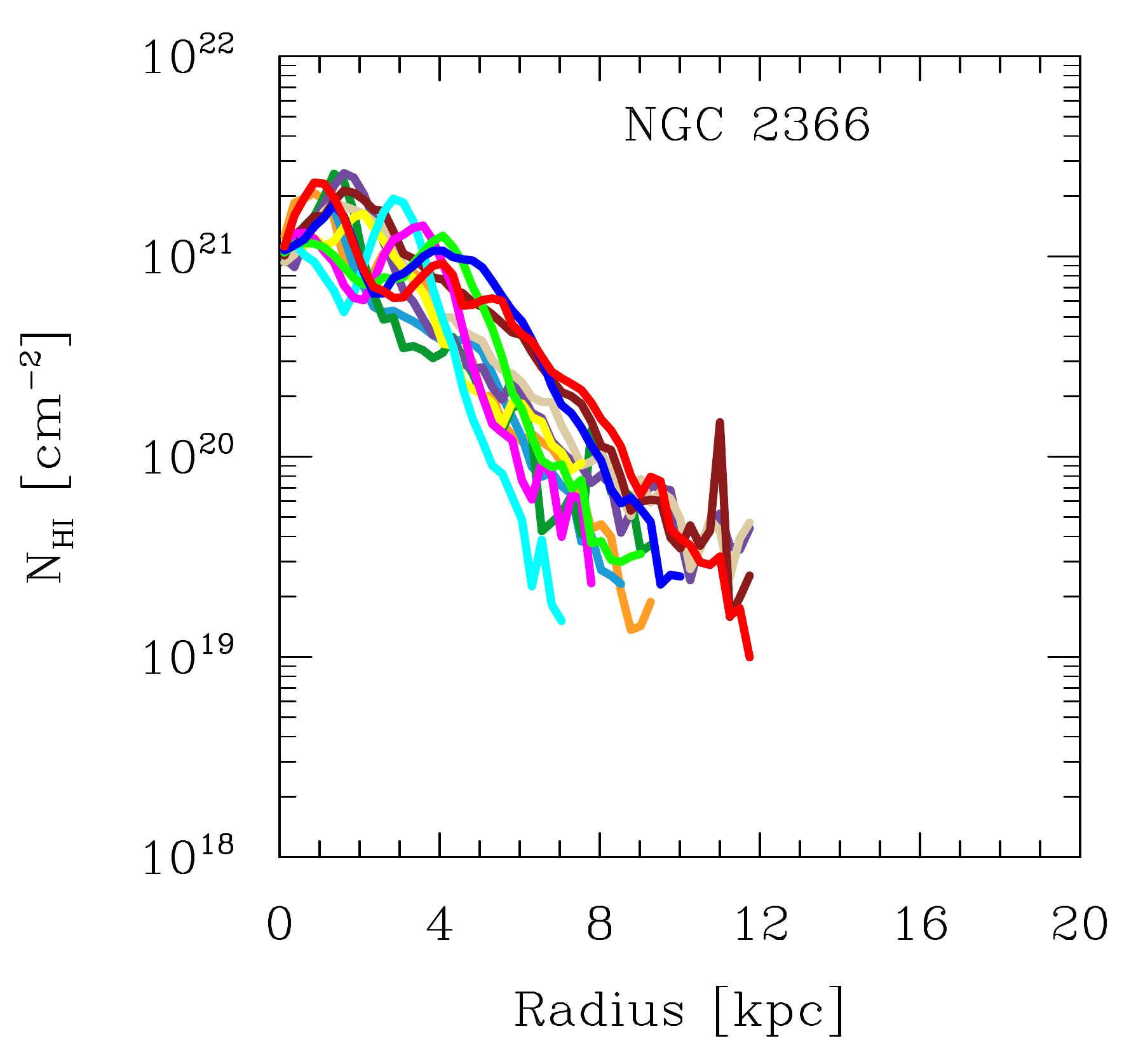}  
    \end{tabular}
    \caption{Odd rows: H\,{\sc i} column density maps overlaid with the annular sectors inside which the radial profiles 
    are derived (only the outermost annular sector is shown). 
    Even rows: the radial H\,{\sc i} column density profiles of the sample galaxies derived inside the different sectors. Different colors 
    represent the different sectors used to derive the radial profiles.} 
    \label{fig:sectors}
\end{figure*}
\setcounter{figure}{0}
\begin{figure*}
    \begin{tabular}{l l l}
       \includegraphics[scale= 0.31]{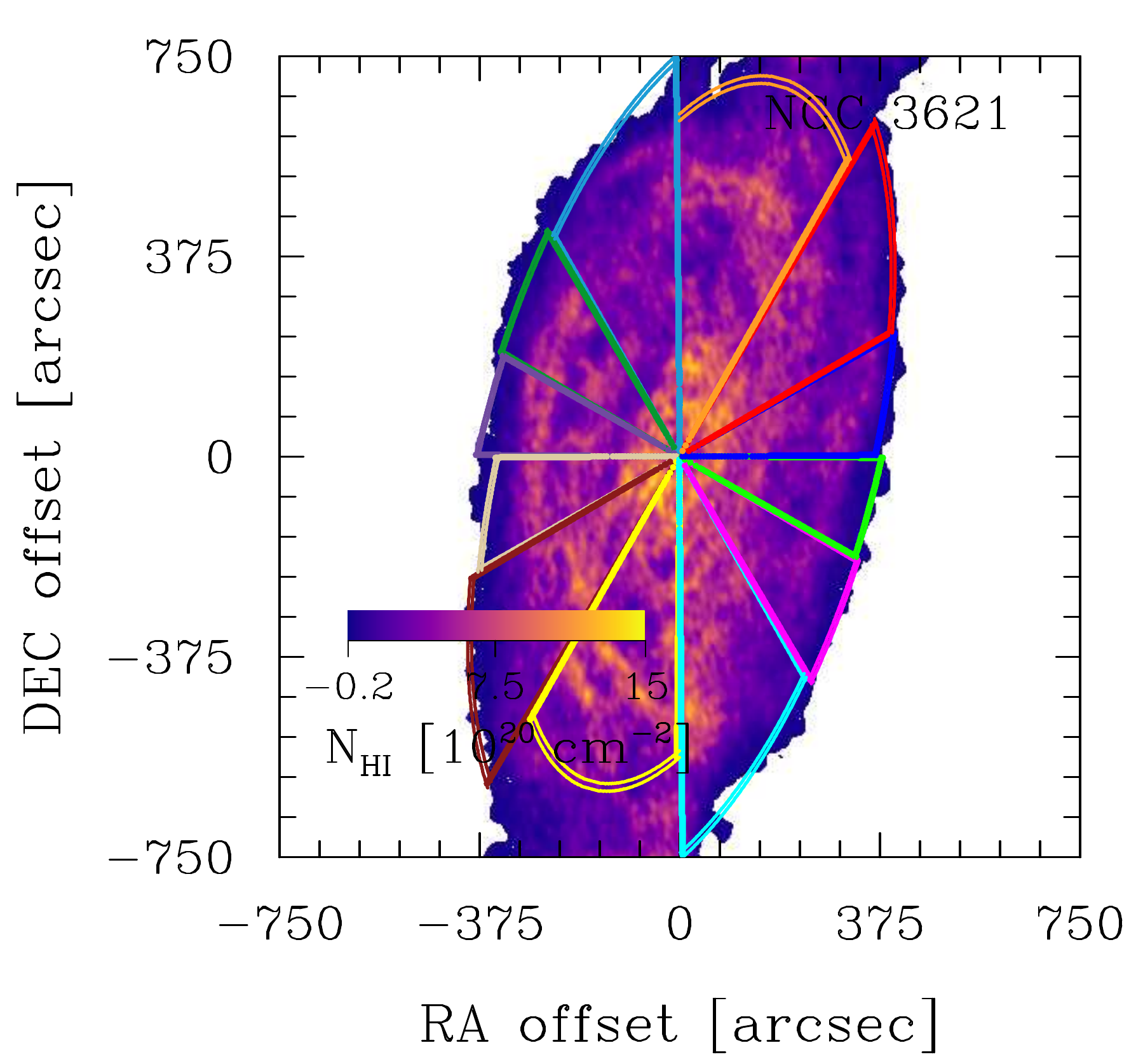} &
       \includegraphics[scale= 0.31]{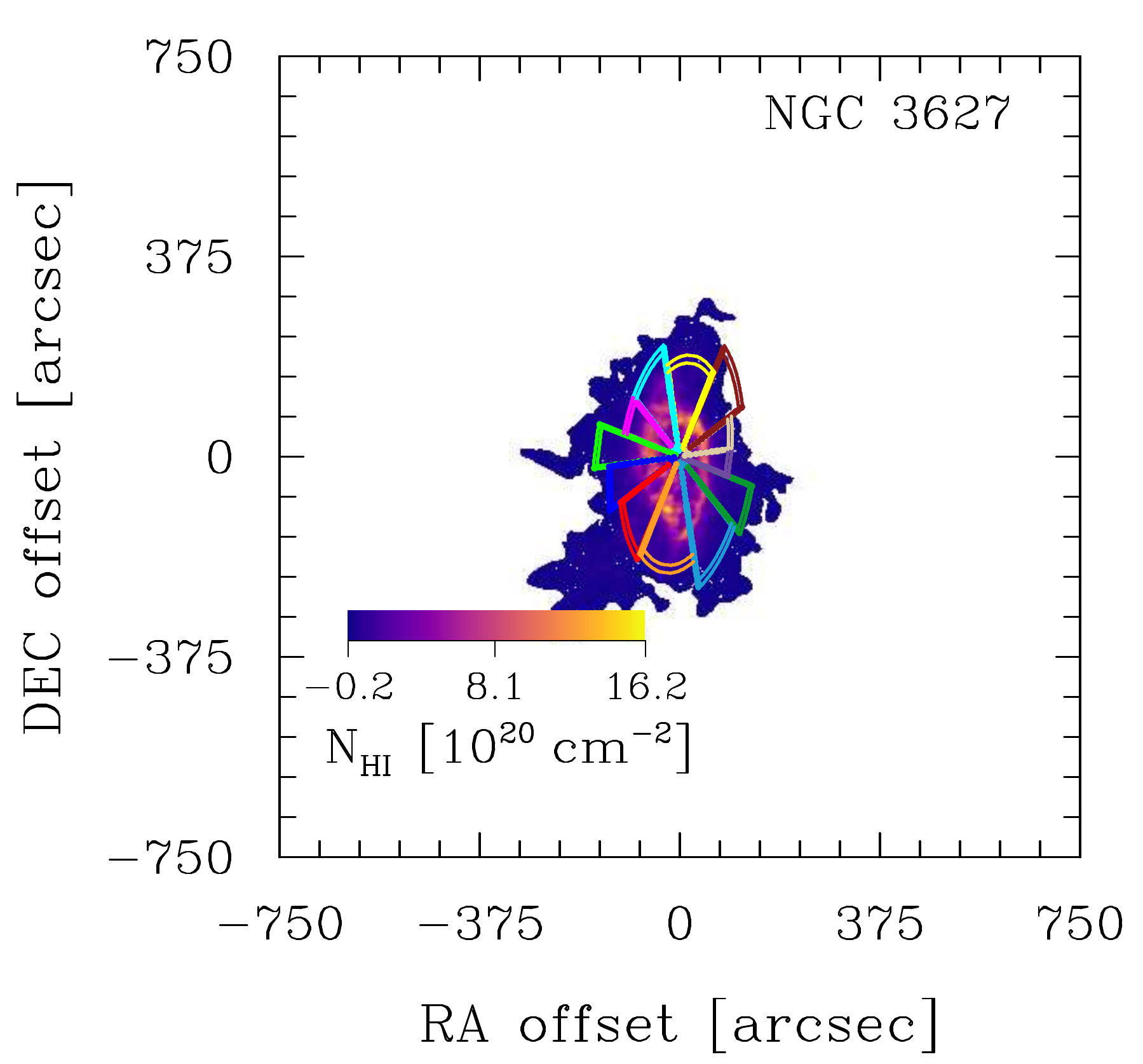} &
       \includegraphics[scale= 0.31]{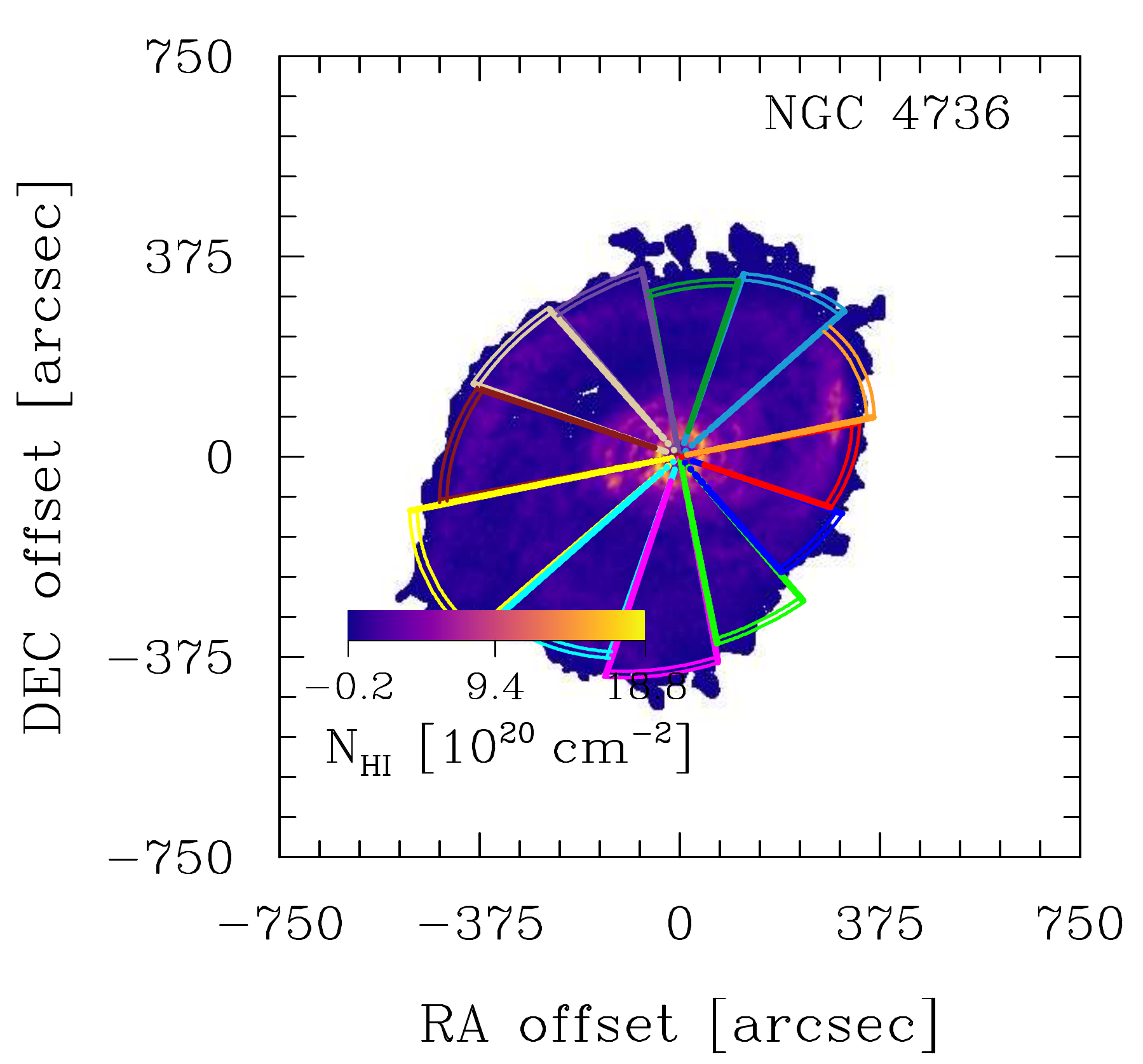} \\
       \includegraphics[scale= 0.31]{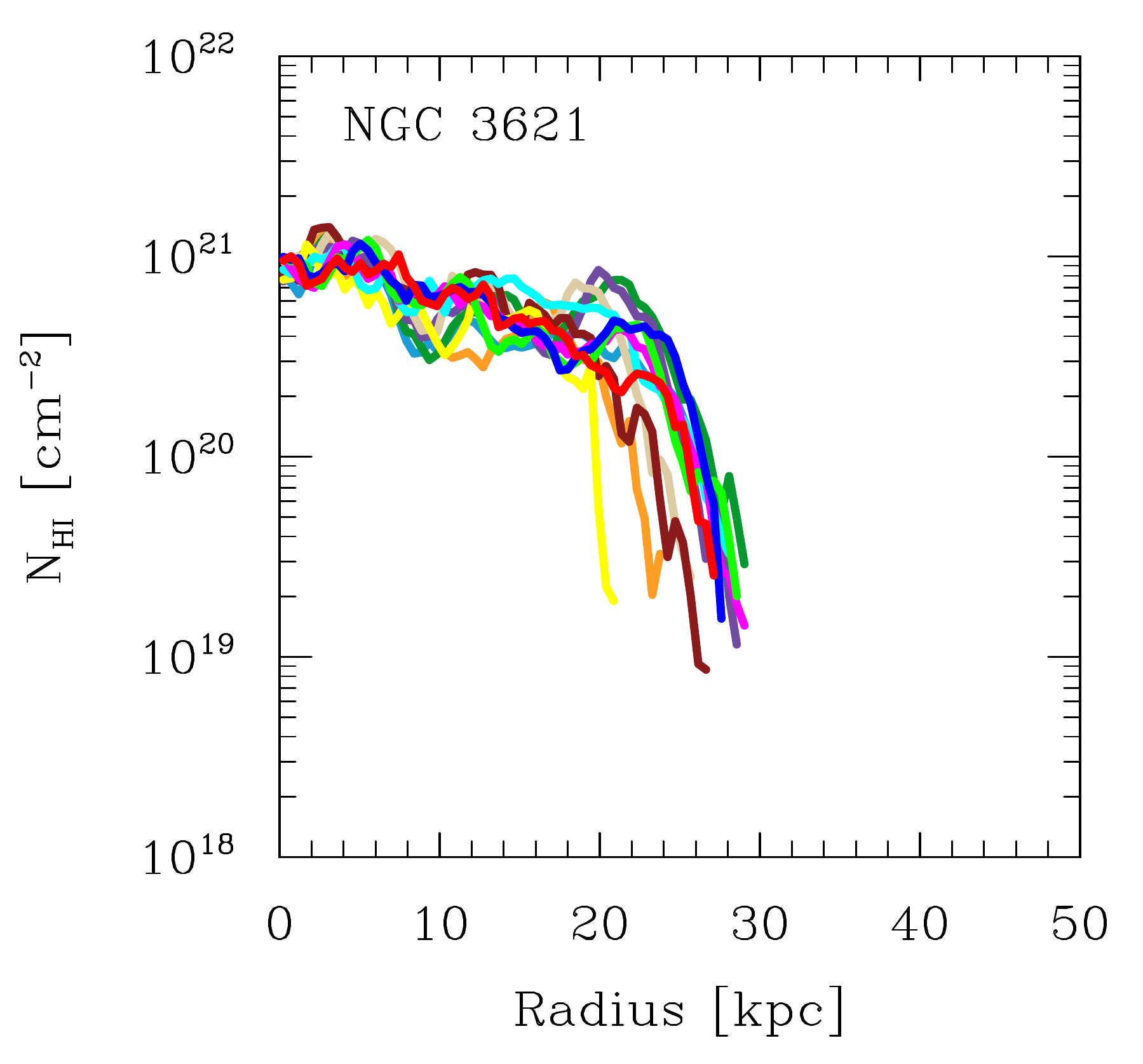} &
       \includegraphics[scale= 0.31]{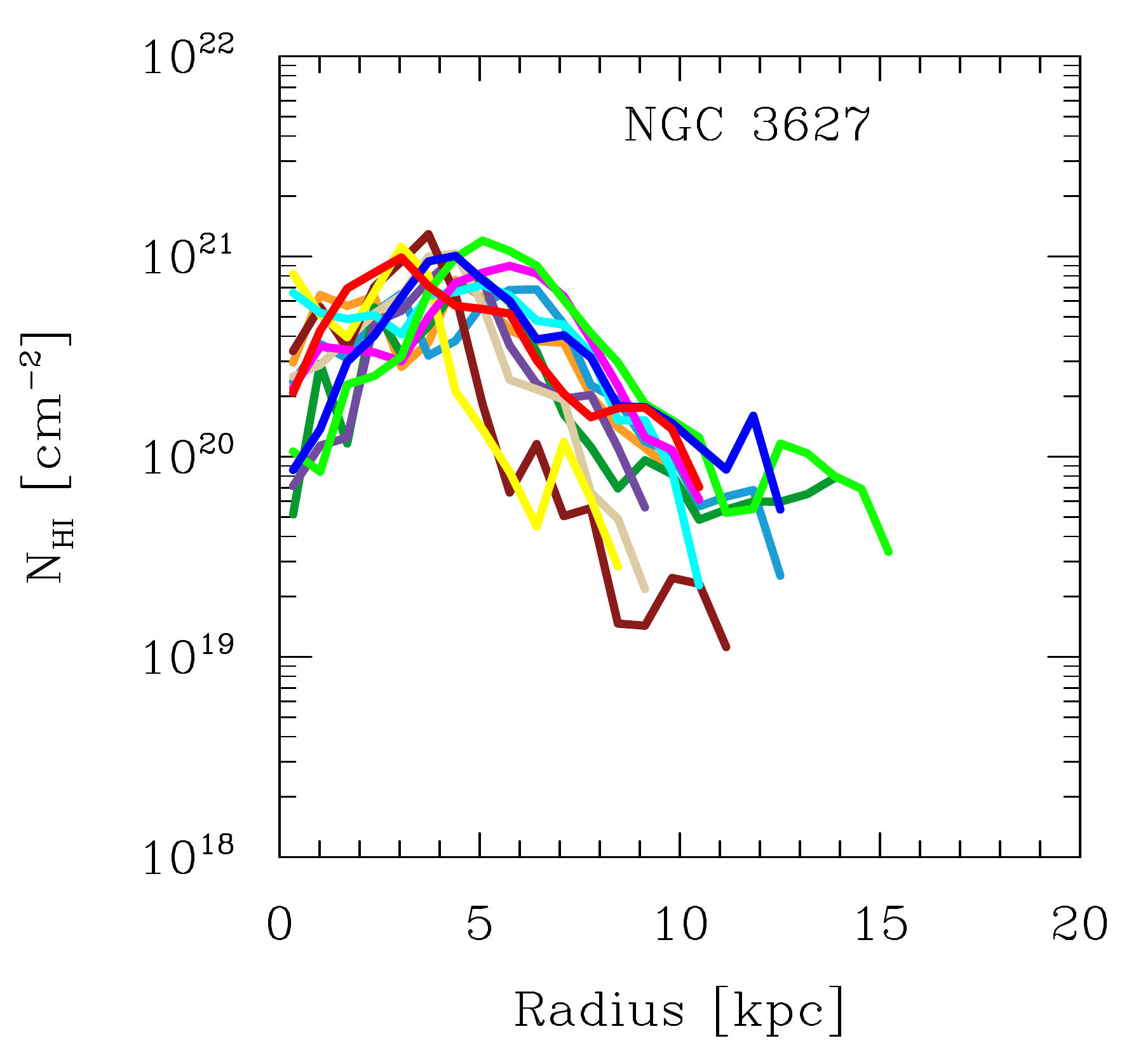} &
       \includegraphics[scale= 0.31]{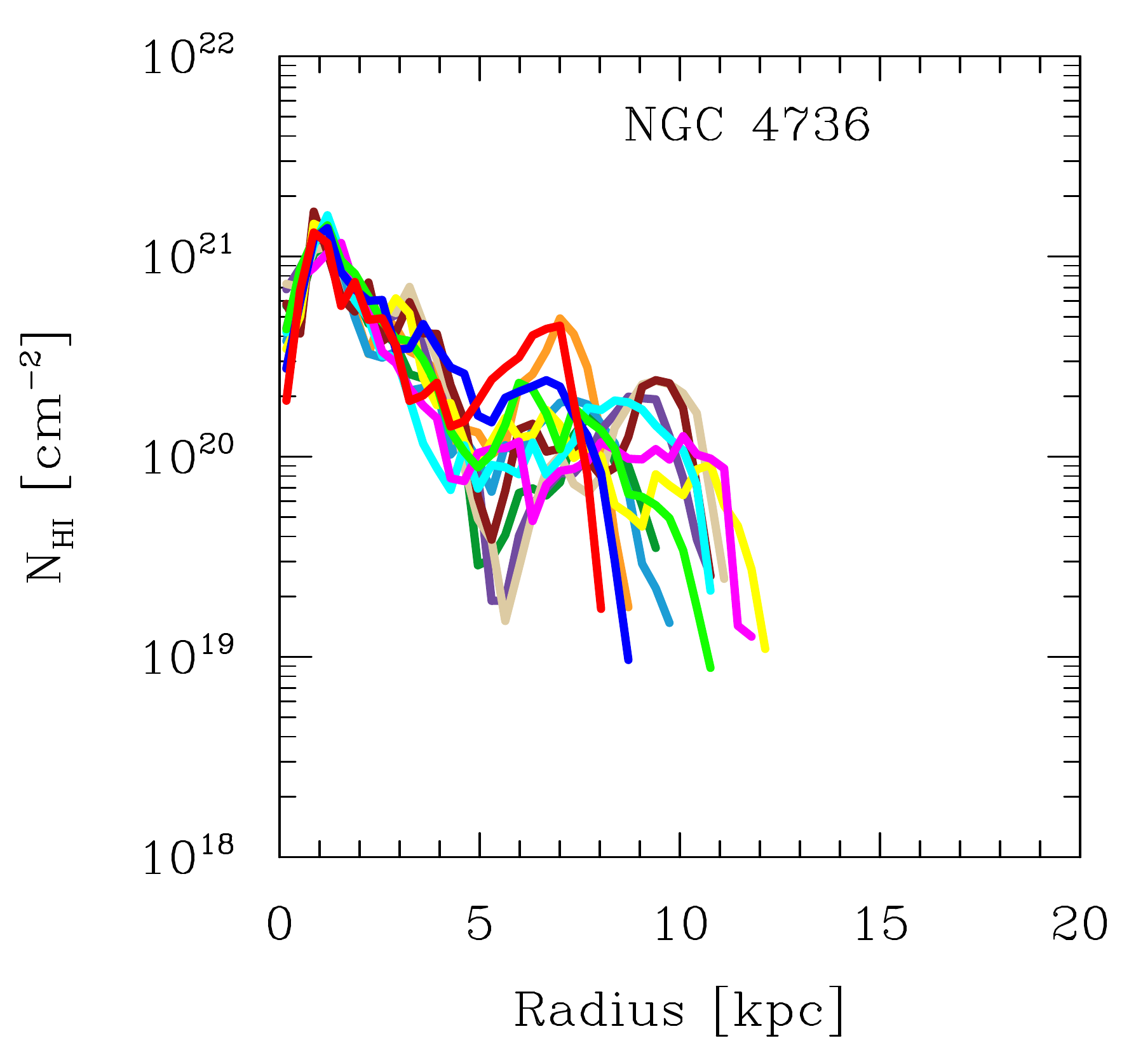} \\
       \includegraphics[scale= 0.31]{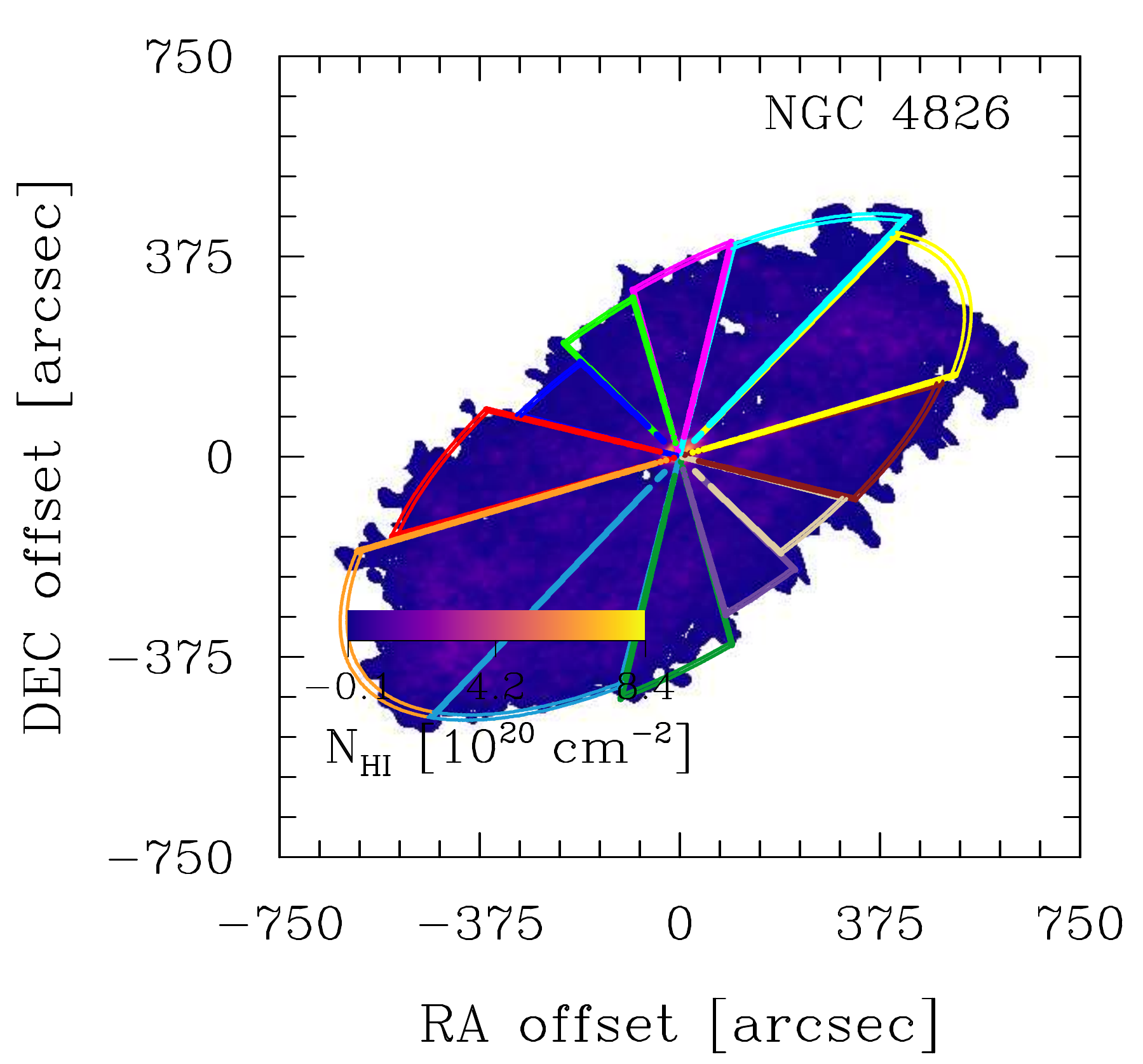} &
       \includegraphics[scale= 0.31]{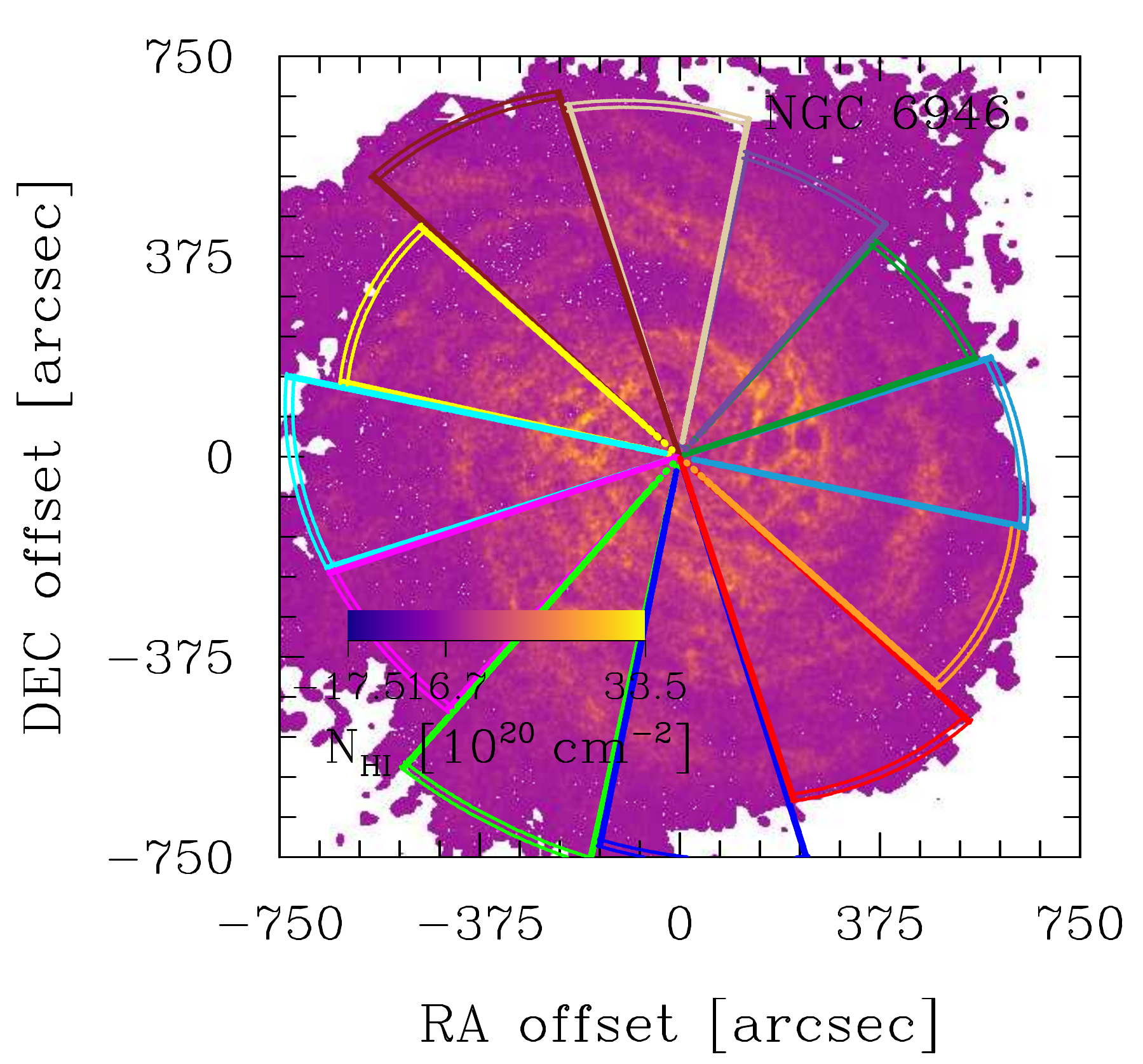} &
       \includegraphics[scale= 0.31]{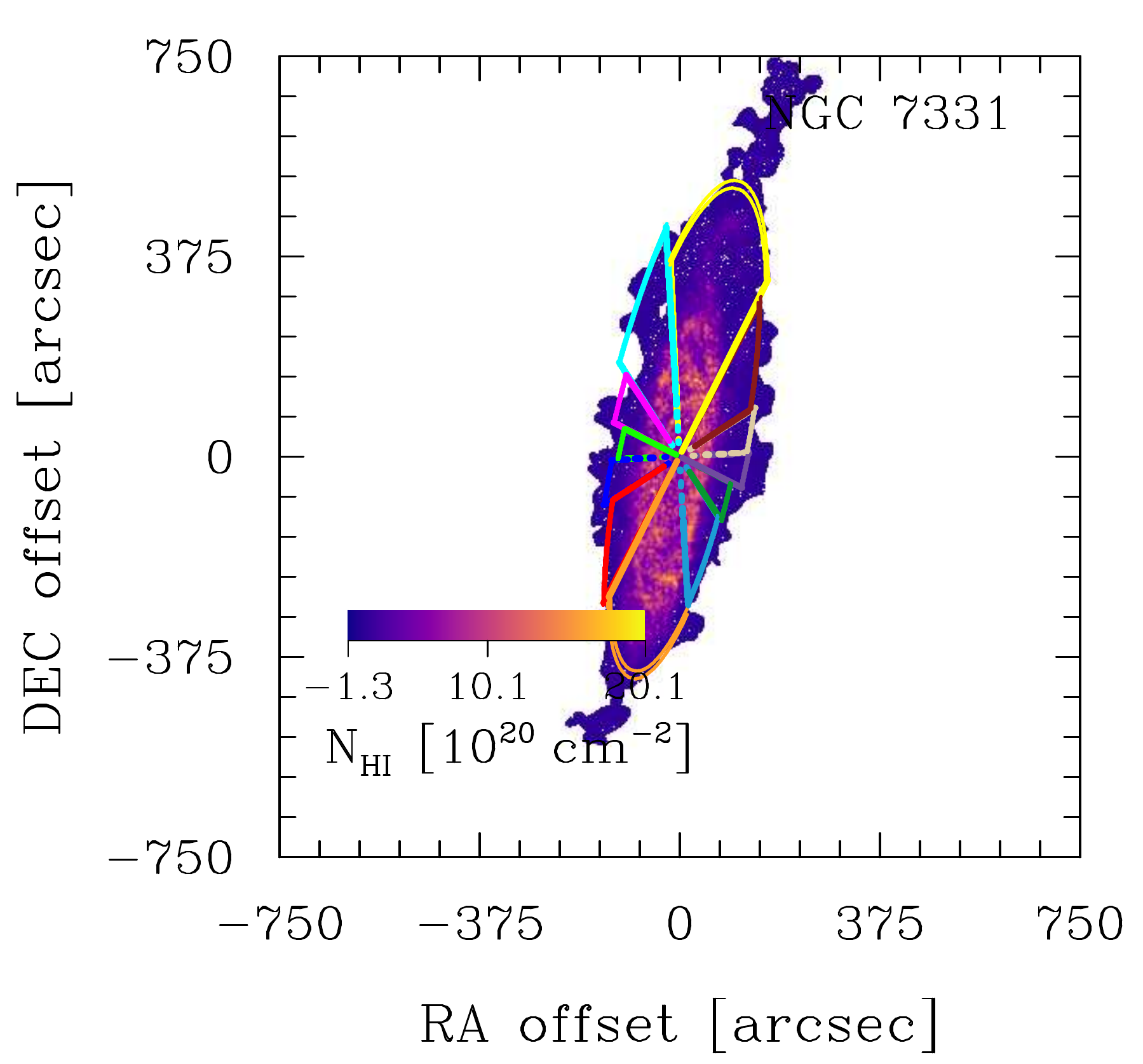} \\
       \includegraphics[scale= 0.31]{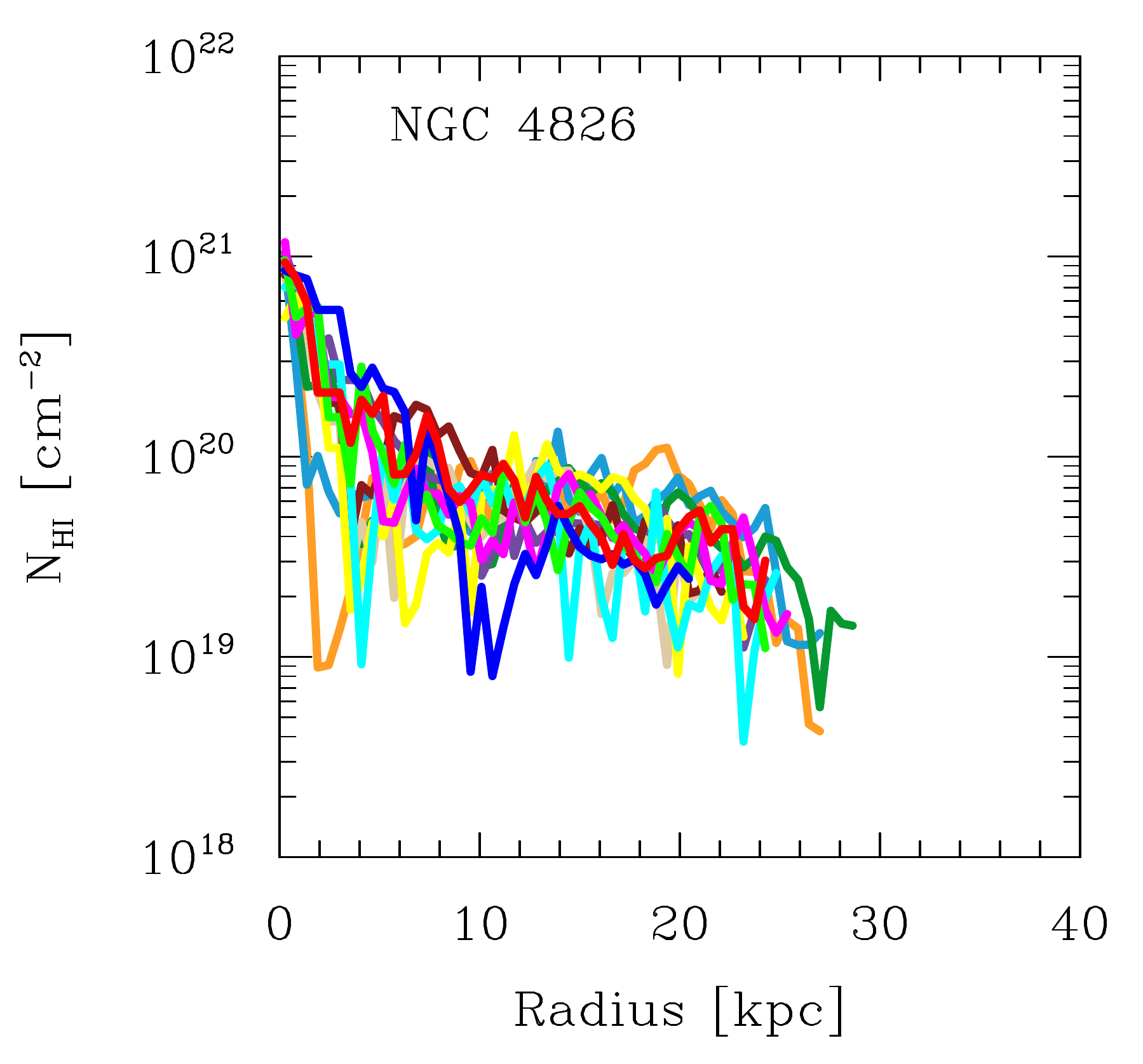} &
       \includegraphics[scale= 0.31]{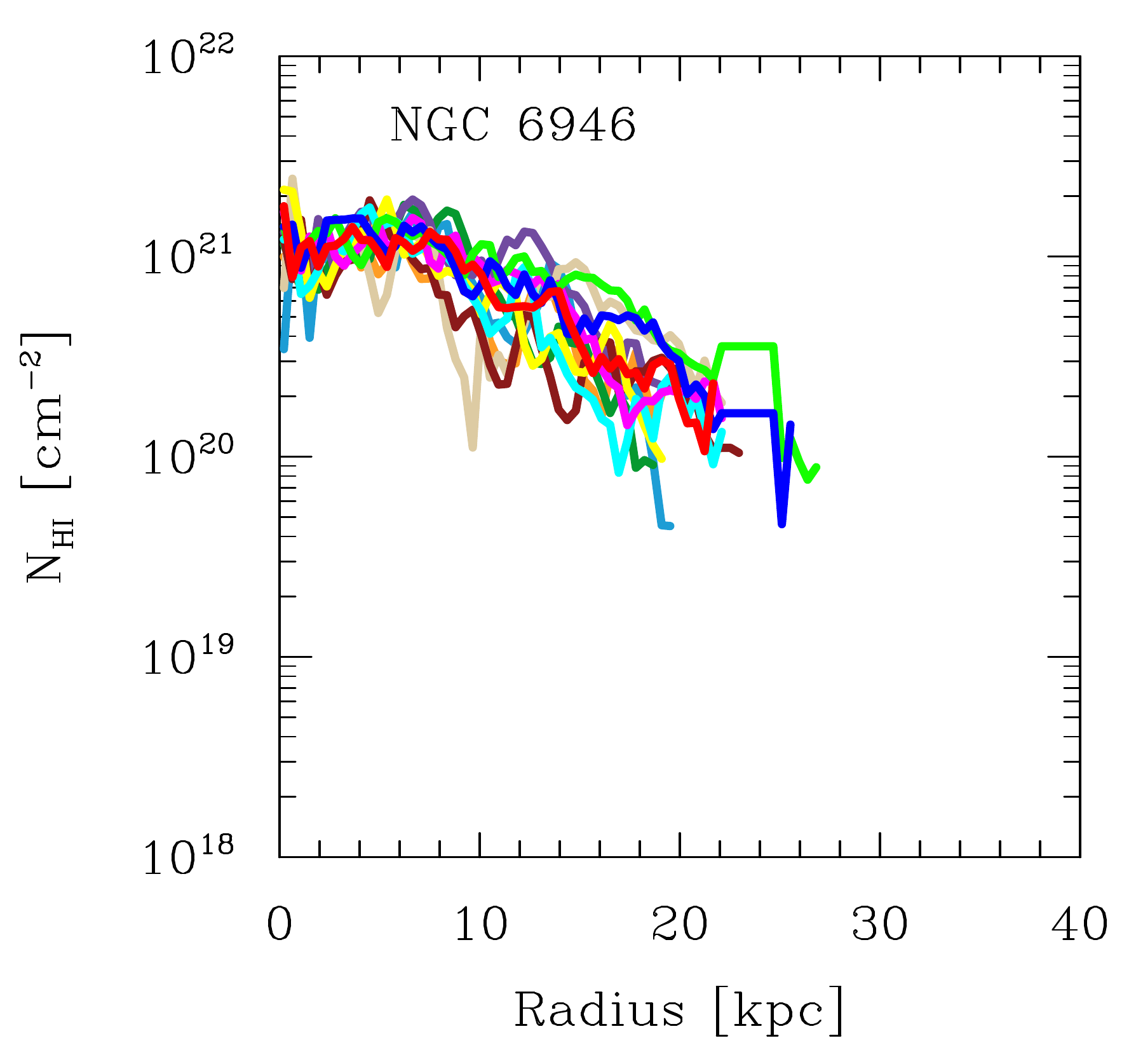} &
       \includegraphics[scale= 0.31]{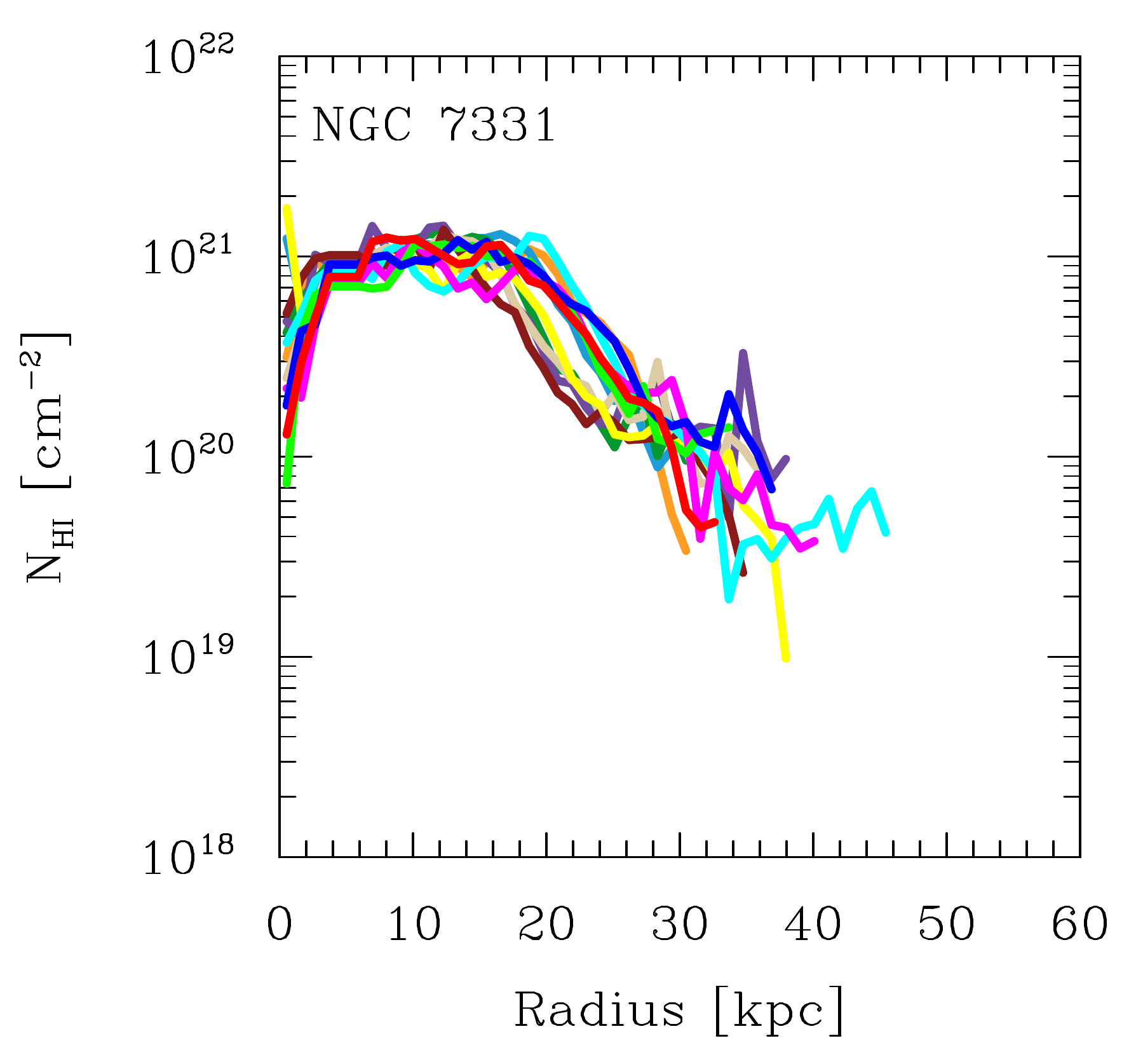} 
    \end{tabular}
    \caption{ Continued. \label{fig:sectors}} 
\end{figure*}
\setcounter{figure}{0}
\begin{figure*}
    \begin{tabular}{l l l}
       \includegraphics[scale= 0.31]{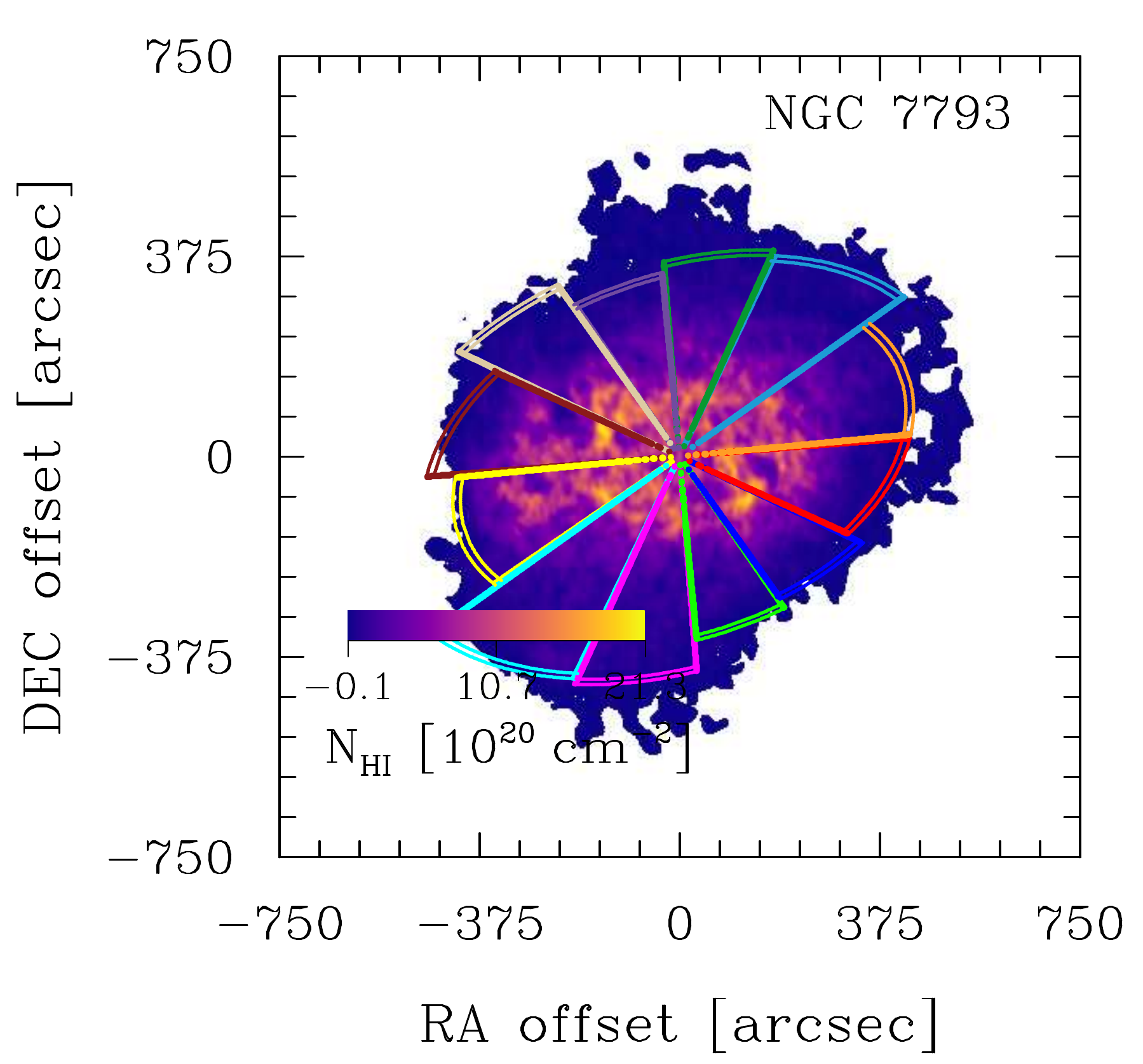} &
       \includegraphics[scale= 0.31]{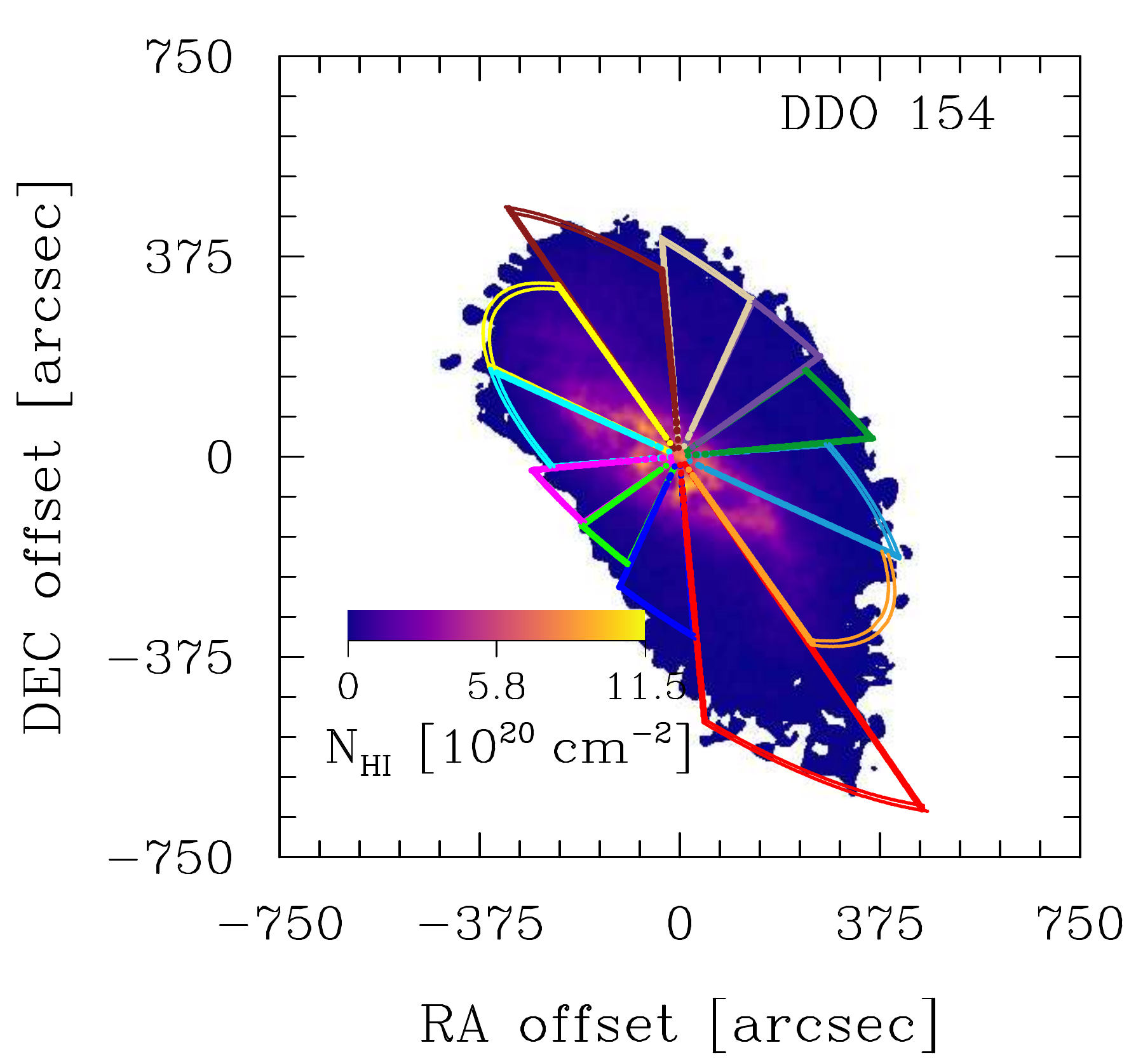} &
       \includegraphics[scale= 0.31]{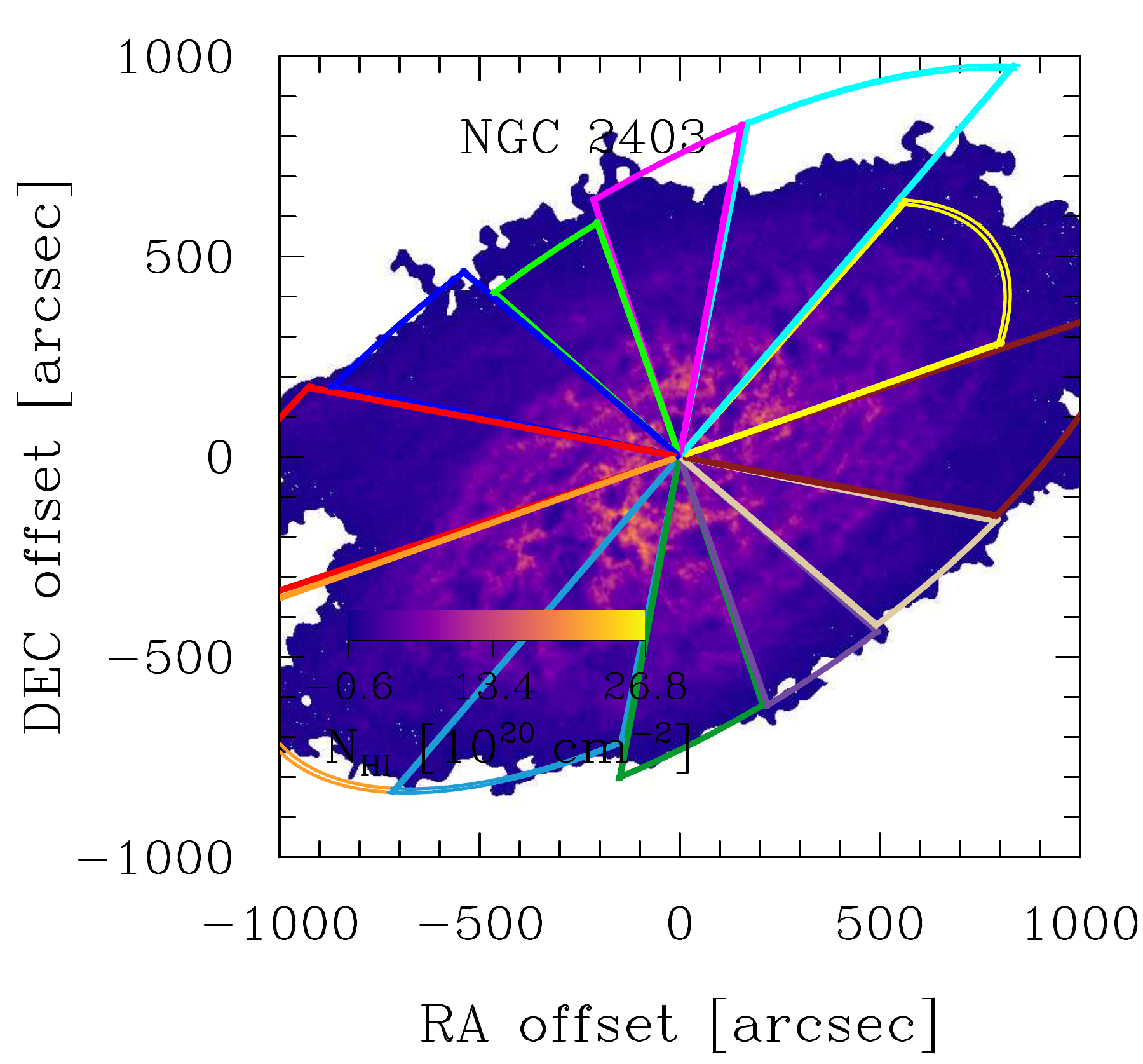} \\
       \includegraphics[scale= 0.31]{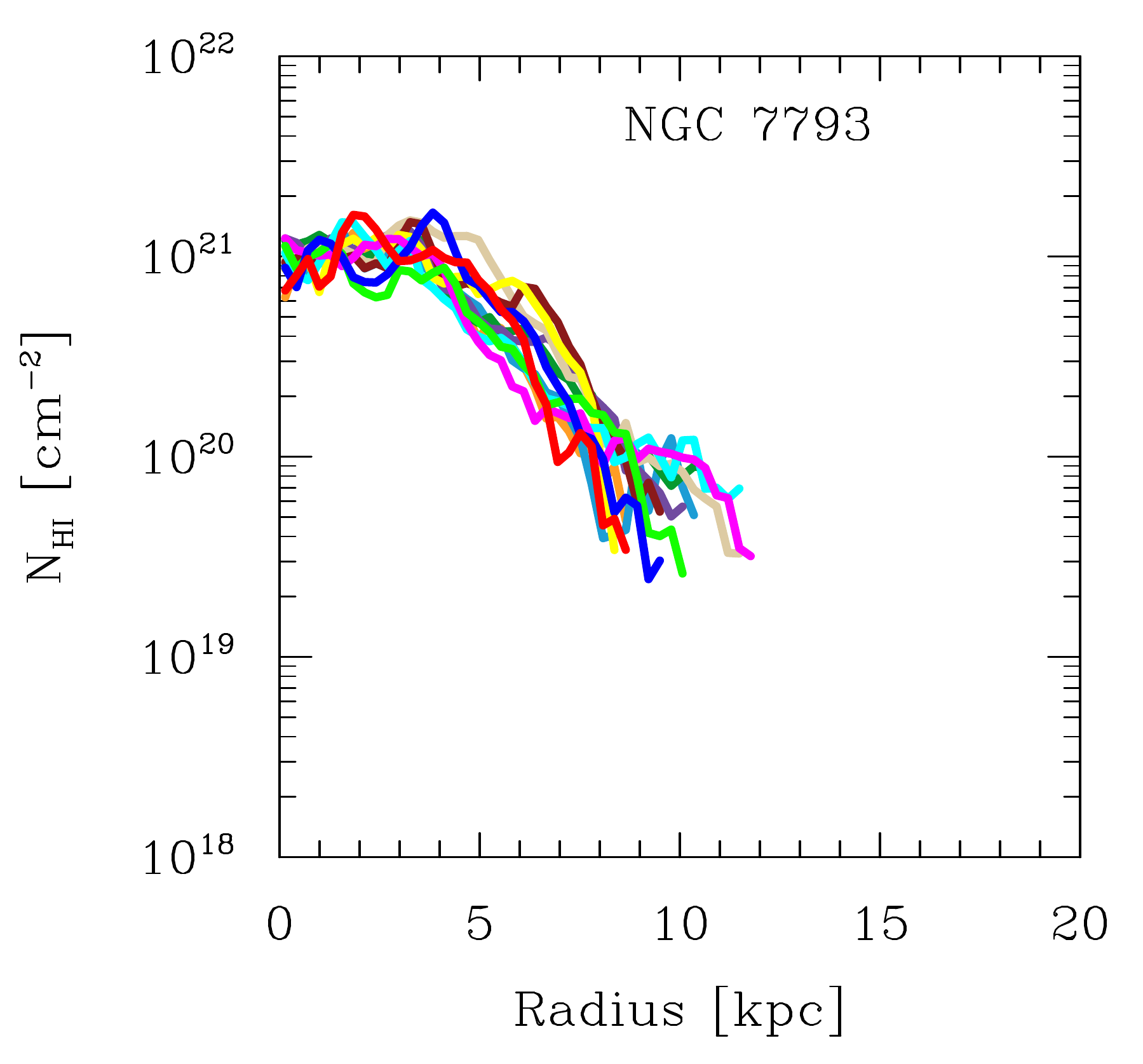} &
       \includegraphics[scale= 0.31]{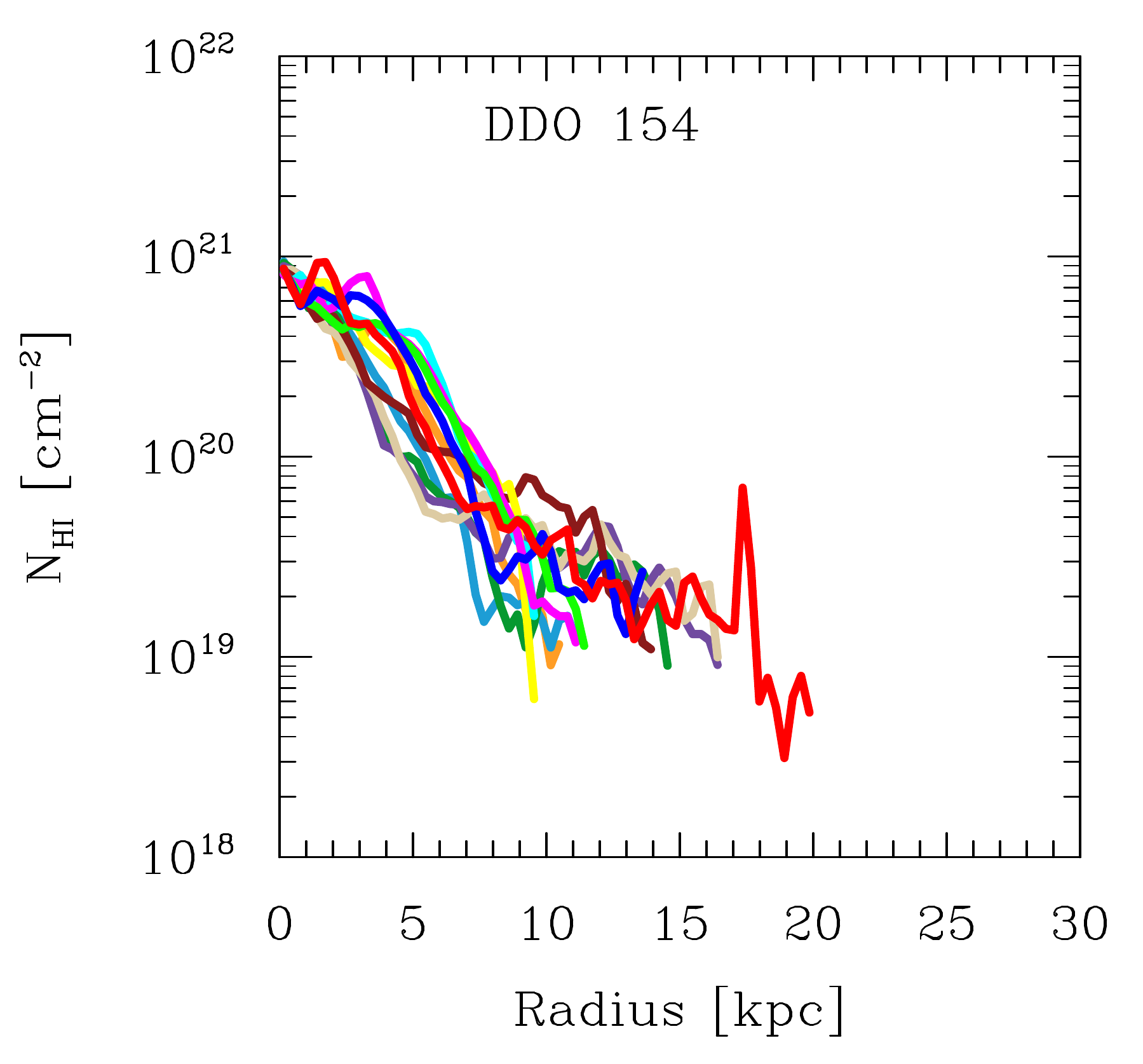} &
       \includegraphics[scale= 0.31]{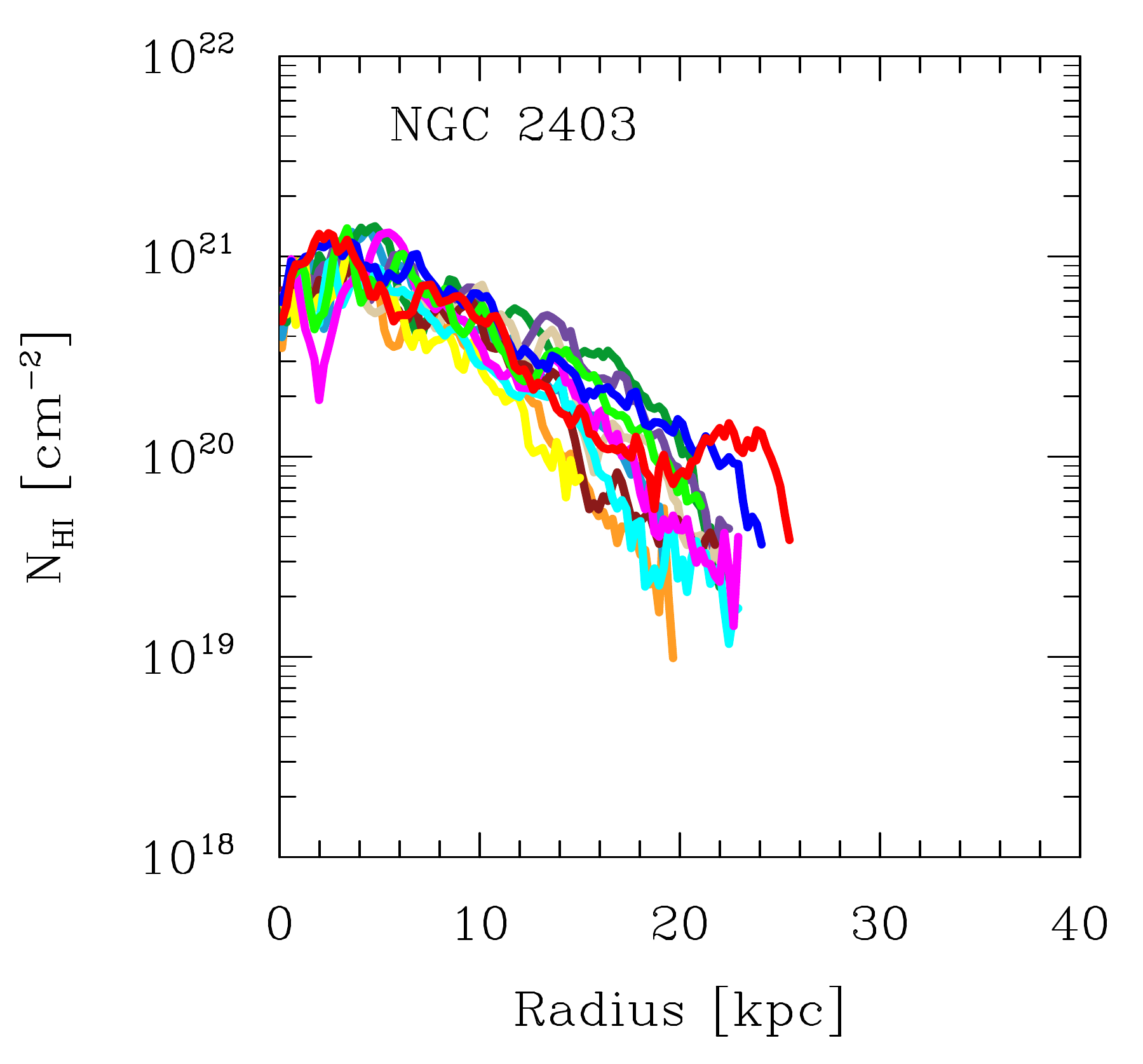} \\
       \includegraphics[scale= 0.31]{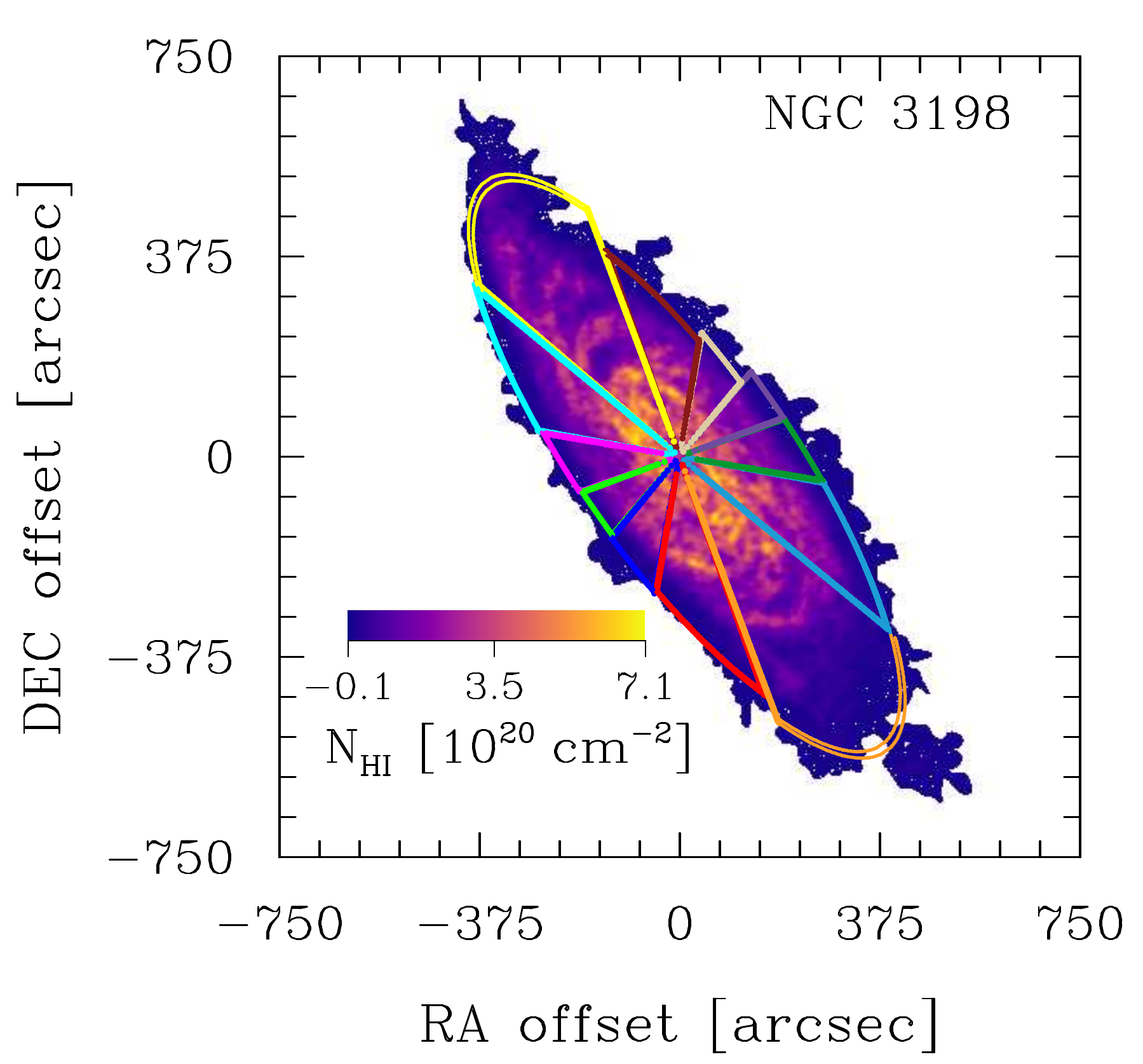} &
       \includegraphics[scale= 0.31]{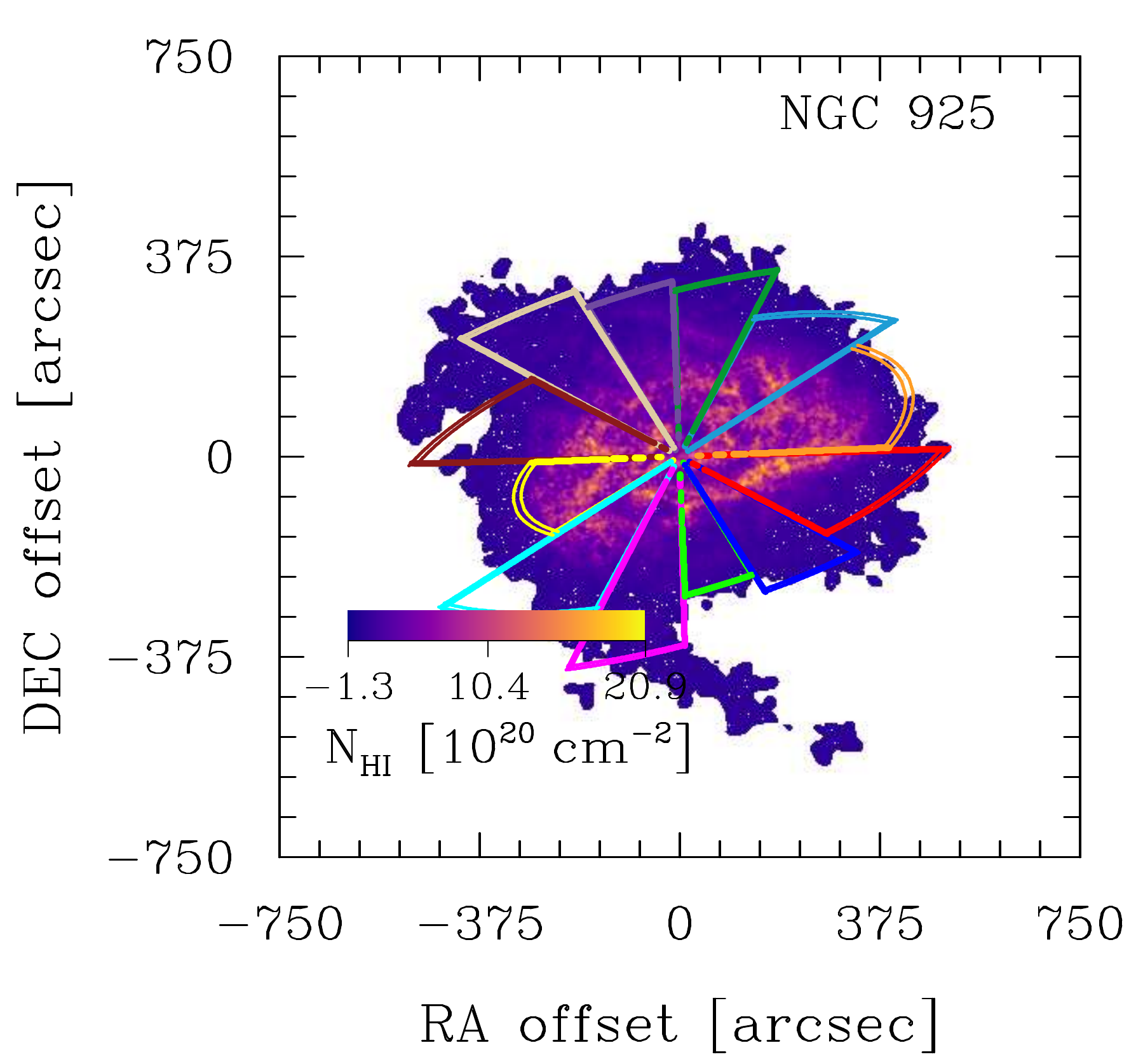} & 
        \includegraphics[scale= 0.31]{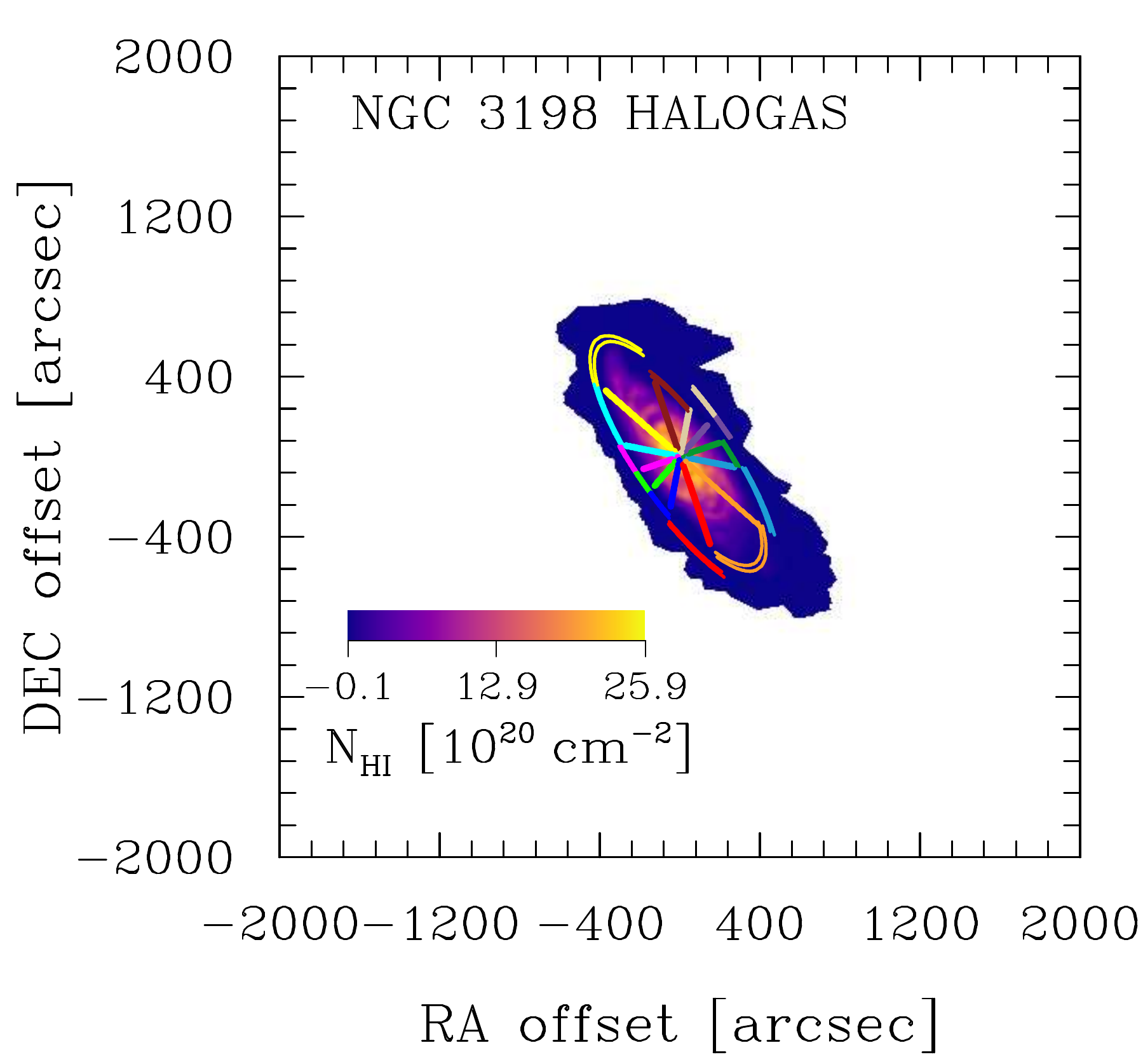} \\ 
       \includegraphics[scale= 0.31]{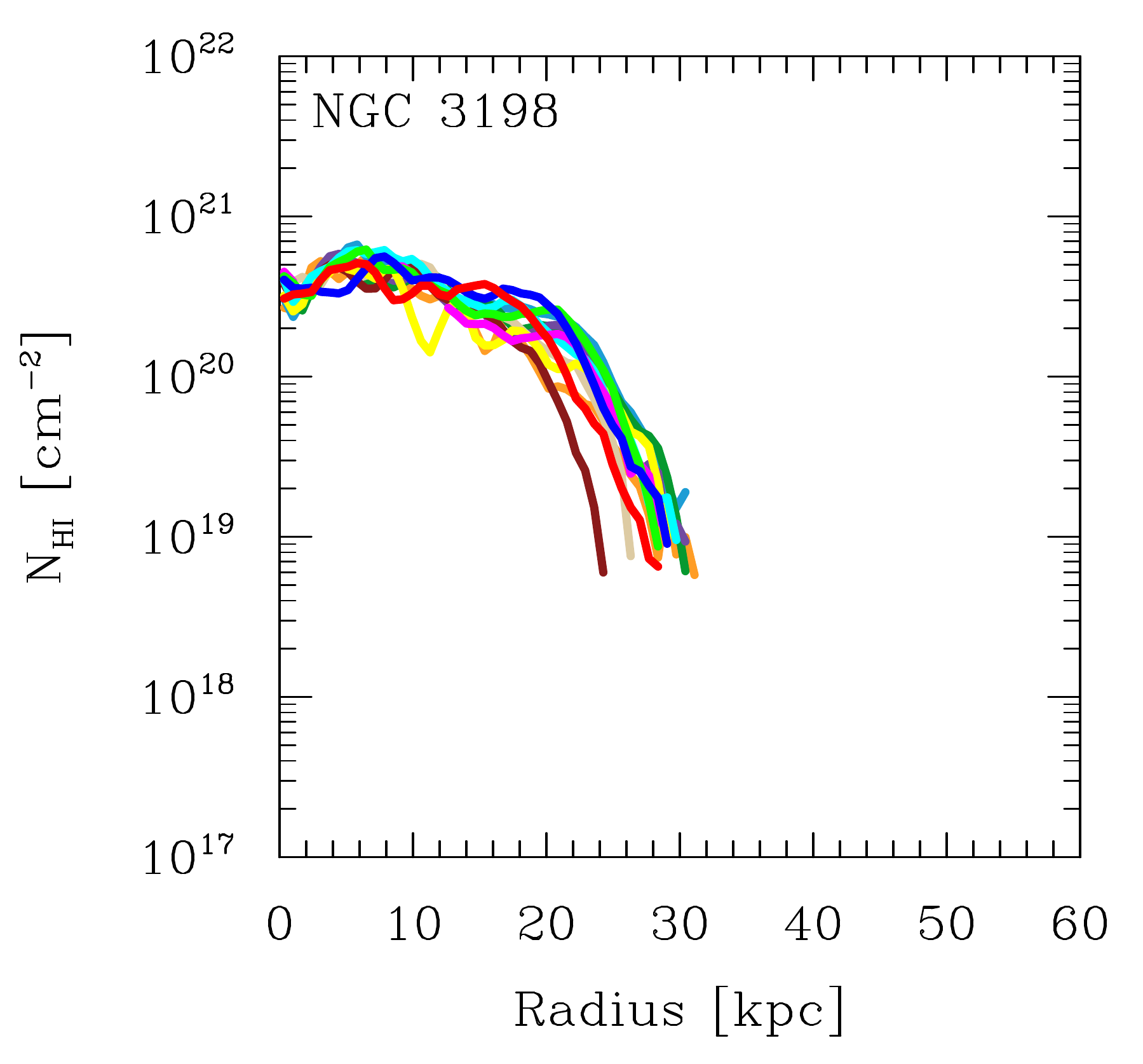} &
       \includegraphics[scale= 0.31]{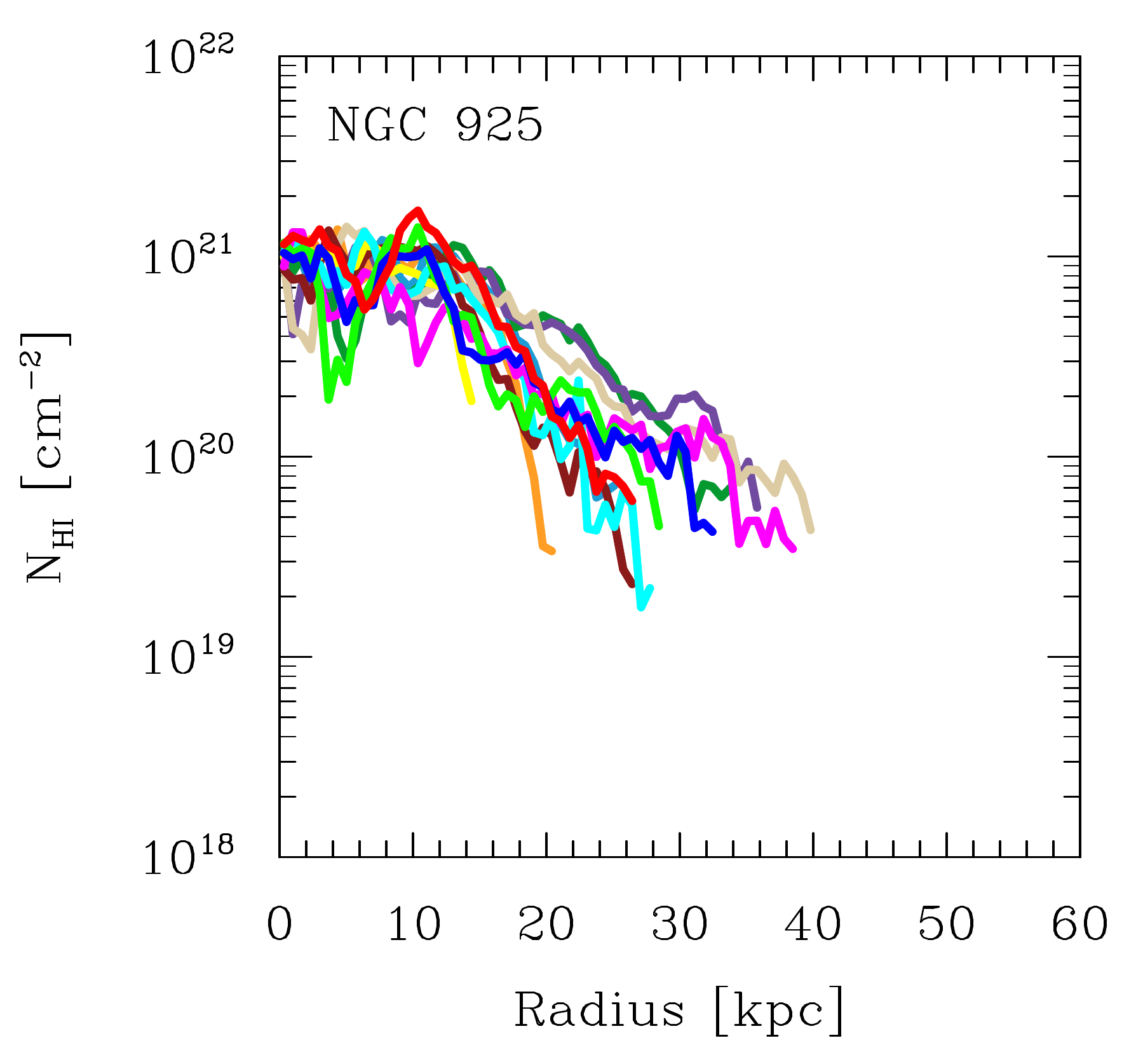} & 
       \includegraphics[scale= 0.31]{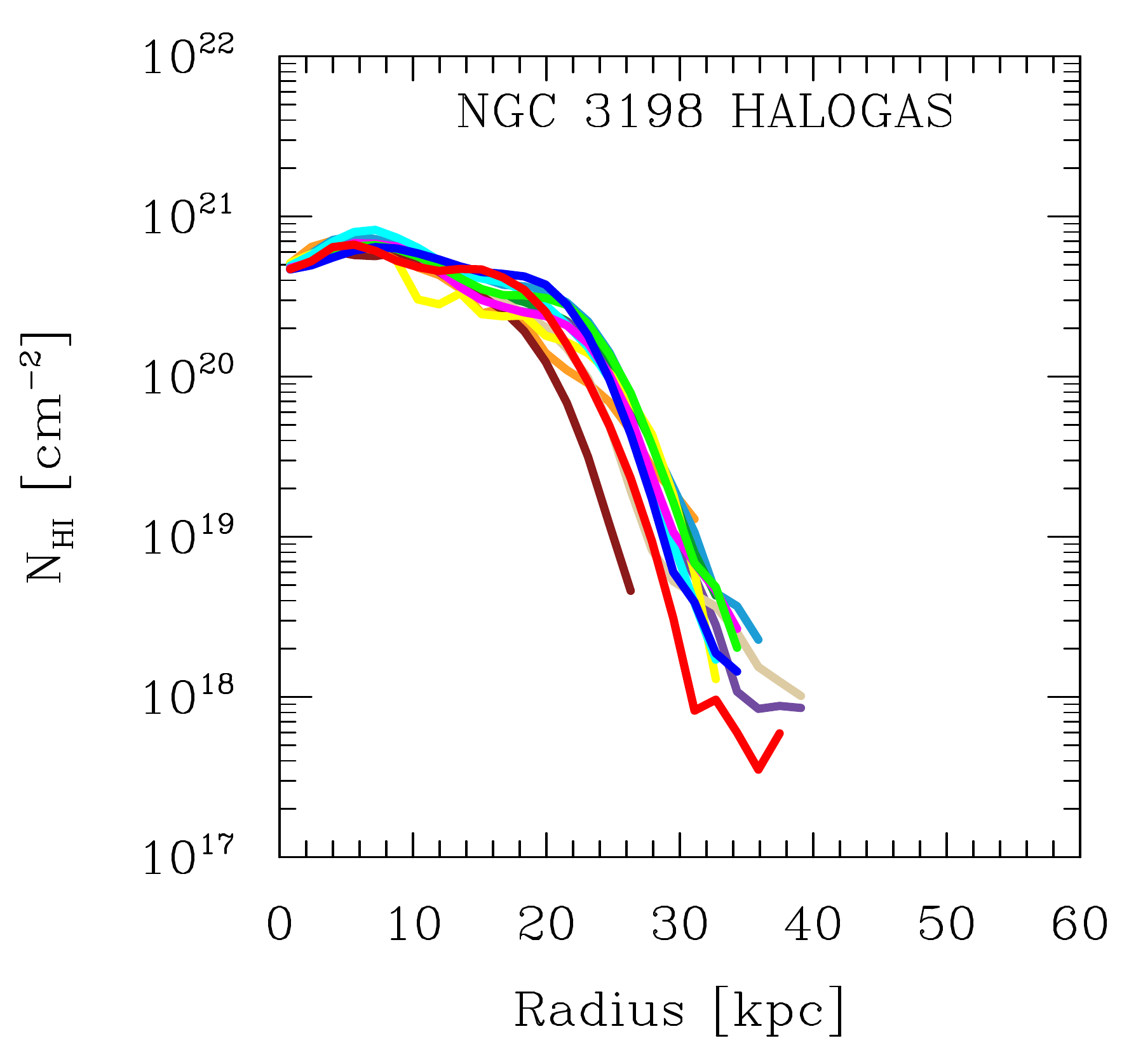} 
    \end{tabular}
    \caption{ Continued. \label{fig:sectors}} 
\end{figure*}
\setcounter{figure}{0}
\begin{figure*}
    \begin{tabular}{l l l}
      \includegraphics[scale= 0.31]{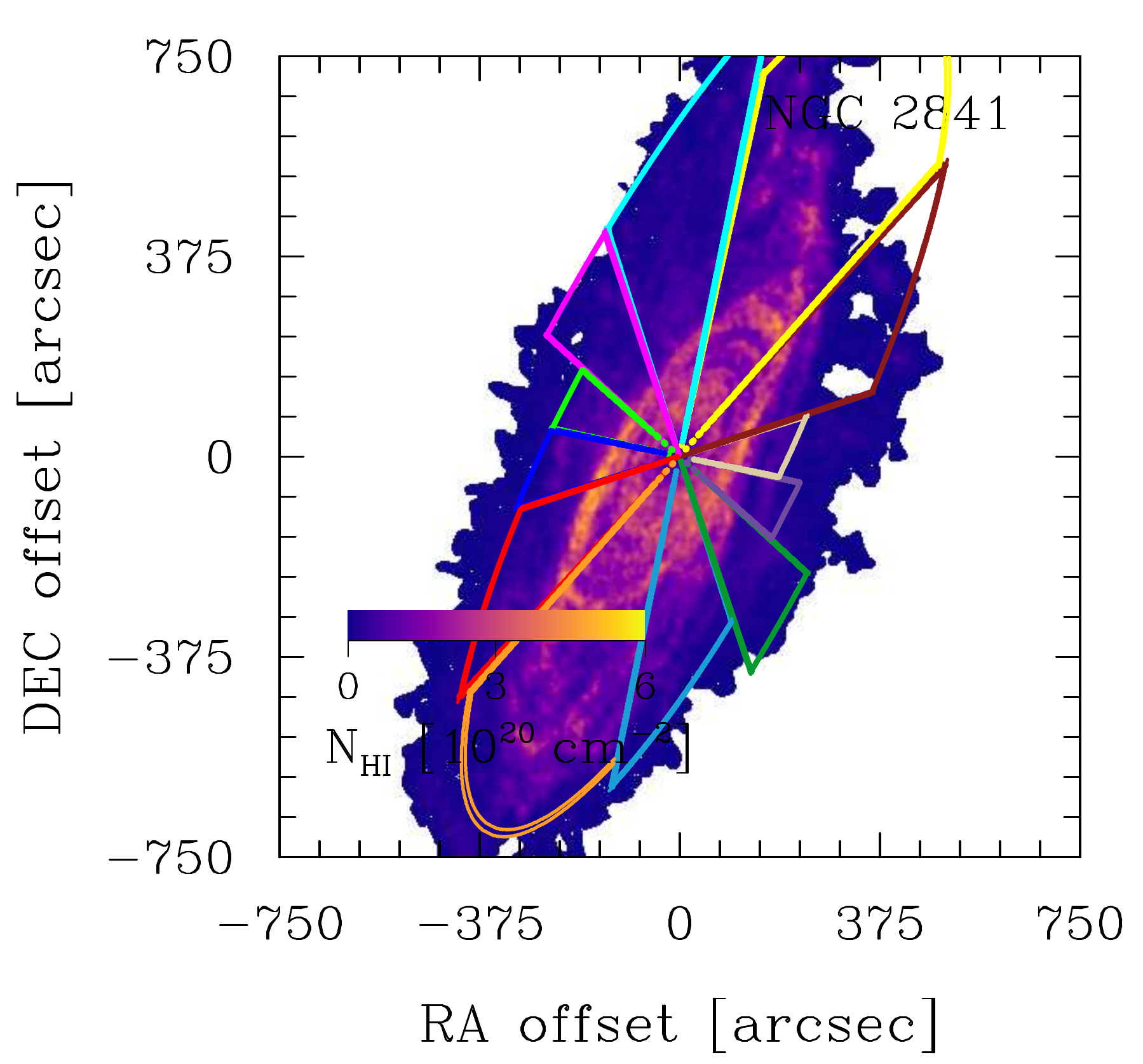} & 
      \includegraphics[scale= 0.31]{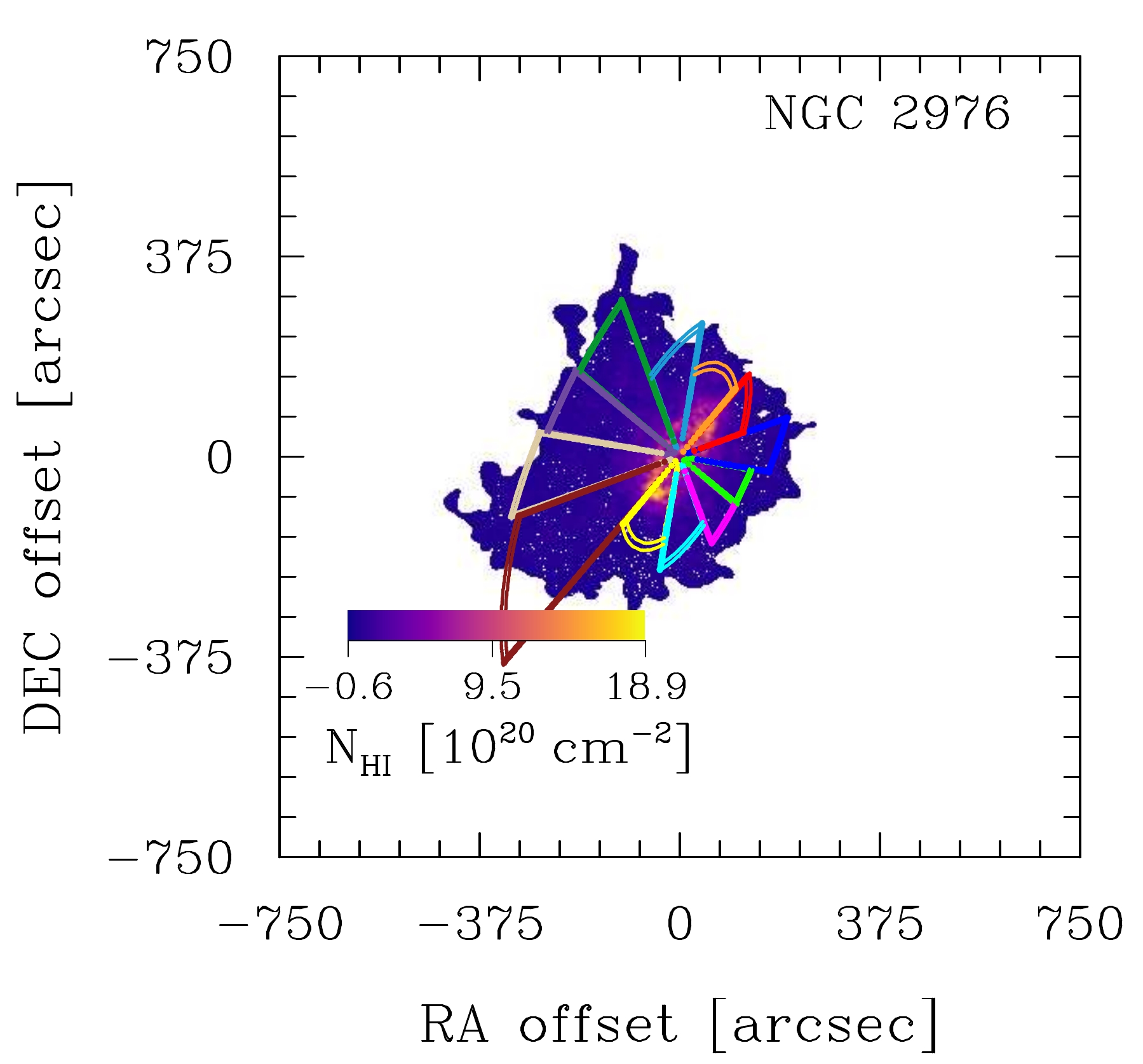} &
      \includegraphics[scale= 0.31]{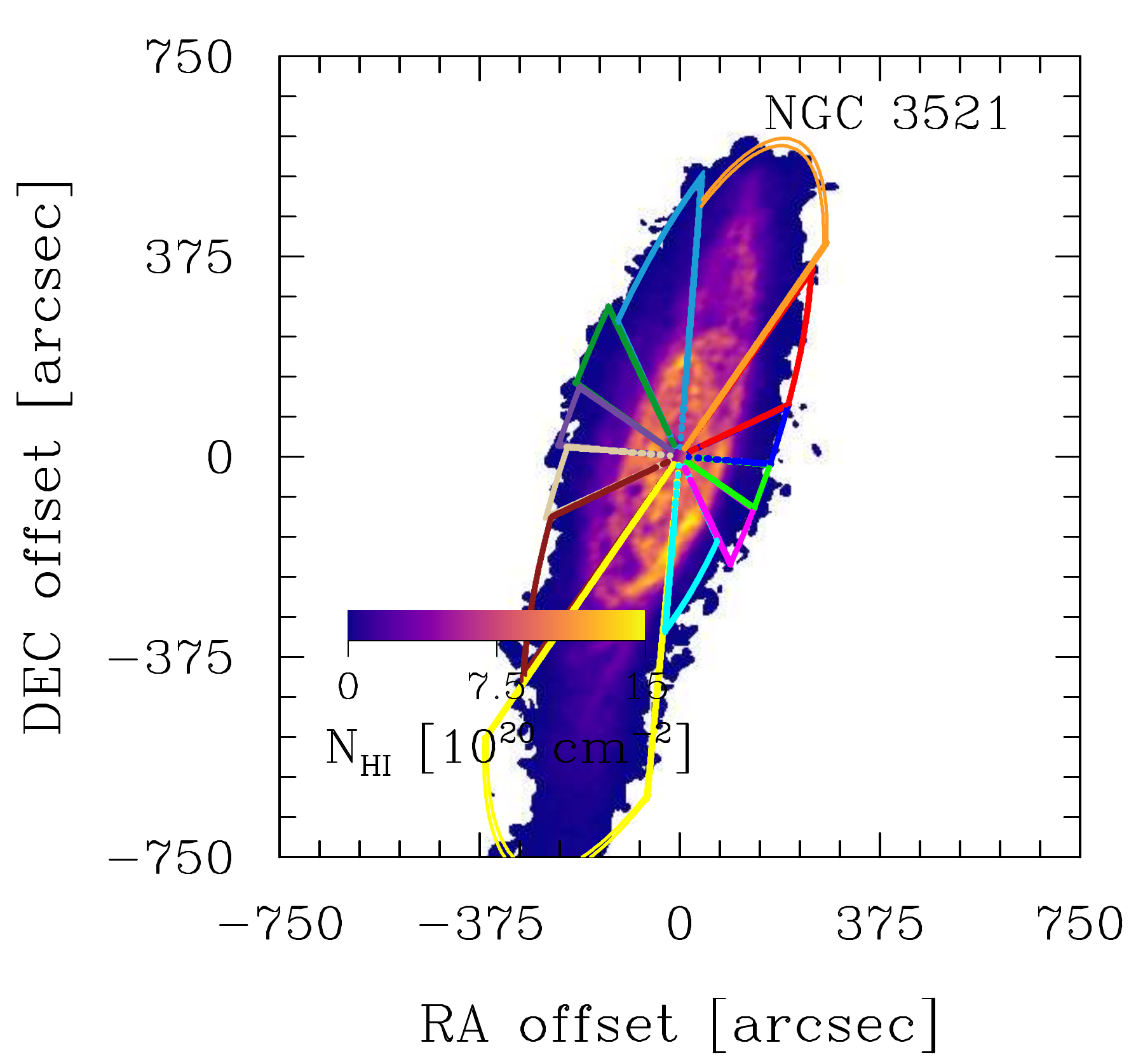}\\
      \includegraphics[scale= 0.31]{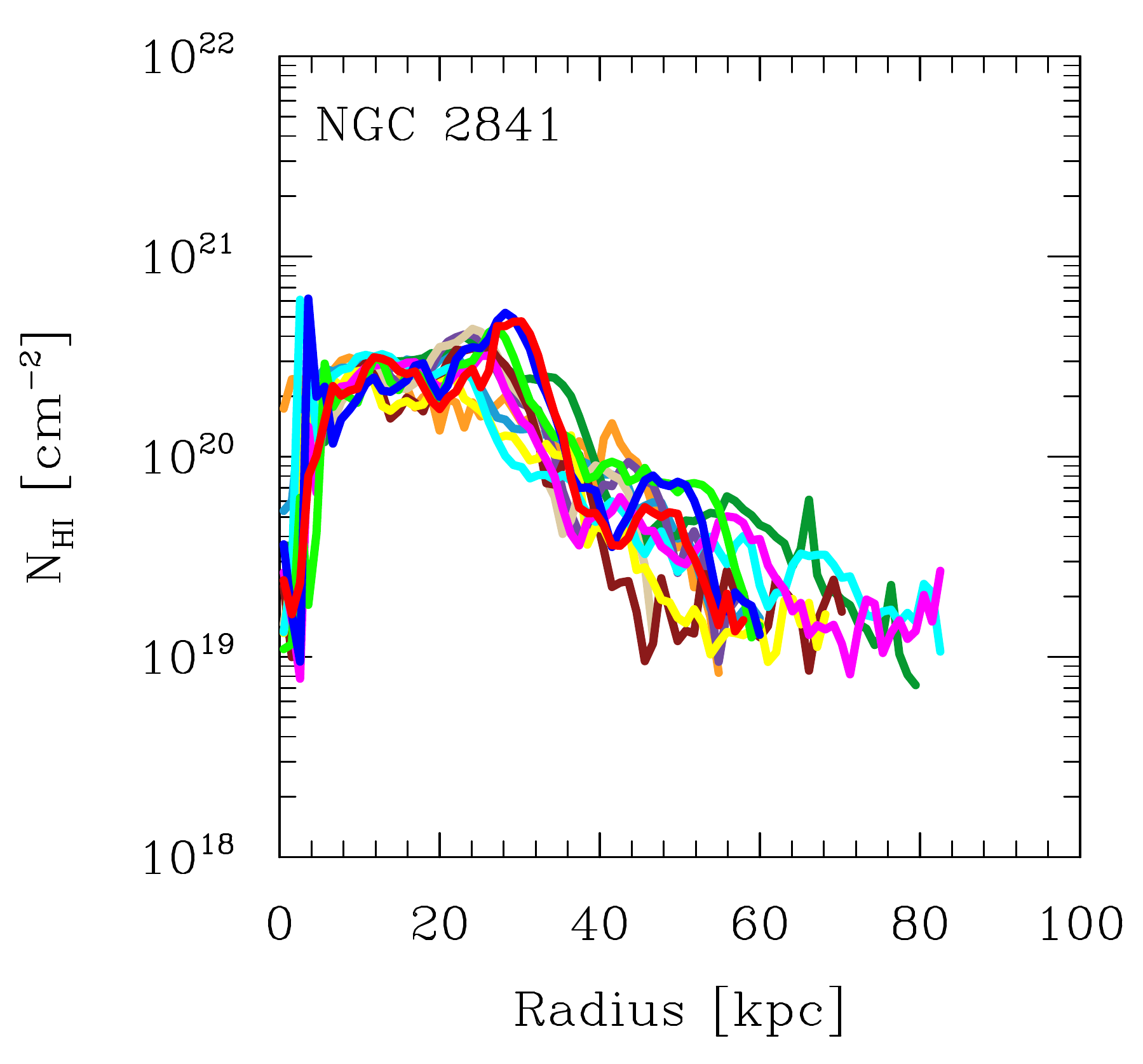} &
      \includegraphics[scale= 0.31]{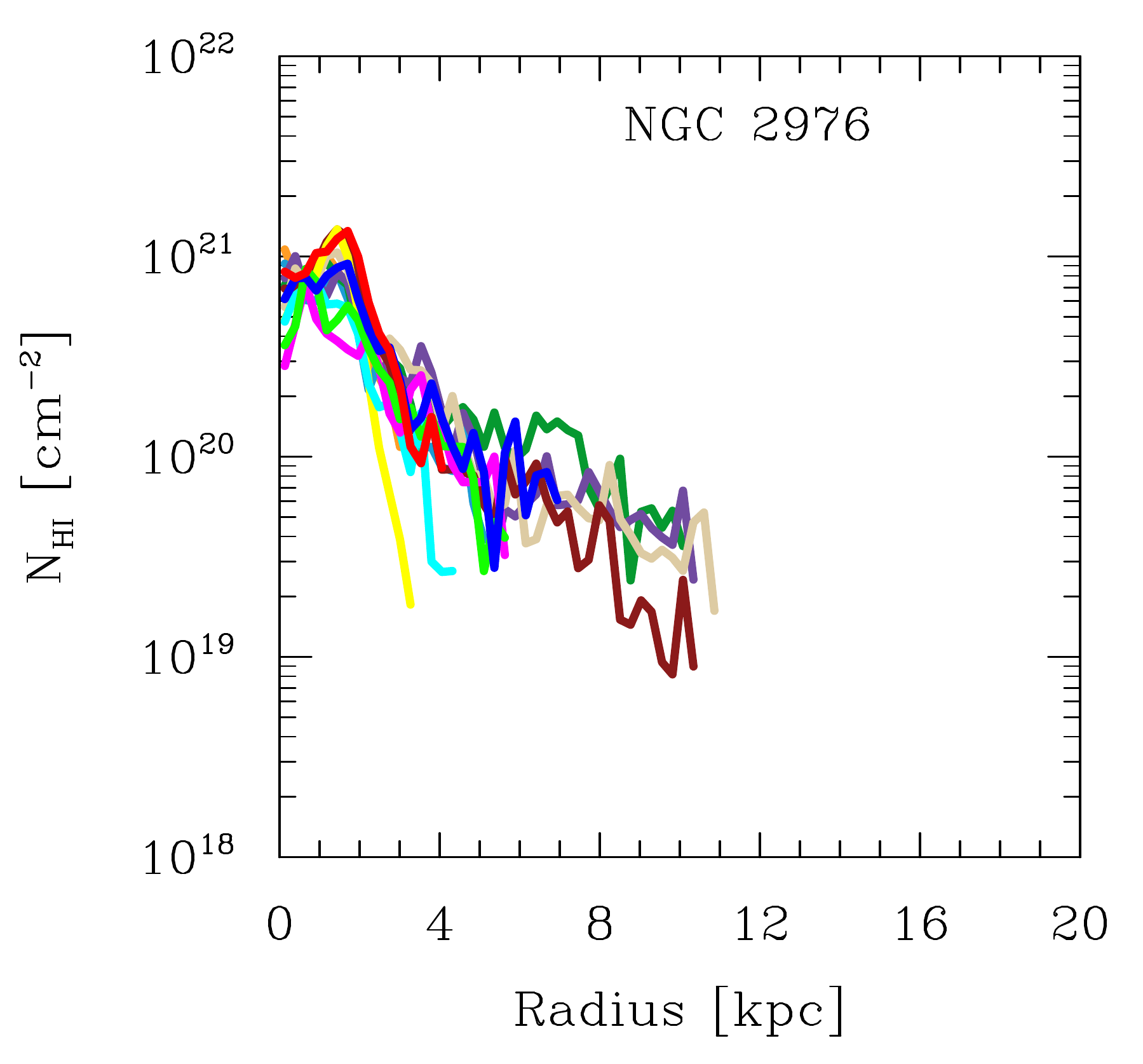} & 
      \includegraphics[scale= 0.31]{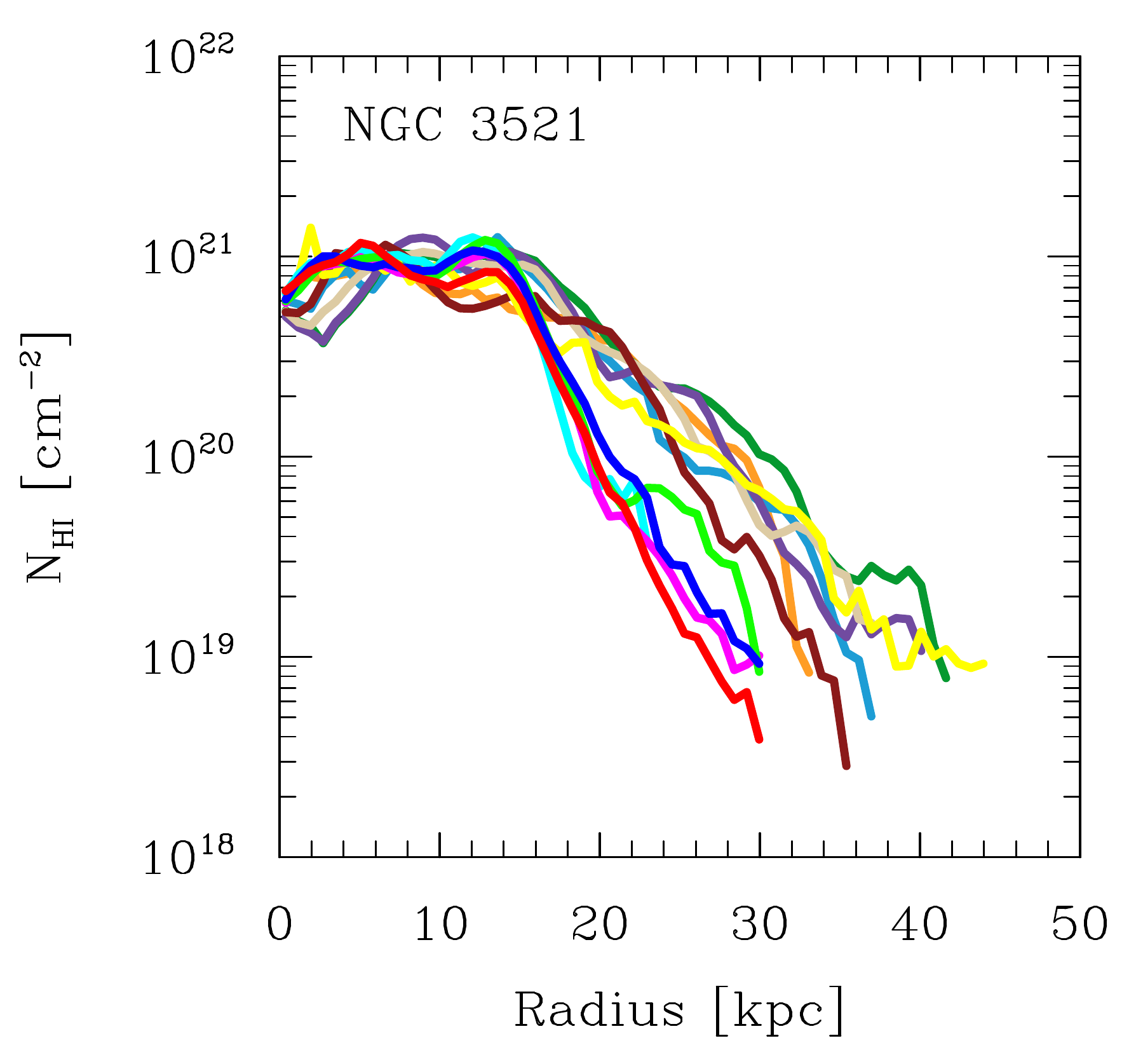}\\ 
       \includegraphics[scale= 0.31]{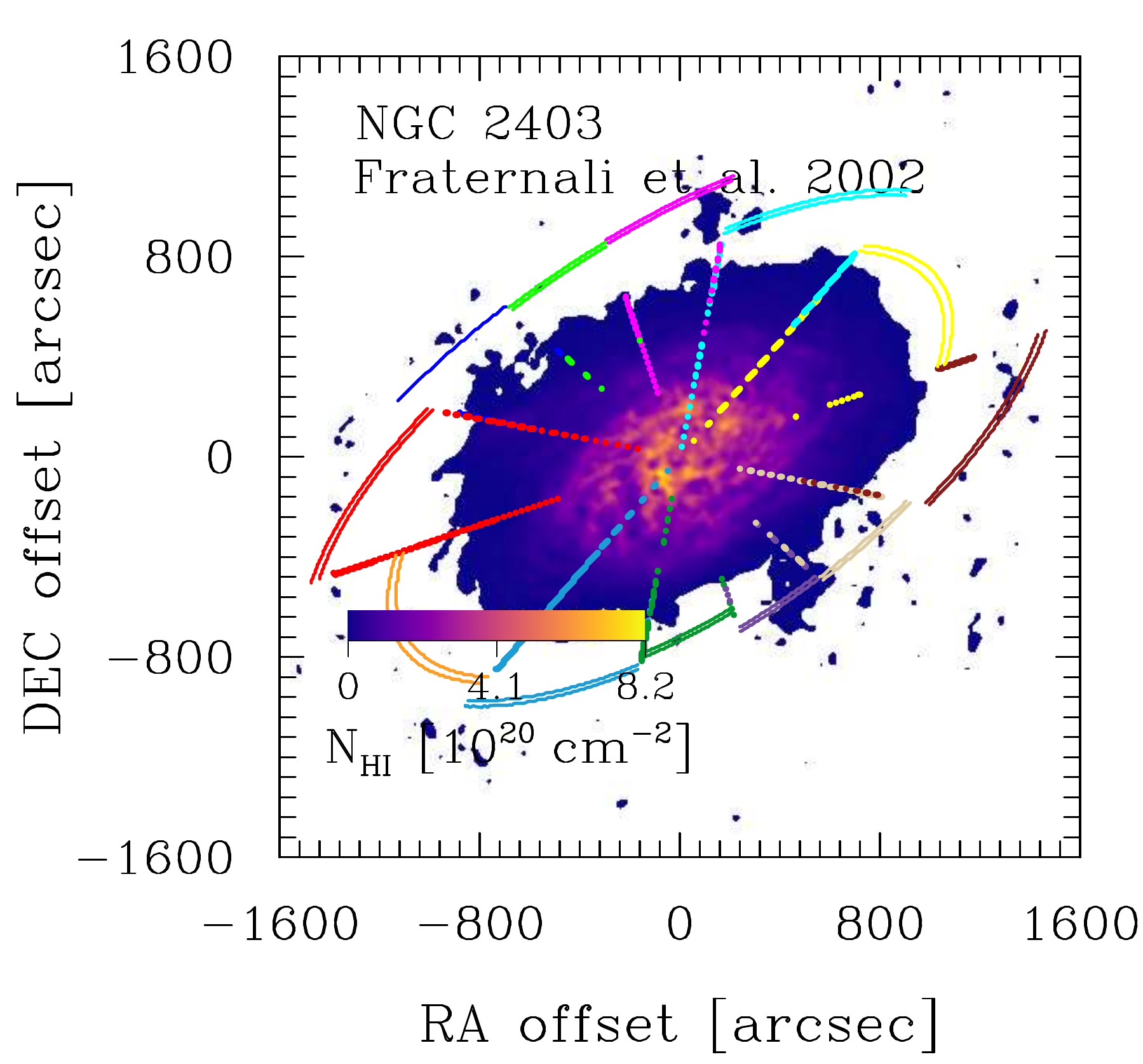} &
       \includegraphics[scale= 0.31]{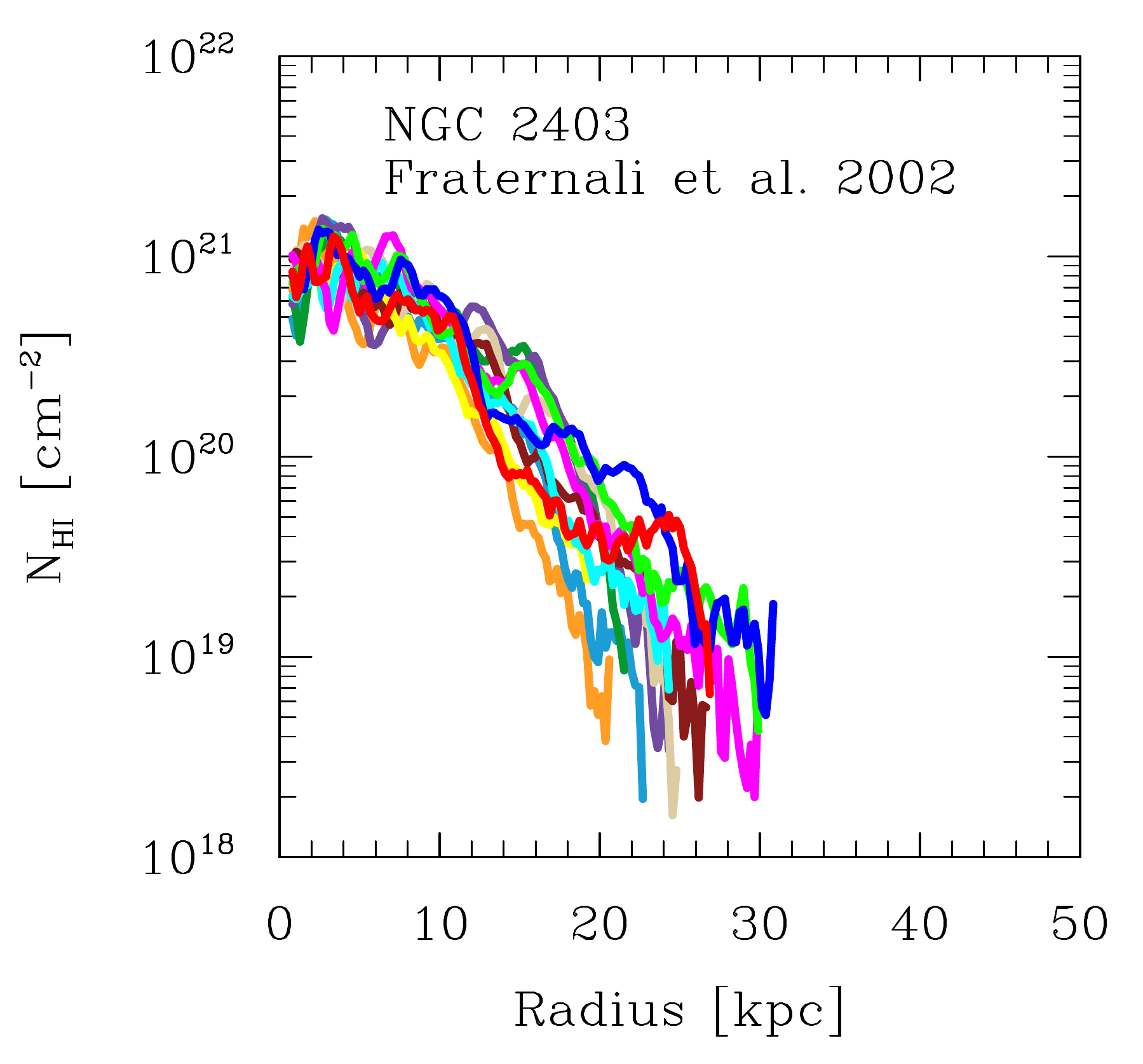} &
    \end{tabular}
    \caption{Continued. \label{fig:sectors}} 
\end{figure*}

      
\end{document}